\pdfoutput=1

\documentclass[12pt,oneside]{article}

\usepackage{hyperref}
\usepackage{ifpdf}

\usepackage{fullpage}
\usepackage{graphicx}

\usepackage{color} 
\usepackage{subfigure}
\usepackage{layout}
\usepackage{epsfig}

\usepackage{amsmath,amssymb,amsfonts}
\usepackage{appendix}

\usepackage{setspace}
\usepackage{array}

\usepackage[hyperpageref]{backref} 

\newcommand\T{\rule{0pt}{1.5ex}}
\newcommand\B{\rule[-1.5ex]{0pt}{0pt}}

\def\chap#1#2{\begin{center}\Huge{\bf{Chapter #1}}\vspace{-0.5cm}\section[#1. #2]{#2}\end{center}}
\def\sub2#1{\vspace{0.2cm}\noindent {\large{\bf#1}}\vspace{0.15cm}}

\begin{document}


\pagestyle{myheadings}

\pagenumbering{roman} {
  \pdfoutput=1

\begin{abstract}

Jefferson Lab Experiment E04-001 used the Rosenbluth technique to measure $R=\sigma_{L}/\sigma_{T}$ 
and $F_{2}$ on nuclear targets. This experiment was part of a multilab effort to investigate
quark-hadron duality and the electromagnetic and weak structure of the nuclei in the nucleon resonance
region. In addition to the studies of quark-hadron duality in electron scattering on nuclear
targets, these data will be used as input form factors in future analysis of neutrino data which
investigate quark-hadron duality of the nucleon and nuclear axial structure functions.
An important goal of this experiment is to provide precise data which to allow a reduction in
uncertainties in neutrino oscillation parameters for neutrino oscillation experiments (K2K,
MINOS).
 This inclusive experiment was completed in July 2007 at Jefferson Lab where the Hall C
High Momentum Spectrometer detected the scattered electron. Measurements were done in the
nuclear resonance region (1 $<$ $W^{2}$ $<$ 4 $GeV^{2}$) spanning the four-momentum transfer range
0.5 $<$ $Q^{2}$ $<$ 4.5 ($GeV^{2}$). Data was collected from four nuclear targets: C, Al, Fe and Cu.
\end{abstract}

}

{\hspace{-0.8cm} \LARGE \bf Acknowledgments \vspace{0.5cm}}\\
I am heartily thankful to my supervisor, Prof. Donal Day, whose encouragement, guidance and support  
enabled me to develop an understanding of the subject. 

I thank the members of University of Virginia research group Dr. Oscar Rondon, Prof. Donald Crabb,  
Prof. Kent Paschke, and Prof. Mark Williams for carefully reviewing my thesis.

I am grateful to Dr. Peter Bosted and Prof. Eric Christy for their continuous involvement and help 
in the analysis of this experiment. 
I am also thankful to graduate students Ibrahim Albayrak and Ya Li for their contribution to this analysis 
and Dr. Patricia Solvignon for organizing analysis meetings.

It is a pleasure to thank those who made this thesis possible such as the authors of the 
E04-001 proposal, Prof. Arie Bodek and Prof. Cynthia Keppel.

{\large{\tableofcontents}}
\vspace{0.7cm}

\newpage





\renewcommand{\arraystretch}{1.2}
\pagenumbering{arabic} 
\section{\hspace{-1.1cm} Introduction}
The study of the strongly interacting particles of nature is the study of hadrons.
The hadrons make up the atomic nucleus and the proton 
and neutron account for almost all the mass in the known universe. 
In contrast to their abundance our knowledge of the force which determines their 
interactions - the force responsible for both holding the nucleus together, 
limiting its maximum size as well as determining its dynamics, is the least well 
understood of the four fundamental forces of nature.
This should not be surprising given the description of the strong interaction in 
terms of Quantum ChromoDynamics (QCD) wherein the nucleons are the lowest 
mass excitations of the complicated blend of quark and gluon 
condensates which form the QCD vacuum. 
The key features associated with the QCD Lagrangian are exhibited by 
the  nucleon are: color confinement, asymptotic freedom and spontaneously broken 
chiral symmetry. 
A study of the nucleon is a study of the strong interaction and QCD.
QCD is a renormalizable field theory based upon the principle of local gauge
invariance under the exchange of color. 
QCD describes the strong force in terms of fermion fields of a given color 
charge, given the name quarks by Gell-Mann, interacting through the exchange
of massless gauge bosons known as gluons. 
The distinctive feature of QCD which sets it apart from the electroweak theory is 
the fact that these gluons possess non-zero color charge (unlike the 
photons in electro-magnetism which carry no electric charge).

Among the main consequences is the property of color 
confinement -- quarks and gluons are not permitted to exist as 
isolated free particles, but must combine to form color-neutral singlets: 
the three-quark baryons, of which the nucleons are the most stable, and the
quark/anti-quark mesons. 
To date, the quantum numbers of all known strongly interacting particles have 
been accounted for using a model based upon the non-Abelian internal $SU(3)_{c}$ 
symmetry of QCD for elementary quarks of six different flavors (these are 
labeled in order of increasing mass as up, down, strange, charm, bottom and top).
Though the static properties of the nucleon can be described in this manner,
the extreme non-linear nature of the strong force means that the dynamical 
properties of its constituents vary significantly depending on the momentum 
scale at which it is studied. 
Practitioners are forced  to confront a profound chasm: large momentum behavior
of the nucleon can accurately be described by the quark and gluon fields of
which it is composed, but this same description fails at low momentum where 
successful models have to rely instead on effective hadronic degrees of freedom.
This is related to the asymptotic freedom of QCD, responsible for the fact that the force between the quarks within
the nucleon becomes weaker as they move closer together.
For high momentum (hard) scattering processes involving the nucleon, perturbative QCD (pQCD) can be used to express the physical 
amplitudes as a perturbative series in the strong coupling constant $\alpha_{s}$, analogous to Quantum Electrodynamics (QED). 
Renormalization of the field theory is required for such a series to 
converge  and is achieved by the introduction of an arbitrary mass scale.
This, in turn, leads to the running of the coupling strength of the strong
interaction. 
Said another way, the effective strength is not constant as in the electromagnetic
case but depends on the momentum scale, decreasing logarithmically at large 
momentum transfers.
Hence an increase in momentum not only implies a shorter distance scale
but a decrease in the effective coupling, allowing the perturbative 
expansion to converge and enabling rigorous predictions based upon the quark
and gluon fields alone.
Asymptotic freedom also means that the relatively large running coupling
strength associated with the low momentum (soft) regime makes a perturbative
expansion in $\alpha_{s}$ of little value at large distance scales, such as typical
hadronic sizes or larger. 
Rigorous solutions are not available for non-perturbative QCD and nucleon 
behavior at comparatively low energies leaving effective theories the only viable alternative. These are based upon identifying appropriate hadronic degrees of freedom for 
a particular momentum range and determining their characteristics.

A variety of methods, each limited in their applicability, have been in use for some time: QCD sum 
rules~\cite{Ioffe:1983ju,Balitsky:1983xk,Nesterenko:1984tk,Belyaev:1992xfa,Radyushkin:2004mt} 
rests on the exploitation of the underlying symmetries of the strong interaction; 
 effective Lagrangians are the basis for many model which approximate 
QCD at low energy;  computationally intensive calculations 
of physical quantities on a discretized space-time lattice  are becoming commonplace.

Common to all these approaches is the special role played by the
non-perturbative vacuum - it links low energy QCD to
the phenomenon of spontaneously broken chiral symmetry. 
Light quarks are approximated as massless (when compared to the nucleon) and the conservation of their left and right-handedness introduces
another layer of symmetry into the effective Lagrangian. 
This chiral symmetry is not exact and can be spontaneously (or dynamically) 
broken by the non-zero mass of confined quarks, leading these quarks
to develop large effective masses through interactions with the vacuum. 
The result is that a nucleon consists not
only of three valence-quarks but includes a complex sea of quark/anti-quarks pairs and
gluons. This sea's characteristics are determined by the momentum scale 
involved.
The electromagnetic response of the 
nucleon has been studied for more than fifty years now with measurements largely focused towards to the extreme upper 
and lower ends of the spectrum of momentum transferred to the nucleon.
The intermediate region involving moderately large momentum transfer, where
there is significant give and take between soft and hard physics and therefore
between quark-gluon and meson-baryon degrees of freedom, has long since been
identified as a prolific testing ground for models of the strong interaction.


Deep inelastic scattering (DIS) has proved to be a powerful tool to study the structure of matter.
The experiments at SLAC~\cite{Coward:1967au,Bloom:1969kc,Breidenbach:1969kd} showed 
for the first time the absence of scale dependence in inelastic electron-proton scattering. 
Later experiments at larger four-momentum transfer revealed logarithmic scaling 
violations, which was explained in terms of QCD and proved QCD as being the correct theory 
of strong interactions. 
In a few GeV energy range where hadronic degree of freedom is dominant the strong coupling  
constant becomes large and pQCD becomes inapplicable. 
This energy range is where the effect of confinement (no free quarks) make strongly-coupled 
QCD highly non-perturbative and it is more easy to work with mesons and baryons. 
Despite the significant difference between low and high energy description of the 
nuclear dynamics a phenomenon of so called quark-hadron duality is observed which can be
though as a link between this two regimes. 
The quark-hadron duality was first observed in data from the early SLAC experiments 
by Bloom and Gilman~\cite{Bloom:1970xb}. 
The observations showed a striking similarity between the $F_{2}$ structure functions measured in
resonance and DIS.
Based on this data Bloom and Gilman proposed that the resonances are not a separate entities 
but are an intrinsic part of the scaling behavior of $F_{2}$. 
In order to analyze the degree of similarity of the $F_{2}$ scaling function, Bloom and Gilman determined a  
Finite Energy Sum Rule~\cite{Collins:1977jy,Donnachie:2002en}  
which allowed one to investigate the phenomenon of quark-hadron quantitatively. 

With the development of QCD in the early 1970s, Bloom-Gilman duality was 
reformulated~\cite{De Rujula:1976tz} in terms of an operator product expansion (OPE) of 
moments of structure functions. 
This has allowed an expansion of the structure function integrals, called moments, in terms of the 
hard scale $1/Q^{2}$, clarified many aspects of quark-hadron duality 
and it's violation. 
However, the OPE is not able to adequately explain why at low energies 
the correlations between partons were suppressed, and how the scaling worked for the 
resonances. 

The advent of high energy, high duty-factor electron accelerators, like the 6 GeV 
electron accelerator at Jefferson Lab, has provided quality data in recent years to 
test and constrain the theoretical calculations in this field. 
One of the significant observation in the JLab data is that Bloom-Gilman duality 
appears to work at $Q^{2}$ values as low as 1 GeV$^{2}$ or even lower.
At low $Q^{2}$ the strong coupling $\alpha_{s}$ is relatively large compared to it's value in DIS, 
but on average the inclusive scattering process appears to mimic the scattering of electrons from 
nearly free quarks. 
These observations led to renewed interest in quark-hadron duality. 
In particular, measuring the $F_{2}$ structure function in the resonance region at high 
Bjorken $x$ ($x>$ 0.7, where $x$ is the longitudinal momentum fraction of 
the hadron carried by a parton in the infinite momentum frame) were of great interest 
since it allowed precision study of the $F_{2}$ structure function in $x>$ 0.7.

The observation of duality in electron scattering is thought to be a fundamental property of quarks and hadrons,  
which raises the question of the existence of duality in neutrino scattering. 
This is expected because scattering takes place from the same constituents as in electron case.
Weak scattering can provide complementary information on the quark structure of hadrons, 
not accessible to electromagnetic probes. 
Neutrino-induced reactions also can provide important consistency checks on the validity 
of duality.
Since the weak interaction mechanism is different from that of electromagnetic one 
the observation of duality may provide additional information.
The MINERvA~\cite{NeutrinoTargets} experiment intends to study duality in neutrino scattering 
and can provide data to answer at what kinematic range duality works, in 
what structure functions and in what reactions. 

Duality in electron scattering have focused mostly on proton. 
There have been some experiments that performed measurements on deuterium and heavy 
nuclei in the high-$x$ and low to moderate $Q^{2}$ region~\cite{Filippone:1992iz,Arrington:1998ps,Arrington:2001ni}.  
The results of these experiments revealed additional information about duality. 
Scaling in nuclei was observed and was interpreted in terms of local duality 
where the averaging over the resonances was accomplished by the Fermi momentum of nuclei 
inside the nucleus.

The primary motivation of this experiment is to study quark-hadron duality in 
electron scattering from nuclear targets in the resonance region as well as to study 
the nuclear dependence of $R=\sigma_{L}/\sigma_{T}$. 
A significant nuclear dependence of $R$ is predicted by Miller~\cite{Miller:2000ta} due to nuclear pions at low values of 
$Q^{2} \sim 0.3$ GeV$^2$. 
The results of this experiment, combined with the results of another experiment, E02-109~\cite{E02-109}, 
will allow the extraction of $\sigma_{L}^{A}$ through a precise Rosenbluth separation~\cite{Rosenbluth:1950yq} and 
test if the pions are really the carriers of the nuclear force as predicted by Miller. 
Also, the data will allow us to model the electron$-$nucleon scattering cross section in the resonance region
which are of great importance for neutrino scattering experiments since the 
neutrino scattering structure functions can be extracted if the vector part is 
provided from electron scattering.

\chap1{Nucleon Structure}
In this section we provide a simple review of the progress made in 
understanding the nucleon, introducing the necessary formalism and key concepts along the way. 
An excellent overview of the field of nucleon structure can be found in Ref.~\cite{Thomas:2001kw}.
\subsection{Inclusive Lepton-Nucleon Reactions}
\begin{figure}[t]
\begin{center}
\includegraphics{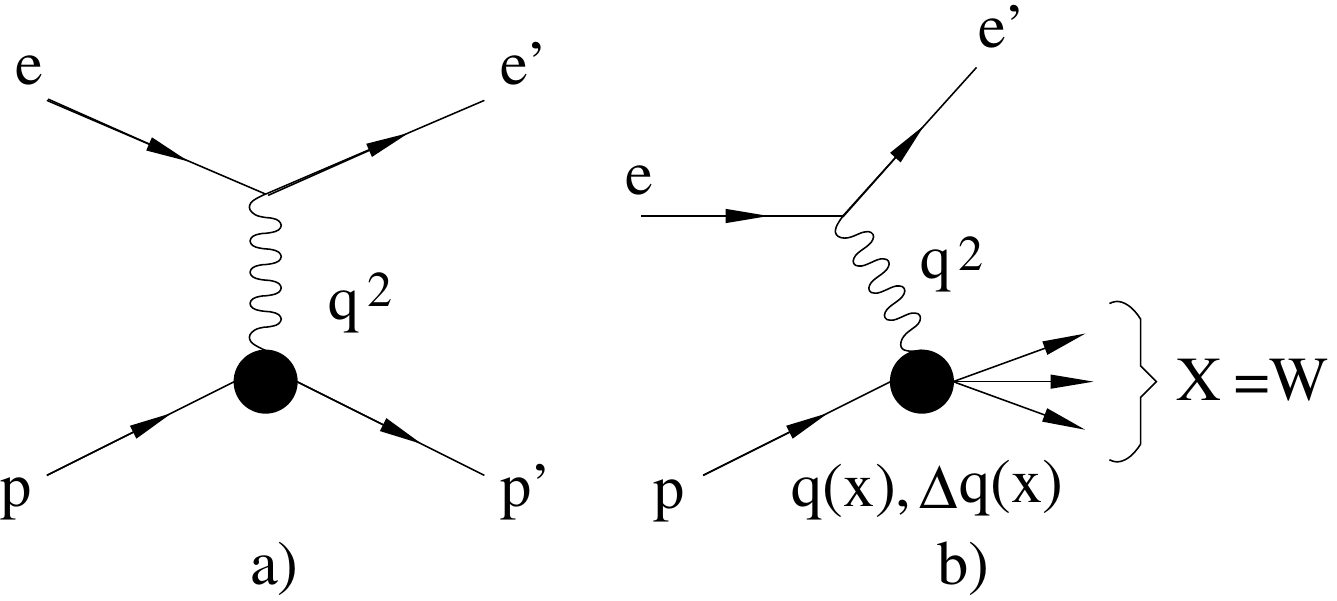}
\end{center}
\caption{ Leading order Feynman diagrams for (a) elastic $ep$ scattering and
(b) deep inelastic scattering. In each respective case, the unknown aspects of
the proton structure are parametrized in terms of the form factors, $F_{1}(Q^{2})$ and
$F_{2}(Q^{2})$, and parton distributions, $q(x)$ and $\Delta q(x)$.
}
\label{fig:FFDG}
\end{figure}

Inclusive lepton-nucleon scattering is an important tool with which to study the structure of the nucleon.
In the energy range accessible to modern accelerators leptons are considered structure-less.
Therefore leptons are ideal to probe the the nucleon since the 
only unknown is the structure of the nucleon. 
In inclusive scattering, \mbox{$eN \rightarrow e'X$}, only the final state electron, $e'$, is registered while 
the hadronic final state $X$ remains undetected. 
In the target rest frame (laboratory frame) the incident electron with energy $E$ scatters from the target, 
the scattered electron have angle $\theta$ relative to the direction of the incident beam and energy $E'$.
In the one photon exchange (Born) approximation, shown in Fig.~\ref{fig:FFDG}, the scattering takes place 
by exchanging a virtual photon (or $W^{\pm}$ or $Z$ boson in neutrino scattering). 
The energy of virtual photon is given by $\nu=E-E'$ and the 3-momentum transfer in $\vec{q}$.
Since the photon is spacelike the virtuality of photon $q^{2}=\nu^{2}-{\vec{q}\hspace{0.1cm}}^{2}$ is negative.
It is convenient to use positive variable $Q^{2}=-q^{2}$, which is related to the initial and final energy and 
angle of the electron.
\begin{equation}
Q^{2}=4EE'\sin^{2}\frac{\theta}{2}
\label{eq:Q2}
\end{equation}
The invariant mass squared of final hadronic state is given by 
\begin{equation}
W^{2}=M^{2}+2M\nu-Q^{2}
\label{eq:W2}
\end{equation}
In general the cross section for inclusive scattering depends on two variables $\nu$ and $Q^{2}$.
In the Deep Inelastic Scattering (DIS) regime, which will be discussed later, 
yet another important variable Bjorken $x$
\begin{equation}
x=\displaystyle\frac{Q^{2}}{2M\nu}
\label{eq:bjx}
\end{equation}
is introduced. 
In the DIS region this variable is the fraction of nucleon momentum carried by the struck parton 
in the ``infinite momentum frame'' $-$ the frame where the electron is in rest while the nucleon 
is speeding toward it. For the nucleon $0\leq x \leq 1$.

Since the electron-nucleon scattering is electromagnetic in nature the one photon exchange 
approximation is a very good approximation. 
Therefore the differential cross section for electron nucleon scattering from unpolarized 
target can be written in as 
\begin{equation}
\frac{ d^{2}\sigma}{d\Omega dE'} = \frac{\alpha^{2}}{Q^{4}} \frac{E'}{E} L^{\mu\nu} W_{\mu\nu}
\label{eq:DiffCSTensor}
\end{equation}
where $\alpha$ is the fine structure constant. 
The leptonic tensor $L_{\mu\nu}$ is calculable from Quantum Electrodynamics and can be written as 
\begin{equation}
L_{\mu\nu}=2\left( k_{\mu}k_{\nu}' + k'_{\mu}k_{\nu} - g_{\mu\nu}k \cdot k'\right) 
\label{eq:LeptonTensor}
\end{equation}
where $k$ and $k'$ are initial and final electron momentum, respectively and $g_{\mu\nu}$ is the metric tensor. 
For an unpolarized initial nucleon the general form of hadronic tensor can be written as 
\begin{equation}
W^{\mu\nu}=W_{1}g^{\mu\nu} + \frac{W_{2}}{M^{2}}p^{\mu}p^{\nu} + \frac{W_{4}}{M^{2}}q^{\mu}q^{\nu} + \frac{W_{5}}{M^{2}}\left( p^{\mu}q^{\nu} + q^{\nu}p^{\mu}\right) 
\label{eq:HadronTensor}
\end{equation}
where $W_{i}$ are unknown functions of $\nu$ and $Q^{2}$. 
The structure function $W_{3}$ and $W_{6}$ do not contribute in the electron scattering, since $W_{3}$ violates 
parity conservation and $W_{6}$'s antisymmetric component is multiplied by a symmetric lepton tensor $L_{\mu\nu}$. 
Based on Lorentz and gauge invariance, together with parity conservation in electron 
nucleon scattering the hadronic tensor can be written as 
\begin{equation}
W^{\mu\nu}=W_{1}(\nu,Q^{2})\left( \frac{q^{\mu}q^{\nu}}{q^{2}}- g^{\mu\nu}\right) + \frac{W_{2}(\nu,Q^{2})}{M^{2}}\left( p^{\mu} + \frac{p\cdot q}{q^{2}}q^{\mu}\right) 
\left( p^{\nu} + \frac{p\cdot q}{q^{2}}q^{\nu}\right) 
\label{eq:HadronTensorTwoW}
\end{equation}
where $W_{1}$ and $W_{2}$ are scalar functions of $\nu$ and $Q^{2}$.
After contracting $L_{\mu\nu}W^{\mu\nu}$ the cross section can be written as 
\begin{equation}
\frac{d^{2}\sigma}{d\Omega dE'} = \sigma_{Mott} \Bigg[W_{2}(\nu, Q^{2})+2W_{1}(\nu, Q^{2})\textrm{tan}^{2}{ \frac{\theta}{2}}\Bigg]
\label{eq:DiffCS}
\end{equation}
where $\sigma_{Mott}$ is the Mott cross section for the scattering from a point particle of charge $e$ and is given by the following formula
\begin{equation}
\sigma_{Mott}=\frac{4\alpha^{2}E'^{2}}{Q^{4}}\cos^{2}\frac{\theta}{2}
\label{eq:SigmaMott}
\end{equation}

The structure functions $W_{1}$ and $W_{2}$ contain all the information about the structure of the nucleon 
and are measured by experiment. 
These can also be written as dimensionless functions as
\begin{equation}
F_{1}(x, Q^{2}) = MW_{1}(\nu, Q^{2}), \hspace{0.5cm} and 
\label{eq:F1SF}
\end{equation}
\begin{equation}
F_{2}(x, Q^{2}) = \nu W_{2}(\nu, Q^{2}).
\label{eq:F2SF}
\end{equation}

The two structure functions $W_{1}$ and $W_{2}$ are related to the photoabsorption cross section.
This is due to the fact that the virtual photon has two states of polarization, transverse (helicity $\pm$ 1) 
and longitudinal (helicty 0). 
Based on this the inclusive cross section can be written in terms of $\sigma_{T}$ and $\sigma_{L}$, 
\begin{equation}
{ d^{2}\sigma \over d\Omega dE^{'} } =  \Gamma \Big[\sigma_{T}(x, Q^{2})+\epsilon \sigma_{L}(x, Q^{2})\Big] 
\label{eq:SigmaPhotoAbsorb}
\end{equation}
where $\Gamma$ is the flux of virtual photons and is given by 
\begin{equation}
\Gamma =   { \alpha K \over 2\pi^{2}Q^{2} } { E'\over E} {1 \over 1-\epsilon}
\label{eq:GammaFlux}
\end{equation}
The factor $K$ is given by 
\begin{equation}
K = {2M\nu - Q^{2} \over 2M} \phantom{l}
\label{eq:KFactor}
\end{equation}
and is defined as ``equivalent photon energy''. 
In other words, $K$ is the energy that would be required to form the final hadron state of the same mass 
that would be created by the real photon with energy $\nu$. 
The ratio of longitudinal to transverse virtual photon polarization given by 
\begin{equation}
\epsilon =   \left[1+2\left(1+{ \nu^{2} \over Q^{2}}\right)\tan^{2}{\theta \over 2}\right]^{-1} \phantom{l}
\label{eq:Epsilon}
\end{equation}
and ranges between $\epsilon=0$ and 1.

The structure functions $F_{1}$ and $F_{2}$ can be written in terms of $\sigma_{T}$ and $\sigma_{L}$ as
\begin{equation}
F_{1}=\frac{KM}{4\pi^{2}\alpha}\sigma_{T}, \hspace{0.5cm} and 
\label{eq:F1SigmaT}
\end{equation}
\begin{equation}
F_{2}=\frac{K}{4\pi^{2}\alpha}\frac{\nu}{(1+\nu^{2}/Q^{2})}\left( \sigma_{L} + \sigma_{T}\right).
\label{eq:F2SigmaTL}
\end{equation}
The ratio of longitudinal to transverse cross section is given as 
\begin{equation}
R \equiv \displaystyle\frac{\sigma_{L}}{\sigma_{T}}=\displaystyle\frac{F_{2}}{2xF_{1}}\left( 1 + \frac{\nu^{2}}{Q^{2}}\right) - 1.
\label{eq:RSigLSigT}
\end{equation}
The structure function $F_{1}$ is purely transverse as it is related to $\sigma_{T}$ only, 
while the $F_{2}$ is a combination of both $\sigma_{L}$ and $\sigma_{T}$. 
A purely longitudinal structure function $F_{L}$ is defined as 
\begin{equation}
F_{L}=\left( 1 + \displaystyle\frac{Q^{2}}{\nu^{2}} \right) F_{2} - 2xF_{1}
\label{eq:FLSF}
\end{equation}
and R can be written as 
\begin{equation}
R = \displaystyle\frac{F_{L}}{2xF_{1}}.
\label{eq:RFLF1}
\end{equation}
Using the ratio $R$, the $F_{2}$ structure function can be extracted from the measured cross sections $\sigma$ 
according to 
\begin{equation}
F_{2}=\displaystyle\frac{\sigma}{\sigma_{Mott}}\nu\epsilon\displaystyle\frac{1+R}{1+\epsilon R}
\label{eq:F2FromCS}
\end{equation}
provided that $R$ is already known.

Inclusive neutrino-nucleon scattering takes place by the exchange of $W$ bosons.
These are the charged current reactions involving neutrinos: \mbox{$\nu N \rightarrow e^{-} X$} or \mbox{$\bar{\nu} N \rightarrow e^{+} X$}.
The cross section can be written as~\cite{Close:1979bt}
\begin{equation}
\displaystyle \frac{ d^{2} \sigma }{d \Omega dE'} = \displaystyle{G^{2}}{(2\pi)^{2}} \left( \frac{m^{2}_{W}}{m^{2}_{W}+Q^{2}}\right)^{2}\frac{E'}{E}L^{(\nu)}_{\mu\nu}W^{\mu\nu} 
\label{eq:NeutrinoTensor}
\end{equation}
where $m_{W}$ is the mass of the $W$ and the tensors correspond to the same definition as in Eq.~\ref{eq:DiffCSTensor}.
After contracting the $L^{(\nu)}_{\mu\nu}W^{\mu\nu}$ the differential cross section is written as 
\begin{equation}
\displaystyle \frac{ d^{2} \sigma^{(\nu,\bar{\nu})} }{d \Omega dE'} = \displaystyle{\frac{G^{2}E'^{2}}{2\pi^{2}}}\left( \frac{m^{2}_{W}}{m^{2}_{W}+Q^{2}}\right)^{2}
\left( 2W_{1}\sin^{2}\frac{\theta}{2} + W_{2}\cos^{2}\frac{\theta}{2} \mp W_{3} \frac{E+E'}{M}\sin^{2}\frac{\theta}{2} \right) 
\label{eq:NeutrinoSigma}
\end{equation}
The difference between electromagnetic and weak scattering is 
the existence of the third structure function. 
This is a result of parity violation in weak interactions.
The structure functions of neutrino scattering can also be written in terms of the photo-absorption cross sections.
\begin{equation}
W_{1}=\displaystyle\frac{K}{\pi G \sqrt{2}}\left( \sigma_{R} + \sigma_{L}\right),
\label{eq:W1Neutrino}
\end{equation}
\begin{equation}
W_{2}=\displaystyle\frac{K}{\pi G \sqrt{2}}\frac{Q^{2}}{Q^{2}+\nu^{2}} \left( \sigma_{R} + \sigma_{L} + 2 \sigma_{S}\right), \hspace{0.5cm} and
\label{eq:W2Neutrino}
\end{equation}
\begin{equation}
W_{3}=\displaystyle\frac{K}{\pi G \sqrt{2}}\frac{2M}{\sqrt{\nu^{2}+Q^{2}}} \left( \sigma_{R} - \sigma_{L} \right).
\label{eq:W3Neutrino}
\end{equation}
In the case of electron scattering parity invariance forces $\sigma_{R}=\sigma_{L}$ 
(L and R stand for left and right, S stands for longitudinal polarization), so there was only one term $\sigma_{T}$. 
One of differences between electron and neutrino scattering is that the electron$-$nuclear structure functions have 
contributions only from vector-vector (VV) terms, while the neutrino structure functions 
are a combination of VV and axial-axial (AA) terms. 
The vector-axial vector interference term is given by $W_{3}$ ($W_{3}$ = 0 in electron scattering since $\sigma_{R}=\sigma_{L}$).
The significance of comparing neutrino scattering to electron scattering
will be discussed later in next section.

\subsubsection{Deep Inelastic Scattering and Quark Parton Model} \label{sec:DISQPM}
\begin{figure}[ht]
\begin{center}
\epsfig{file=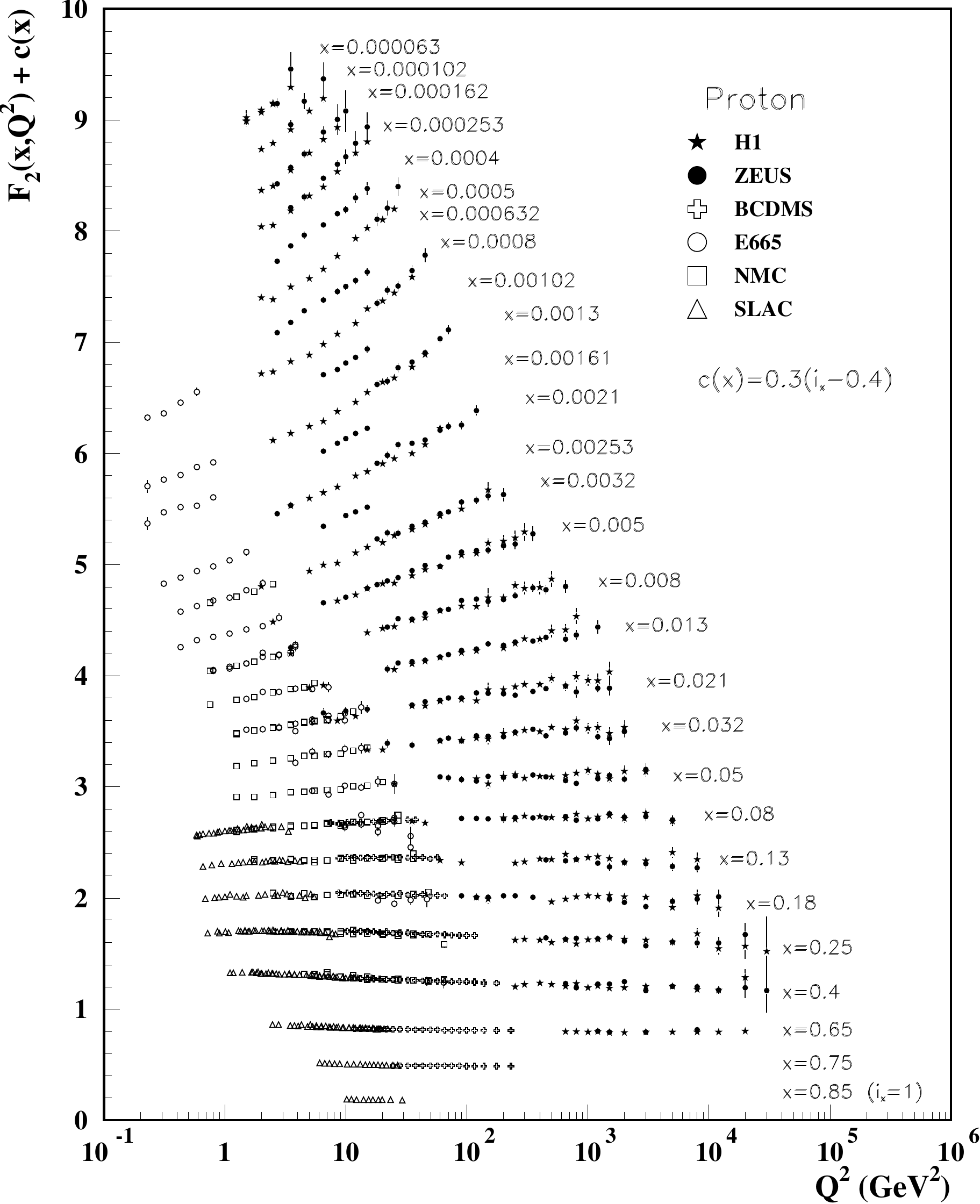, width=4.0in}
\caption{ World data on the structure function $F_{2}(x, Q^{2})$ of the proton from the Particle Data Group~\cite{Hagiwara:2002fs}.}
\label{fig:WorldF2DIS}
\end{center}
\end{figure}

Elastic electron-nucleon scattering happens when the struck nucleon recoils 
without disintegration and the final hadronic mass $W$ is equal to the nucleon mass, $W=M_{N}$.
Nuclear resonances are excited states of the nucleon and can be seen in the spectrum of inelastic 
electrons when $W < 2$ GeV and Q$^2 < $ 4.5 GeV$^2$.  
At fixed $W$ nuclear the structure functions go to zero when $Q^{2} \rightarrow \infty$,  
\begin{equation}
\left. \begin{array}{l}
 \mbox{ $m_{N}W_{1}(W,Q^{2})$} \\
 \mbox{ $\nu W_{2}(W,Q^{2})$ }
       \end{array} \right\} \xrightarrow{Q^{2} \rightarrow \infty} 0.
\label{eq:StrcFuncElast}
\end{equation}
If instead of keeping $W$ fixed, the variable $x$ (defined in Eq.~\ref{eq:bjx}) is kept constant 
for $Q^{2}>1$ GeV$^2$ and $W>2$ GeV the remarkable phenomenon of Bjorken scaling~\cite{Bjorken:1968dy} is observed, 
see Fig.~\ref{fig:WorldF2DIS}.
The structure functions $W_{1}$ and $W_{2}$ become independent of the mass scale.
It was the measurement of $W_{1}$ and $W_{2}$ at SLAC~\cite{Breidenbach:1969kd} that uncovered the 
first evidence that the nucleon consists of structure-less particles, and the cross section was an 
incoherent sum of individual elastic scattering cross sections from these 
constituents. 
The data showed that at $Q^{2}$ values above a few GeV, the structure functions
depend only on this new variable $x$:
\begin{displaymath}
m_{N}W_{1}(x,Q^{2})\rightarrow F_{1}^{qpm}(x),
\end{displaymath}
\begin{equation} \label{eq:eqw1w2}
\nu W_{2}(x,Q^{2})\rightarrow F_{2}^{qpm}(x),
\end{equation}
Recalling that the structure function for any object with internal 
structure must be $Q^{2}$-dependent, this observation implies that 
the virtual photons in DIS must be scattering from point-like, structure-less 
objects inside the proton. 
These internal particles were given the name of partons, it was 
only later that they were associated with the quarks and gluons of QCD. 
The $Q^{2}$-independent behavior is the property of Bjorken 
scaling and arises as a consequence of the asymptotic freedom associated with QCD at short distances  
and implies independence of the absolute resolution scale and hence point-like substructure. 
The physical significance of $x$ becomes clear in the infinite momentum frame~\cite{Close:1979bt}  
where nucleon is moving with a momentum approaching to $\infty$ in the z-direction. 
In such a frame, relativistic time dilation implies that during the interaction 
with the virtual photon, the nucleon can be considered as a collection of 
non-interacting partons (the impulse approximation~\cite{Close:1979bt}), each with
different fractions of the total nucleon longitudinal momentum. 
For one of these partons to absorb a virtual photon of energy $\nu$ and mass 
$Q^{2}$, it must carry exactly the momentum fraction $x$. 
It is then clear that DIS, in addition to providing evidence for internal 
structure of the proton, can be used to measure the momentum distribution 
of the partons inside the nucleon (for a review see Ref.~\cite{Roberts:1990ww}). 
The unpolarized parton distribution $q(x)$ for both 
spin-1/2 and spin-0 partons can be related to the nucleon
structure functions of Eq.~\ref{eq:eqw1w2} by the relation:
\begin{equation} \label{eq:eqquarkdis2}
F_{2}^{qpm}(x)=x\sum_{i}e_{i}^{2}f_{i}(x)
\end{equation}
where the sum is over the various species of partons of charge $e^{2}_{i}$  and $f_{i}(x)$ is the 
probability that the parton has momentum in the interval $x \rightarrow x + dx$.
The $F_{1}(x)$ structure function is related to parton distribution functions by the following formula
\begin{equation} \label{eq:eqquarkdis1}
F_{1}^{qpm}(x)=F_{2}(x)/2x=\frac{1}{2}\sum_{i}e_{i}^{2}f_{i}(x).
\end{equation}
This relation is known as Callan-Gross relation.
From Eq.~\ref{eq:FLSF} and Eq.~\ref{eq:RFLF1} the $R$ defined in Eq.~\ref{eq:RFLF1} can be written as 
\begin{equation} \label{eq:RQPM}
R=\frac{Q^{2}}{\nu^{2}}=\frac{4M^{2}x^{2}}{Q^{2}}.
\end{equation}
The above equation shows that when $Q^{2} \rightarrow \infty$, $R \rightarrow 0$.
Since the virtual photons have helicity $\pm$ 1, they can only be absorbed by partons with  
non-zero spin (this follows from conservation of angular momentum) otherwise after absorbing a photon the parton will still have 
zero angular momentum, violating conservation of angular momentum, since the total angular momentum of initial state was not zero). 
This prediction, that $R$ is small at large $Q^2$ is in agreement with data \cite{Whitlow:1990gk,Dasu:1988ru,Aubert:1985fx,Benvenuti:1989rh,Berge:1989hr}, 
but the predicted quadratic reduction of $R$ at small $x$ is not observed. 
$R$ being small supports the idea of partons having spin $\frac{1}{2}$.

All of the above discussion assumes that partons don't have mass or transverse momentum.
Including these effects~\cite{Close:1978fk} $R$ can be written as 
\begin{equation} \label{eq:RQPMMassEffect}
R=\frac{4k_{T}^{2}+m^{2} \pm \Delta}{Q^{2}}
\end{equation}
where $k_{T}$ is transverse momentum of partons, $m$ is the mass of partons,
$\Delta$ is binding energy of partons. 
This form is significant since it predicts large values for $R$ in $Q^{2}=$1-5 GeV range. 
A very nice review about QPM can be found in Ref.~\cite{Close:1978fk}.

The quark and gluon parton distributions have been accurately measured over a 
wide range of kinematics, and have led to important revelations concerning 
the internal nature of the nucleon. 
The most significant of these discoveries is the fact that the sum of
quark momenta only amounts to around \mbox {45 \%} of the nucleon's total momentum.
If instead of considering the spin-averaged structure functions one introduces
the spin-dependent structure function $g_{1}$, it is possible to define 
the polarized parton distributions $\Delta q(x)$ and $\Delta \bar{q}(x)$ in a 
similar way:
\begin{equation} \label{eq:eqg1}
g_{1}=\sum_{q}e_{q}^{2}(\Delta q(x)+\Delta \bar{q}(x)).
\end{equation}
The polarized parton distributions are defined as the difference between 
quark distributions with spin parallel/anti-parallel to the proton 
spin $\Delta q(x)=q(x)^{\uparrow}-q(x)^{\downarrow}$.
Measurement of these entities led to another 
discovery: the spin of the quarks only contribute about 25 \% to the 
total spin of the nucleon. 
This spin crisis was discovered at CERN in the EMC 
experiment~\cite{Ashman:1989ig} and has led to the question: if the quarks only carry 
a quarter of the nucleon's intrinsic spin, where does the remainder come from?
To answer this, the nucleon spin $S_{N}$ can be decomposed into three components:
\begin{equation} \label{eq:eqspin}
S_{N}=\frac{1}{2}\Delta \Sigma + \Delta G +L_{z}=\frac{1}{2}
\end{equation}
where $\Delta \Sigma$ is the total spin of the quarks, $\Delta G$ the total 
spin of the gluons and $L_{z}$ is the quark's orbital angular momentum. 
Recent results of DIS data~\cite{COMPASS} showed that contribution of $\Delta G$ is compatible with zero.

\subsubsection{Scaling Violation in DIS} \label{sec:ScalingViol}
In the simple quark-parton model partons are assumed to be asymptotically free at high $Q^{2}$.
In addition if one assumes that number of partons do not change with $Q^{2}$ the scaling should not be 
violated at all. 
However, it can be seen from Fig.~\ref{fig:WorldF2DIS} that scaling is violated since function $F_{2}$
has some $Q^{2}$ dependence. 
The observed scaling violation is due to interactions between partons and a change in their numbers.
At low values of $x$ the structure function $F_{2}$ rises as $Q^{2}$ increase
while at large $x$ it decreases.
The scaling violation at low $x$ is caused by gluon splitting to quark-antiquark pairs. 
As a consequence, the number of quarks with low values of $x$ increase, which is equivalent of $F_{2}$ increasing 
as $Q^{2}$ increase.
At high values of $x$ where valance quarks are dominant, gluon radiation shifts quarks to lower 
$x$ causing $F_{2}$ to fall with increasing $Q^{2}$. 

This scaling violation can also be explained in terms of the resolution effects of the virtual photon. 
At some value of $Q_{1}^{2}$ virtual photon can probe structure of $\lambda \sim  1/\sqrt{Q^{2}_{1}}$.
At low $x$, where sea quarks and gluons carry most of the momentum of the nucleon, only 
part of them can be seen by the virtual photon. 
At $Q_{2}^{2} > Q_{1}^{2}$ resolution of the virtual photon is higher and is able 
to resolve more quarks and gluons increasing the value of $F_{2}$.
At the same time, the higher resolution of virtual photon at high $x$ (valance quarks are dominant) means that 
instead of seeing quarks with high $x$ the virtual photon can see quarks which radiate a gluon and 
have their $x$ lowered by some amount, effectively decreasing $F_{2}$.

The structure functions are not calculable in the framework of QCD.
However, it is possible to calculate the $Q^{2}$ evolution of structure functions in perturbative QCD.
This evolution is done according to Dokshitzer-Gribov-Lipatov-Altarelli-Parisi equations (DGLAP)~\cite{Gribov:1972ri,Dokshitzer:1977sg,Altarelli:1977zs}. 



Besides scaling violations caused by the self-coupling of gluons, which has a $\log Q^{2}$
dependence, there exist scaling corrections of the form $(1/Q^{2})^{n}$.
These are called power corrections. 
These power corrections to perturbative QCD are due to the non-vanishing target nucleon mass and higher twist effects. 
The higher twist effects are caused by the interactions between the 
struck quark in the electron-quark scattering process and the other quarks in the nucleon (Final State Interactions). 
At large $Q^{2}$, the target mass and higher twist effects are negligible. 
But, these effects may be significant at low $Q^{2}$. 
In order to study QCD scaling violations in $F_{2}$ structure function Nachtmann introduced a scaling variable~\cite{Nachtmann:1973mr},
\begin{equation} \label{eq:eqnach}
\xi = \displaystyle\frac{2x}{1 + \sqrt{1 + \displaystyle\frac{4M^{2}x^{2}}{Q^{2}}}}.
\end{equation}
As it can be seen from the expression of Nachtmann scaling variable, as $Q^{2}$ increase $\xi \rightarrow x$. 
Using the Nachtmann scaling variable $\xi$ instead of $x$ the scaling of the $F_{2}$ structure function can be extended to lower $Q^{2}$.
The lower limit of $Q^{2}$ at which the QCD is still applicable is not exactly known. 
However, a comparison of the data \cite{Gluck:1995yr} to a QCD analysis suggests that $Q^{2}$ evolution work 
well for $Q^{2}$ values as low as 0.3~(GeV/c)$^{2}$. 

\subsection{\texorpdfstring{$F_{2}$}{F2} and \texorpdfstring{$R$}{R} in the Nuclear Resonance Region}
In the nucleon resonance region $W^{2} < 4$ and Q$^{2}=$ 0.5-4.5 GeV$^2$, the
non-perturbative higher twist and target mass effects may be quite significant. 
This is expected since the resonances are bound states of quarks
and gluons and necessarily involve higher twists.
Therefore, the behavior of $F_{2}$ and $R$ in the resonance region may be different from that in the DIS region.
Even at high values of $Q^{2}$, $R$ may not be small due to strong gluon binding effects (see Ref.~\cite{Close:1978fk}). 

\begin{figure}[ht]
\begin{center}
\epsfig{file=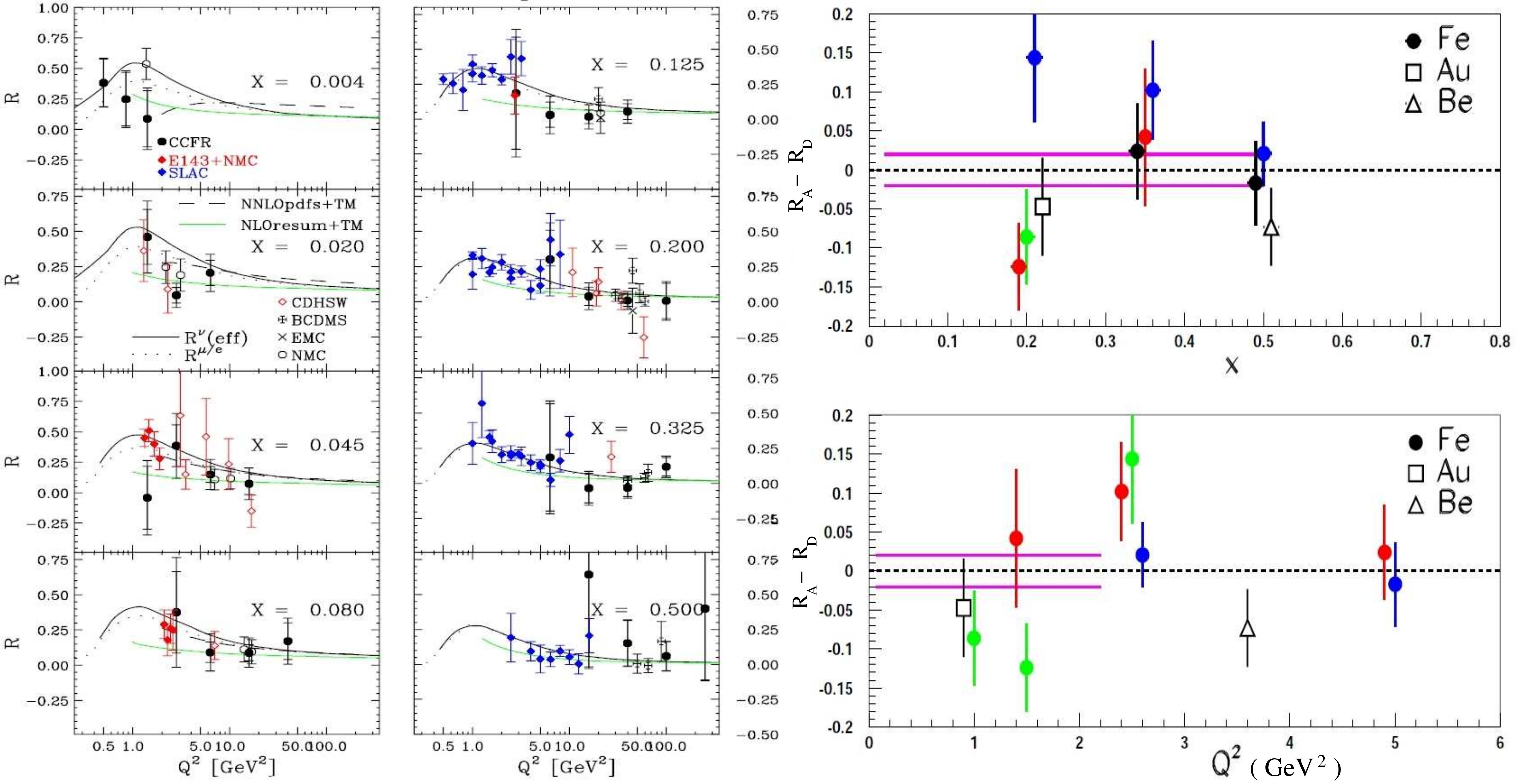, width=6.0in}
\caption{ Left: A compilation of the world's published data on $R$ at high $Q^{2}$ in the DIS for both nucleons and nuclei.
Right: The SLAC E140 data on the nuclear dependence of $R$ in the DIS region presented in the form $R_{A} - R_{D}$ (The magenta lines in the figure 
on the right represent the projected uncertainties of this experiment.) }
\label{fig:WorlDISR}
\end{center}
\end{figure}

The nuclear dependence of $R$ in the resonance region has never been determined experimentally.
The only data that exists is in the DIS region.
In Fig.~\ref{fig:WorlDISR} the world data on $R$ is shown for all nuclear targets.
The left plot in Fig.~\ref{fig:WorlDISR} shows the world's data on $R$ for all nuclear targets in the DIS region.
The right plot in Fig.~\ref{fig:WorlDISR} shows all available data on the nuclear dependence of $R$ in the DIS region. 
The errors are very large and none of the data are in the resonance region. 
Note that the data in Fig.~\ref{fig:WorlDISR} (right) were taken at high values of $W^{2}$ in the DIS region. 
At these high W and $Q^{2}$ values, the overall value of $R$ is generally quite small.
Therefore, it is difficult to discern any nuclear effects. 
The measurements of the present experiment are at low $Q^{2}$, where $R$ is larger and a nuclear dependence should be 
easier to discern. 
The lines in Fig.~\ref{fig:WorlDISR} (right) represent the small projected uncertainties of the current experiment.

Precision measurements of $R$ in the resonance region at moderate to low $Q^{2}$ 
will greatly aid efforts to develop a reliable global description of $R$. 
The new data on $R$, $F_2$, $F_1$, and $F_L$ in the resonance 
region will also help efforts to develop global models to describe all unpolarized cross
sections, covering the range from large $x$, low $Q^{2}$ to smaller $x$, low $Q^{2}$. 
(These models are useful for electron$-$nuclear scattering model development, 
and the extraction of parton distribution functions.)
A detailed study of $F_{2}$ and $R=\sigma_{L}/\sigma_{T}$ on nuclear targets in the resonance
region is an important ingredient in forming an integrated description of charged lepton and 
neutrino scattering cross-sections. 
High rate neutrino beams now under construction or planned at Fermilab and J-PARC will allow the first 
precision experimental comparisons of electron and neutrino cross sections and present and planned 
future neutrino oscillation experiments will use these results to predict event rates.

The full program of studies requires first additional precise electron scattering data, in particular 
$\sigma_{L}$ and $\sigma_{T}$ (or equivalently $F_{2}$ and $R$) on nuclear targets 
(materials suited for future neutrino oscillation detectors, including water~\cite{Itow:2001ee}, 
hydrocarbons~\cite{NUMI}, liquid argon and steel), where
the most precise high energy neutrino cross sections have been measured~\cite{Yang:2001xc}) 
in the relevant kinematic regime. 
Later, as the new generation of high rate neutrino beams at Fermilab and J$-$PARC become available, 
the approach can be directly validated with comparisons to data from high rate neutrino cross$-$section 
experiments on the same targets~\cite{NeutrinoTargets}.

Additionally, $R$ will be used to investigate the phenomenon of quark$-$hadron duality (discussed in the next section), 
which suggests that, on average, electroproduction structure functions in resonance region scale the same way as in the 
DIS region.

\subsection{Quark Hadron Duality}
At high energies perturbative QCD methods fully describe experimental results 
and at low energies chiral perturbation theory allows the study of low-energy 
dynamics of QCD. 
A phenomenon of quark-hadron duality, first discovered by Bloom and Gilman, 
is observed connecting these two regimes. 
In intermediate kinematics, a wide variety of reactions can be described 
simultaneously by single particle (quark) scattering, and by exclusive 
resonance (hadron) scattering.
\subsubsection{Bloom-Gilman Duality} \label{sec:BGDuality}
In the early 1970’s, when analyzing inclusive electron-proton scattering data from SLAC, 
Bloom and Gilman observed~\cite{Bloom:1970xb, Bloom:1971ye} a remarkable connection between 
structure function $\nu W_{2}(\nu,Q^{2})$ in the nucleon resonance region and in the deep 
inelastic region. 
It was found that the resonance structure function averaged over the scaling variable $\omega'$ 
was approximately equal to deep inelastic structure function.
The scaling variable is given by the following formula
\begin{equation} \label{eq:eqomegaprime}
\omega' = \displaystyle\frac{2M\nu + M^{2}}{Q^{2}}=1 + \displaystyle\frac{W^{2}}{Q^{2}}=\omega + \displaystyle\frac{M^{2}}{Q^{2}}
\end{equation}
where M is the nucleon mass.
The choice of $\omega'$ over $\omega=1/x$ was not clear at the time it was used. 
If one looks at the Nachtmann variable defined in Eq.~\ref{eq:eqnach} it will be clear that 
using $\omega'$ for scaling parameter is somewhat equivalent of using $1/\xi$. 
This new variable $\omega'$ was found without theoretical guidance but an extraction of the Nachtmann variable $\xi$ 
revealed a similarity.
The Bloom and Gilman scaling variable $\omega'$ included additional $Q^{2}$ dependence that incorporates the 
finite target mass effects.
Using the variable $\omega'$, Bloom and Gilman were able to make the first quantitative observations 
of quark-hadron duality in inclusive electron scattering.

\begin{figure}[ht]
\begin{center}
\epsfig{file=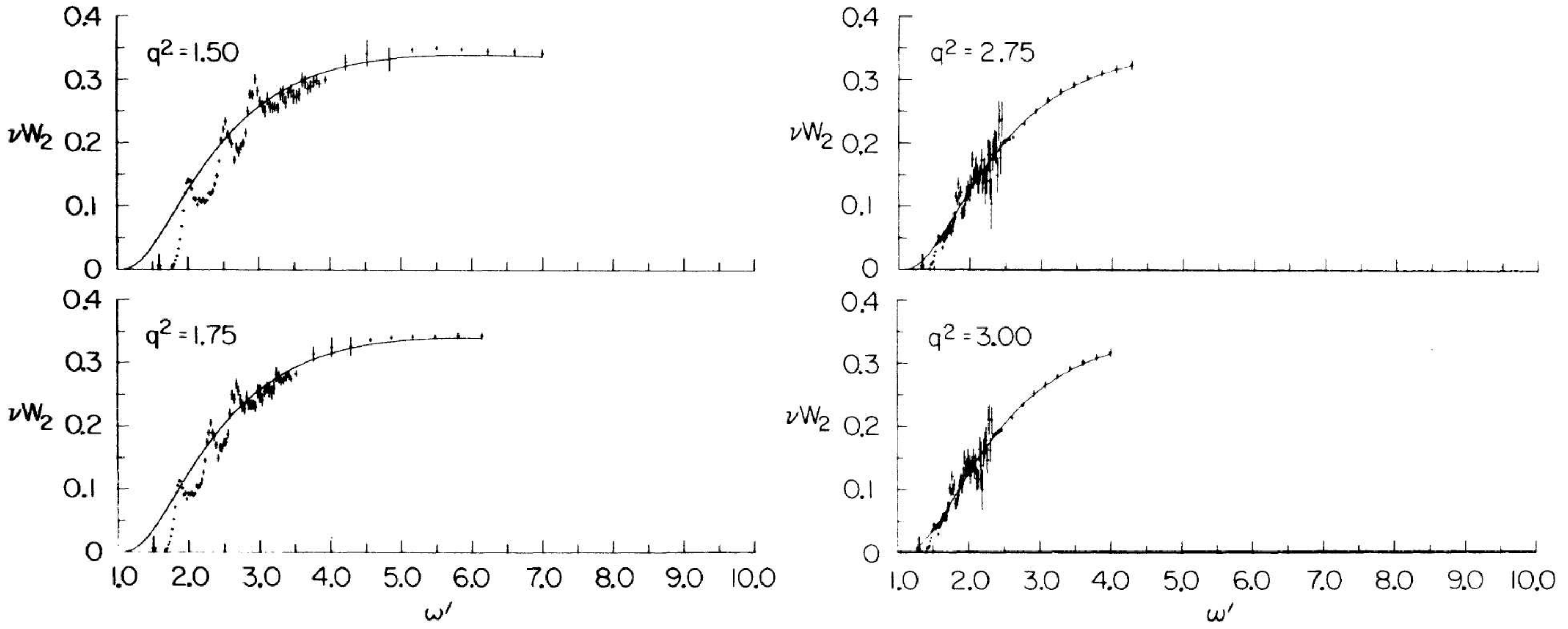, width=6.0in}
\caption{ Early proton $\nu W^{2}$ structure function data in the resonance region, as a function of $\omega'$,
compared to a smooth fit to the data in the scaling region at larger $Q^{2}$. 
The resonance data were obtained at $Q^{2} = $ 1.5, 1.75, 2.75, 3.0 GeV$^2$, for the longitudinal
to transverse ratio R = 0.18. (Adapted from Ref.~\cite{Bloom:1971ye}.)}
\label{fig:BloomDuality}
\end{center}
\end{figure}

The original data on the proton $\nu W^{2}(\nu,Q^{2})$ structure function in the resonance 
region are illustrated in Fig.~\ref{fig:BloomDuality} for several values of $Q^{2}$ from 1.5 to 3.0 GeV$^{2}$. 
In the resonance region there are about twenty nucleon resonances, but only three of them are visible in 
measured inclusive electron-proton cross section. 
The first peak corresponds to a single resonance P$_{33}$(1232), while other 
resonances are composed of overlapping states. 
The second resonance region comprises from the S$_{11}$(1535) and D$_{13}$(1520) resonances.
The data presented in Fig.~\ref{fig:BloomDuality} are from inclusive measurements, and may contain 
tails from heavier resonances as well as some nonresonant components. 
The $\nu W^{2}(\nu,Q^{2})$ structure function data were extracted from the measured cross sections 
using a fixed value of the longitudinal to transverse cross section ratio, $R=\sigma_{L}/\sigma_{T}$ = 0.18.

The scaling curve shown in Fig.~\ref{fig:BloomDuality} is a parametrization of the high-W (high-$Q^{2}$) data 
available in the early 1970s~\cite{Miller:1971qb}.
It can be seen that the resonance data are clearly seen to oscillate about, and average to, the scaling curve.
Based on the similarity of the structure functions in the resonance and DIS region Bloom and Gilman 
concluded that the resonance data are, on average, equivalent to the scaling curve. 
Also, the agreement of resonance region data gets better with increasing $Q^{2}$.
These observations led Bloom and Gilman to conclude that resonances are not a separate entity 
but are an intrinsic part of the scaling behavior of $\nu W_{2}(\nu,Q^{2})$~\cite{Bloom:1971ye}.

In order to quantify the similarity of scaling functions, Bloom and Gilman wrote a finite energy sum 
rule~\cite{Bloom:1971ye} at fixed $Q^{2}$ for $\nu W_{2}(\nu,Q^{2})$ .
They equated the integral over $\nu$ of $\nu W_{2}$ in the resonance region, to the integral over 
$\omega'$ of the scaling function:
\begin{equation} \label{eq:DualBGsumrule}
\displaystyle\frac{2M}{Q^{2}}\int^{\nu_{m}}_{0} \nu W_{2}(\nu,Q^{2})d\nu=\int^{1+W_{m}^{2}/Q^{2}}_{1} \nu W_{2}(\omega')d \omega'.
\end{equation}
In this equation the upper limit on the $\nu$ integration, $\nu_{m}= ( W^{2}_{m} - M^{2} + Q^2 )/2M$, 
corresponds to the maximum value of $\omega'=1 + W^{2}_{m}/Q^{2}$, where $W_{m} \sim$ 2 GeV, 
so that the integral of the scaling function covers the same range in $\omega'$ as the resonance region data. 
The finite energy sum rule allows the area under the resonances in Fig.~\ref{fig:BloomDuality} to be compared 
to the area under the smooth curve in the same $\omega'$ region to determine the degree to which the 
resonance and scaling data are equivalent. 

A comparison of both sides in Eq.~\ref{eq:DualBGsumrule} for $W_{m}=$ 2 GeV showed that the relative 
differences ranged from $\sim$ 10\% at $Q^2=$ 1 GeV$^2$, to $\sim$ 2\% beyond $Q^2=$ 2 GeV$^2$~\cite{Bloom:1971ye}.
This demonstrates the equivalence on average of the resonance and deep inelastic regimes. 
Using this approach, Bloom and Gilman$'$s quark-hadron duality was able to qualitatively describe the 
data in the range $1 < Q^{2} < 10$ GeV$^2$.
The resonances in inclusive inelastic electron-proton scattering do not vanish 
with increasing $Q^2$ relative to the ``background'' beneath them, but instead fall at about 
the same rate with increasing $Q^2$. 
The nucleon resonances are therefore strongly correlated with the scaling behavior of $\nu W_{2}$.

\subsubsection{Duality in Nuclei}
Duality studies have been studied mostly on protons. 
There are handful of experiments in the high$-x$ and low to moderate $Q^{2}$ region that tried to 
study duality on deuterium and heavy nuclei~\cite{Arrington:2001ni,  Arrington:1998ps, Filippone:1992iz}.
Inclusive electron-nucleus experiments at SLAC designed to probe the $x > 1$ region in the $F^A_{2}$ structure
function concluded that the data began to display scaling indicative of local duality~\cite{Filippone:1992iz}.
An explanation for the origin of $\xi$ scaling was proposed by the Benhar and Liuti~\cite{Benhar:1995rh}. 
The authors suggested that the observed scaling might instead come from an accidental cancellation of $Q^{2}$ 
dependent terms, and would occur only for a limited range of momentum transfers (up to $Q^{2} \sim$ 7 GeV$^2$).

\begin{figure}[ht]
\begin{center}
\epsfig{file=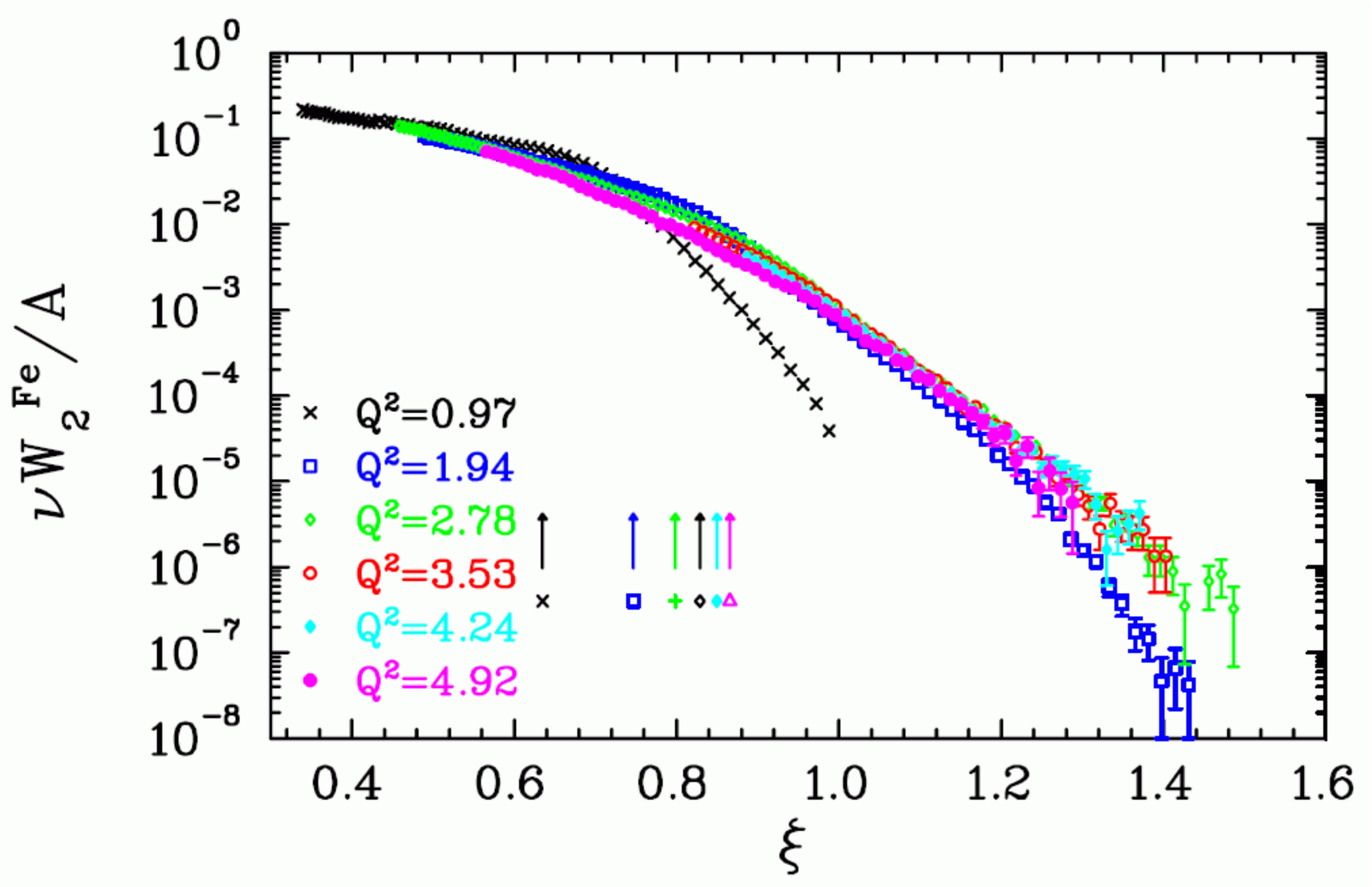, width=4.5in}
\caption{ The $\nu W_{2}^{Fe}=F_{2}$ structure function for iron (per nucleon) as a function of $\xi$. 
The data were obtained at fixed electron scattering angle, and the quoted $Q^2$ (in units of GeV$^2$) are
the values for $x = 1$. 
The arrows indicate the values of $\xi$ corresponding to the quasi-elastic peak for each setting.
(Figure from Ref.~\cite{Arrington:2001ni})}
\label{fig:NuclearFilipone}
\end{center}
\end{figure}

In order to study this effect and confirm the observation of the experiments done at SLAC an experiment was done 
at Jefferson Lab~\cite{E89008}.
A plot shown in Fig.~\ref{fig:NuclearFilipone} is shown to illustrate the results of that experiment. 
Here, $F_{2}^{Fe}/A$ is plotted as a function of $\xi$ at several values of $Q^{2}$ ($x = 1$).
An important thing to realize from the first glance is the absence of resonance structure, which is clearly 
observed for the free nucleon. 
Also, the quasi-elastic peak is not visible. 
The absence of a distinct quasi$-$elastic and resonance structure is a result of Fermi smearing. 
It was discussed in the previous section that structure functions averaged over the nuclear resonance region 
resembles the structure function of DIS. 
In the case of nuclei, it appears that averaging is done by Fermi motion of nucleons inside the nuclei. 
As the resonance structure is smeared by Fermi motion the scaling can be observed at all $\xi$ and 
it is impossible to differentiate DIS and resonance regimes other than by calculating kinematics. 
It can be seen from Fig.~\ref{fig:NuclearFilipone} that all data except the lowest $Q^2$ fall on a smooth scaling curve. 

Qualitatively, the nuclear effects in the resonance region appear to be similar to those in
the deep inelastic region. 
The nuclear dependence of the scaling structure functions are not expected to be the same 
as the nuclear dependence of resonance production.
This is somewhat surprising, but perhaps it is another hint that quark-hadron duality 
can be applied in inclusive nuclear scattering too. 
Based on the data of $\xi$-scaling in nuclei, it can be implied that Fermi motion is 
averaging the nucleon electromagnetic response over a finite energy range 
(for the proton this was done using Eq.~\ref{eq:DualBGsumrule}).

There is limited data available in the nuclear resonance region and at the same time 
the heavy model dependence at large $x$ does not allow duality studies for nuclei at the same 
level as it is done for proton.
The results of the current experiment will complement the existing data and allow precise studies of  
the nuclear dependence of quark-hadron duality. 

\subsection{Theoretical Basis of Duality}
At the time when Bloom-Gilman duality was observed QCD was not yet fully developed.
In order to give theoretical explanation to it, phenomenological models or 
models based on hadronic degree of freedom were developed.
After the development of QCD and the recognition that it is a real theory of
strong interactions, the phenomenon of duality was reanalyzed using the Operator Product Expansion (OPE). 
This is discussed in the next section.

\subsubsection{Moments of Structure Functions}
Before starting a discussion of the Operator Product Expansion let's introduce Cornell-Norton~\cite{Cornwall:1968cx}
moments first. 
The n-th moments of the spin-averaged $F_{1}$, $F_{2}$ and $F_{L}$ structure functions are defined as: 
\begin{equation} \label{eq:M1Moment}
M_{1}^{n}(Q^{2})=\displaystyle\int^{1}_{0}dx\hspace{0.2cm}x^{n-1}F_{1}(x,Q^{2})
\end{equation}
\begin{equation} \label{eq:M2Moment}
M_{2,L}^{n}(Q^{2})=\displaystyle\int^{1}_{0}dx\hspace{0.2cm}x^{n-2}F_{2,L}(x,Q^{2}).
\end{equation}
In this definition the $n = 1$ moment of the $F_{1}$ structure function in the parton model 
counts quark charges, while the $n = 2$ moment of the $F_{2}$ structure function corresponds to 
the momentum sum rule. 
In the Bjorken limit, the moments of the $F_{1}$ and $F_{2}$ structure functions are related via the
Callan-Gross relation, see Eq.~\ref{eq:eqquarkdis1}, as $M_{2}^{(n)}=2M_{1}^{(n)}$.
The Cornwall-Norton moments defined in terms of the Bjorken $x$ scaling variable are 
appropriate in the region of kinematics where $Q^{2}$ is much larger than typical 
hadronic mass scales, where corrections of the type $M^{2}/Q^{2}$ can be neglected. 
In the next section it will be clear how the structure function moments can be utilized 
to test the validity of quark-parton duality at small $Q^{2}$ where the contributions from 
higher twist effects are not negligible.

\subsubsection{Operator Product Expansion}
The theoretical basis for describing Bloom-Gilman duality in QCD is the operator product
expansion (OPE) of Wilson~\cite{Wilson:1969zs}.
Based on OPE, sum rules are derived which can be used to analyze the structure functions and their moments.
The important feature of these sum rules is model independence. 
They are derived using only some very general results from quantum field theory. 

The OPE allows one to evaluate a products of operators in the asymptotic limit. 
The product of two operator-valued fields, $A(x)$ and $B(y)$, can be expressed as 
an infinite series 
\begin{equation} \label{eq:OPEExpention}
A(x)B(y)=\displaystyle\sum_{n} C_{n}(x-y)O_{n}(x)
\end{equation}
where $C_{n}(x-y)$ are analytic functions of $x-y$ and $O_{n}(x)$ are local fields. 
This expansion is valid as long as $x-y$ is small enough or equivalently, $Q^{2}$ is 
large enough compared to relevant mass scale. 
At any fixed values of $x-y$ only finite number of terms contribute in the above expansion.

According to the OPE, at large $Q^{2} >> \Lambda^{2}_{QCD}$, the moments of the structure 
functions can be expanded in powers of $1/Q^{2}$. 
The coefficients in the expansion are matrix elements of quark and gluon operators corresponding 
to a certain twist, $\tau$, defined as the mass dimension minus the spin, n, of the operator. 
For the $n$-th moment of the $F_{2}$ structure function, $M_{2}^{(n)}$, for example (see Eq. (43)), 
one has the expansion
\begin{equation} \label{eq:OPEMomentExpen}
M_{2}^{(n)}(Q^{2})=\displaystyle\sum_{\tau=2,4,...}^{\infty}\displaystyle\frac{A_{\tau}^{(n)}(\alpha_{s}(Q^{2}))}{Q^{\tau-2}}, \hspace{1cm} n = 2,4,6...
\end{equation}
where $A_{\tau}^{n}$ are the matrix elements with twist less than $\tau$.
The $Q^{2}$ dependence of the matrix elements can be calculated perturbatively, 
with $A_{\tau}^{n}$ expressed as a power series in $\alpha_{s}(Q^{2})$.

The moments of structure functions defined in Eq.~\ref{eq:M2Moment} can be used in a QCD analysis. 
It must be emphasized that OPE analysis is perturbative since the structure functions 
are expanded in terms of a hard scale, $1/Q^{2}$. 
The usefulness of OPE is that it allows a separation of ``soft'', nonperturbative physics 
contained in parton correlation functions, from the ``hard'' scattering of the probe 
from the partons~\cite{Feynman}.

The relation between the higher-twist matrix elements and duality in electron scattering
was explained in the paper of De Rujula, Georgi and Politzer~\cite{De Rujula:1976tz, Georgi:1976vf}.
They explained the empirical observation of Bloom-Gilman duality in terms of the twist 
expansion of the structure function moments in QCD. 
Examining Eq.~\ref{eq:M2Moment} they saw that the lowest moment corresponding to $n=2$ 
is the Bloom-Gilman integral defined in Eq.~\ref{eq:DualBGsumrule}.
Scaling ($Q^{2}$ independence of the $F_{2}$ structure function) is only observed at high enough 
values of $Q^{2}$, which means that at low $Q^{2}$ duality should be violated. 
At high $Q^2$, where the scaling of the structure function is observed, moments become independent of 
$Q^2$. 
In order for the integrals of the structure functions at higher momentum to be equal to the integral of the structure functions
at low $Q^{2}$, the higher-twist contributions must be small or cancel. 
This is equivalent to saying that duality can exist if the interactions between 
the scattered quark and the spectator system is suppressed. 

\subsection{The Nuclear Dependence of \texorpdfstring{$R$}{R}}
Pions are thought to be the carrier of the nuclear force at nucleon distance scale 
and hence are the dominant agent responsible for the binding of nuclei. 
The Yukawa interaction can be used to describe the strong nuclear force between nucleons  
carried by pions. 
This was strengthened via the observation of very significant effects of pion exchange 
currents in a variety reactions involving electromagnetic probes of nuclei. 
In the EMC experiment~\cite{Arneodo:1992wf} it was discovered that the structure function $F_{2}$ 
depends on the mass number $A$ of the target. 
This effect has been studied by comparing $F_{2}$ measured on bound nucleons in nucleus $A$ and deuterium. 
This implies that there is a significant difference between the parton distributions of free 
nucleons and nucleons in a nucleus. 
There were several attempts to understand the observed effect in terms of nuclear pions but all of them failed 
to explain the experimental data at all kinematics.
These failures to observe the influence of nuclear pions caused a crisis for nuclear theory~\cite{Bertsch:1993vx}.

\begin{figure}[h]
\begin{center}
\epsfig{file=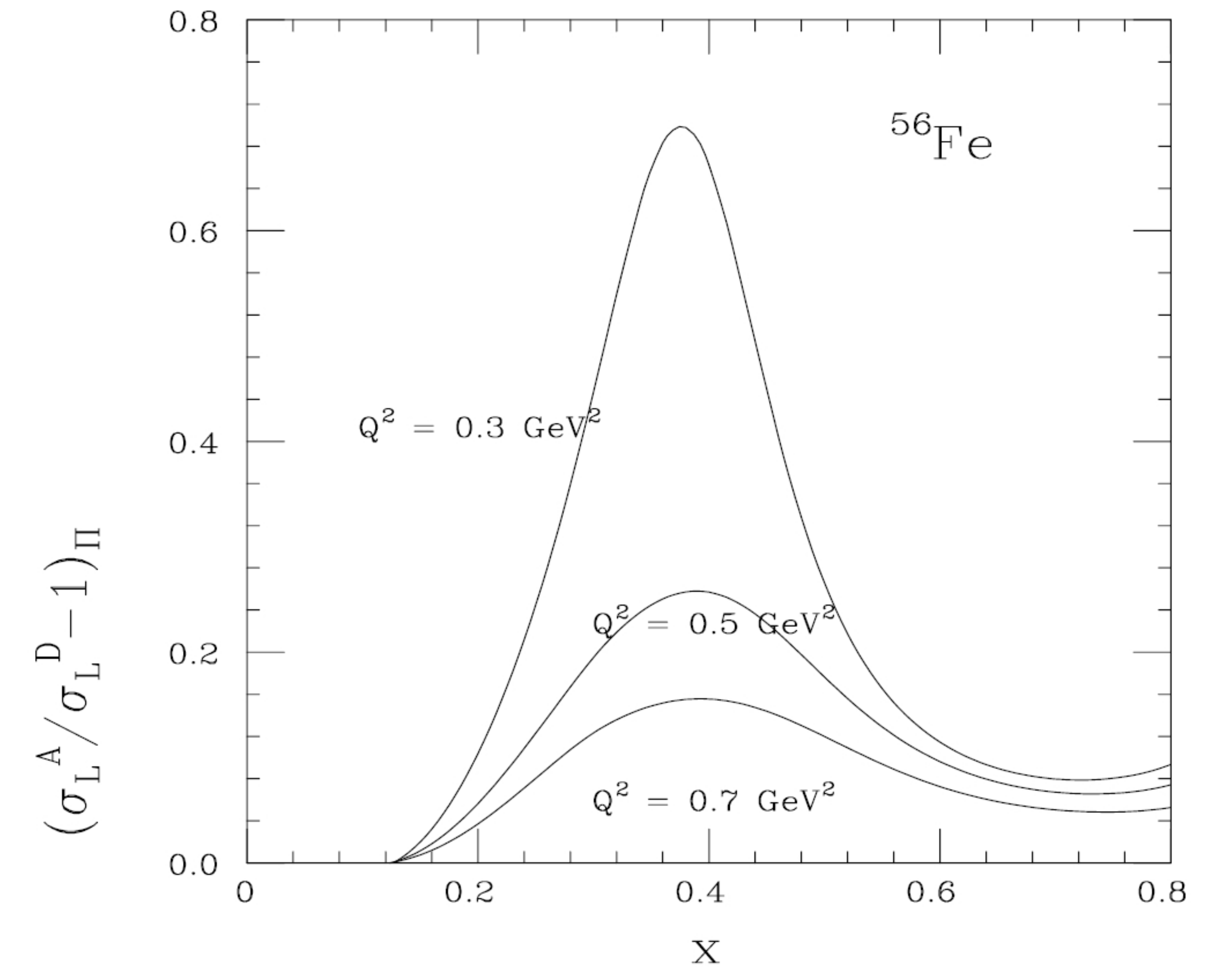,height=3.0in}
\end{center}
\caption { Plot from G. A. Miller~\cite{Miller} showing the predicted sensitivity of
the inclusive longitudinal cross section ratio of iron to deuterium due to pion excess. }
\label{fig:miller}
\end{figure}

One possibility for verification arises from measurements of the ratio $R$, of scattering
of virtual photons in a longitudinal or transverse polarization state , $R=\sigma_{L}/\sigma_{T}$ 
from nuclei. 
A large value of $\sigma_{L}$, and the corresponding violation of the Callan-Gross relation 
Eq.~\ref{eq:eqquarkdis1}, indicates the presence of nuclear bosons as fundamental constituents 
of nuclei~\cite{Miller:2000ta}. 
A measurement of $R$ in the region covered by this experiment combined with the data of experiment 
E02-109~\cite{E02-109} can provide information about the dynamics of the internucleon force 
inside nuclei as predicted by Miller~\cite{Miller}.
The calculations performed by Miller~\cite{Miller} predict a significant pion excess 
enhancement in the $\sigma_{L}^{A}/\sigma_{L}^{D}$ ratio at low $Q_{2}$ and moderate $x_{Bj}$, 
as shown in {Fig.~\ref{fig:miller}}.
The phenomenology is described in the reviews \cite{Piller:1999wx,Frankfurt:1988nt,Arneodo:1992wf}. 
The main point is that the quark and anti-quark distributions can be given as convolution of 
the $q,\bar{q}$ distributions in a given nuclear hadronic constituent with the light cone 
distribution functions: $f_{\pi/A}(y),f_{\pi/N}(y)$, the probability to find an excess pion 
in the nucleus (A) or nucleon (N), with a plus-momentum given by $ym_{N}$.
The function $f_{\pi/N}(y)$ is constrained by the data on the nucleon sea which restrict 
the $\bar{u}$ and $\bar{d}$ distributions to be similar~\cite{Thomas:1983fh,Frankfurt:1989wq}.

The longitudinal cross sections for nuclear and deuterium targets are related by the formula~\cite{Miller}
\begin{equation} \label{eq:MillerAD}
\displaystyle\frac{\sigma_{L}^{A}}{\sigma_{L}^{D}}=1+x\frac{2}{3}f_{\pi}(\xi)\frac{\nu^{2}}{(Q^{2}+\nu^{2})}\frac{F_{\pi}^{2}}{F_{2}^{D}R_{D}}(1+R_{D})
\end{equation}
where $F_{\pi}^{2}$ is the pion form-factor. 
This equation is plotted shown in Fig.~\ref{fig:miller} for iron an target. 
The largest sensitivity to pions is predicted at $Q^{2}=0.3$ GeV$^2$ and quickly vanishes with the increase of
the $Q^{2}$ due to the falloff of the pion form-factor.
The results of this experiment, combined with the results of the E02-109~\cite{E02-109} experiment, 
will allow the extraction of $\sigma_{L}^{A}$ and $\sigma_{L}^{D}$ ($Q^{2}$ $\approx$ 0.3 GeV$^2$, $W^{2} <$ 4.0 GeV$^2$). 
The expected enhancement in the $\sigma_{L}^{A}/\sigma_{L}^{D}$ ratio at low $Q^{2}$ 
will be regarded as an indication of pion excess in heavy nuclei as suggested by Miller~\cite{Miller}.

\subsection{Neutrino Scattering}
Neutrino interactions with nuclei are important for the determination
of the internal structure of the nuclear matter and 
complement deep inelastic electron scatterings experiments. 
Electron scattering experiments are sensitive to the average of the square 
of the electric charge of the nucleon constituents, while neutrino experiments 
measure their weak structure. 
In particular, neutrino(antineutrino)$-$nucleus scattering can be used to
understand the flavor composition of nuclei as neutrino scattering 
is sensitive to specific flavors of quarks. 

It is also interesting to test if Bloom-Gilman duality is observed in neutrino scattering. 
It was argued in Ref.~\cite{Carlson:1993wy} that duality should also exist in weak structure functions. 
This can only be confirmed by experiment and there is not enough data in the resonance region 
to draw a final conclusion. 
The interaction of neutrinos with the nuclear matter is weak and cross sections 
are small which requires the use of nuclear targets. 
Also nuclear effects in DIS have been studied using only muon and electron beams. 
Neutrino scattering experiments were only done using heavy nuclear targets such 
as iron target-calorimeters. 
Results of these experiments indicate that the nuclear corrections for $e-A$ and $\nu-A$ are different. 
Among these differences is evidence for quark-flavor dependent nuclear effects~\cite{Gran:2008zz}.

The MINERvA~\cite{Gran:2008zz} experiment seeks to measure low energy neutrino interactions both in support 
of neutrino oscillation experiments and also to study the strong dynamics of the 
nucleon and how the nuclear medium affects these interactions.
\subsubsection{Structure Functions in DIS}
Deep inelastic neutrino structure functions are directly related to the parton distribution functions 
and the same parton distribution functions determine the charged lepton scattering.
For electron scattering the unpolarized parton distribution $q(x)$ for both 
spin-1/2 and spin-0 partons is related to the nucleon structure functions according to 
Eq.~\ref{eq:eqquarkdis2}
where the sum is over the various species of partons of charge $e_{i}$  
and $f_{i}(x)$ is the probability that the parton has momentum in the interval 
$x \rightarrow x + dx$. 
Assuming that partons have quantum numbers of quarks and using the Eq.~\ref{eq:eqquarkdis2}, 
$F_{2}$ electron scattering structure functions of proton and neutron can be written as: 
\begin{equation} \label{eq:F2QPMProton}
\displaystyle\frac{1}{x}F_{2}^{ep}=\frac{4}{9}[u^{p}(x)+\bar{u}^{p}(x)]
+\frac{1}{9}[d^{p}(x)+\bar{d}^{p}(x)]
+\frac{1}{9}[s^{p}(x)+\bar{s}^{p}(x)]
\end{equation}
\begin{equation} \label{eq:F2QPMNeutron}
\displaystyle\frac{1}{x}F_{2}^{en}=\frac{4}{9}[u^{n}(x)+\bar{u}^{n}(x)]
+\frac{1}{9}[d^{n}(x)+\bar{d}^{n}(x)]
+\frac{1}{9}[s^{n}(x)+\bar{s}^{n}(x)]
\end{equation}
where proton is indicated as p and neutron as n.
Since u, d quarks and proton, neutron both form isospin doublets one can 
see that $u^{p}=d^{n}$ (call this $u$), $d^{p}=d^{n}$ (call this $d$), 
$s^{p}=s^{n}$ (call this $s$). 
Here the contribution of other quarks is small and is not considered.
Taking into account the isospin symmetry described above the structure functions can
be written as: 
\begin{equation} \label{eq:F2QPMProton1}
\displaystyle\frac{1}{x}F_{2}^{ep}=\frac{4}{9}(u+\bar{u})+\frac{1}{9}(d+\bar{d}+s+\bar{s})
\end{equation}
\begin{equation} \label{eq:F2QPMNeutron1}
\displaystyle\frac{1}{x}F_{2}^{ep}=\frac{4}{9}(d+\bar{d})+\frac{1}{9}(u+\bar{u}+s+\bar{s}).
\end{equation}
For neutrino nucleon scattering $F_{2}$ can also be written in terms of quark 
distributions. 
Charged current neutrino scattering is different from the electron scattering since it only couples to
specific flavors of quarks, in particular for neutrino proton scattering only 
$d,s,\bar{u},\bar{c}$ quarks contribute, and in neutrino-neutron scattering only
$u,c,\bar{d},\bar{s}$ quarks contribute. 
Neutrino scattering structure function can be written as 
\begin{equation} \label{eq:F2QPMNeurino}
F_{2}(x)=x\sum_{i}f_{i}(x),
\end{equation}
where only specific quarks contribute as discussed above.
At energies where the charm quark production is absent or negligible and 
the Cabibo angle~\cite{Cabibbo:1963yz} (The Cabibo angle is the probability that one flavor of quark will 
change into other flavors under the action of the weak force) is zero, the proton structure function 
can be written in terms of quark distributions as 
\begin{equation} \label{eq:F2QPMNeurinoProton}
\displaystyle\frac{1}{x} F_{2}^{\nu P}(x)  =2[d(x)+\bar{u}(x)]
\end{equation}
\begin{equation} \label{eq:F2QPMNeurinoNeutron}
\displaystyle\frac{1}{x} F_{2}^{\nu N}(x) =2[u(x)+\bar{d}(x)]
\end{equation}
where the factor 2 comes from the fact that weak current contains both
vector and axial parts. 
Using equations~\ref{eq:F2QPMProton1}, \ref{eq:F2QPMNeutron1}, \ref{eq:F2QPMNeurinoProton}, 
\ref{eq:F2QPMNeurinoNeutron} 
the structure functions of electron and neutrino scattering can be related to each other as 
\begin{equation} \label{eq:F2QPMElecToNeutrino}
\left[ \displaystyle\frac{ F_{2}^{eN} + F_{2}^{eP} }{ F_{2}^{\nu N} + F_{2}^{\nu P} } \right](x)=
\frac{5}{18} \left[  \frac{u + \bar{u} + d + \bar{d} + 2/5( s + \bar{s})}{u + \bar{u} + d + \bar{d}}\right].
\end{equation}
If the contribution of strange quarks is negligible ($x>0.2$) the structure functions are related 
by a factor of $5/18$, known as the 5/18ths rule. 
The observation that electron scattering and neutrino scattering structure functions 
are approximately related by a factor of $5/18$, was a significant triumph for the QPM.
An interesting and very useful review about this QPM can be found in~\cite{Close:1979bt}.
\subsubsection{Neutrino Scattering in Resonance Region}
In the previous section relating the structure functions of electron and neutrino scattering 
from the nucleon in DIS region was rather straightforward. 
The same can not be said in the nuclear resonance region. 
In the resonance region higher order QCD effects are not negligible as in DIS 
and it is not possible to consider nucleons as non-interacting group of 
partons.  
Since the electromagnetic and weak interactions are sensitive to different types of 
partons (quarks) the relationship between the structure functions in these reactions can only be 
established if specific conditions are satisfied. 
For example a neutrino beam can convert a neutron into a proton, but it cannot convert a 
proton into a neutron (and vice versa for an antineutrino beam). 
From the conservation of the vector current, the vector structure functions measured in 
electron scattering can be related to their counterparts in neutrino scattering 
for only specific isospin final states.
The axial structure functions in neutrino scattering can only be determined 
if the vector part is provided from electron scattering.

In order to give a general idea how the structure functions in electron scattering can be 
related to their counterparts in neutrino scattering, electron scattering cross section
is written in terms of helicity amplitudes,
\begin{equation} \label{eq:CSHelicityAmpl}
\displaystyle\frac{1}{\Gamma_{T}} \frac{ d^{2} \sigma }{d \Omega dE'}=
\frac{1}{2}\left( \sigma_{T}^{1/2} + \sigma_{T}^{3/2}\right) + \epsilon_{L}\sigma_{L}
\end{equation}
where $\Gamma_{T}$ is the virtual photon flux, $\epsilon_{L}$ describes the degree
of longitudinal polarization of the photon, $\sigma_{T}^{1/2}$
is the total transverse absorption cross section with helicity $1/2$
for the photon-nucleon system, and $\sigma_{T}^{3/2}$ is the helicity $3/2$ cross section.
Here $\sigma_{L}$ is the total cross section for the absorption of a longitudinal (scalar) photon.
Analyses of electroproduction data give numerical values for cross
sections at the peak of each resonance~\cite{Burkert:2002zz, Aznauryan:2004jd},
\begin{equation} \label{eq:CSHelicitySigT}
\sigma_{T}(W=M_{R})=\displaystyle\frac{2m_{N}}{\Gamma_{R}M_{R}}\left( A_{1/2}^{2} + A_{3/2}^{2} \right) 
\end{equation}
\begin{equation} \label{eq:CSHelicitySigL}
\sigma_{L}(W=M_{R})=\displaystyle\frac{2m_{N}}{\Gamma_{R}M_{R}}\frac{Q^{2}}{q^{2}_{z}}S_{1/2}^{2}
\end{equation}
where $\Gamma_{R}$ is the width of the resonance, and $M_{R}$ is the mass of the resonance. 
Using these formulas one can relate electromagnetic to weak form
factors using isotopic symmetry. 
Photons can have two isospin states $|I,I_{3}>=|1,0>$ and $|I,I_{3}>=|0,0>$.
The isovector component belongs to the same isomultiplet with the vector
part of the weak current. 
Each of the amplitudes $A_{3/2}$, $A_{1/2}$, $S_{1/2}$ can be further decomposed into 
three isospin amplitudes~\cite{Lalakulich:2006sw}.
A general helicity amplitude on a proton ($A_{p}$) and neutron ($A_{n}$) target has the
decomposition 
\begin{equation} \label{eq:ProtNeutrIsospinDecomp1}
A_{p}=A_{p}(\gamma p \rightarrow R^{+})=b - \sqrt{\displaystyle\frac{1}{3}}a^{1}+\sqrt{\displaystyle\frac{2}{3}}a^{3}
\end{equation}
\begin{equation} \label{eq:ProtNeutrIsospinDecomp2}
A_{n}=A_{n}(\gamma n \rightarrow R^{0})=b + \sqrt{\displaystyle\frac{1}{3}}a^{1}+\sqrt{\displaystyle\frac{2}{3}}a^{3}.
\end{equation}
For the weak current we have only an isovector component of the vector current, therefore the 
$b$ amplitude never occurs in weak interactions.
The $a^{1}$ and $a^{3}$ can be found from weak interactions according to the following formulas
\begin{equation} \label{eq:ResonanceA1A3}
A(W^{+} n \rightarrow R^{(1)+})=\frac{2}{\sqrt{3}}a^{1},
\end{equation}
\begin{equation} \label{eq:ResonanceA1A3_1}
A(W^{+} p \rightarrow R^{(3)++})=\sqrt{2}a^{3},
\end{equation}
\begin{equation} \label{eq:ResonanceA1A3_2}
A(W^{+} n \rightarrow R^{(3)+})=\sqrt{\frac{2}{3}}a^{3}.
\end{equation}

Comparing Eq.~\ref{eq:ResonanceA1A3} with Eq.\ref{eq:ProtNeutrIsospinDecomp1} it 
can be seen that for the isospin $1/2$ resonance the weak amplitude satisfies 
the equality $A(W^{+} n \rightarrow R^{(1)}+)=A_{n}^{1}-A_{p}^{1}$.
Since the amplitudes are linear functions, the weak form-factors are 
related to electromagnetic form-factors with a simple relation
\begin{equation} \label{eq:ResonanceAvAnAp}
I=1/2: \hspace{1cm}   C_{i}^{V}=C_{i}^{n}-C_{i}^{p}.
\end{equation}
For the isospin $3/2$ resonances the 
$A_{n}^{3}(W^{+} n \rightarrow R^{(3)+})=A_{p}^{2}(W^{-} p \rightarrow R^{(3)0})=\sqrt{2/3}a^{3}$. 
The weak form-factors are 
\begin{equation} \label{eq:ResonanceAvAnAp1}
I=3/2: \hspace{1cm}   C_{i}^{V}=C_{i}^{n}=C_{i}^{p}.
\end{equation}

The above description of relating the weak and electromagnetic structure functions 
in nuclear resonance region is thoroughly described in Ref.~\cite{Lalakulich:2006sw}.
The article is self-contained and facilitates writing simple programs to reproduce the 
cross sections.

The MINERvA~\cite{NeutrinoTargets} experiment will perform a high statistics 
neutrino-nucleus scattering experiment using a fine-grained detector specifically designed to 
measure low-energy neutrino-nucleus interactions accurately. 
The high-luminosity NuMI beam line at Fermilab will provide energies spanning 
the range $\sim$ $1-15$ GeV, over both the resonance and deep inelastic regimes, 
making MINER$\nu$A a potentially very important facility to study quark-hadron duality in 
neutrino scattering.
The results of this experiment will be used in future analysis of neutrino data (including MINERvA) 
in order to investigate quark hadron-duality in the axial structure functions of nucleons and nuclei.

\chap2{Experimental Apparatus}
\subsection{Overview}
The Jefferson Laboratory Hall C experiment E04-001~\cite{e04001} 
was performed in May-June 2007. 
The purpose of this experiment is to measure the longitudinal-transverse (L-T) 
separated structure functions $F_{2}$ and $R=\sigma_{L}/\sigma_{T}$ from nuclear 
targets in the resonance region. The targets used include carbon-12, aluminum-27, 
iron-56 and copper-64. The data were taken at 2.1, 3.2, 3.3, 4.1, 5.1 GeV beam energies 
and electron scattering angle ranging from 12$^{o}$ to 76$^{o}$. The scattered 
electron was detected in the High Momentum Spectrometer (HMS). The kinematic coverage 
of the experiment is shown in {Fig.~\ref{fig:kinem}} and is listed in Appendix.

\begin{figure}[!ht]
\begin{center}
\epsfig{file=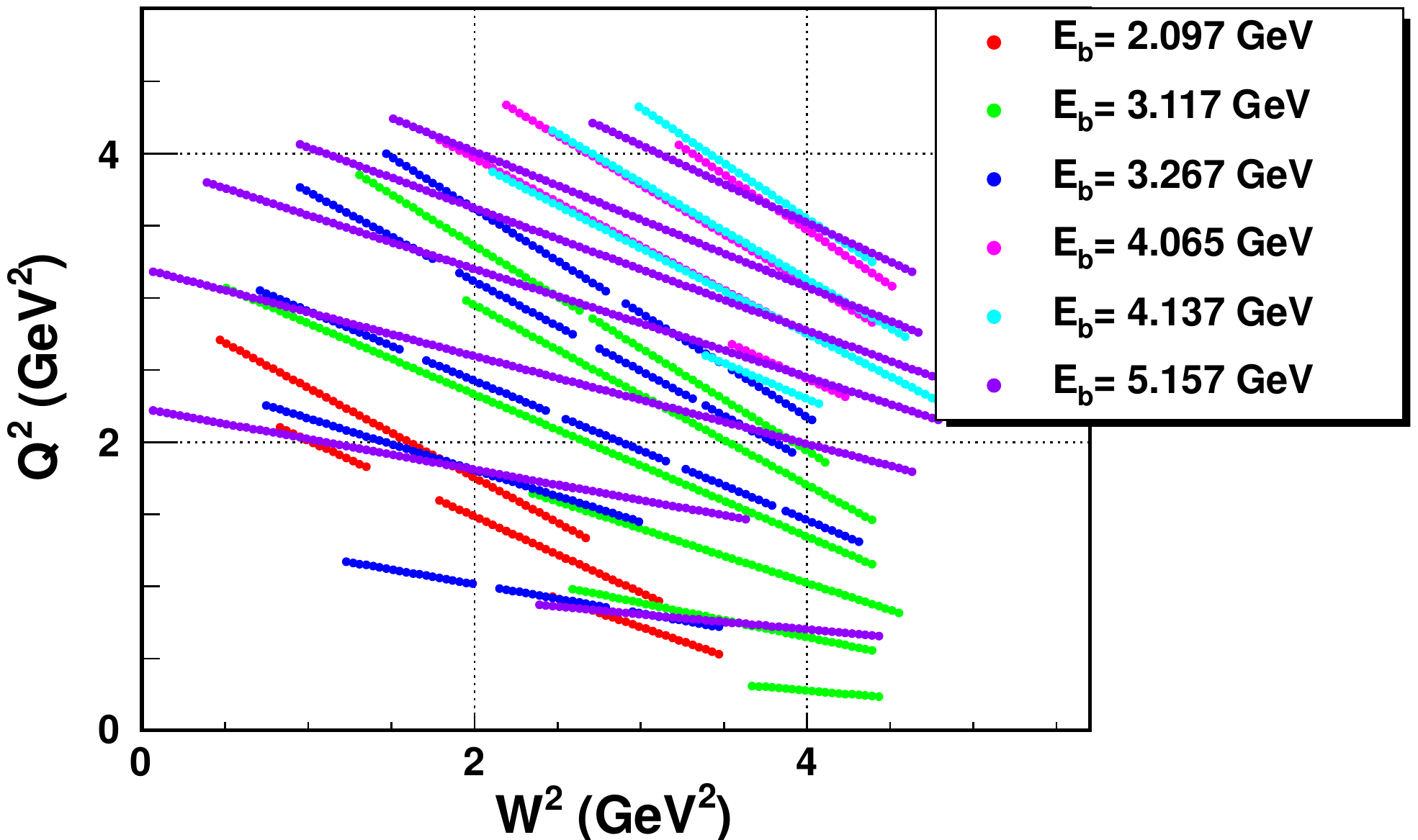,height=3.0in}
\end{center}
\caption{ Kinematic coverage of E04-001 experiment.}
\label{fig:kinem}
\end{figure}

\subsection{The Thomas Jefferson Accelerator Facility}
\begin{figure}[ht]
\begin{center}
\epsfig{file=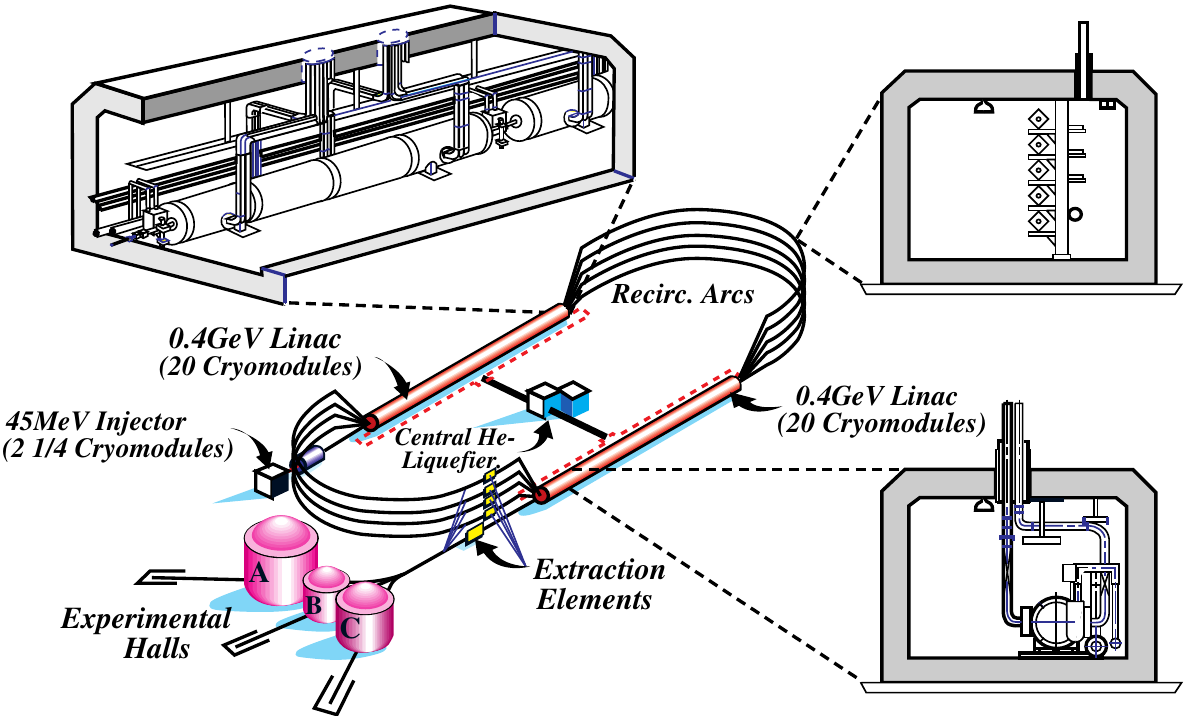,height=2.5in}
\end{center}
\caption{ Lay-out of the CEBAF facility. 
The electron beam is produced at the injector by illuminating a photocathode 
and the produced photoelectrons are then accelerated to 67 MeV.
The beam is then further accelerated in each of two superconducting linacs, 
through which it can be recirculated up to five times.
The beam can be extracted simultaneously to each of the three experimental 
halls.}
\label{fig:cebaf}
\end{figure}
The Continuous Electron Beam Accelerator (CEBAF) of JLab~\cite{Leemann:2001dg} 
was designed to deliver a continuous beam of electrons simultaneously to 
three experimental end-stations. 
A diagram of the racetrack shaped accelerator is given in 
\mbox{Fig.~\ref{fig:cebaf}}.
The source of the injector is a 100 kV photocathode guns~\cite{Kazimi:2000gg} with a 
maximum beam current of a few mA.
Next, the beam is incident on a chopping aperture which contains slits of 
different sizes, one for each of the experimental Halls A, B and C. 
The width of these slits determines the beam current that is delivered to 
each Hall. 
The chopper sweeps the beam over the slits with a rotating electric field 
with a frequency of 1497 MHz. 
The beam then enters the first superconducting accelerator section, 
where it is accelerated to 67 MeV and then injected into the 
North Linac.
The North Linac is a string of 20 cryomodules, with each cryomodule 
containing eight super-conducting niobium cavities. 
These cavities are kept super-conducting by 2 K Helium coolant from the 
Central Helium Liquefier (CHL). 
Klystrons excite the cavities and establish electric fields which accelerate the electrons 
as they travel through them. 
At the end of the North Linac, the electron beam has a nominal energy of 
600 MeV, although 
by careful tuning of the accelerating electric field of the cavities, 
this energy can be raised or lowered. 
Next, the beam enters the east magnetic recirculation arcs, where it is 
bent in a semi-circle to the South Linac. 
Here again the beam is accelerated through a string of 20 cryomodules. 
At the end of the South Linac, the beam can be extracted for use in any 
of the experimental halls or it can proceed around the west recirculation
arcs for another pass around the accelerator. 
The beam can traverse the accelerator a maximum of five times, gaining a 
nominal 1200 MeV of energy with each pass around the machine. 
In the North and South Linacs the different energy beams, resulting from 
each pass around the accelerator, travel in the same beam-line. 
However, the different energy beams require different bending fields in the 
recirculation arcs. 
When the beams reach the arcs they are separated by momentum and each one 
goes through a different arc. 
At the end of the arcs, the beams are recombined into the same beam-line 
again. 
When the beam is of the energy requested by the experimental halls, it is
extracted from the accelerator to the Beam Switch Yard (BSY). 
There the three sets of beam bunches are separated into the appropriate 
experimental hall beam-line by deflecting cavities operating at 499 MHz. 
Each hall receives a short (1.67 ps) train of pulses with a frequency of 
499 MHz.
The beam has a very small transverse size ( $\geq$ 200 $\mu$m (FWHM) at 
845 MeV). 
The fractional energy spread ($\Delta$E/E) is at the 10$^{-4}$ level.
\subsection{Beam-line Apparatus}
\begin{figure}[ht]
\begin{center}
\epsfig{file=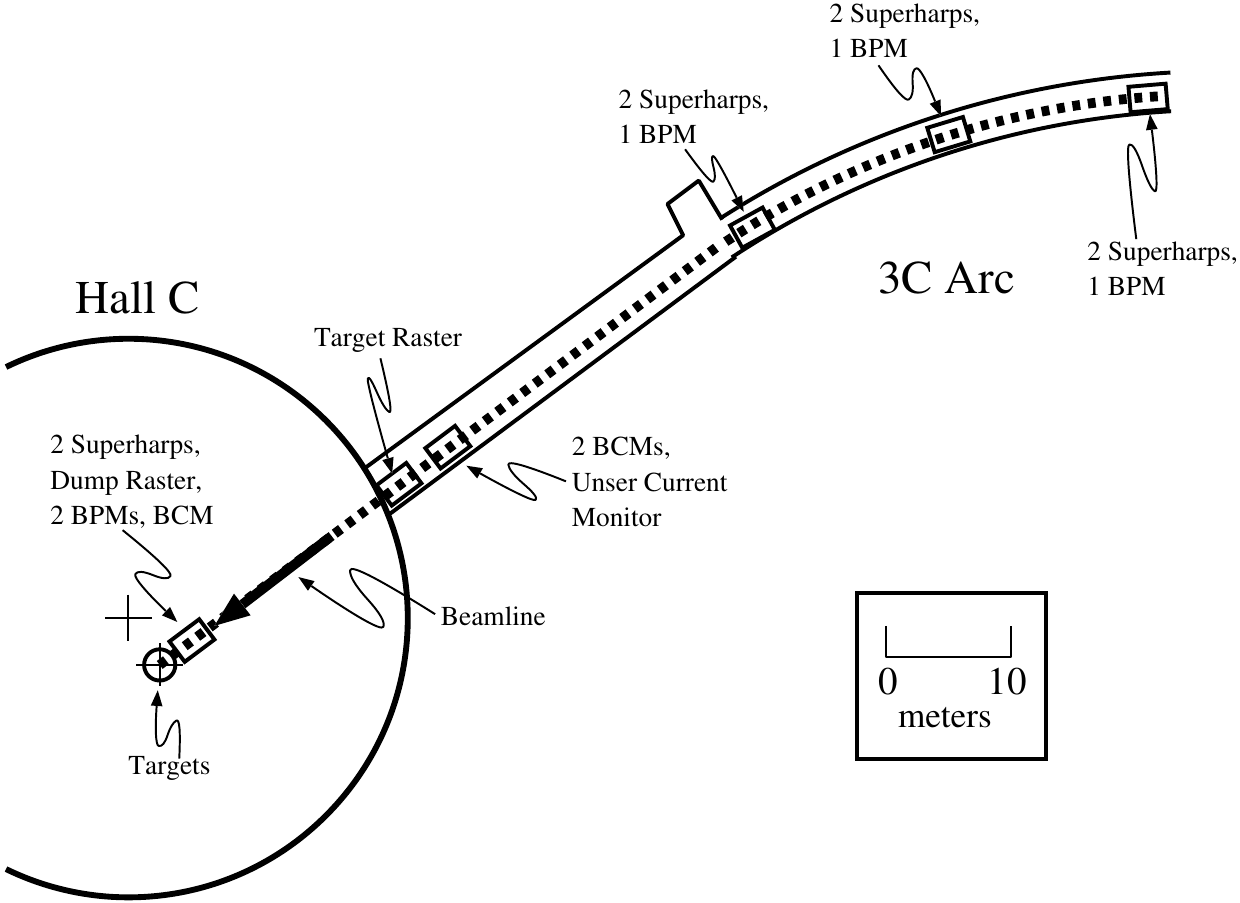,height=3.5in}
\end{center}
\caption{
  Schematic lay-out of Hall C, indicating the location of the 
  the raster, the beam energy measurement system, the beam current monitors (BCM) and 
  the beam position monitors (BPM) upstream of the target.
}
\label{fig:whallc}
\end{figure}

Electron beam parameters are measured by several devices 
(see Fig.~\ref{fig:whallc}) situated along the beam-line, 
upstream of the target, near the entrance to the hall. 
Measurements of position, current, energy and longitudinal 
polarization 
are possible, with some intentional redundancy to allow cross 
checking. 
Some typical beam parameter values for the present experiment and 
the uncertainties associated with their measurement 
are given in Table~\ref{tab:beam}
\begin{table}[htbp!]
\begin{center}
\begin{tabular}{| l | c | c | c | }\hline
Beam Parameter   & Beam-line  & Measured & Accuracy(absolute)\\
                 & Device(s) & Value &                      \\
\hline\hline
Position x (at target)  	 &  BPM/Superharp & - &  $\leq$200$\mu$m\\
Position y (at target)  	 &  BPM/Superharp & -  & $\leq$200$\mu$m \\
\hline
Current 	 & BCM     & 30-80 $\mu$A & $\leq$3$\times$10$^{-3} \mu$A \\
\hline
Energy 	         & ARC  & - & 2$\times$10$^{-4}$ GeV\\
\hline
\end{tabular}
\end{center}
\caption{
  Typical values measured for the various electron beam properties 
  described in the text, together with the associated accuracies.
}
\label{tab:beam}
\end{table}

To measure the absolute value of the energy of the electrons in 
the beam, eight dipole magnets between the beam switch-yard and the 
hall entrance deflect the electrons through a nominal angle of 
34.3$^{\circ}$.
Wire scanners before and after the magnets accurately determine
this angle, which along with the measured arc field integral
$\int{B dl}$ (magnetic field integral over the path of the electron beam), 
is used to calculate the electron momentum. 
\subsubsection{Beam Position Monitors}
So-called superharps make a destructive but extremely precise measurement of 
the beam position and profile. A superharp consists of three tungsten wires, 
two vertical that measure the horizontal beam profile and one horizontal wire that 
measures the vertical beam profile, mounted in a frame which is connected to an arm 
that can be moved in and out of the beam.
Analog-to-Digital converters (ADCs) connected to each wire read the signals on the wires 
as the frame is moved in and out of the beam, while a position encoder determines where 
the wire intercepts the beam. With the position information and the ADC measurements 
the position and profile of the beam can be measured. More detailed information about 
the superharps can be found in Ref.~\cite{Yan:1995uu}.

\begin{figure}[h!]
\begin{center}
\epsfig{file=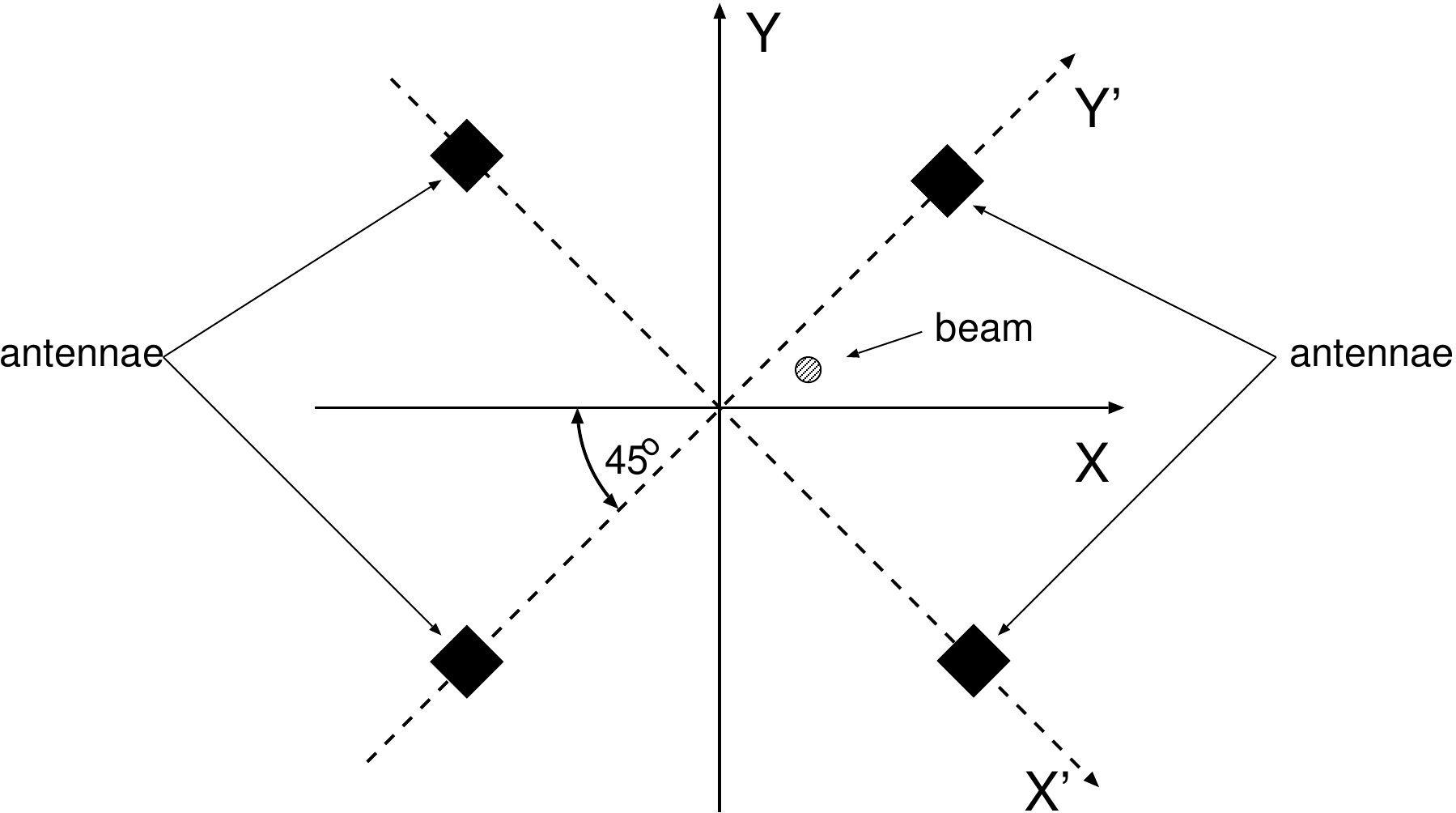,height=8cm,angle=-90}
\caption{Diagram of the orientation of the BPM antennae.}\label{fig:bpm}
\end{center}
\end{figure}

The beam position is also measured with three Beam Position Monitors (BPMs) which provide 
non-destructive beam position information along the beam line which are used during 
data taking; see Fig.~\ref{fig:bpm}. A BPM is a cavity containing two pairs of antennae 
perpendicular to each other and inclined by $\pm$45$^{\circ}$ with respect to the horizontal plane. 
When the beam passes through the cavity 
parallel to the symmetry axis both pairs of antennae pick up the frequency of the beam. 
Each antennae gives a signal proportional to the distance that the antennae is positioned from 
the beam. The beam position is derived from the difference over the sum of the signals 
from antennae on opposite sides of the beam. This method provides a position measurement 
independent of beam current. Since this is a position relative to the central axis of the 
beam line these measurements must be compared to the Superharp measurements to determine 
the absolute position of the beam during data taking. The final accuracy of the beam 
position measurements is $\pm$1.0 mm, with a relative position uncertainty of 0.1-0.2 mm.
More detailed information about the BPMs can be found in Ref.~\cite{gue2}.

\subsubsection{Beam Current Monitors}

The Hall C beam line has four devices installed to measure the beam current: three microwave cavity 
Beam Current Monitors (BCMs), and a parametric DC current transformer (Unser monitor).

The BCMs measure the integrated beam current in two second intervals, and their output voltage, as
well as the output voltage of the Unser, is converted into frequency with a voltage-to-frequency 
converter, and read out by a scaler. 
A BCM consists of a cylindrical wave guide, mounted in the beam line 
so that the beam travels along the axis of the cylinder. The dimensions of the cylinder
were selected so that the 499 MHz structure of the beam excites the 1497 MHz TM$_{010}$ mode
in the wave guide. The resonance frequency is picked up by wire loop antennae, and converted
to a DC voltage through a RMS-to-DC converter.
The cavities have a stable gain and offset, and a high signal/noise ratio, but cannot measure 
the absolute current, because the output power as a function of the measured beam current depends 
on the cavity impedance, quality factor, and the signal cable attenuation. 
The absolute calibration is done with the Unser monitor.

The Unser monitor is installed between BCM1 and BCM2.
It cannot be used for charge measurements because it is sensitive to thermal
fluctuations, resulting in large drifts in its zero-offset. However, because the 
gain is stable and well measured, the Unser monitor is used to calibrate the 
gain of the BCMs. More information about the Unser monitor can be found in Refs.~\cite{uncer1, Unser:1981fh}.



\subsubsection{Beam Raster}

The electron beam generated at CEBAF is a high current beam with a small 
transverse size ($\le$ 200 $\mu$m FWHM). It can deposit a large amount of power in the
target. In order to prevent local boiling in the cryotargets or melting of the solid targets the beam is 
rastered before striking the target. For that purpose the beamline is equipped with a pair of fast 
raster magnets, located 25 meters upstream of the target. 
The first set rasters the beam vertically and the second 
horizontally. The current driving the magnets is varied sinusoidally, at 17 kHz in 
the vertical direction, and 24.2 kHz in the horizontal direction. The frequencies are chosen
to be different so that the beam motion does not form a stable (Lissajous) figure at the target, but it
moves over a square area.
The amplitude of the raster pattern on the target was $\pm$1.0 mm in both directions during 
the present experiment. 
More details about fast raster can be found in Ref.~\cite{Yan:1995uu}.
\subsection{Experimental Hall C}
Hall C is the second largest of the three experimental halls at Jefferson Lab.
It contains two spectrometers the High Momentum Spectrometers (HMS) and the Short Orbit Spectrometer (SOS) 
located on either side of the electron beam-line.
The spectrometers can be moved clockwise or counter-clockwise about the 
target position over a wide range of angles.
The HMS has a minimum angle of 10.5$^{\circ}$ with 
respect to the beam-line and a maximum central angle of 165$^{\circ}$. 
The SOS has now been dismantled. 
Between the beam switch-yard and the target (which sits at the center 
of the hall) there are various devices to monitor and measure 
beam current, beam position, beam energy and beam polarization.
\begin{figure}[ht]
\begin{center}
\epsfig{file=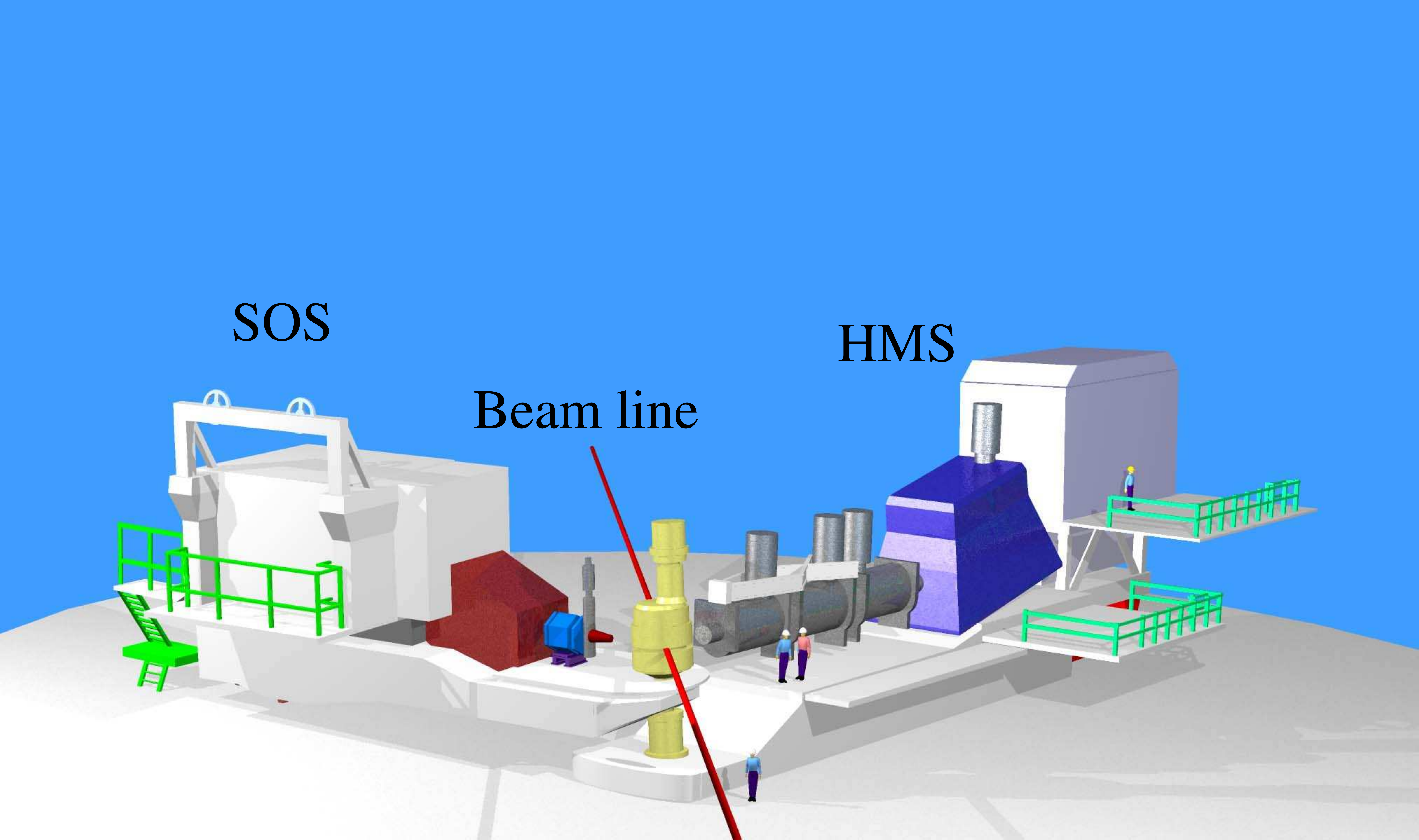,height=3.5in}
\end{center}
\caption{Schematic layout of Hall C.}
\label{fig:hallc}
\end{figure}
A schematic of \mbox{Hall C} is shown in Fig.~\ref{fig:hallc}.
At the pivot point of the two spectrometers sits the cryotarget, 
encased within a cylindrical aluminum scattering chamber. 
The electron beam is incident on the target through the beam$-$line.
The targets used for this experiment were deuterium, carbon, aluminum, 
iron and copper. Scattered particles were detected by HMS detector.
The primary beam continues along the path of the beam-line to the 
beam dump.
\subsection{Target}
The targets are located in a cylindrical aluminum scattering chamber, which is installed at 
the spectrometer pivot. The scattering chamber has an inner radius of 61.6 cm and a height of 150 cm. 
The beam exit windows are made of 0.4 mm and 0.2 mm thick aluminum foils on the sides facing the  
HMS and SOS spectrometers, respectively.
\begin{figure}[h!]
\begin{center}
\epsfig{file=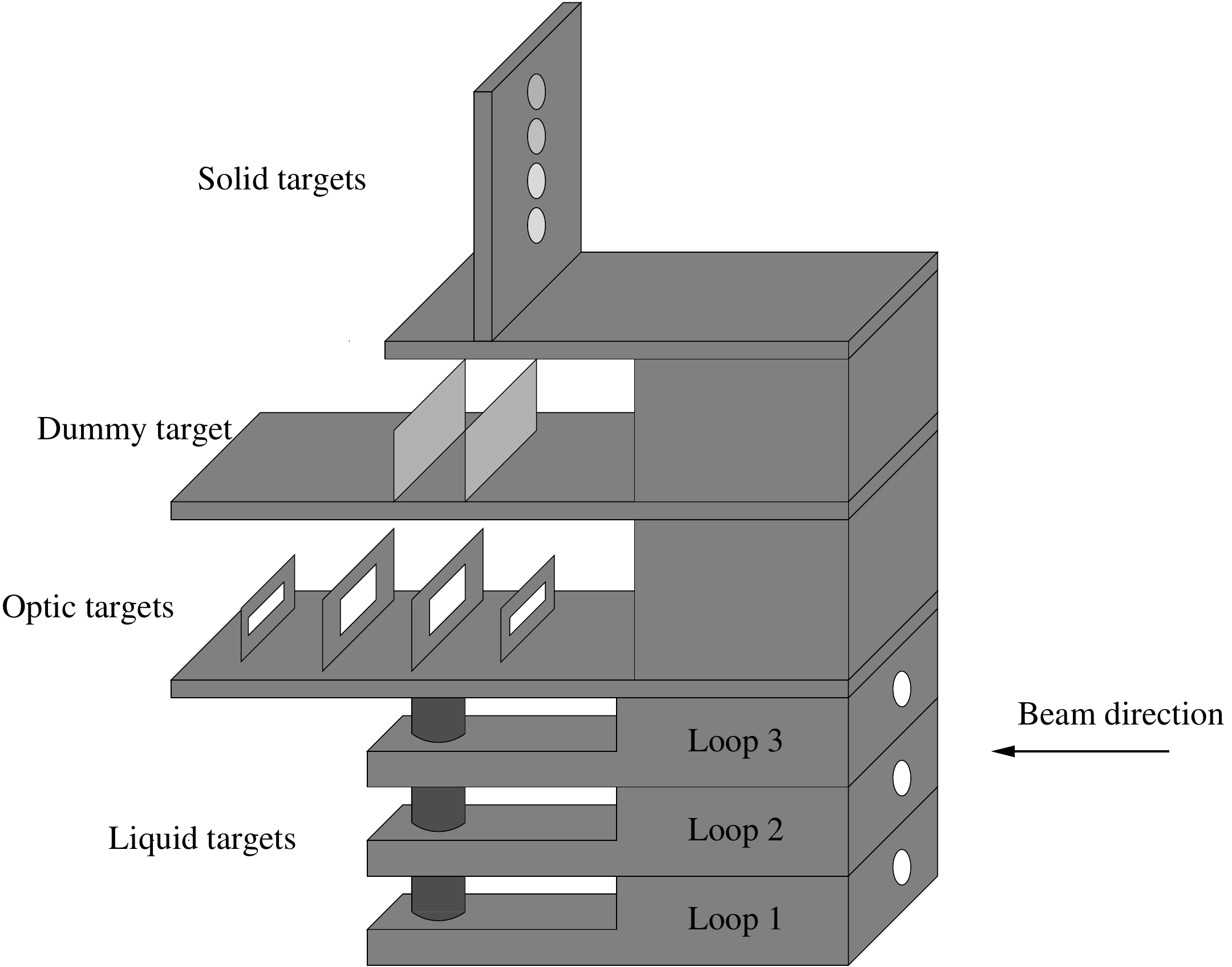,height=4.0in}
\caption{Hall C target ladder.} \label{fig:cham}
\end{center}
\end{figure}     
The target ladder (see Fig.~\ref{fig:cham}) is located inside the scattering chamber.
The solid targets are BeO, carbon, copper, iron targets (See Table~\ref{tab:tar_sol}).
The targets are mounted on the target ladder and can be exchanged in 
few minutes by the lifter mechanism. The mechanism permits accurate, reproducible 
positioning of any of the targets at beam height. 

\begin{table}
\begin{center} 
\begin{tabular}{|c|c|c|c|}
\hline
Target & Purity & Thickness (g/cm$^{2}$) & Radiation Length (\%)\\
\hline
BeO    & 99.00\% & 0.2918  $\pm$ 0.00030 & 0.45 \\
\hline
Carbon & 99.95\% & 0.35525 $\pm$ 0.00030 & 1.06 \\
\hline
Copper & 99.95\% & 0.17775 $\pm$ 0.00015 & 1.81 \\
\hline
Iron   & 99.95\% & 0.11870 $\pm$ 0.00014 & 1.00 \\
\hline
\end{tabular}
\end{center}
\caption{Thicknesses and radiation lengths of the solid targets.}\label{tab:tar_sol}
\end{table}

\begin{figure}[h!]
\begin{center}
\epsfig{file=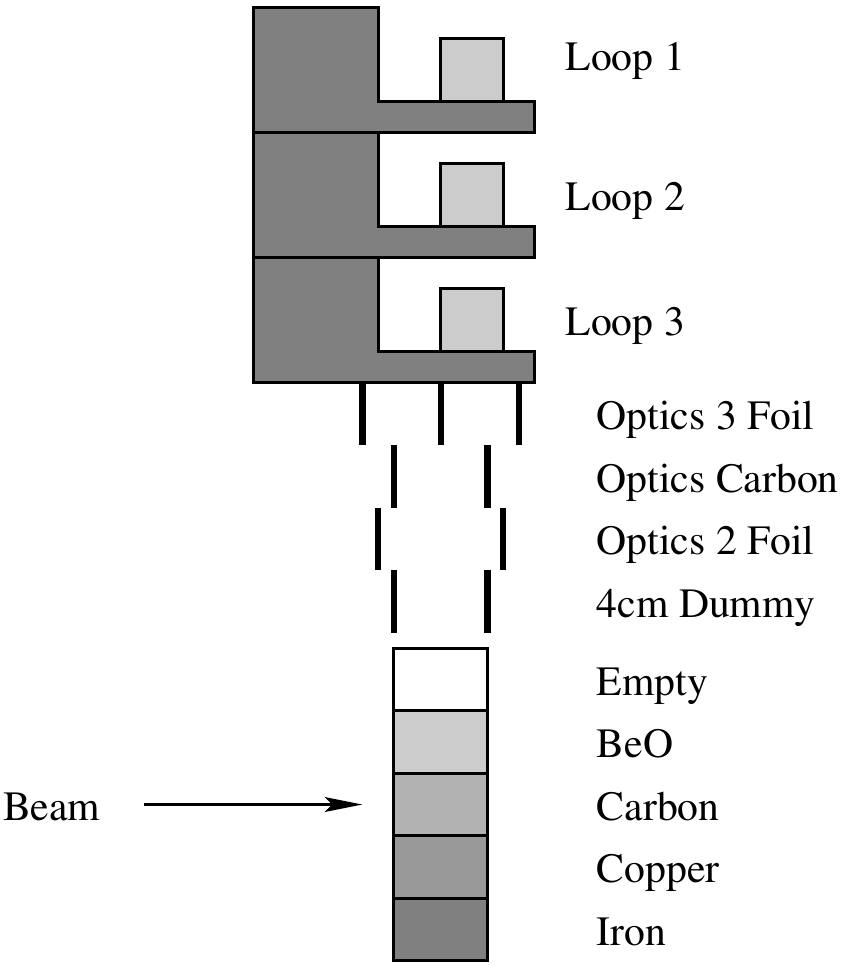,height=4.0in}
\caption{Hall C target ladder.}\label{fig:sol}
\end{center}
\end{figure}

The cryotarget consists of three loops for circulating cryogenic liquids, see Fig.~\ref{fig:sol}.

This experiment and E06-009~\cite{E06009} experiment ran simultaneously and used the same target ladder 
shown in Fig.~\ref{fig:sol}. 
The E06-009 used loop 1 filled with deuterium, while the E05-017 experiment, which also ran 
during the same time period, used loop 1 filled with liquid hydrogen. 
The liquid hydrogen (LH$_{2}$) was cooled down to 19.0 K and held at a density of 0.0723 $\pm$ 0.0004 g/cm$^{3}$. 
The liquid deuterium (LD$_{2}$) was cooled to a temperature of 22.0 K and a density 
of 0.1670 $\pm$ 0.001 g/cm$^{3}$. 
The temperature and pressure of the liquid targets were monitored by target control software which was
programmed to sound alarms when the temperature or pressure in the targets changed beyond the predetermined 
limits.

The liquid target cells are made of 0.0127 cm aluminum. The 4 cm dummy target (Al6061-T6) consists of 
two dummy endcaps to simulate an empty target for target wall background measurements.
The thicknesses and radiation lengths of the targets in the cryotarget ladder are listed
in Table~\ref{tab:tar_crio}.

\begin{table}
\begin{center} 
\begin{tabular}{|c|c|c|c|}
\hline
Foil Number   & Position &  Thickness (g/cm$^2$)          &  Radiation Length (\%)\\
\hline
Foil 1        &  Upstream foil    &  0.2658 $\pm$ 0.0035  & 1.10\\
\hline
Foil 2        &  Downstream foil  &  0.2549 $\pm$ 0.0034  & 1.06\\
\hline
\end{tabular}
\end{center}
\caption{Dummy target thickness and positions. The alloy is Al6061-T6. }\label{tab:tar_dummy}
\end{table}

\begin{table}
\begin{center} 
\begin{tabular}{|c|c|c|}
\hline
Target &Thickness (g/cm$^{2}$) & Radiation Length (\%)\\
\hline
4 cm LD$_{2}$          & 0.6570 $\pm$ 0.0039 & 0.56 \\
Cell Walls ($^{27}$Al) & 0.0340 $\pm$ 0.0035 & 0.14 \\
4 cm Dummy ($^{27}$Al) & 0.5210 $\pm$ 0.0017 & 2.11 \\
\hline
\end{tabular}
\end{center}
\caption{Thicknesses and radiation lengths of the targets in the cryotarget ladder. Cell walls 
represent entrance and exit foils.}\label{tab:tar_crio}
\end{table}

\subsection{High Momentum Spectrometer}
The High Momentum Spectrometer (HMS) was one of the two standard spectrometers in Hall C.
It was designed to have moderately large acceptance, good position and angular resolution 
in the scattering plane, an extended target acceptance, and a large angular range.
During the current experiment the spectrometer was situated on the right-hand side of 
the beam-line (with respect to incoming beam direction) and was used to detect the 
scattered electron. 

\subsubsection{HMS Optics Design}
\begin{figure}[ht]
\begin{center}
\epsfig{file=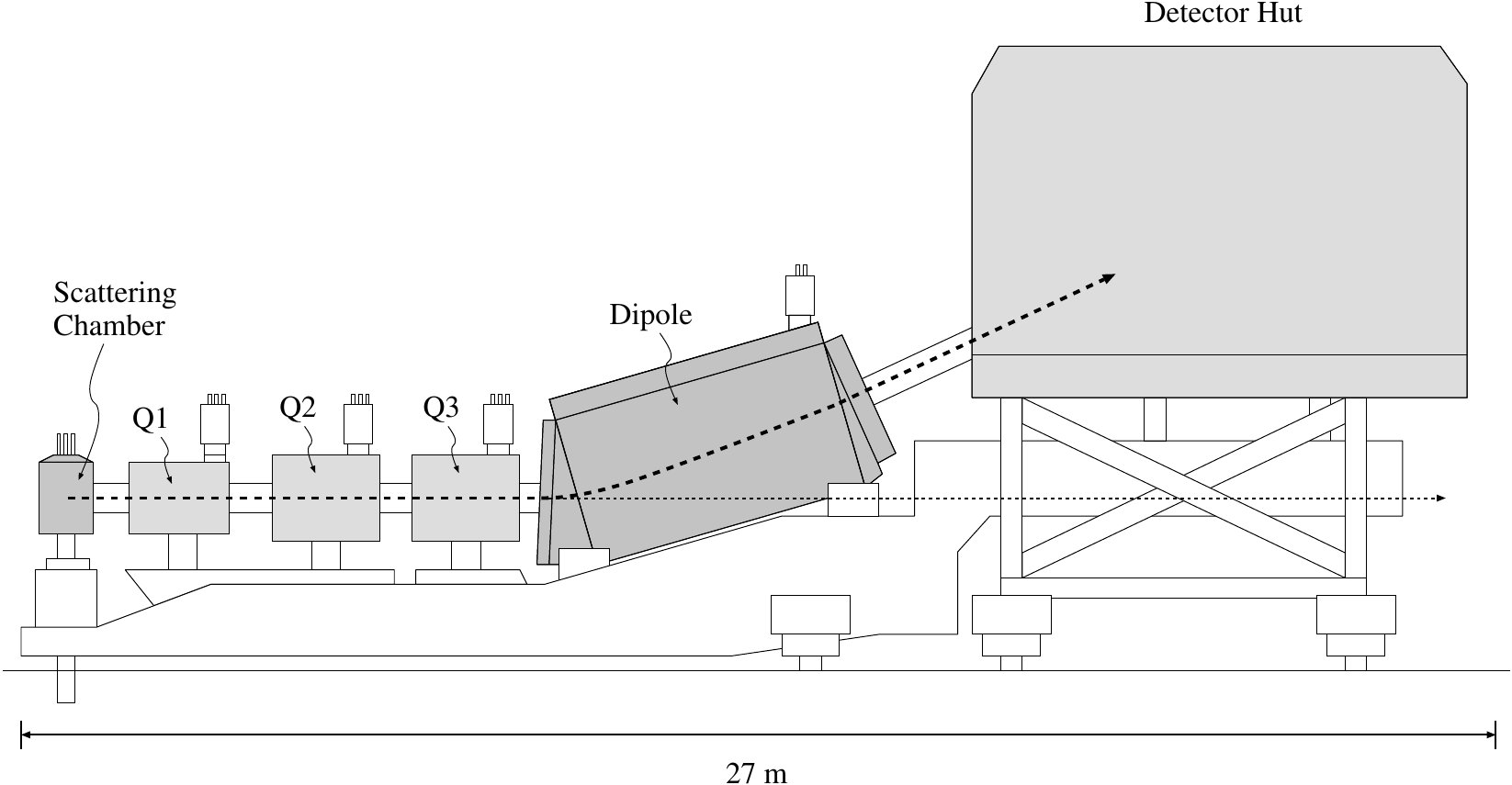,height=3.0in,angle=0.0}
\end{center}
\caption{Schematic lay-out of a HMS magnetic elements, showing the geometrical 
configuration of the three quadrupole and the dipole magnets. }
\label{fig:hmslayout}
\end{figure}

The magnetic elements of the HMS consist of three superconducting 
quadrupoles and a superconducting dipole, 
in a QQQD arrangement (Fig.~\ref {fig:hmslayout}). 
The dipole magnet is a superconducting, cryostable magnet. Its basic parameters 
are an effective length of 5.26 m, a bend radius of 12.06 m and a gap width of 42 cm. 
It was designed to achieve a 25$^{o}$ bending angle for 7.4 GeV/c momentum particles.
It provides defocusing in the vertical plane to achieve good momentum resolution.
Quadrupole Q$_{2}$ focuses in the horizontal plane, whereas Q$_{1}$ and Q$_{3}$ both 
provide vertical focusing. 
Horizontal focusing of Q$_{2}$ provides a large momentum bite, solid angle and 
extended target acceptance. The HMS characteristics are given in Table~\ref{tab:hms}.

\begin{table}
\begin{center} 
\begin{tabular}{|c|c|}
\hline
Maximum central momentum                   & 7.4 GeV/c \\
Momentum acceptance                        & $\pm$10\%    \\
Momentum resolution                        & $<$0.1\%   \\
Solid angle				   & 6.7 msr   \\
Scattering angle acceptance                & $\pm$40.0 mr \\
Out-of-plane angle acceptance              & $\pm$80.0 mr \\
Extended target acceptance                 & 10.0 cm     \\
In plane angle resolution                  & 0.4 mr    \\
Out of plane angle resolution              & 0.9 mr    \\
Useful target length                       & 10.0 cm   \\
Vertex Reconstruction Accuracy             & 2.0 mm     \\

\hline
\end{tabular}
\end{center}
\caption{HMS performance characteristics.}\label{tab:hms}
\end{table}
\subsection{HMS Detector Package}
\begin{figure}[h!]
\begin{center}
\epsfig{file=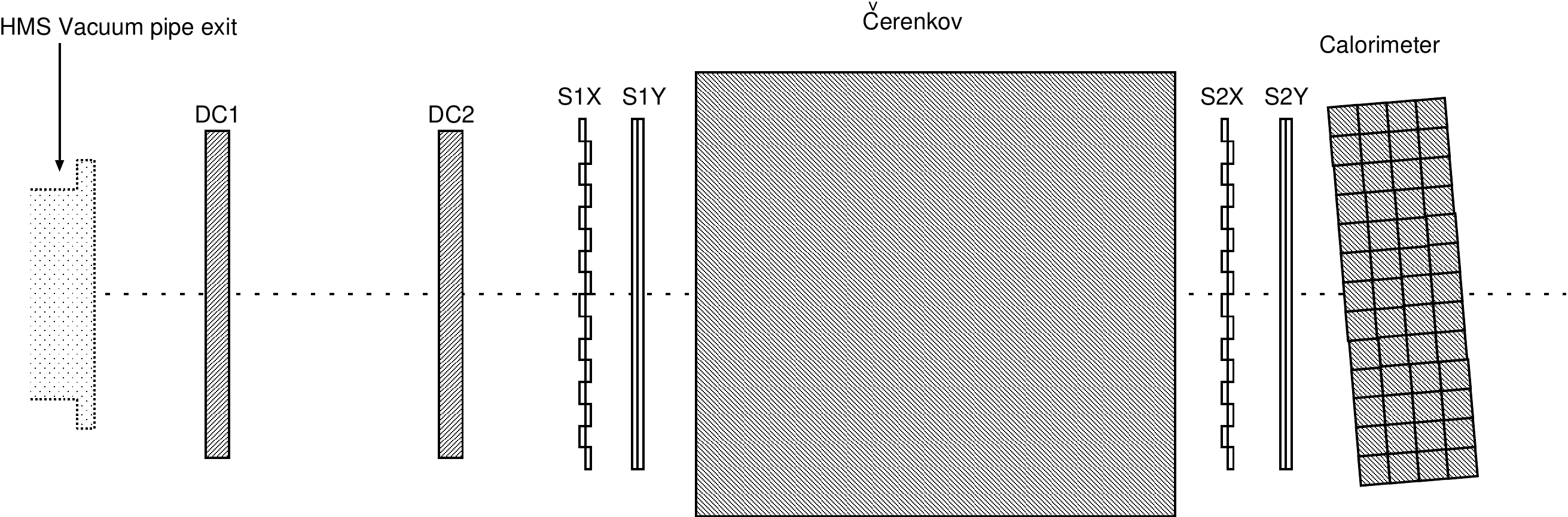,width=5cm,height=14cm,angle=-90}
\caption{Schematic side view of the HMS detector package.}
\label{fig:det}
\end{center}
\end{figure}

The detector package of the HMS consists of two drift chambers, DC1 and DC2, two pairs
of scintillator hodoscopes, S1 and S2, a gas \u{C}erenkov detector and 
a lead-glass calorimeter. The detectors are mounted on frames that connect to the carriage 
that supports the magnets. This design insures that the detector package and magnets 
stay unmoved relative to each other. The detector package has the following functions: 
triggering, tracking and particle identification.
A schematic view of the HMS detector package is shown in Fig.~\ref{fig:det}.
\subsubsection{Drift Chambers}
A drift chamber is a particle tracking detector that measures the drift time 
of ionization particles in a gas to calculate the hit coordinates of
ionizing particle. Combined with knowledge of the optical transfer properties of the 
spectrometer, the drift chamber hit coordinates are used to reconstruct the particle 
trajectory, reaction vertex and momentum at the target. 

\begin{figure}[t]
\begin{center}
\epsfig{file=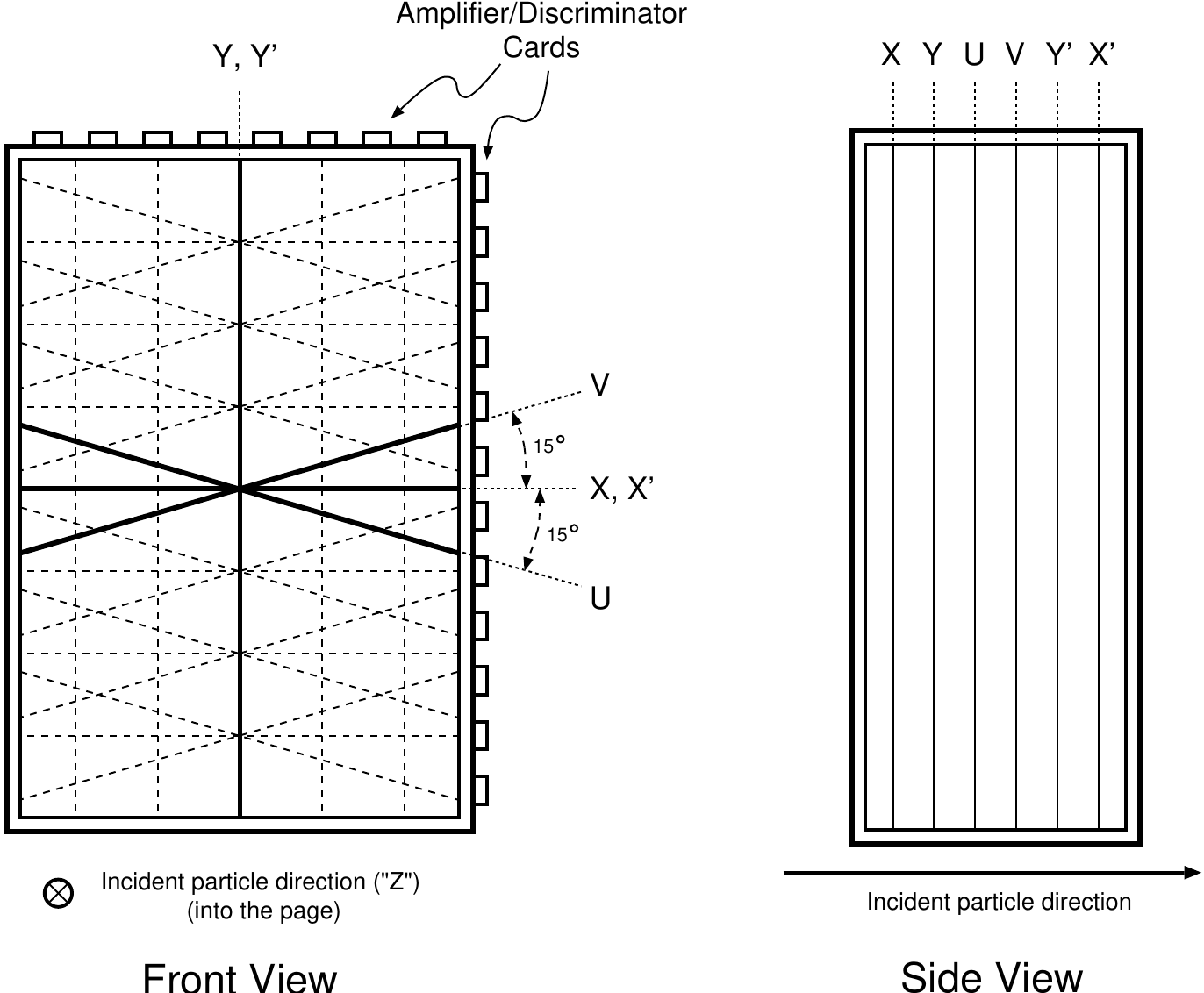,height=3.5in}
\end{center}
\caption{ Drift chamber}
\label{fig:dc}
\end{figure}

The drift chambers are spaced 81.2 cm apart, and each has an active area of about 
113 cm ($x$) by 52 cm ($y$). 
They consist of six separate planes of sense wires (anodes) of 
25 $\mu$m diameter gold-plated tungsten spaced 1.0 cm apart in a gas mixture of argon and ethane,
and field wires (cathodes) of 150 $\mu$m gold-plated copper beryllium wires.
The planes are spaced 1.8 cm apart. Between each sense wire is a field wire which is held 
at a negative potential (-1800 V to -2500 V). On both sides of the sense planes is an additional 
plane of field wires, held at the same negative potential. 
The planes are ordered X, Y, U, V, Y$^{'}$, X$^{'}$ as seen by incoming particles. 
The X and X$^{'}$ planes provide two measurements of position of the particles in the 
dispersive direction. The Y and Y$^{'}$ planes offer two measurements of  
position in the transverse direction  and the U and V are rotated $\pm$15$^{\circ}$ 
from the X (X$^{'}$) planes to prevent the right-left ambiguity, see Fig.~\ref{fig:dc}. 

When a charged particle passes through one of the chambers, it ionizes
the gas atoms and produces a trail of electrons and ions. 
Far from a sense wire, where the electric field between wire and 
cathode is uniform and parallel, electrons drift toward the wire with a 
constant velocity. 
As they get closer, the field has a stronger radial nature, causing 
the electrons to accelerate and produce a secondary electron avalanche. 
This avalanche generates a negative pulse on the wire. 
The signals from each wire are amplified and discriminated on the 
cards attached directly to the drift chambers
and then sent to the TDCs (Time-to-Digital Converters) located in the 
back of the detector hut. 
The measured drift time and known electron drift velocity can 
then be used to calculate the perpendicular distance between wire 
and particle track. A typical track generates signals in about five 
wires per plane. The position resolution at the focal plane is approximately 
280 $\mu$m per plane.
More information about the HMS drift chambers can be found in Ref.~\cite{Baker:1995ky}.

\subsubsection{Hodoscopes}

\begin{figure}[h!]
\begin{center}
\epsfig{file=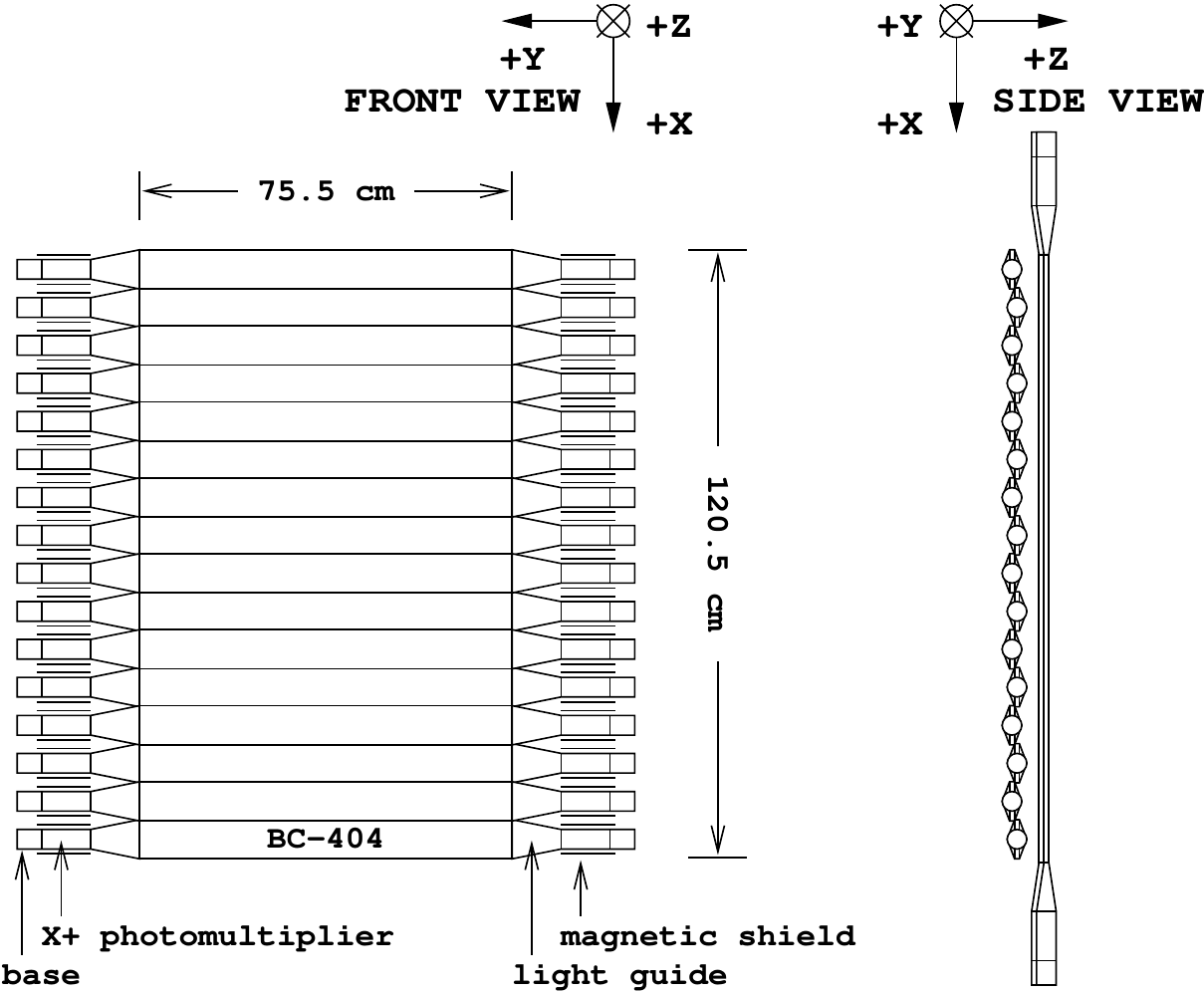,height=3.5in}
\caption{Hodoscope geometry.}\label{fig:hod}
\end{center}
\end{figure}

The HMS has 4 hodoscopes, which provide the trigger for detector read-out and
allows the identification of heavy particles through time-of-flight (TOF), 
though this was not used in the present experiment. 
The hodoscopes are paired in two horizontal-vertical X-Y sets (S1X, S1Y and S2X, S2Y).
The sets are separated by 220 cm. Each X hodoscope consists of 16 horizontally oriented 
scintillators (paddles), while the Y hodoscopes consist of 10 vertically oriented 
scintillators. The paddles of the X and Y hodoscopes are all 1 cm thick and 8 cm wide, 
but they have different lengths. The X paddles are 75.5 cm long, and the Y paddles 
are 120.5 cm long. Each scintillator paddle is wrapped in one layer of light-tight 
aluminum foil and two layers of Tedlar. At both ends of each paddle are Photomultiplier Tubes 
(PMTs) attached to the paddle through lucite light guides.  
The paddles are staggered in the beam direction with 0.5 cm overlap between the paddles to avoid 
gaps, see Fig.~\ref{fig:hod}.
When charged particles pass through the paddles they excite the atoms of the scintillators.
These atoms emit light as they return to their ground state.
The light is detected by PMTs at the ends of the paddles.
The light that is not emitted along the length of the paddle is reflected internally
through the scintillator and ultimately also detected by the PMTs.    

The signal pulses from the PMTs are sent to the counting house where they run through
the splitter, giving two signals with 1/3 and 2/3 of the amplitude of the original input 
signal. The smaller signal goes to Analog-to-Digital Converters (ADCs) that measure the 
integral of the signal. The other part of the signal is discriminated and one set
of outputs is sent to TDCs (for timing information). The other set of outputs is 
sent into a logic module.  
The logic module first generates the logical OR of all the discriminated signals
from the tubes on one side of a given plane, for example: S1X$+$ $\equiv$ (S1X1$+$ OR S1X2$+$ OR ... S1X16$+$). 
There are equivalent sets of signals for the $-$ side of each plane.
Then, these sets of signals are combined into six outputs:

a) S1X $\equiv$ (S1X$+$ AND S1X$-$) and analogously for S1Y, S2X, S2Y. These four output logic signals indicate
which of the hodoscope planes are active and make a new logic signal in case at least three of them have fired 
which is called SCIN.

b) The X$-$Y pairs are further combined to form
S1 $\equiv$ (S1X OR S1Y), and S2 $\equiv$ (S2X OR S2Y). These signals indicate whether the pairs are active
and make a logic signal called STOF.

These logical outputs are then sent to the main trigger logic
and to the scalers to be recorded. 
A more detailed description of the hodoscopes can be found in Ref.~\cite{Arring}.

\subsubsection{Gas \v{C}erenkov Detector} \label{subsubsection:Cerenkov}
\begin{figure}[h!]
\begin{center}
\epsfig{file=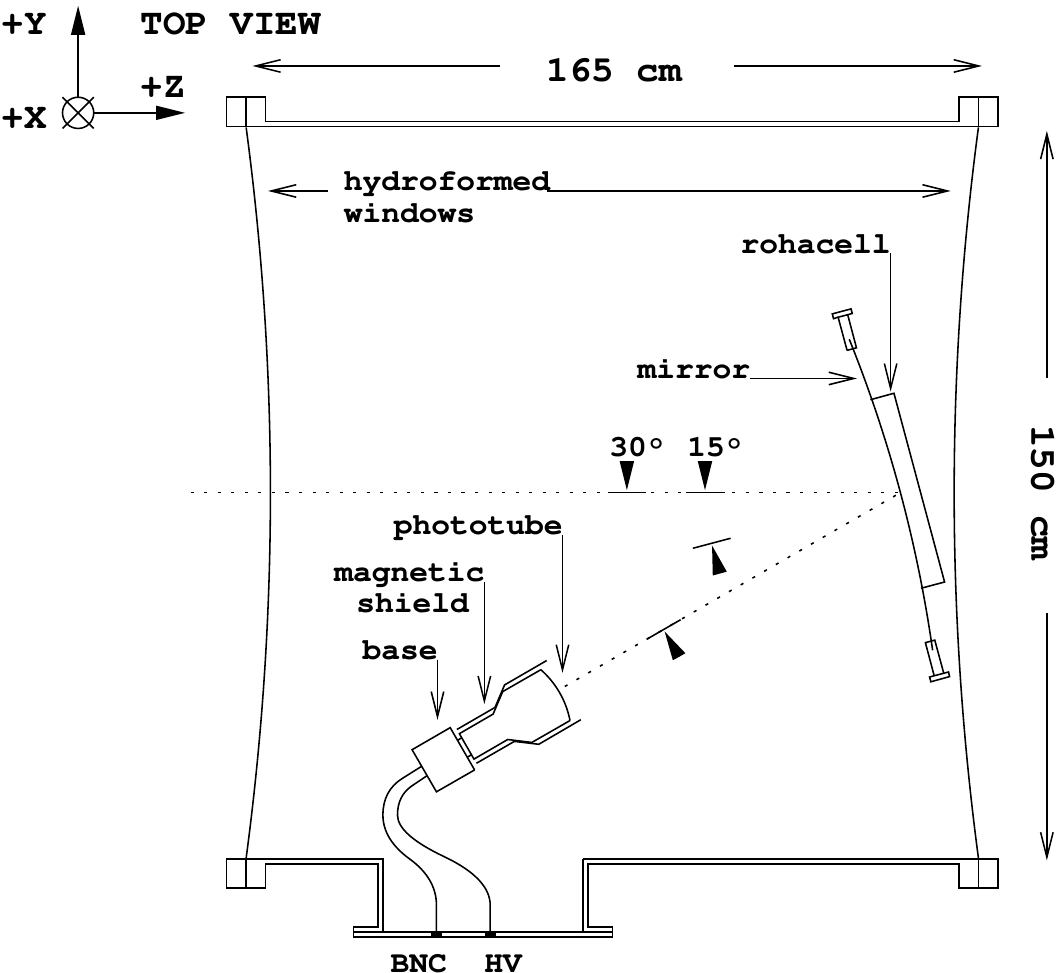,height=3.6in}
\caption{\u{C}erenkov geometry.}
\label{fig:cer}
\end{center}
\end{figure}

\begin{figure}[!ht]
\begin{center}
\epsfig{file=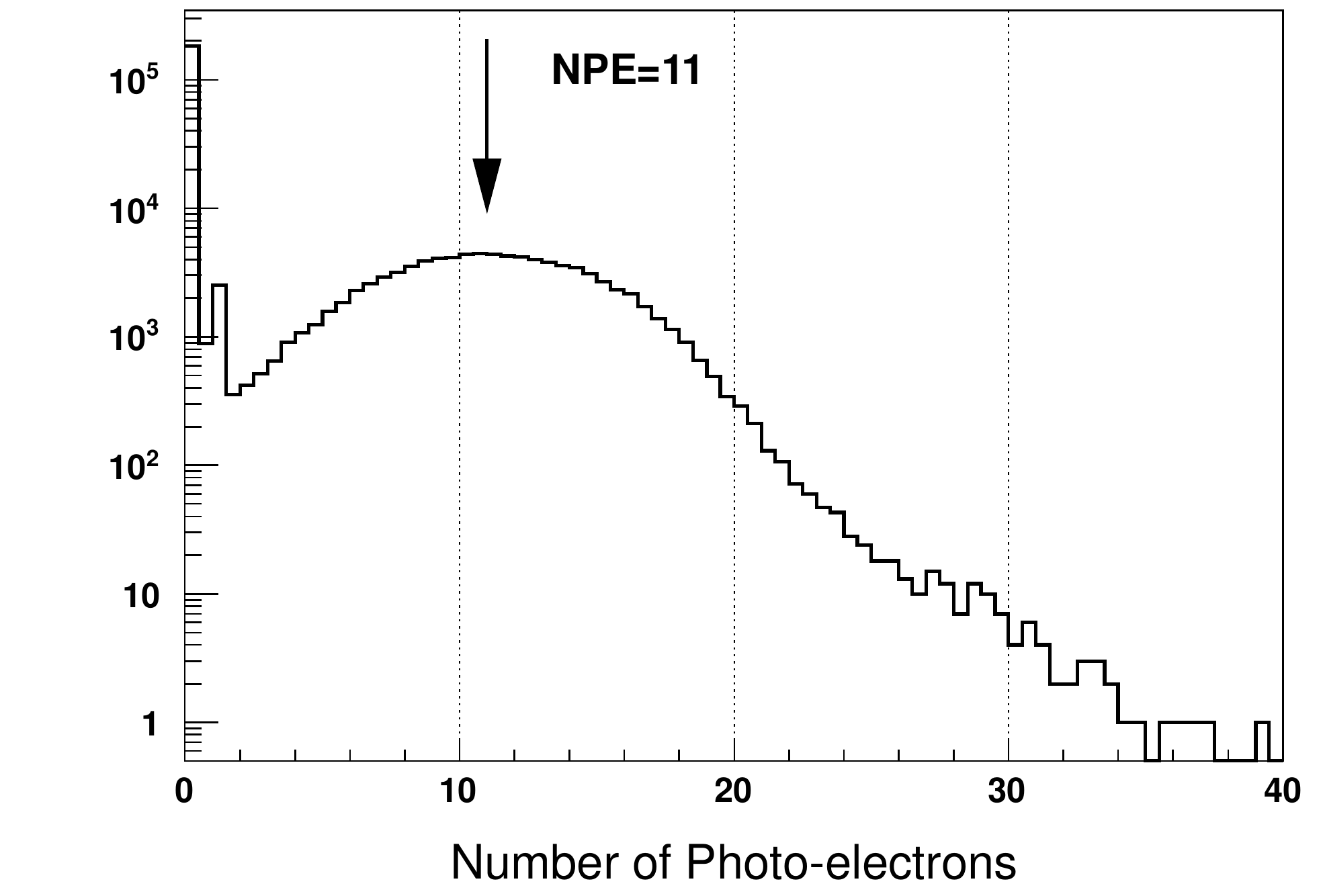,height=3.5in}
\caption{HMS \u{C}erenkov spectrum.
Most of the pions appear at zero photoelectrons.}\label{fig:cerenkov}
\end{center}
\end{figure}

The HMS gas \u{C}erenkov detector provides particle identification
by operating as a threshold detector.
It consists of a large cylindrical tank (diameter of 150 cm, length of 165 cm)
situated in the middle of the detector stack between the hodoscope pairs S1 and 
S2, see Fig.~\ref{fig:det}.
A pair of front reflecting spherical mirrors mounted vertically with 1 cm overlap at the rear 
of the detector is rotated over 15 degrees to focus the light on a pair of PMTs, see Fig.~\ref{fig:cer}.
The detector is filled with C$_{4}$F$_{10}$ gas.
The \u{C}erenkov detector measures the light emitted when a charged particle travels 
through the gas with a velocity above the speed of light in the gas. 
This is known as \u{C}erenkov radiation. The light will be emitted with an angle
$\rm{cos}(\theta) = 1/n\beta$, where $\beta$ is the velocity of the particle relative to the speed of light and
$n$ is the index of refraction of  the material. If $n\beta < 1$ no light will be emitted. 
The light is reflected from focusing mirrors to PMTs, which generate a signal
proportional to the number of \u{C}erenkov photons. 
The gas (and here the index of refraction) is chosen  
such that electrons at the spectrometer momentum will emit \u{C}erenkov radiation and pions will not.  
The refractive index of the $C_{4}F_{10}$ gas at 1 atm is n = 1.0006 and this gives a pion threshold above 4 GeV/c
and an electron threshold of about 15 MeV/c. The average measured signal from an electron is
about 10 photoelectrons, see Fig.~\ref{fig:cerenkov}. However, it is still possible for a pion to be misidentified 
as an electron when it produces a knock-on $\delta$-electron that fires the \u{C}erenkov detector.

The signal from each PMT is sent to the counting house where each signal is split in a fashion similar
to the hodoscope signals. One pair of signals is sent to the ADC and the other pair is summed
and put through the discriminator to give signals for the TDC and trigger logic.
More information about the \u{C}erenkov detector can be found in Ref.~\cite{Arring}.

\subsubsection{Electromagnetic Calorimeter} \label{subsubsection:Calo}
\begin{figure}[h!]
\begin{center}
\epsfig{file=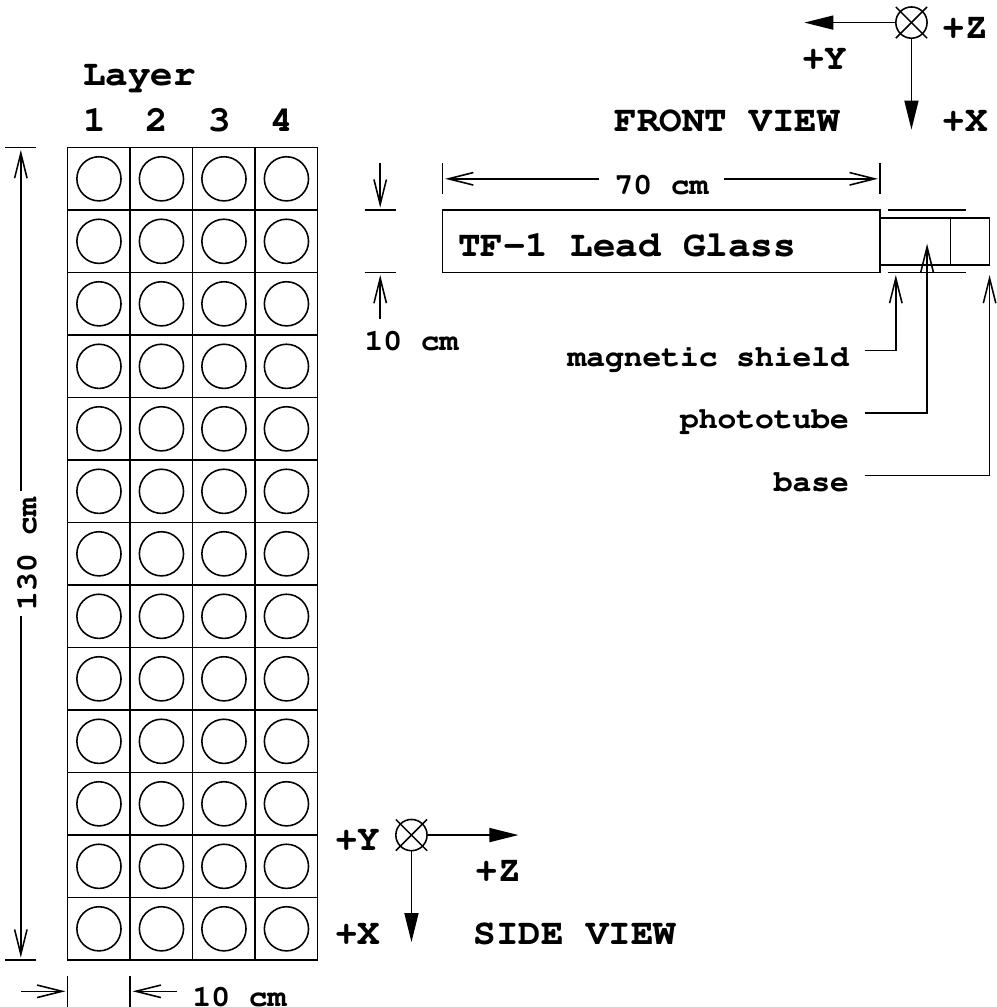,height=3.0in}
\caption{Calorimeter geometry.}\label{fig:calo}
\end{center}
\end{figure}

\begin{figure}[h!]
\begin{center}
\epsfig{file=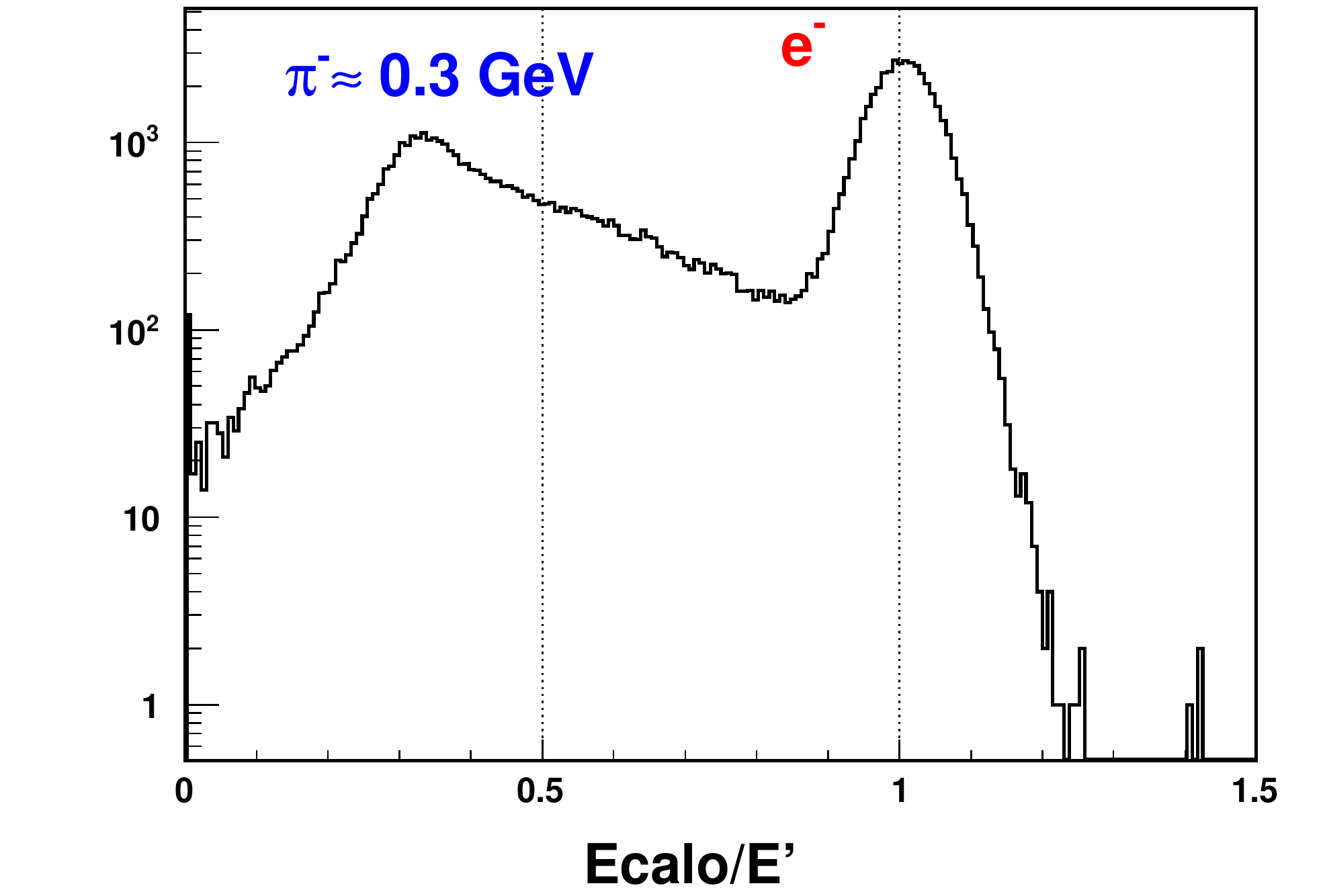,height=3.0in}
\caption{Calorimeter energy distribution divided HMS momentum (\mbox{$P_{HMS}$=1.0 GeV/c}). }\label{fig:calo_spec} 
\end{center}
\end{figure}

The HMS has a lead-glass calorimeter which is used to discriminate between electrons and 
pions. Fig.~\ref{fig:calo} depicts the calorimeter which consists of 52 TF1 type lead glass blocks 
10 cm $\times$ 10 cm $\times$ 70 cm, with a PMT on one end. 
The blocks are arranged in four layers with 13 blocks per layer, giving a total thickness 
of the calorimeter along the direction of particle motion of 16 radiation lengths. 
As shown in Fig.~\ref{fig:det} the calorimeter is rotated 5$^{\circ}$ with respect to 
the dispersive plane to prevent particles from passing between the blocks. 
Electrons interacting with the lead glass radiate photons in the calorimeter, 
which in turn produce electron-positron pairs (when the photons are energetic enough). 
These pairs in turn also radiate photons and so a shower of particles $(e^{+},e^{-},\gamma)$ is 
produced in the calorimeter. 
The charged particles produce \u{C}erenkov
radiation which is detected by photomultiplier tubes. The produced signal 
is proportional to the total track length of the particles in the 
calorimeter which is in turn proportional to the energy of the 
initial electron.  
Electrons (positrons) entering the calorimeter deposit their entire 
energy, and thus the ratio of deposited energy of electrons (positrons) in the 
calorimeter to the detected energy derived from the particle bending in the spectrometer 
is unity, see Fig.~\ref{fig:calo_spec}.

Pions normally deposit about 300 MeV through ionization in the calorimeter so 
a peak in the $E_{calo}/E^{'}$ spectrum can be observed at 0.3 GeV/$E'$. 
However, pions can have a charge-exchange reaction and produce a neutral pion,
which in turn decays into two photons, the full energy of which will be deposited 
in the calorimeter. This leads to a high-energy tail for pions, 
which can result in pion misidentification.
   
The signals from the calorimeter PMTs are sent to the counting house where they are split 50/50. 
One half is sent to the ADC and the other half to the linear modules to be summed. 
The sum in the first layer (PRSUM) and the sum of the entire calorimeter (SHSUM) are 
discriminated to give three logic signals for the trigger. 
PRSUM is the sum of all signals from the first layer of the calorimeter.  
The SHSUM signal, obtained by summing the signals from all lead-glass blocks, represents the
total energy deposited in the calorimeter. 
PRSUM are used to form the high (PRHI) and low (PRLO) thresholds on the energy in the 
the first layer of calorimeter. 
SHLO is a cut on the total energy in the calorimeter.
\subsection{Data Acquisition System}
The standard Hall C DAQ system consists of a variety of CAMAC, NIM, 
Fastbus and VME electronics, as well as modules custom built by the 
JLab electronics group. 

The data acquisition is handled by the CODA (CEBAF Online Data Acquisition)~\cite{coda} 
software package running on a PC. 
Data for each run are written directly to rotating memory and consists of three types of events:
1) detector information handled by the ADCs and TDCs, 2) scaler information, 
and 3) information from the EPICS~\cite{Dalesio:1994qp} database.
The ADCs and TDCs are read-out for each 
event, while the scalers are read every 2 seconds. 
The EPICS database contains information such as magnet settings, beam position and target 
temperature and pressure. These quantities are read out every 30 seconds. 
More information about data acquisition can be found in Ref.~\cite{coda}.

\subsubsection{Triggers}
Triggers generated by the PMT signals from the scintillator planes and from calorimeter
blocks are sent to DAQ.

The physical information of interest is ultimately obtained also from raw data (digital numbers)
read out from hundreds of ADC and TDC channels.
\begin{figure}[ht]
\begin{center}
\epsfig{file=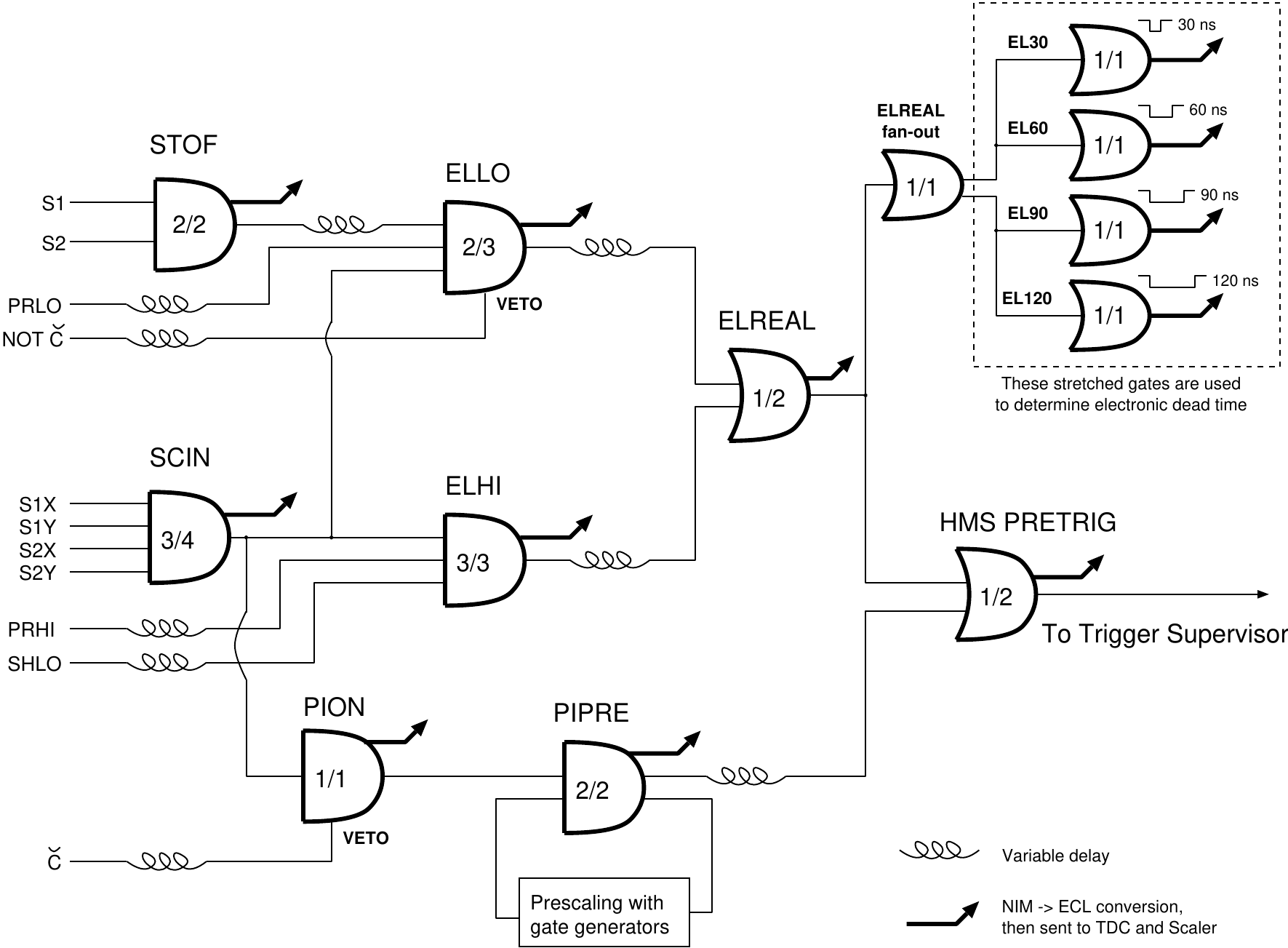,height=4.0in,angle=0.0}
\end{center}
\caption{Schematic diagram of the HMS trigger logic.}
\label{fig:rcstrg}
\end{figure}

Charged particles passing through the spectrometer produce triggers in one or more of the detectors 
described earlier. 
Certain combinations of these triggers are used to form the pretrigger. 
An electron trigger (ELREAL) can be produced in two ways:

1) The low $-$ level electron trigger (ELLO) requires a \u{C}erenkov signal (\u{C}), plus at least two 
out of three of the following conditions:
a) at least one of the two scintillator layers of each hodoscopes has fired (STOF $\equiv$ S1 AND S2).
b) at least three of the four scintillator layers of both hodoscopes have fired (SCIN).
c) There is a (PRLO) signal from calorimeter. 

2) A high $-$ level electron trigger (ELHI) requires that all of the following signals are present:
a) The (SCIN) signal. 
b) The (PRHI) signal from the calorimeter.
c) The (SHLO) signal from the calorimeter. 
The high $-$ level electron trigger (ELHI) does not use the \u{C}erenkov signal.

The electron trigger (ELREAL) is (ELLO) OR (ELHI).
The reason for using two electron triggers (ELLO) and (ELHI) is to reduce the trigger inefficiency and 
to provide the most efficient electron-hadron separation by the \u{C}erenkov and the calorimeter detectors.

There is an additional pion trigger (PION) which requires the (SCIN) signal, and no 
\u{C}erenkov signal ($\bar{\hat{C}}$). The (PION) signal is then combined with a prescaling circuit 
to form (PIPRE), which ensures that a low-rate sample of pions is sent along to the data acquisition system.
Two copies of (ELREAL) are generated: one is fanned out into four
logic units with dead times set between 30 and 120 ns 
in order to allow a measurement of the electronic dead time
and another (ELREAL) OR (PIPRE) forms the signal (PRETRIG), which is forwarded to the 
Trigger Supervisor (TS)~\cite{tsup}. 
 
When a run is started first 1000 pedestal triggers are generated by (PED PRETRIG) 
and the data acquisition system records the read-outs from the ADCs. 
Then the data acquisition system begins to record physics events. 
When a physics event is signaled by (PRETRIG),
the data acquisition system records the read-outs of the detectors ADCs and TDCs.  
The TS sets TS BUSY while the data acquisition system records the read-outs. After the data 
acquisition system has completed recording the event information, TS BUSY is set off 
to allow for the next physics event. Both trigger and pretrigger signals are written 
into scalers providing information about the computer dead time. 
   

\chap3{Data Analysis}
The goal of data analysis is to perform and apply detector calibrations and extract cross sections 
from experimental data.
In this chapter the calibration of detectors, along with background subtraction and cross section extraction methods are described.
\subsection{BCM Calibrations} \label{sec:bcm}

\begin{figure}[ht]
\begin{center}
\epsfig{file=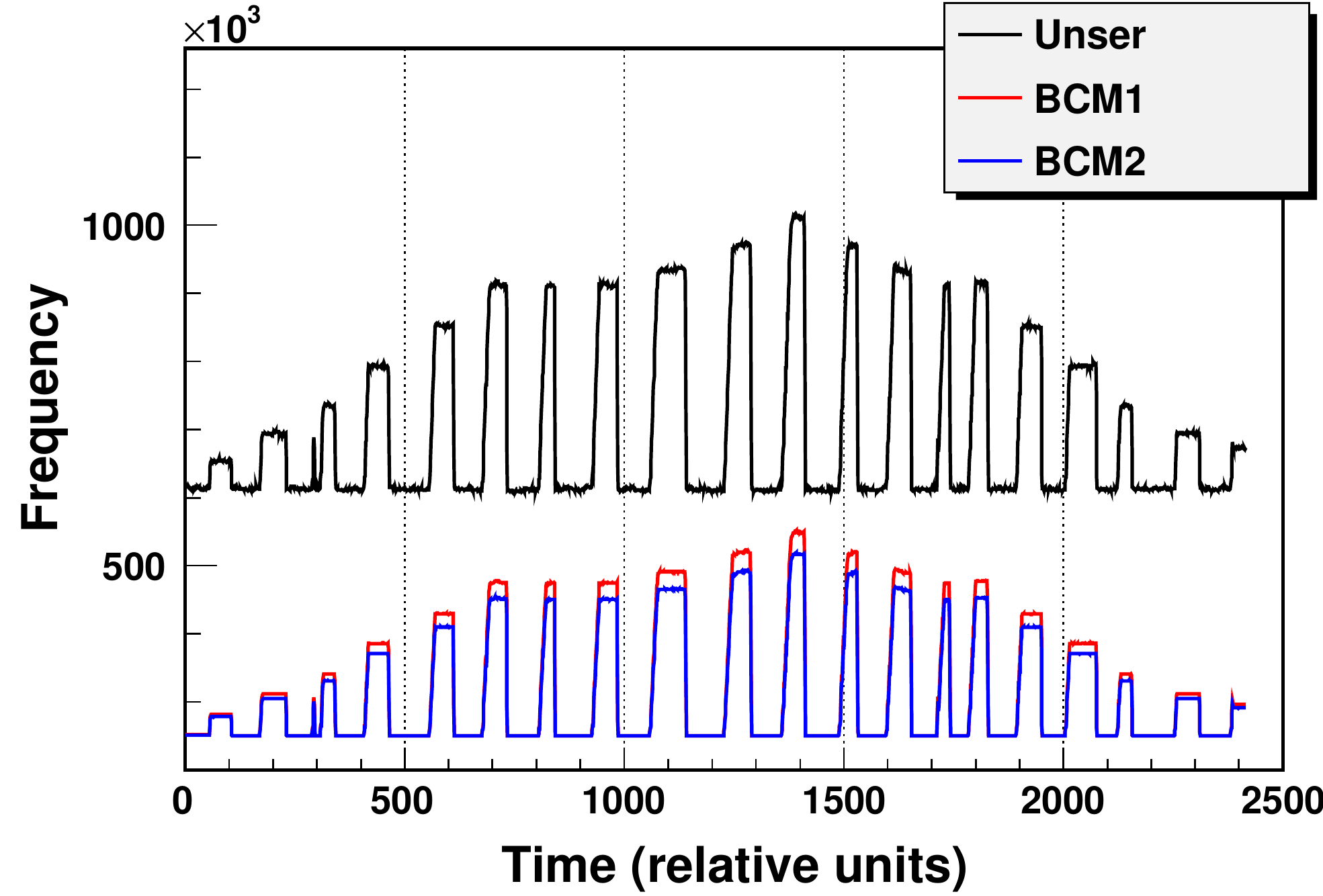,height=3.0in}
\end{center}
\caption { Unser and BCM frequency versus time. }
\label{fig:unser}
\end{figure}

The BCMs measure the integrated beam current in two second intervals, an example is shown in Fig.~\ref{fig:unser}.
The Unser monitor is used to calibrate the BCMs since it has very well measured and stable gain.
Despite its stable gain, the Unser's offset is very unstable and can change depending on temperature 
and other physical conditions. This is the reason it is not used for an absolute beam current measurement.
The BCMs have very stable offsets but their gains can change over time and need to be calibrated a few times
during this experiment. 

During a BCM calibration run the accelerator delivers beam to the hall in current steps lasting 
2 minutes, first increasing up to the current at which data are taken, 
then decreasing, with equal time intervals of no beam between the steps.
\begin{figure}[ht]
\begin{center}
\epsfig{file=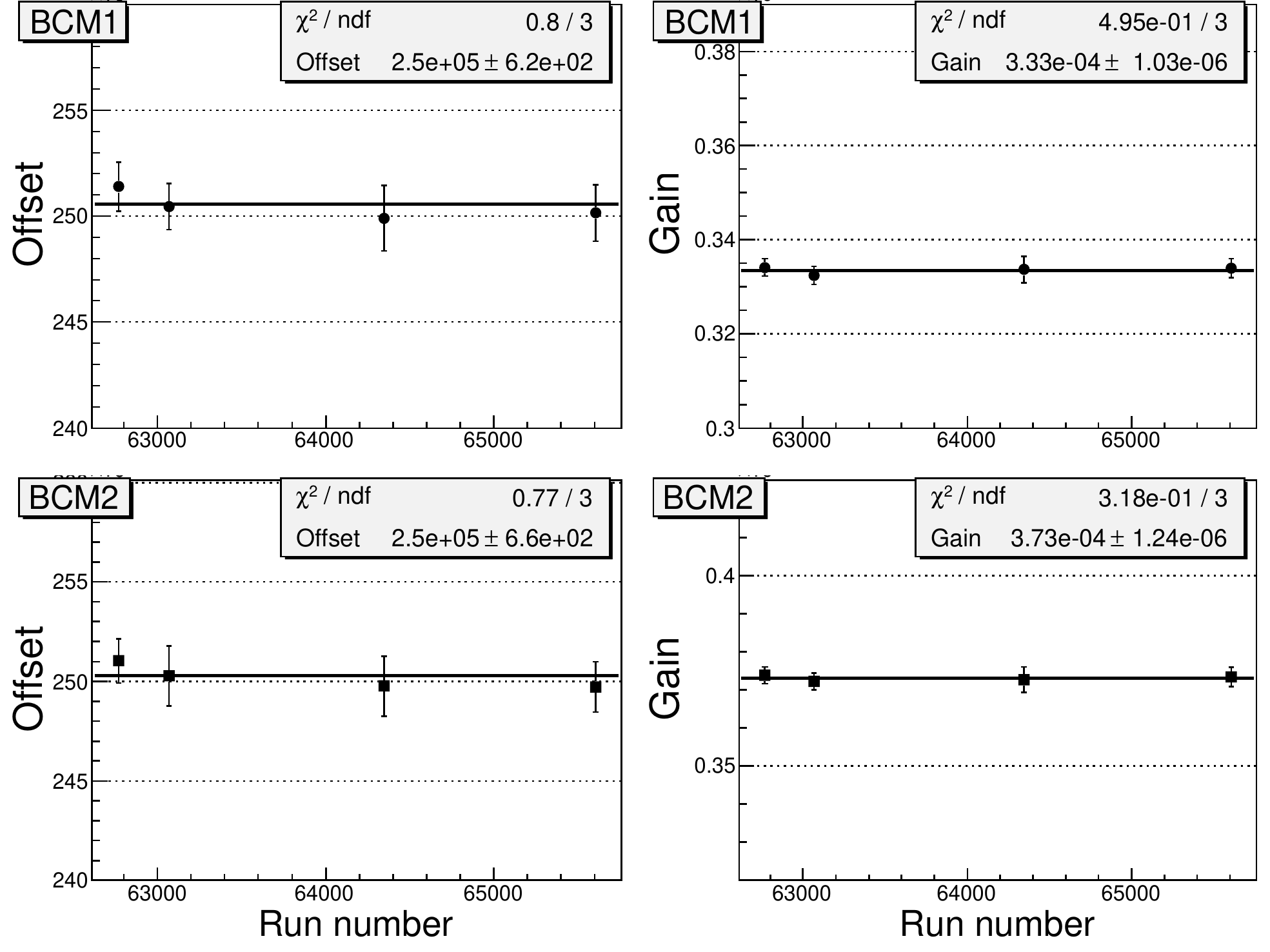,height=4.0in}
\end{center}
\caption { BCM calibration results for the entire period of the experiment. The time interval between first and last
measurement is two months. This shows that BCM gains and offsets were very stable.}
\label{fig:bcmcalib}
\end{figure}
During the beam off intervals the zero offsets of the Unser monitor are measured by calculating the average 
frequency of the Unser before beam was on and after beam was off (before 2 minutes interval and after it). 
This allows a very accurate determination of the Unser monitor's offset during the 2 minutes beam on interval.
The beam current for each beam on period is calculated using the measured Unser offset and gain.
Plotting the BCMs average frequency, during the beam on period, versus the Unser current and doing a linear fit
one can extract the BCMs' gains and offsets. These gains and offsets are used in the replay engine to calculate
the beam current.

\begin{figure}[ht]
\begin{center}
\epsfig{file=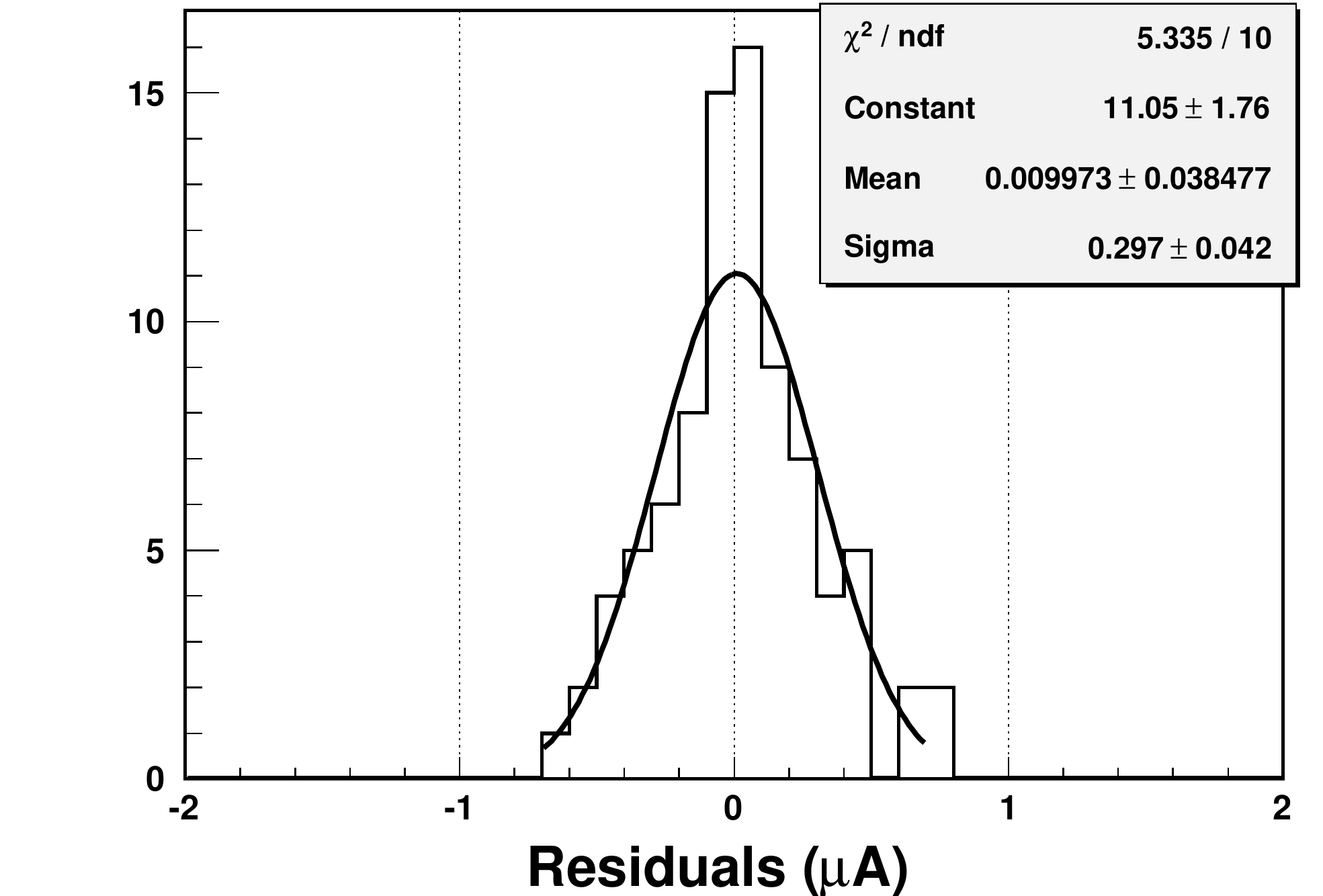,height=3.0in}
\end{center}
\caption { BCM calibration residuals. }
\label{fig:bcmresid}
\end{figure}

In the E04-001 experiment the beam current for the iron target was limited to maximum of 40~$\mu$A, 
while for carbon, aluminum and copper up to 80~$\mu$A. 
For production runs the beam current was always higher than 35~$\mu$A to minimize the uncertainty in the charge.

The systematic error from the calibration procedure, due to noise of the Unser monitor signal and 
uncertainty of the gain of the power meter signal used to measure the beam current signal from the BCM, 
yields on overall absolute systematic uncertainty on the current measurement of 0.3~$\mu$A.
This was estimated by analyzing carbon runs taken with same kinematics but with beam currents 
ranging from 10~$\mu$A to 100~$\mu$A. 
All runs were corrected by a factor $1.0/(1+0.3/I)$ to take account this offset.

\subsection{Detector Calibrations}
For all previous Hall C experiments, the software package HALL C Engine has been 
used for data analysis. 
It is a code based on GNU FORTRAN 77, which utilizes many of the data 
manipulation and display tools that are part of the widely-used CERNLIB package.  
It involves techniques for the HMS data analysis which, as a result of 
copious use in recent years, are stable and well-tested. 
The principal element is drift chamber (DC) tracking and the closely related magnetic 
reconstruction of the scattered electron momentum, direction and its reaction 
vertex at the target. 
In this chapter calibration of the DC, the \u{C}erenkov, and the Calorimeter will be 
discussed.
\subsubsection{Drift Chamber Calibration}

Drift chambers provide the tracking information, coordinates and angles, 
for particles entering the HMS. 
The time difference between the fast START signal from the hodoscope counter and 
the STOP signal from the drift chamber is used to calculate the track position.
In order to avoid negative drift times the overall offset between the 
times measured by the drift chamber and the times measured by the hodoscope 
is removed by doing hodoscope timing calibration. 
The time taken for electrons produced by ionization along a electron track to 
reach the anode wire is measured by TDC's and  converted into a perpendicular 
distance from the wire. 
Since the hit time in a particular TDC depends on specific cable lengths 
and signal processing times, that may differ from channel to channel, 
a reference time ($t_{\circ}$) is found for each wire. 
\begin{figure}[th]
\begin{center}
\epsfig{file=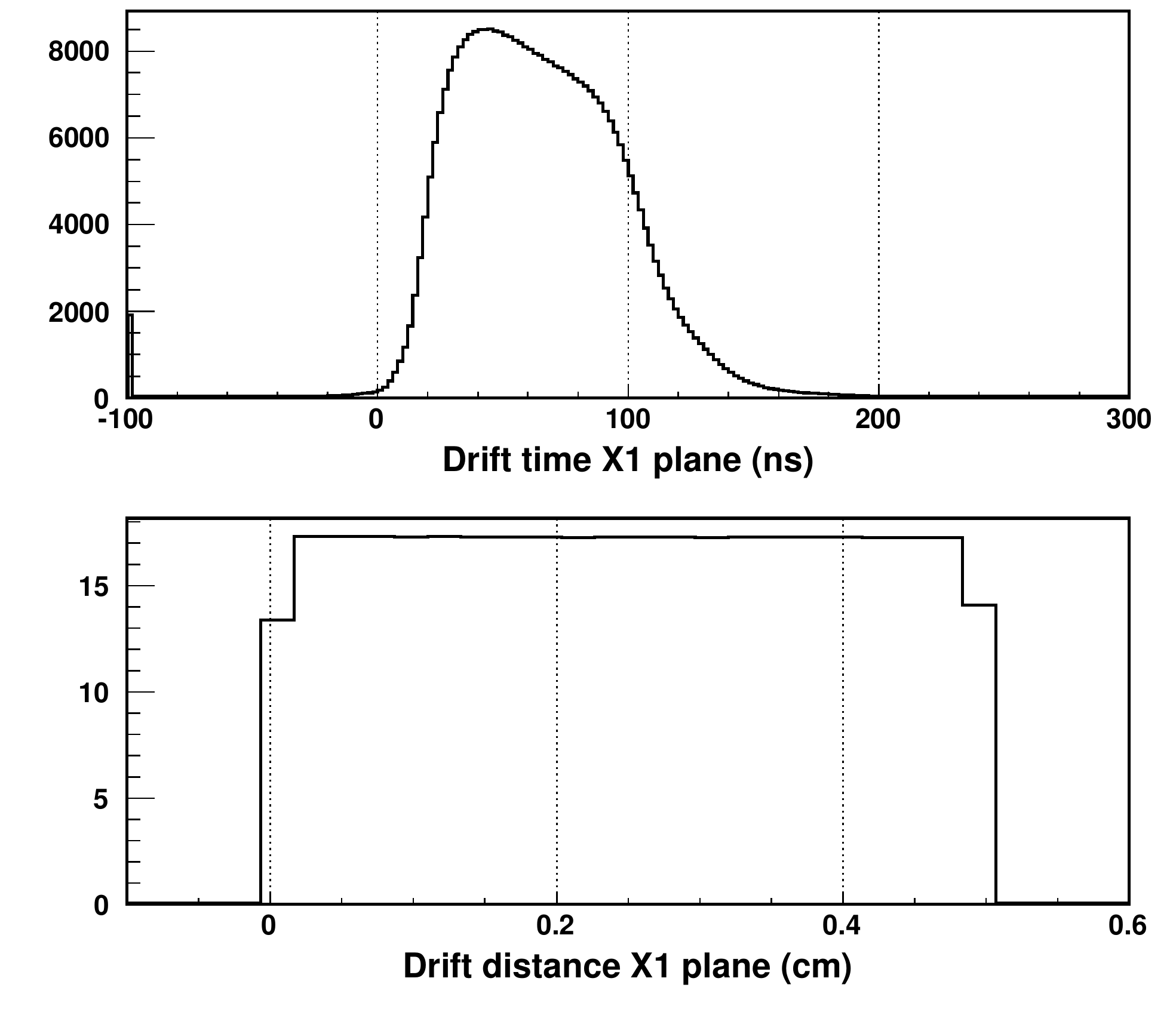,width=5.0in}
\end{center}
\caption{ Drift time and distance distributions for the HMS drift chamber plane X1. 
These distributions are the sum of all production runs.}
\label{fig:dc_calib}
\end{figure}
This is used to extract the drift time from the actual TDC time. 
In order to determine how far the track was from the wire we generated a 
time-to-distance map using the following procedure. 
About 200000 events are analyzed to obtain the drift time spectrum. 
Assuming that hits are distributed evenly around the sense wire one can obtain the
drift time by the following formula:
\begin{equation} \label{eq:DriftTime}
D(t)=D_{o}\frac{\displaystyle\int_{t_{_{min}}}^{t}F(\tau)d\tau}{\displaystyle\int_{t_{_{min}}}^{t_{max}}F(\tau)d\tau}
\end{equation}
where $D$ is the calculated distance from the wire, $D_{o}$ is the drift cell size (5 mm), $T$ is the
TDC time, $t_{min}$,  $t_{max}$ are time limits corresponding START to STOP signals (250 ns), 
$\displaystyle F(t)$ is the measured drift time distribution. 

At high rates there are many accidental events which can force the tracking 
algorithm to fail and therefore introduce a tracking inefficiency. 
In order to reduce the number of these accidental events a narrow  
HMS drift chamber TDC time window from 2400 to 2900 channels 
(before it was 1800-3300 channels, 1 channel 0.5 ns) was used. 

\subsubsection{\u{C}erenkov Calibration}
The \u{C}erenkov calibration is accomplished by finding the minimum ADC signal corresponding to 
the production of one photoelectron in the PMT. 
\begin{figure}[ht]
\begin{center}
\epsfig{file=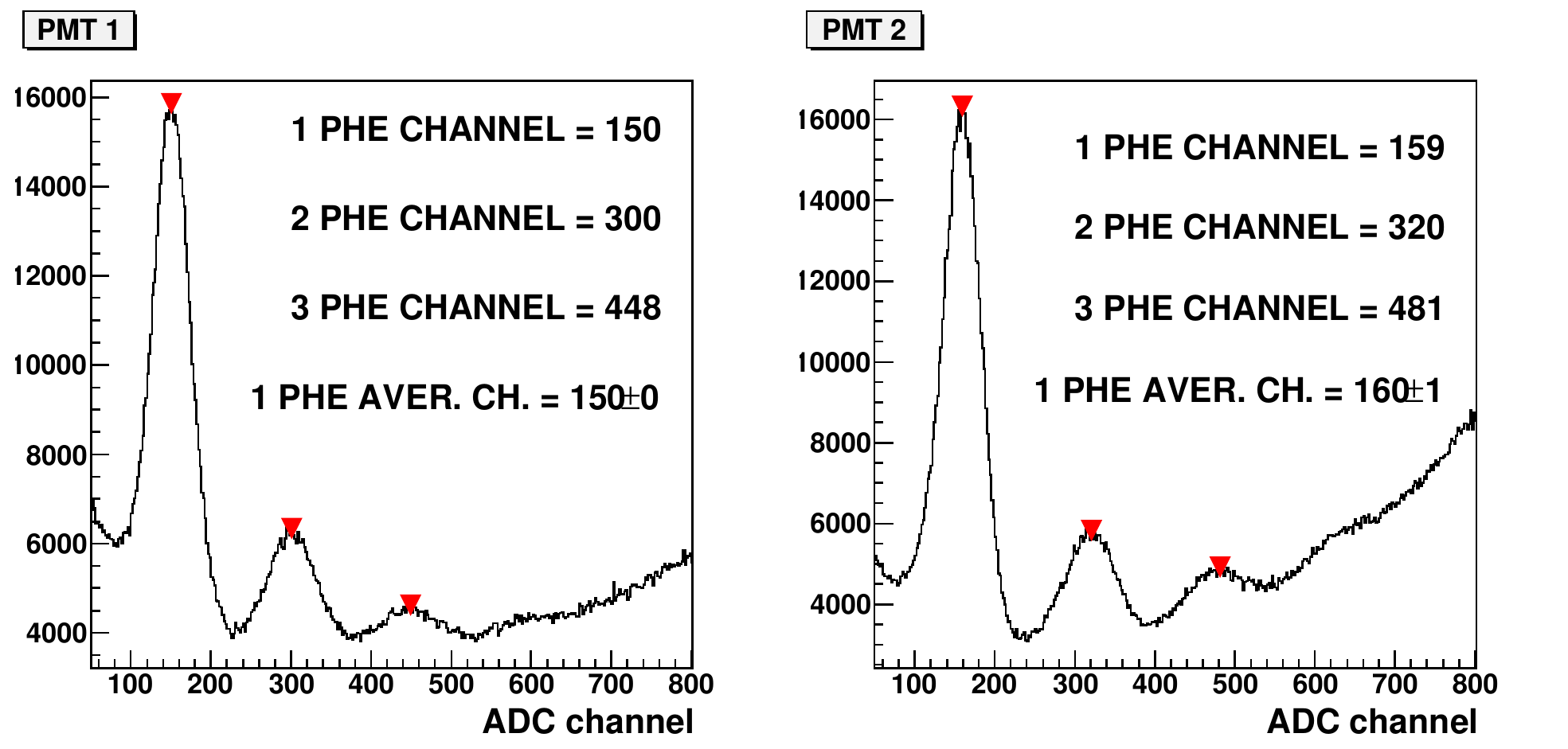,height=2.8in,angle=0.0}
\end{center}
\caption{ADC spectrum of the \u{C}erenkov detector, PMT1 and PMT2.}
\label{fig:ceronephe}
\end{figure}
To see the one photoelectron peak one should select events that originate from less populated 
regions of the HMS acceptance to allow a minimal amount of light into the PMT. 
By doing so, it is possible to see the one, the two and even the three photoelectron peaks. 
The results of the calibration are shown in Fig.~\ref{fig:ceronephe} where 
three different peaks are clearly visible. For PMT1 the 
one photoelectron peak is located at ADC channel 150, the two at 300 and a third peak three at  
448 indicating that the calibration is done correctly (the pedestal peak is centered at the zero ADC channel), 
since two and three photoelectron peaks are multiples of the one photoelectron peak. 
The same is true for PMT2.
The ADC spectrum shown in Fig.~\ref{fig:ceronephe} is the sum of all production runs. 
During the entire experiment the PMT pedestals had widths about 30$-$40 channels and were very stable.
The calibrated signals from two PMTs are summed to get the total number of photoelectrons produced 
for each event. 

As described in Section~\ref{subsubsection:Cerenkov} the \u{C}erenkov detector 
has two mirrors. 
Due to engineering difficulties there is a gap at the region where the mirrors meet. 
The gap causes photon detection inefficiency around the central part of the $\delta p$ 
acceptance of the HMS.
\begin{figure}[ht]
\begin{center}
\epsfig{file=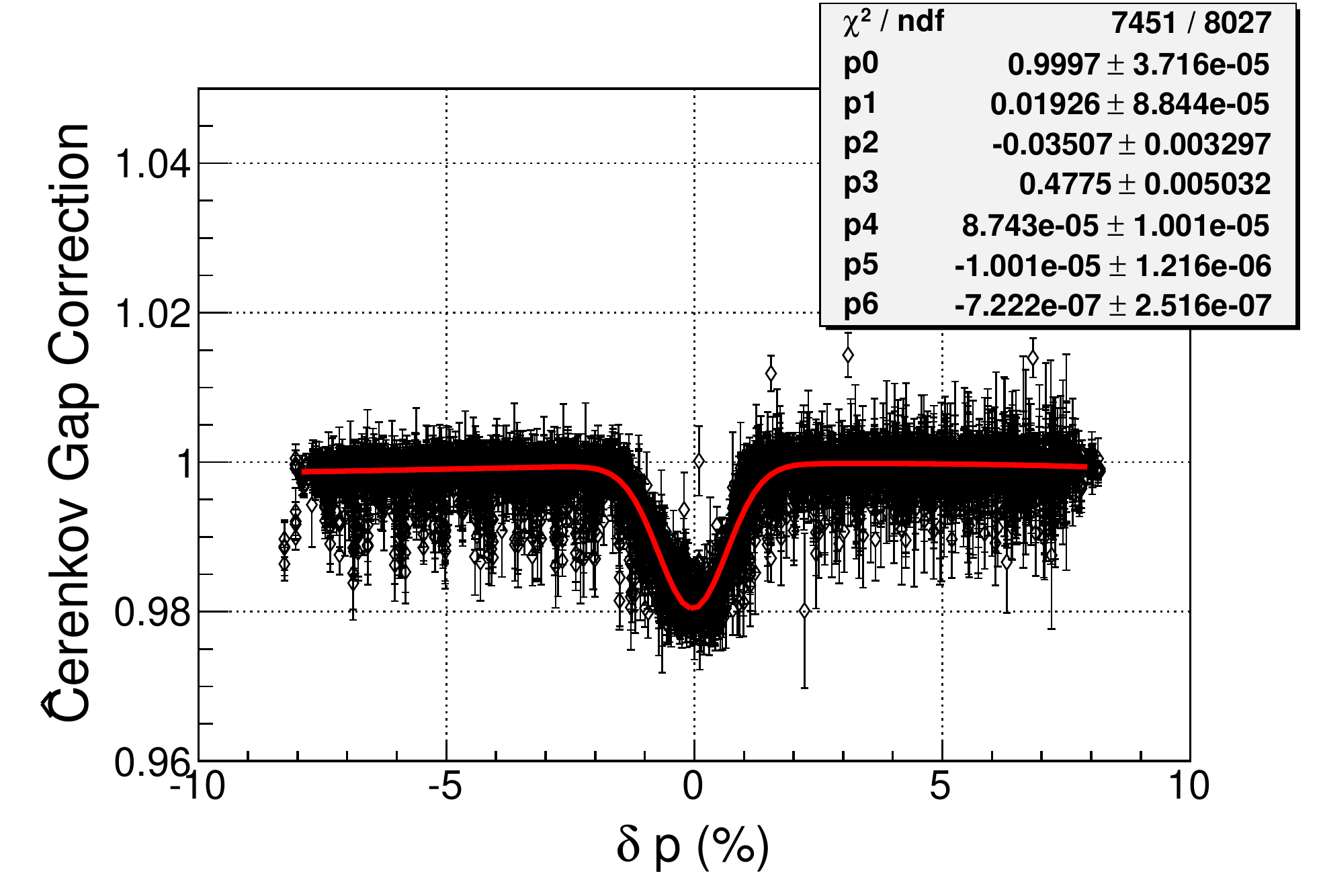,width=5.0in,angle=0.0}
\end{center}
\caption{ Inefficiency of the central part of the \u{C}erenkov due to a gap between the mirrors. 
Inefficiency is parametrized in $\delta p$ and applied to all cross sections on bin-by-bin basis.
All production runs are used to calculate the inefficiency.}
\label{fig:cergapcorr}
\end{figure}
In order to estimate the size of the inefficiency and correct the cross sections 
the following procedure is done: 
the ELHI (see Section~\ref{sec:trigeff}) trigger is required to fire in addition to all cross section cuts, 
the experimental yield is calculated with a \u{C}erenkov cut with the number of photoelectrons 
greater than 0 and greater than 2 (see Sec.~\ref{sec:PIDcuts} for \u{C}erenkov cuts) and the ratio 
of the yields is plotted versus $\delta p$.
The ratio is scaled by a constant to remove the effect of the \u{C}erenkov cut 2 on pions 
(this exposes the inefficiency only due to the gap). 
As it can be seen from the Fig.~\ref{fig:cergapcorr} the \u{C}erenkov cut 2 causes up to 
2\% inefficiency at the central part of the $\delta p$ spectrum. 
The ratio is parametrized in $\delta p$ and is applied to the cross sections (for $|\delta p| < 2$\% range) on bin-by-bin basis 
using the following formula 
\begin{equation} \label{eq:cergap}
\displaystyle\sigma(W^{2}_{i},\theta)_{corr}=\sigma(W^{2}_{i},\theta)\Big/\int_{{W^{2}_{i}}_{min}(\delta p)}^{{W^{2}_{i}}_{max}(\delta p)} f(\delta p)d\delta p.
\end{equation}
It is important to mention that except for the inefficiency caused by the gap no position dependent ( $\delta p$ ) 
inefficiency is found. 
The point-to-point uncertainty (fit residuals) due to this parametrization is estimated to be 0.15\%. 
\subsubsection{Calorimeter Calibration}

Shower detector counters measure the energy deposited by the incoming particle.
While passing  through the lead glass of the shower counter, photons and electrons
produce secondary photons and electrons, thus leading to an electromagnetic shower.
The secondary electrons emit \v{C}erenkov light, which after being reflected from
the inner surface of the shower blocks, is collected on the cathodes of the 
photomultiplier tubes(PMT), mounted at the end of each block.
The registered light is linearly proportional to the energy deposited by
the incoming particle. 
\begin{figure}[p]
\begin{center}
\epsfig{file=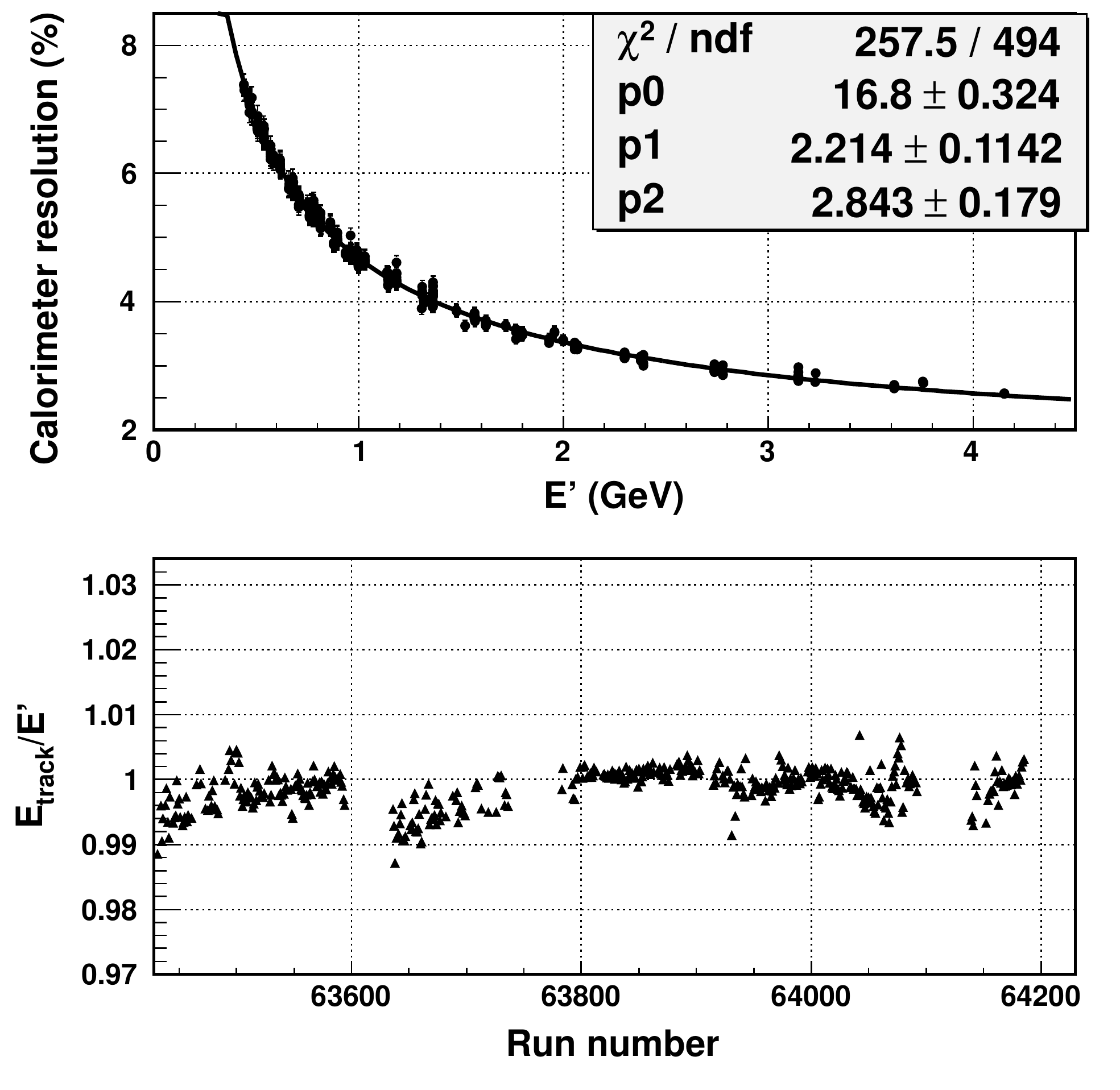,width=5.0in}
\end{center}
\caption{Calorimeter resolution. The top plot shows calorimeter resolution for all production
runs versus HMS momentum. 
In order to find the energy resolution of the calorimeter the distribution of ratio of 
calorimeter track energy to the HMS momentum is fitted by a gaussian function. The width of the 
gaussian function gives the calorimeter resolution. 
The bottom plot shows the center of the gaussian function versus run number. }
\label{fig:calocalib}
\end{figure}
Light flashes on the PMT cathode produce electrical signals, which are sent to
ADC's and are stored in the data stream for further analysis.
The purpose of the shower cluster reconstruction in a shower detector is to 
calculate energy and in some cases the position of the particle.
For this, at first shower cluster is identified.
The cluster in a shower detector is determined as a group of adjacent shower 
counters, where an electromagnetic shower has developed, i.e. an energy 
deposition was detected. 
The block that has the maximum energy deposition is called the central counter of the cluster.

The parameters of the shower --- the energy and the coordinates --- are calculated.
The energy $E$ is calculated as a sum of energy deposition in all of counters 
included into the cluster by the formula
\begin{eqnarray}
\label{form:SHe}
  E=\sum_{i \in M}E_i\ ,
\end{eqnarray}
where:\\
\begin{tabular}{lcl}
\hspace*{1cm} $i$  &---& number of shower counter, included into the cluster;\\
\hspace*{1cm} $M$  &---& set of counters numbers, included into the cluster;\\
\hspace*{1cm} $E_i$ &---& energy deposition in the $i$-th counter;\\
\end{tabular}
\\
\\
For the shower parameters reconstruction, the shower counters gain 
calibration is required. 
The first step involves using the HMS electron energy E$_{e}$.

The purpose of calibration is to define a coefficient for transformation of 
the ADC amplitude to the energy deposition for every shower counter.
i.e., define coefficients $C_i$ such that, 
\begin{eqnarray}
  \label{form:SHedep}
  E_i=C_i \cdot (A_i-P_i)\ ,
\end{eqnarray}
where: \\
\begin{tabular}{lcl}
\hspace*{1cm} $A_i$ &---& ADC amplitude;\\ 
\hspace*{1cm} $P_i$ &---& pedestal of the amplitude;\\ 
\hspace*{1cm} $C_i$ &---& calibration coefficient;\\ 
\hspace*{1cm} $E_i$ &---& energy deposition in the $i$-th counter.\\
\end{tabular}

The calibration coefficients $C_i$ are calculated by minimization of the functional
\begin{eqnarray}
  \label{form:SHmin}
  \chi^2 = \sum_{n=1}^N \Big[ 
  \sum_{i \in M^n} C_i \cdot (A_i^n-P_i) - E_e^n \Big]^2\ .
\end{eqnarray}
Here: \\
\begin{tabular}{lcl}
\hspace*{1.0cm} $n=1 \div N$ &---& Number of event;\\
\end{tabular}

\begin{tabular}{lcl}
\hspace*{1.0cm} $i$     &---& Number of blocks, included in the cluster;\\
\hspace*{1.0cm} $M^n$   &---& Set of counters numbers in the cluster;\\
\hspace*{1.0cm} $A_i^n$ &---& Amplitude in the $i$-th counter;\\
\hspace*{1.0cm} $P_i$   &---& Pedestal of the $i$-th channel;\\
\hspace*{1.0cm} $E_e^n$ &---& Known energy of a particle;\\
\hspace*{1.0cm} $C_i$   &---& Shower counters calibration coefficients to be 
                              fitted.\\\\
\end{tabular}

After obtaining the calibration coefficients, the energy deposition $E$ is calculated by the formula 
(Eq.~\ref{form:SHe}).

\subsection{Target Coordinate Reconstruction} \label{subsection:TargetCoordRec}  

In the transport coordinate system the trajectory of the particle at 
the focal plane, at the target and through the HMS magnetic elements 
is described by a vector (t), which expresses the track relative to 
the central reference trajectory. 
This vector is characterized by five components:
\begin{equation}
\mathbf{}\vec{t}=
\left[\begin{array}{c}
x \\
x' \\
y \\
y' \\ 
\delta \\ 
\end{array}\right]
\end{equation}
where:
\begin{itemize}
\item 
{
  x is the dispersive (vertical) displacement from the central 
  trajectory expressed in meters.
}
\item 
{
  x$'$ is the angle the trajectory makes in the dispersive (vertical) plane 
  relative to the central trajectory $(dx/dz)$, expressed in radians.
}
\item 
{
  y is the displacement in the non-dispersive (horizontal) plane.
}
\item 
{
  y$'$ is the angle the trajectory makes in the non-dispersive (horizontal) plane 
  relative to the central trajectory $(dy/dz)$, expressed in radians.
}
\item
{
$\delta$ is the fractional deviation of the momentum from the central
  value \mbox{$(p-p_{\circ})/p_{\circ}$}, expressed in percent.
}
\end{itemize}
In a first-order approximation, a transport matrix can be defined 
to relate the measured focal plane coordinates to their counterparts 
at the target: 
\begin{equation}
\mathbf{
\mathbf{}
\left[\begin{array}{c}
\delta \\
x' \\
y \\
y' \\ 
\end{array}\right]_{\it{tar}}
}=
\left[\begin{array}{cccc}
<\delta|x> & <\delta|x'>&0 &0\\
<x'|x> & <x'|x'>&0 &0\\
0 & 0&<y|y> &<y|y'>\\
0 & 0 &<y'|y> &<y'y'>\\ 
\end{array}\right]
\mathbf{}
\left[\begin{array}{c}
x \\
x' \\
y \\
y' \\ 
\end{array}\right]_{fp}
\label{firstorder}
\end{equation}

This matrix involves only four unknown parameters since $x_{tar}$ is 
known from the BPM data.
In practice, the expansion of the focal plane coordinates is 
performed up to the fifth order. 
The reconstruction is performed with this formula:
\begin{equation} \label{eq:ytg}
 x^{i}_{tar}={\displaystyle\sum}^{N}_{j,k,l,m}M^{i}_{jklm}(x_{fp})^{j}(y_{fp})^{k}(x'_{fp})^{l}(y'_{fp})^{m} \hspace{0.2cm} for \hspace{0.2cm}  (1 \leq j + k + l +m \leq N) \\
\end{equation}
where $M^{i}_{jklm}$ denote the elements of the reconstruction matrix.

\cleardoublepage
\subsection{Extraction of the Differential Cross Section} \label{Sec:CSEC}

The inclusive electron-nucleus cross section is extracted using the following formula:
\begin{center}
\begin{equation}
\displaystyle\frac{d^{2}\sigma}{d\Omega dW^2} = {
PS\times N_{events}\over Acc\times \Delta  \Omega \times \Delta W^{2} \times N_{in}\times Eff \times LiveTime \times N_{tg}} 
\phantom{l}
\end{equation}
\end{center} 
where
\begin{itemize}
\item { PS is the prescale factor.}
\item { N$_{events}$ is the number of scattered electron observed in solid angle d$\Omega$ and dW$^2$ range. }
\item { Acc is the acceptance of the d$\Omega$ and dW$^2$ bin.}
\item { $\Delta$W$^{2}$ is the width of W$^2$ bin.}
\item { $\Delta  \Omega$ is the solid angle.}
\item { $N_{in}=Charge/Q_{electron}$ is the number of incident electrons.}
\item { Eff is the product of all efficiencies, trigger, particle identification, and tracking,  } 
\item { LiveTime is the product of electronic and computer live times.}
\item { N$_{tg}$=$\rho N_{A}d/A$ is the number of scattering centers per cm$^2$. }
\end{itemize}

\subsubsection{Acceptance Cuts}

The Hall C event reconstruction program reconstructs track parameters at the focal plane 
and using an optic model of HMS finds the target parameters using the Eq.~\ref{eq:ytg}. 
At the focal plane the track parameters are the $x$ and $y$ coordinates, and slopes of the 
track are $x'_{fp}$ and $x'_{fp}$. 
These parameters are defined in Section~\ref{subsection:TargetCoordRec}.
Cuts are applied to the reconstructed target quantities in order to eliminate events 
that are outside of the spectrometer acceptance but end up in the detectors 
after multiple scattering in the magnets or shielding. 
The acceptance cuts are given in the Table~\ref{tab:TabAccepCut}.
\begin{table}[ht]
\begin{center} 
\begin{tabular}{|c|}
\hline
HMS acceptance cuts   \\
\hline
$|x'_{tar}| <$ 0.08 rad       \\
\hline
$|y'_{tar}| <$ 0.04 rad       \\
\hline
$|\delta| <$ 8\%       \\
\hline
\end{tabular}
\end{center}
\caption{Cuts on HMS reconstructed tracks.} \label{tab:TabAccepCut}
\end{table}
The angle cuts are applied to eliminate events coming from outside the HMS acceptance.
These events are the result of multiple scattering in HMS magnets and shielding material. 
The angle cuts are selected to be large enough to allow most events.
The HMS momentum cut is applied to limit the HMS momentum acceptance since the 
reconstruction matrix elements could provide reliable tracking reconstruction only 
within these limits.

For the HMS, the $x'_{tar}$ and $y'_{tar}$ cuts  typically rejected less than 
0.3\% of the total tracked events, and never more than 1.5\%. 
For those rejected events, most come from events that are outside of the spectrometer 
acceptance but end up in the detectors after multiple scattering. 
The loss of these events is compensated by including multiple scattering effects in the acceptance calculation.
Also, these events are mostly lost from the very edge of the $\theta$ bins.
In the cross section calculation this edge bin is eliminated. 
The percentage of events lost due to not using the edge bins depends on kinematic setting and 
is in 0.1-0.5\% range.

\subsubsection{Particle Identification Cuts} \label{sec:PIDcuts}

As in all experiments there are events coming from reactions other then 
the one the experiment wants to study. These events are regarded as a background. 
\begin{figure}[!ht]
\begin{center}
\epsfig{file=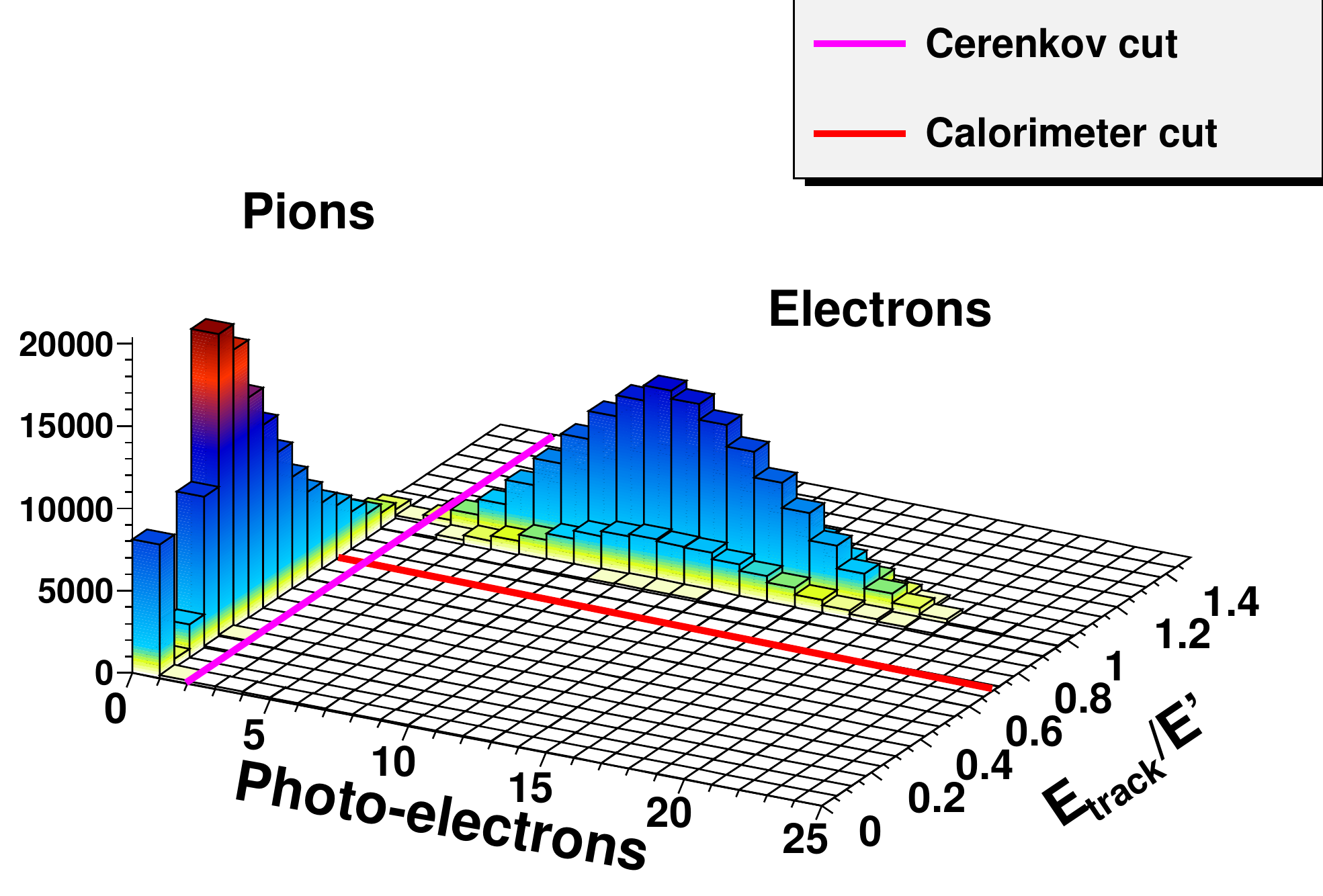,width=5.0in}
\end{center}
\caption{HMS calorimeter track energy hsshtrk/E$'$ vs number of \u{C}erenkov 
photoelectrons. The magenta line shows the \u{C}erenkov cut, the red line shows the Calorimeter cut.}
\label{fig:PID3D}
\end{figure}
Different type of detectors are constructed to decrease the number of background 
events in order to achieve the experiment's specific precision goals. 
For this experiment there are two detectors used for particle identification, 
a threshold \u{C}erenkov detector described in Section~\ref{subsubsection:Cerenkov}, 
and an electromagnetic lead-glass calorimeter described in Section~\ref{subsubsection:Calo}.
The main background in this experiment is from negative pions produced by charge 
exchange reactions. The ratio of pions to electrons varied from 0.1 to 30 for all runs.
In this experiment the \u{C}erenkov cut required that number of photoelectrons be bigger 
than 2 ($hcer\_npe>2$), and the ratio of the Calorimeter track energy divided to particle energy (determined from HMS)
be greater than 0.7 ($hsshtrk/hse>0.7$).
Fig.~\ref{fig:PID3D} gives event distribution versus the number of photoelectrons 
on one axis and the track energy against the other.
The pions and electrons are clearly identified.

\begin{figure}[ht]
\begin{center}
\epsfig{file=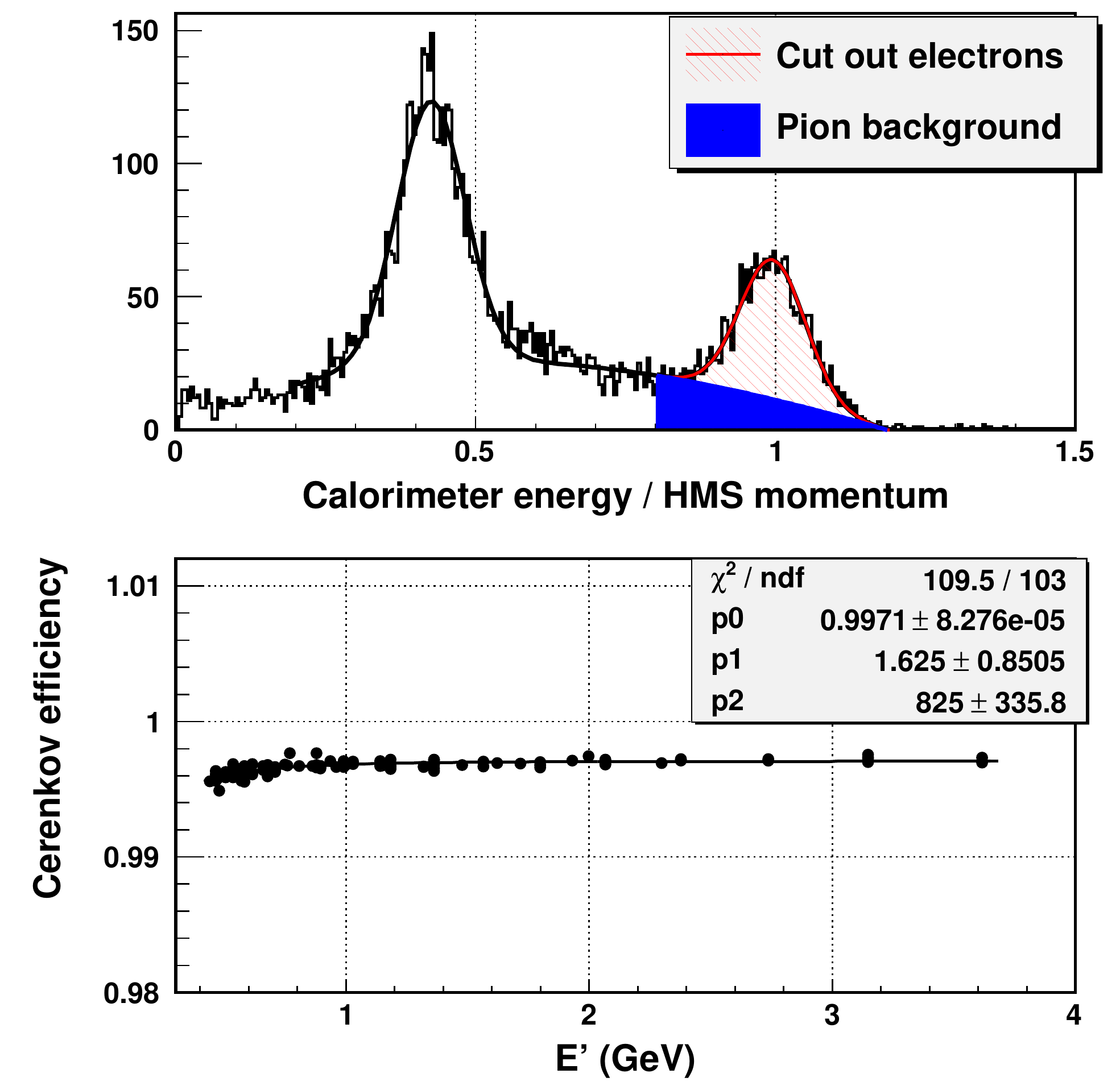,width=5.0in}
\end{center}
\caption{First plot: HMS calorimeter total energy hcal\_et/E$'$ distribution when number of photoelectrons 
are higher than 0 but less than 2. Second plot: The \u{C}erenkov cut efficiency as a function of scattered energy.}
\label{fig:CerEff}
\end{figure}

\begin{figure}[!ht]
\begin{center}
\epsfig{file=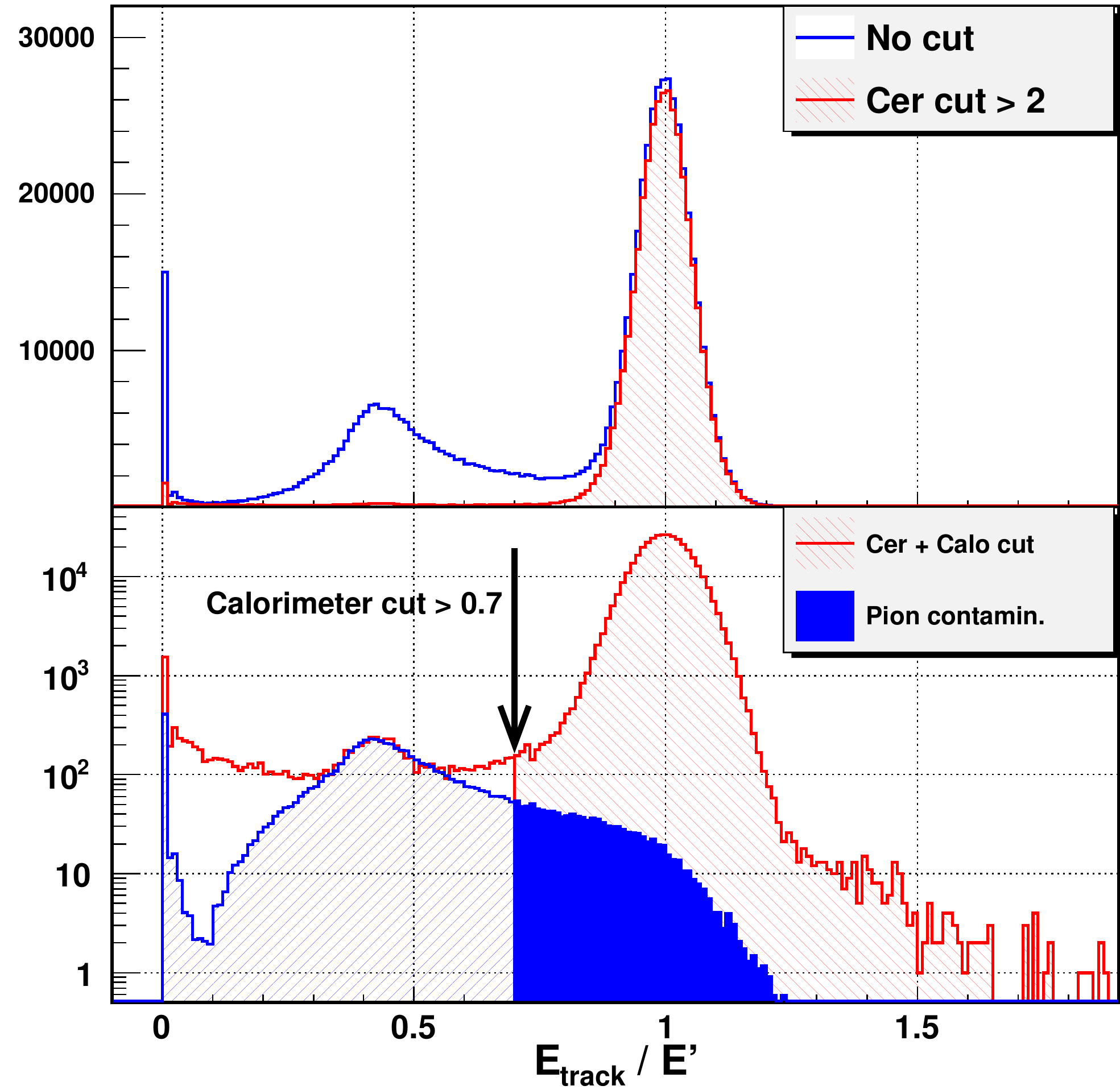,width=5.0in}
\end{center}
\caption{HMS central momentum is 0.71 GeV. Top plot: HMS calorimeter track energy $E_{track}/E'$ (hsshtrk/hse) 
distribution without \u{C}erenkov cut (the blue line) and with \u{C}erenkov cut $>$ 2 (the red line). 
Bottom plot: The $E_{track}/E'$ distribution after \u{C}erenkov cut $>$ 2 and $E_{track}/E'>$ 0.7 cut (the red hatched area). 
The solid blue area is the pion contamination. 
}
\label{fig:CaloPID}
\end{figure}

{\hspace{-0.8cm} \bf \u{C}erenkov Cut Efficiency:}
Fig.~\ref{fig:cerenkov} shows a HMS \u{C}erenkov spectrum. The mean HMS signal is around 
10 photoelectrons and most of the pions have zero photoelectrons. 
The majority of the pions that have a signal with more than 2
photoelectrons are pions that produce delta electrons at the front window or in the gas of the \u{C}erenkov.  
These delta electrons emit \u{C}erenkov light and the pion is misidentified as electron.
In order to reduce the number of pions, a cut on \u{C}erenkov signal is applied.
The cut requires that number of photoelectrons to be greater than 2. 
This number is chosen to eliminate most of the pions and at the same time
allows the majority of the electrons to pass. Not all electrons can be
eliminated due to this cut so it is necessary to estimate how many. 
It is possible to estimate the number of electrons lost to this cut by looking at the 
total energy distribution in the calorimeter when applying all cross section cuts 
except the \u{C}erenkov cut. 
This \u{C}erenkov cut requires that number of photoelectrons to 
be in the range 0 to 2, where 0 is excluded to reduce number of pions 
in the energy distribution. 
In the top plot of Fig.~\ref{fig:CerEff} the ratio (hcal\_et/hsshtrk) of total Calorimeter energy 
to HMS momentum is shown after applying the aforementioned cut.
The blue shaded area is the pion background estimated by a polynomial function.
The parameters of the polynomial are obtained by fitting the distribution of hcal\_et/hsshtrk ratio 
with a \u{C}erenkov photoelectron cut equal to zero. 
The red hatched area is the estimated number of electrons that were cut by the requirement that the 
number of photoelectrons was greater than 2 and are the source of the \u{C}erenkov inefficiency. 
The number of these events divided by the total number of electrons is the inefficiency of the 
\u{C}erenkov cut and is shown in the bottom plot of Fig.~\ref{fig:CaloPID}. 
The \u{C}erenkov cut efficiency is 99.7\% and drops a little at the lowest scattered energy
(most likely caused by high pion/electron ratio).
The electron cut efficiency is parametrized in E$'$ and used to correct the cross section.
The systematic error of the \u{C}erenkov cut is estimated to be about 0.1\%.

{\hspace{-0.8cm} \bf Calorimeter Cut Efficiency:} The \u{C}erenkov cut alone is not enough to 
achieve the necessary pion rejection power at low energies ( 0.4 $-$ 1.0 GeV). 
An additional cut, $hsshtrk/hse>0.7$, on the Calorimeter energy is applied to reduce the number of pions that
pass the \u{C}erenkov cut. 
The distribution of $hsshtrk/hse$, where hsshtrk is the track energy in the calorimeter and 
hse is the energy of the particle measured in the HMS, has electrons peaked at 1 and pions peaked 
at $0.3/E_{track}$, see the bottom plot of Fig.~\ref{fig:CaloPID}.
The $hsshtrk/hse>0.7$ cut will remove most of the pions, but it will also 
remove some good electrons, which for some reason have less energy in the calorimeter than expected. 
It will cause an inefficiency (Calorimeter cut inefficiency) and needs to be estimated. 
There is another source of inefficiency associated with the calorimeter cut called Wrong 
Tracking Efficiency (WTE). 
The WTE is not specific to calorimeters, but is rather a result of imperfect tracking algorithm.
The WTE is absorbed into the calorimeter inefficiency since it is the result of the calorimeter cut. 

At the lowest scattered electron energy as this experiment, 0.44 GeV, the Calorimeter resolution is the worst and is equal to 
7\% (see Fig. \ref{fig:calocalib}). 
Even in this case the $hsshtrk/hse>0.7$ cut is 3$\sigma$ away from the peak and 
is more than 99.9\% efficient.
However, contribution from calibration errors, light collection inefficiency, and pedestal drift 
can leave some electrons below the the 3$\sigma$ threshold and will be cut by the $hsshtrk/hse>0.7$ cut. 

To estimate the number of this electrons the following procedure is followed: 
\begin{itemize}
\item{ 
First, it is necessary to find the lowest value of $hsshtrk/hse$ for good electrons. 
A few elastic runs with several E$'$ are analyzed to estimate this value. 
It was found that small number of electrons, after a strict cut ( $>$ 10 ) on the \u{C}erenkov, can have $hsshtrk/hse$ as low as 0.3.}
\item{ Second, the cut efficiency is calculated with this formula 
\vspace{-1.0cm}
\begin{center}
\begin{displaymath}
\epsilon_{cal}^{cut}= {N^{events}(E_{calo}^{total}/E'>0.7, hcal\_e\textit{1}/hse>0.2, Cerenkov>10) \displaystyle\over N^{events}(E_{calo}^{total}/E'>0.3), hcal\_e\textit{1}/hse>0.2, Cerenkov>10) }
\end{displaymath}
\end{center} 
where $hcal\_e\textit{1}$ is the total energy deposited in the first layer of the calorimeter. Since pions are less likely to deposit significant 
energy in the first layer (the first layer acts as preshower) this cut further reduces their number. 
In addition to the cuts in the above equation, acceptance cuts are also applied both in numerator and denominator.
}
\end{itemize}

The procedure described above will count some pions as electrons even with strict cuts on the \u{C}erenkov and on the first layer of the calorimeter.
This means that it will underestimate the calorimeter cut efficiency. 
For each momentum setting there are a few runs with different angles, and therefore with different $\pi/e$ ratio. 
If there are no pions after $E_{calo}^{total}/E'>0.3$ cut one would expect that the Calorimeter cut efficiency to be the
same. 
In order to estimate the systematic error these pions introduce, the efficiencies from runs with the same momentum but
different angles are averaged and the average is used as the cut efficiency for that momentum.
The error bar is the standard deviation from the average. 

There are some good events (having passed the acceptance and the \mbox{\u{C}erenkov $>$ 2 cuts}) for which a cluster in the calorimeter 
is not found.
These events can be seen in the spectrum of $hsshtrk/hse$ around zero, as seen in the bottom plot of Fig.~\ref{fig:CaloPID}.
For these events the tracking algorithm was unable to predict the correct position of the track on the front face of the Calorimeter,  
that is a wrong track energy reconstruction.
However, the value of the total deposited energy in the calorimeter divided by the energy of the detected particle 
is still around 1 ($E_{calo}^{total}/E'$ $\approx$ 1).
The loss of these good events is compensated by calculating the efficiency due to the wrong tracking, denoted as $\epsilon^{wrong}$.
The $\epsilon^{wrong}$ is calculated using this formula
\vspace{-1.0cm}
\begin{center}
\begin{displaymath}
\epsilon_{cal}^{wrong}= \frac{N^{events}(E_{calo}^{track}/E'>0.7, Cerenkov>10, ntracks=1)}  {N^{events}(E_{calo}^{total}/E'>0.7, Cerenkov>10, ntracks=1) }.
\end{displaymath}
\end{center}
The requirement of having one track is essential to ensure that there is only one cluster in the calorimeter.
At high counting rates it is possible to have two clusters produced by two high energy 
electrons. 
These two electrons will be counted as one and the lost electron will contribute as electronic dead time.
\begin{figure}[th]
\begin{center}
\epsfig{file=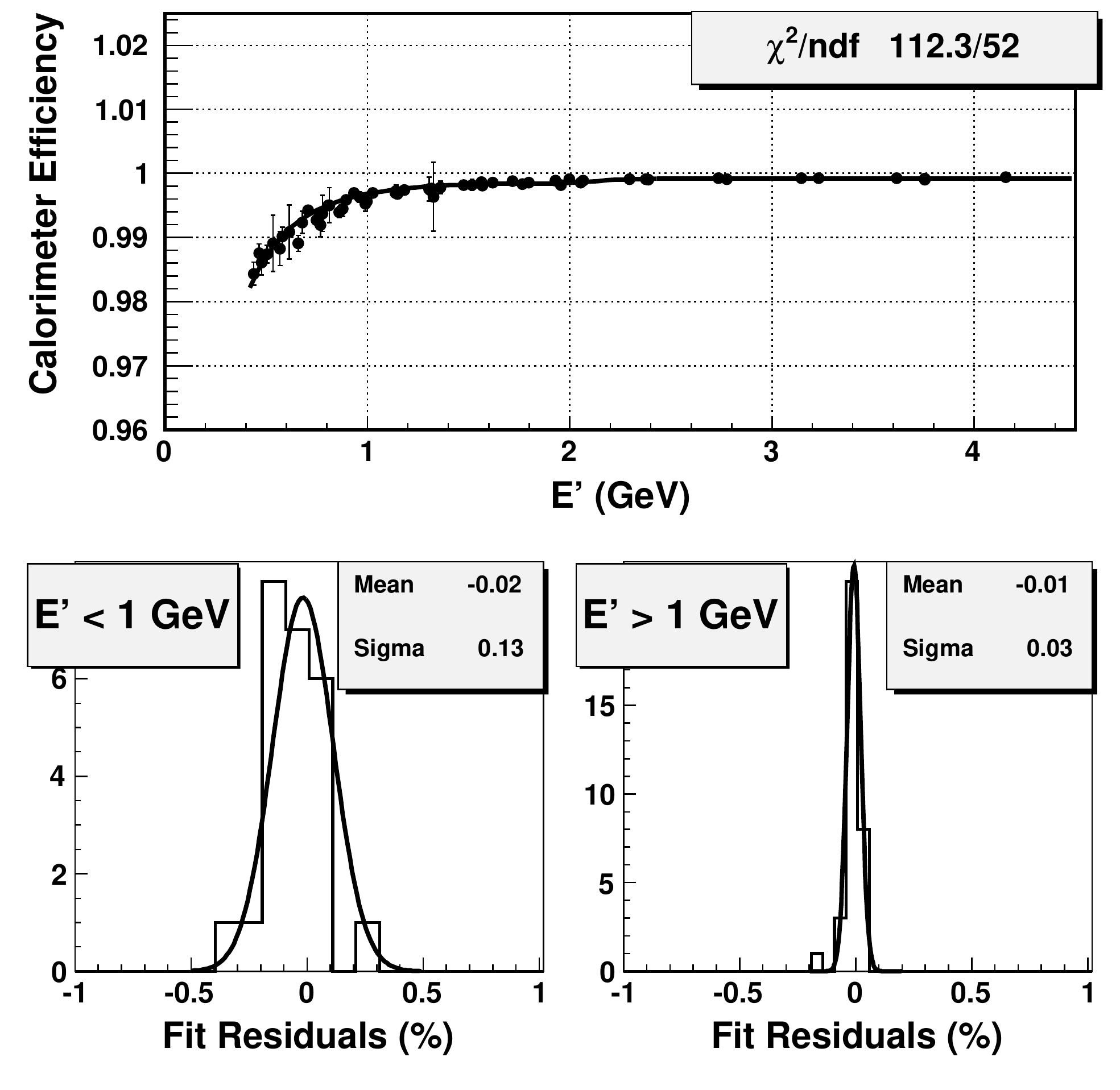,width=5.0in}
\end{center}
\caption{ Upper plot: Calorimeter efficiency versus E$'$. 
Second and third plots: Residuals of the fit when E$'<$1 GeV (left) and E$'>$1 GeV (right).}
\label{fig:CaloCutEff}
\end{figure}
In addition to the cuts in the above equation, acceptance cuts are also applied both in numerator and denominator.

The total Calorimeter efficiency is the product of WTE and Calorimeter cut efficiency and is shown in Fig.~\ref{fig:CaloCutEff}.
The top plot is the Calorimeter total efficiency vs E$'$. It is parametrized as a function of
E$'$ and used in the cross section analysis. 
The second and third plots are the residuals of the parametrization for E$'<$ 1 GeV and E$'>$ 1 GeV. 
The systematic uncertainty is estimated as the $\sigma$ of the residuals (left and right plots) and is used in cross 
section analysis. 
Systematic uncertainty is about 0.13\% for E$'<$ 1 GeV and 0.03\% for E$'>$ 1 GeV. 
The difference is the result of higher $\pi/e$ ratio at low energies.

\subsubsection{Background from the target walls}

Backgrounds are unfortunate part of all nuclear physics experiments. 
Depending on the goals of the experiment they must be taken into account by subtracting 
background events or by estimating them and assigning a systematic uncertainty to the final measurement. 
In this experiment all backgrounds that can contribute more than half a percent are estimated and
subtracted to obtain the final result. 
\begin{figure}[ht]
\begin{center}
\epsfig{file=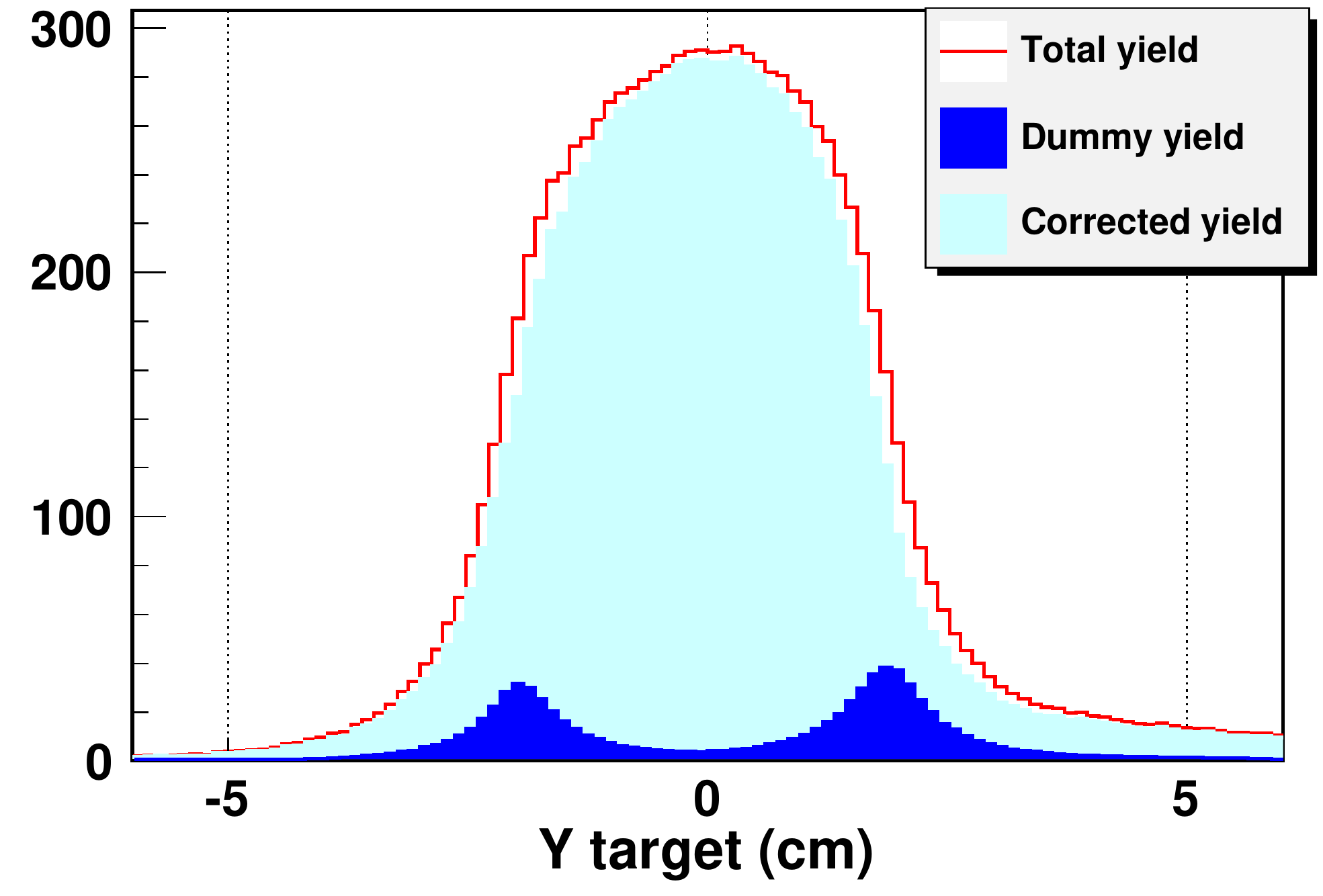,width=4.0in}
\caption{Normalized yield distributions for deuterium target. The solid red line is the total normalized yield, the blue solid
area is the contribution of estimated events (from dummy target) produced by aluminum walls of the liquid deuterium target, the cyan 
shaded area is the corrected deuterium yield. }
\label{fig:al_wals}
\end{center}
\end{figure}

The cryotarget liquid targets have aluminum walls which can be source of background 
induced by scattering on nucleons of the aluminum nuclei.
In order to determine the magnitude of this background, data on the aluminum dummy 
target have been taken at exactly the same kinematic settings as the deuterium data.
The size of this background is estimated by the following formula
\vspace{-1.0cm}
\begin{center}
\begin{equation}
Y_{corrected} = {Y}_{Cryot.}-Y_{Alum.}{T_{Walls}\over T_{Alum.} }{R_{Alum.}^{ext}\over R_{Walls}^{ext}} \phantom{l}, 
\end{equation}
\end{center}
where, $Y$ are the yields for the cryo and aluminum dummy targets, $T$ is the thickness of the aluminum walls in the 
cryotarget and aluminum dummy target, and $R$ is the external radiative correction for the aluminum dummy target or the 
aluminum walls in the cryotarget.
Yield for one of the deuteron runs without aluminum background subtraction is shown in the 
Fig. \ref{fig:al_wals}. 
In the same figure the yield for the target wall background and the corrected deuterium yield are shown. 

The size of the background from the aluminum target walls is generally about 10\%.
The background has been calculated for each run and the subtraction has been done on a bin-by-bin basis 
in ($W^{2}, \theta$).

\subsubsection{Charge Symmetric Background} \label{sec:CSB}

In this experiment inclusive electrons are scattered from nuclear targets.
The probability of production of $\gamma$ and $\pi^{o}$ particles in the target is comparable 
to the cross section of inclusive electron scattering cross section in the kinematic range of
this experiment. 

The $\pi^{o}$ decays predominantly to two photons which produce electron-positron 
pairs when passing through the target or any material between target and the drift chambers.
In the beam energy range of the current experiment bremsstrahlung of $\gamma$ 
particles (virtual and real) is the dominant electron energy loss mechanism. 
The $\gamma$ particles will produce electron-positron pairs in the field of a nucleon (Bethe-Heitler process).
The resulting electrons produced through the reaction mechanisms described above is kinematically 
indistinguishable from inclusive electrons and passes all particle identification
cuts. 
These electrons are a background and need to be subtracted from real inclusive electrons scattered from the target.
Since the normal configuration of the experiment is unable to remove these electrons, 
another approach is used. 
From the production mechanism of background electrons it can be seen that there should be equal 
number of positrons and electrons (that is why this is called charge symmetric background (CSB)). 
Thus, the polarity of the HMS spectrometer is reversed and positron data are taken at the 
spectrometer settings where the charge symmetric background is large. 
This background is significant for larger scattering angles and small final electron energies 
of the electrons. 
\begin{figure}[!ht]
\begin{center}
\epsfig{file=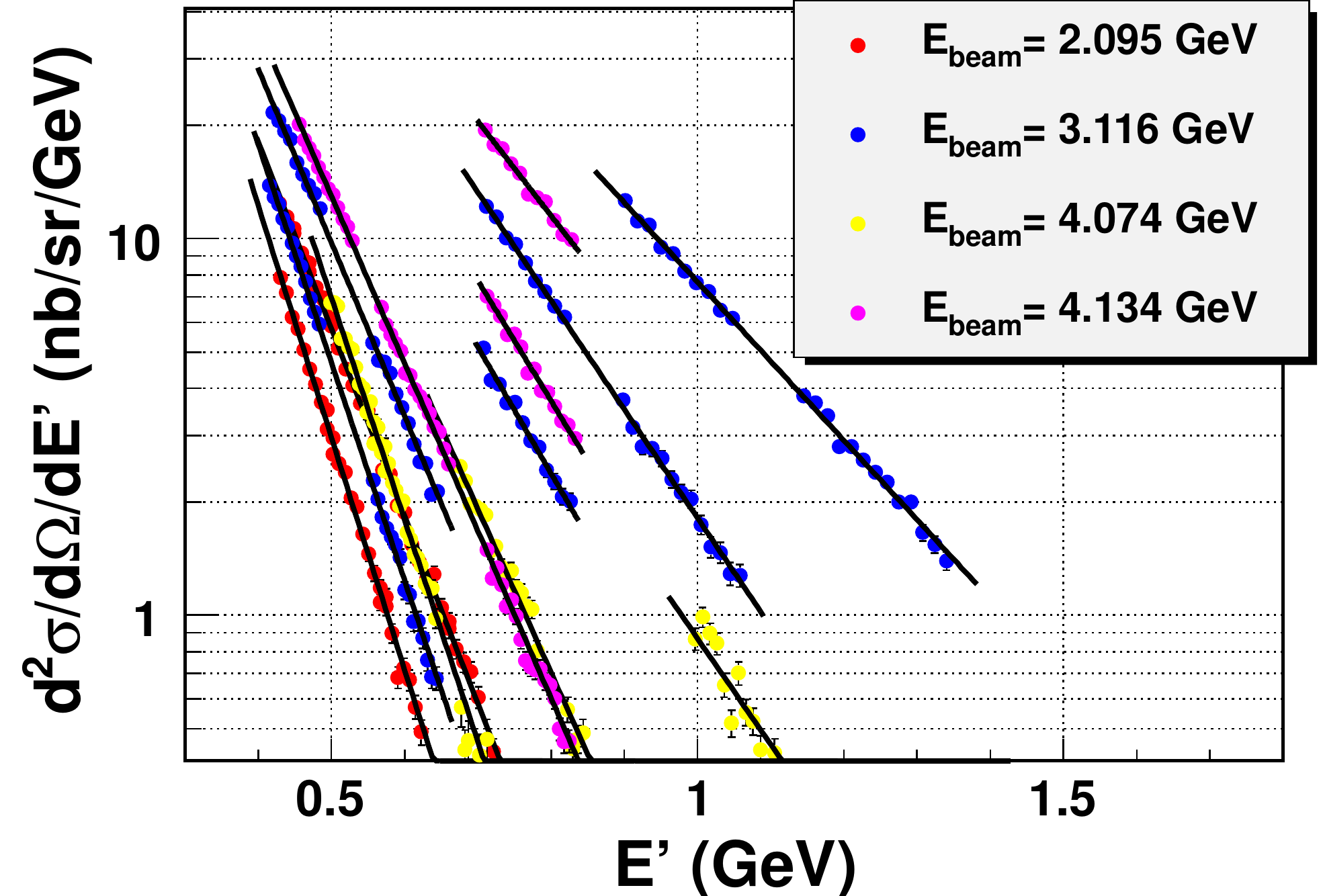,width=5.0in}
\end{center}
\caption{ Cross section of Charge Symmetric Background for the carbon target.}
\label{fig:CSB_CS}
\end{figure}

\begin{figure}[!ht]
\begin{center}
\epsfig{file=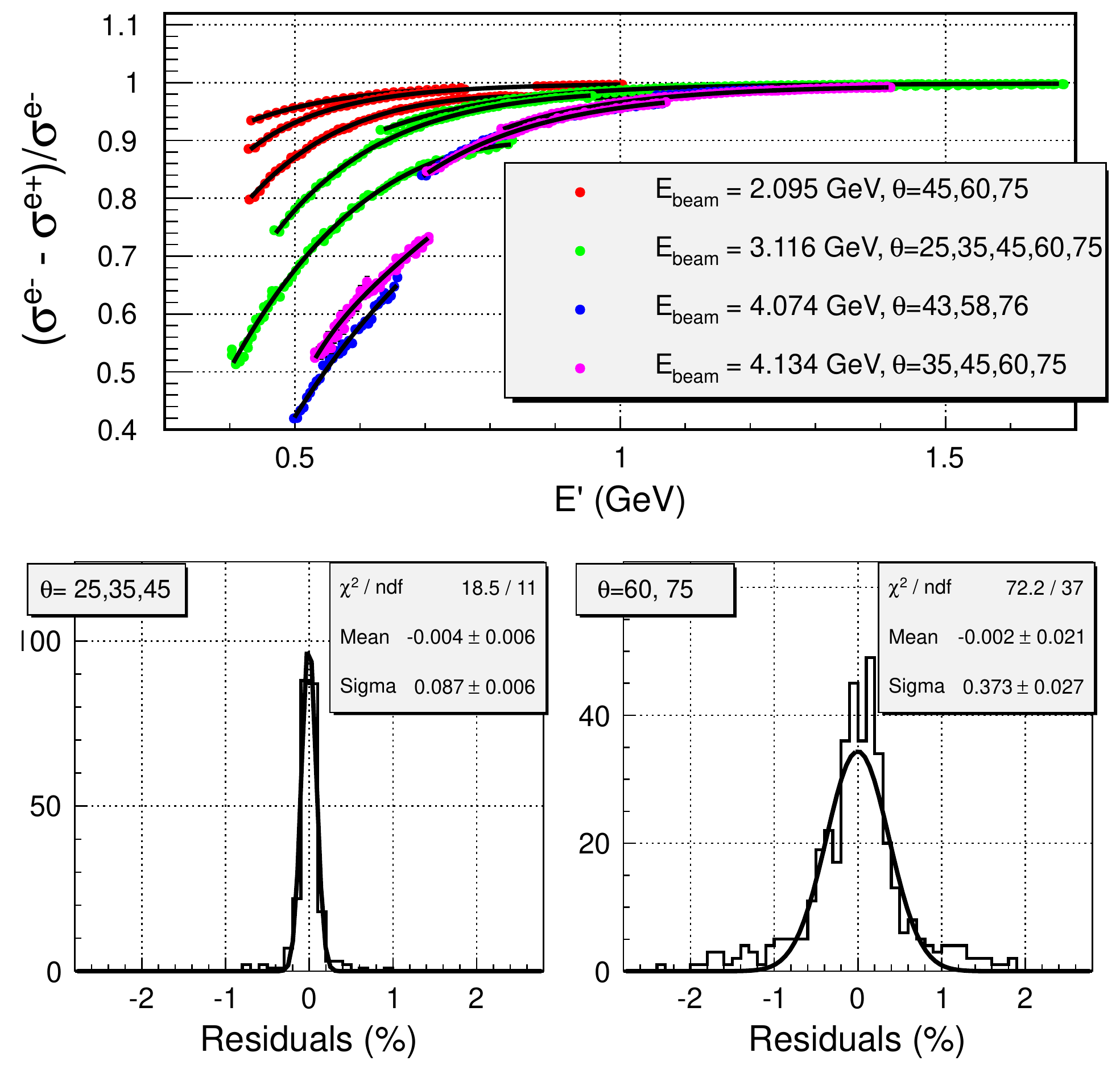,width=5.0in}
\end{center}
\caption{ First plot: Ratio of CSB cross section to raw electron cross section versus E$'$ for all carbon runs. 
Second plot: The residuals of CSB fit for $\theta$=25,35,45 degrees (left) and $\theta$=60,75 degrees (right).}
\label{fig:CSB_SYSTEM}
\end{figure}

The background contribution from kaon decay is negligible and is not taken into account, for more details see Ref.~\cite{liang}.
Runs with reversed HMS polarity were taken where the CSB was considered to have significant contribution. 
The CSB cross section is calculated for all these runs and then extrapolated by the  
following function: 
\vspace{-1.0cm}
\begin{center}
\begin{displaymath}
F\left( E' \right) = e^{C}\left( e^{S\left( E_{beam}-E'\right) }-1\right)  \phantom{l},
\end{displaymath}
\end{center}
where E$'$ is the electron energy. $C(\theta)$ and $S(\theta)$ are free parameters determined from fitting the cross sections 
for each beam energy and angle as shown in Fig.~\ref{fig:CSB_CS}. 
The form of the function is chosen to go to zero at E$'$ equal to the beam energy and have smallest possible number
of parameters. 
After fitting the CSB cross sections at each energy and angle, the $C(\theta)$ and $S(\theta)$ parameters are interpolated in theta by a 
polynomial function of second degree.
The parameters of the polynomials are stored in a lookup table and are used in the cross section analysis. 
The CSB is subtracted from the total electron cross section for each bin ($\theta,W^{2}$).

The CSB reaches 60\% at the smallest values of $E'=0.44$ GeV and the largest angles, $\theta=$76$^o$
and about 50\% for $\theta=$75$^o$, as can be seen in the top plot of Fig.~\ref{fig:CSB_CS}. 
The point to point uncertainty has been estimated as the deviation of the measured cross section 
from the parametrization and is equal to 0.1\% for angles less than 45$^o$ and 0.4\% for angles
60$^o$ and 75$^o$. See the bottom plots of Fig.~\ref{fig:CSB_SYSTEM}.

\subsubsection{Pion Contamination} \label{sec:pioncontam}

The \u{C}erenkov and the calorimeter cuts can't fully remove all pions.
Most of the pions do not give a signal in the \u{C}erenkov detector, 
but some pions can produce knock-on electrons of high enough energy ( more than 20 MeV) to emit \u{C}erenkov light.
These pions will pass the \u{C}erenkov cut.
In the calorimeter, pions can be registered due to ionization losses. 
This is visible as a wide peak around 0.3/$E'$(GeV) in the normalized energy spectrum of the 
calorimeter.
Also pions can undergo charge-exchange reactions and produce a neutral pion which decays into 
two photons.
These photons will deposit their full energy in the calorimeter which leads 
to a high energy tail in the normalized energy spectrum of the calorimeter and can go beyond $E_{cal}/E'=0.7$ ( the calorimeter cut). 
As a result both \u{C}erenkov and calorimeter cuts will fail to reject some of the pions and they will  
be accepted as electrons, resulting in pion contamination in the electron sample.
In the kinematic range of the current experiment the pion contamination is estimated to be up to 3\%. 
Investigation showed no rate dependence of pion contamination.

\subsubsection{Electronic and Computer Dead Times} \label{sec:DeadTime}

\begin{figure}[!ht]
\begin{center}
\epsfig{file=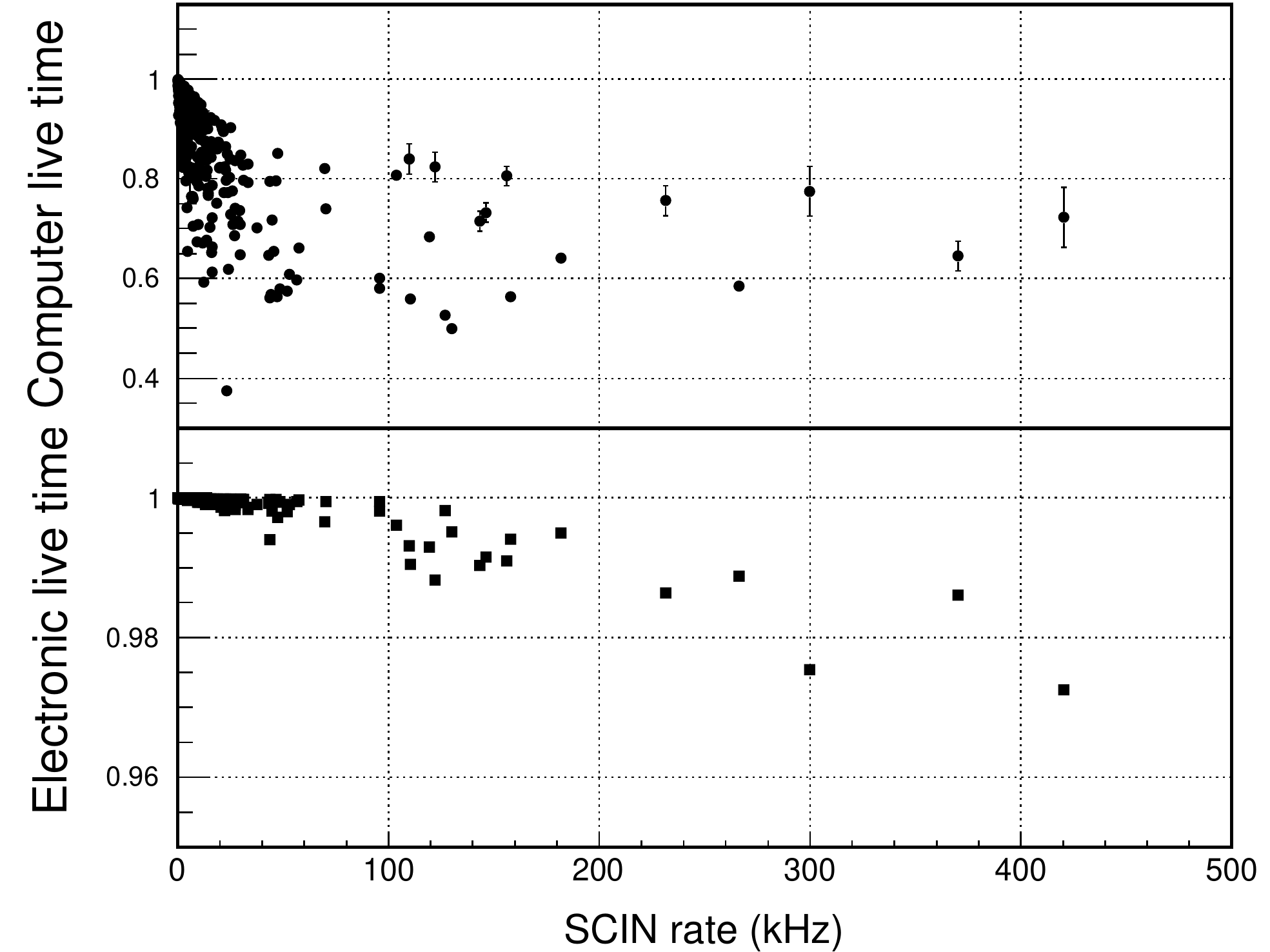,width=5.0in}
\end{center}
\caption{ First plot: Computer live time as a function of rate. Second plot: Electronic live time as a function of rate.}
\label{fig:DEADTIME}
\end{figure}

In this experiment electronic dead time was low (less than 3\% ) since the highest counting rate is less than 500 kHz. 
Computer dead time reached up to 60\% for very few runs and is generally lower than 30\%; see Fig.~\ref{fig:DEADTIME}. 

{\hspace{-0.8cm} \bf Electronic Deadtime:}
Electronic dead time (EDT) is caused when a trigger is missed because the hardware 
is busy when an event that should generate a trigger comes in.
Detector dead time occurs when a detector is unable to respond to an 
event because it is still responding to a previous event.

For an average event rate $R$ the probability of finding $n$ counts in a time interval $\tau$
is given by the Poisson distribution, 
\vspace{-1.0cm}
\begin{center}
\begin{equation}
P(n) = { (Rt)^{n}e^{-Rt} \over n! } \phantom{l}.
\end{equation}
\end{center}
The live time is the probability of no events occurring in the interval $\tau$ is $P(0) = e^{-R\tau}$.
For small $R\tau$ this can be approximated by $P(0) \approx 1-R\tau$.
An event will be missed when it arrives within a time $\tau$ of an event accepted by the gate,
where $\tau$ is the gate width of the logic signal. 
Therefore, the fraction of the measured events is equal
to the probability that the time between events will be greater than $\tau$, 
\vspace{-1.0cm}
\begin{center}
\begin{equation}
{ N_{measured} \over N_{total}} \approx 1-R\tau \label{eq:el_time} \phantom{l}.
\end{equation}
\end{center}
In order to determine $N_{total}$, scalers with four different gate widths ($t$=50, 100, 150, 200 ns)
have been used and the number of events recorded in each of them have been measured 
($N_{50}$, $N_{100}$, $N_{150}$, $N_{200}$).
A linear extrapolation back to zero gate width gives $N_{total}$.
\begin{figure}[ht]
\begin{center}
\epsfig{file=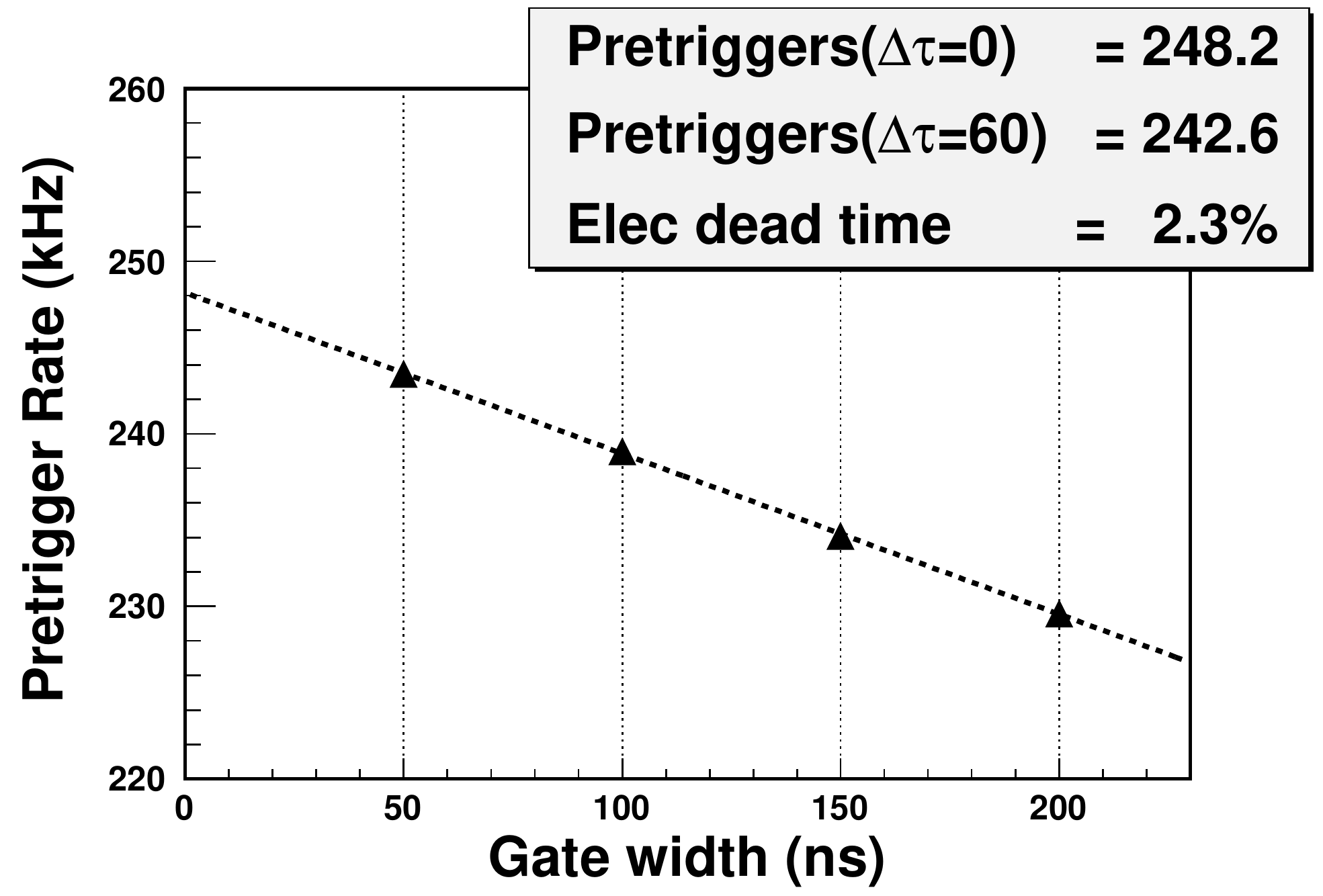, width=3.5in}
\caption{ Rates with different gate widths for run 63915.}\label{fig:el_dt}
\end{center}
\end{figure}
In Hall-C ENGINE the electronic dead time is calculated using only two pretriggers,
PRE100 and PRE150. 
The following formula is used to get the final electronic dead time (EDT) for each run.
\vspace{-1.0cm}
\begin{center}
\begin{equation}
EDT \approx \frac{(PRE100-PRE150)}{PRE100} \frac{\Delta\tau}{50} \phantom{l}.
\end{equation}
\end{center}
where $\tau$ is the real electron trigger gate width (60 ns).

All runs are corrected on a run by run basis, and no systematic uncertainty is assigned to the electronic dead time.
For more information about electronic dead time see Ref.~\cite{Arring}.

{\hspace{-0.8cm} \bf Computer Deadtime:}
The computer dead time occurs as a result of the data acquisition
computers being busy processing an event and not being available to process new events.  
In that case a new event is lost.
Events recorded by the electronics have been logged as pretriggers. The pretriggers that have been
processed successfully by the trigger supervisor are recorded as triggers.
The ratio of triggers to pretriggers gives the computer live time (see Fig.~\ref{fig:DEADTIME}),
\begin{displaymath}
L=\displaystyle\frac{N_{trigger}}{N_{pretrigger}}.
\end{displaymath}
The computer live time is calculated for each run and applied on a run-by-run basis. 
The systematic uncertainty for the computer live time has been studied by comparing 
experimental yields at the same spectrometer kinematic setting
but varying the prescale factors and therefore, computer live times~\cite{liang}. 
These studies show that the yields agree within 0.2\%.

\subsubsection{Trigger Efficiency} \label{sec:trigeff}
Triggers are used to reduce the rate of useful events into a range manageable by the data acquisition equipment.
Triggers also provide timing signals to the various detector parts. 

Some events can be lost due to inefficiencies of the detectors which are used to form different triggers of the HMS spectrometer.
The schematics of the single arm trigger logic is shown in Fig.~\ref{fig:rcstrg}. 
It can be seen that the two electron triggers (low - level and high - level) are strongly correlated. 
For example, if the (SCIN) signal (at least three of the four scintillator layers of both hodoscopes have fired) 
is present for an event, there must also be a (STOF) signal (which requires one front panel and one back panel).
If (PRHI) is present then there must also be a (PRLO) signal (high (PRHI) and low (PRLO) threshold on 
the energy in the first layer of the calorimeter).
Thus, assuming that the \u{C}erenkov signal (\u{C}) is always present, the efficiency for the 
low-level electron trigger can be calculated as $\epsilon_{PRLO} \times \epsilon_{STOF}$. 
The small inefficiency of the \u{C}erenkov is covered by the ELHI trigger which does not require the \u{C}erenkov signal. 
If the calorimeter has an inefficiency, an electron trigger still could be produced if there is a (SCIN) signal. 
Accordingly the electron trigger efficiency for the HMS can be calculated (estimated) by the following formula, 
\vspace{-1.0cm}
\begin{center}
\begin{equation}
\epsilon_{trg} = \epsilon_{PRLO} \times \epsilon_{STOF} + (1 - \epsilon_{PRLO}) \times \epsilon_{3/4}\label{eq:tr} \phantom{l}, 
\end{equation}
\end{center}
where $\epsilon_{3/4}$ is defined in Eq.~\ref{eq:eps34}.

\begin{figure}[!ht]
\begin{center}
\epsfig{file=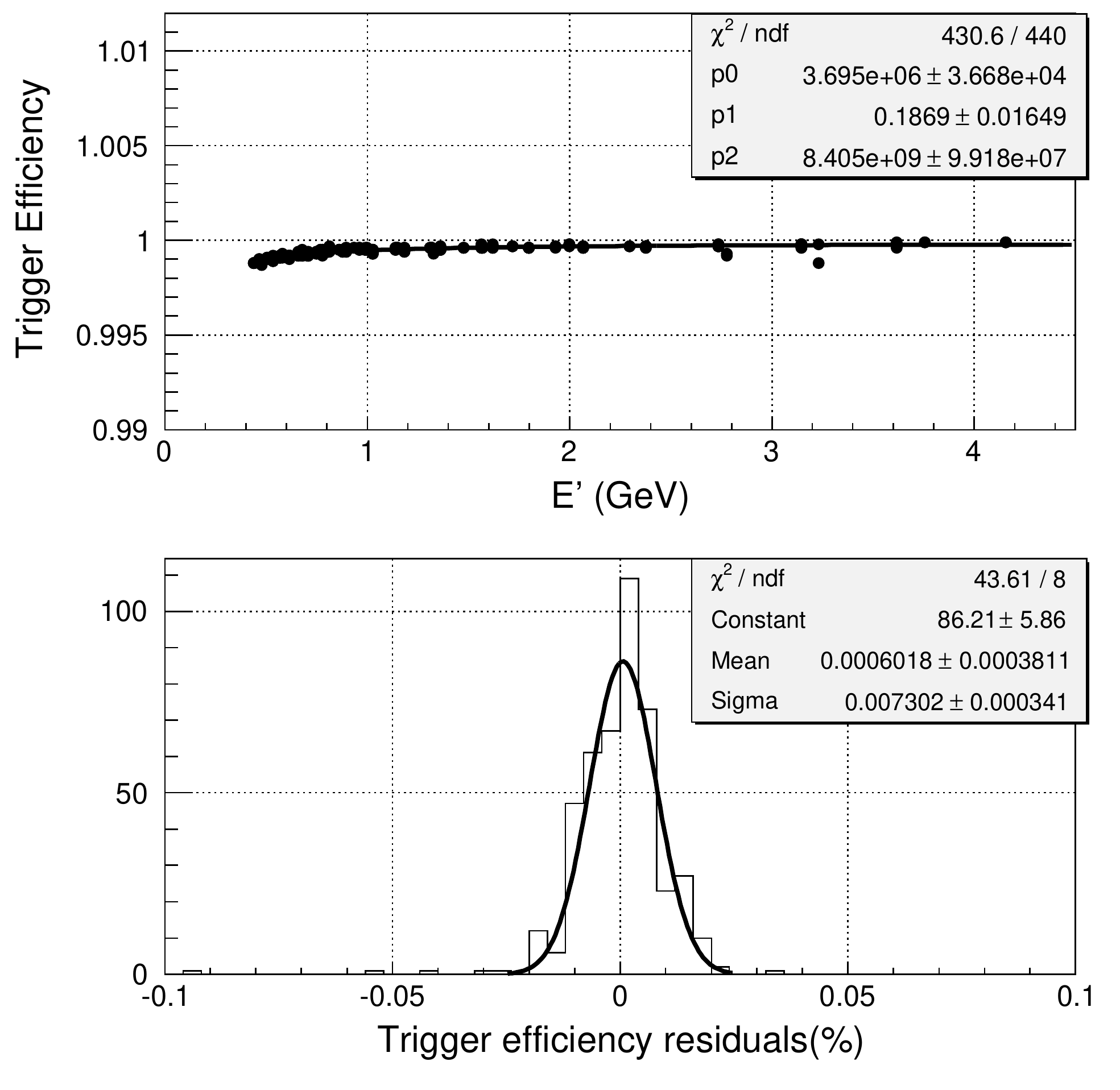,width=5.0in}
\caption{HMS electron trigger efficiency as a function of the scattered electron energy.} \label{fig:trigeff}
\end{center}
\end{figure}

The scintillator 3/4 efficiency ($\epsilon_{3/4}$) has been calculated by summing over all possible combination that 
would satisfy the 3/4 trigger
\begin{center} 
\begin{equation}
\epsilon_{3/4} =  \prod_{i = 1, 4}\epsilon_{i} + \sum_{j = 1, 4}(1-\epsilon_{j})\prod_{i \ne j}\epsilon_{i} \phantom{l}, \label{eq:eps34}
\end{equation}
\end{center}
where $i$, $j$ are the plane numbers and $\epsilon$ is the corresponding efficiency for each plane.
The average scintillator 3/4 efficiency ($\epsilon_{3/4}$) is about 0.983. 

The (PRLO) efficiency ($\epsilon_{PRLO}$) has been calculated as the ratio of the events that have
a (PRLO) signal and a signal from the \u{C}erenkov over the events that have a signal from the \u{C}erenkov, 
\vspace{-1.0cm}
\begin{center}
\begin{equation}
\epsilon_{PRLO} = { PRLO \textrm{ \& }   \textrm{\u{C}}ER > 0.5  \over \textrm{\u{C}}ER > 0.5  } \phantom{l}.
\end{equation}
\end{center}
The average (PRLO) efficiency is about 0.98.

The (STOF) efficiency ($\epsilon_{STOF}$) has been calculated by using the efficiency of each 
scintillator ($\epsilon_{1}$, $\epsilon_{2}$, $\epsilon_{3}$, $\epsilon_{4}$,) by the
following formula
\vspace{-1.0cm}
\begin{center}
\begin{equation}
\epsilon_{STOF} =  \Big(1 - (1-\epsilon_{1})*(1-\epsilon_{2})\Big) \times \Big(1 - (1-\epsilon_{3})*(1-\epsilon_{4})\Big) \phantom{l}.
\end{equation}
\end{center}
The average (STOF) efficiency is about 0.999.

Finally, the total electron trigger efficiency for the HMS is calculated by Eq.~\ref{eq:tr} 
and the result is shown in Fig.~\ref{fig:trigeff}.
The trigger efficiency is always higher than 0.999. 
The trigger efficiency has been parametrized as a function of the scattered electron energy and
during the data analysis this parametrization has been used on a run-by-run basis. 
The systematic uncertainty is estimated as the spread of the measured trigger efficiency 
from the parametrization and is equal to 0.007\%. A detailed description of trigger efficiency calculation can be 
found in Ref.~\cite{vlad}.
\subsubsection{Tracking Efficiency} \label{sec:TrackEff}
Sometimes track reconstruction can fail even when there is a legitimate 
track passing through the detector system. 
These lost events should be taken into account during cross section calculations. 
\begin{figure}[ht]
\begin{center}
\epsfig{file=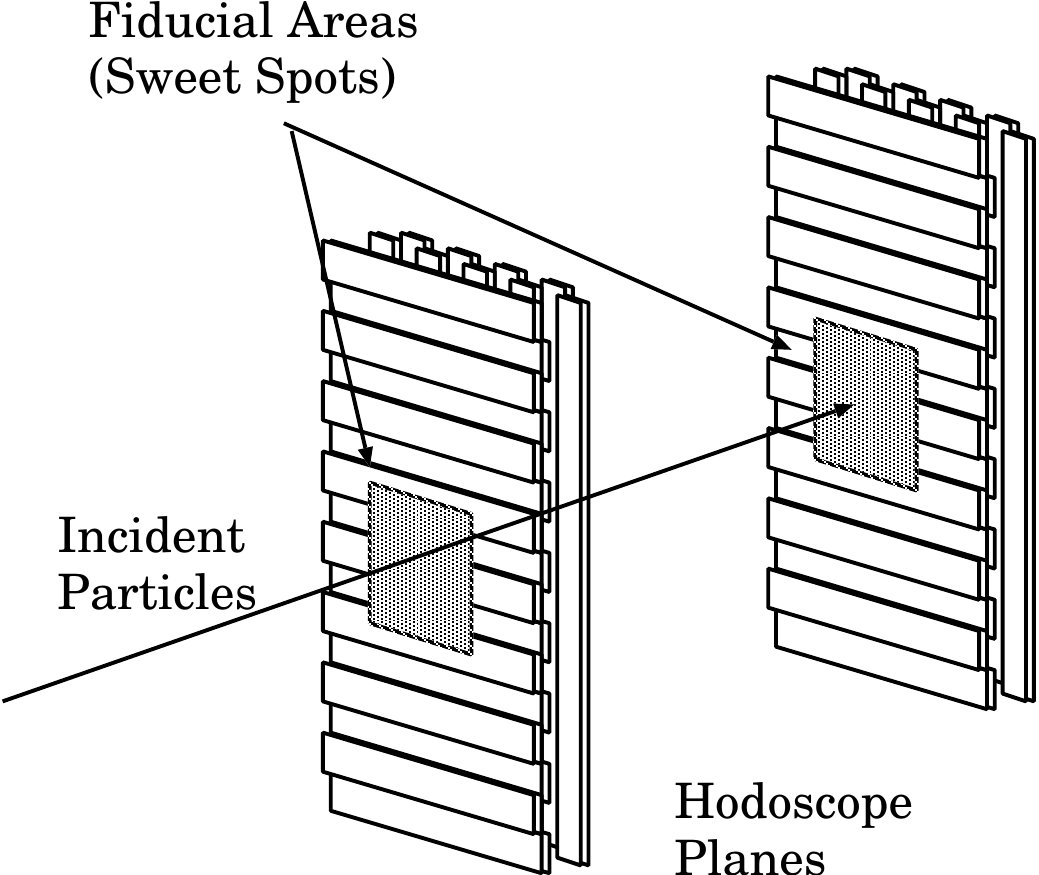, width=3.5in}
\caption{Illustration of the fiducial areas on the scintillator hodoscopes used for determination of the tracking efficiency.}\label{fig:swit_spot}
\end{center}
\end{figure}
Lost tracks happen when the tracking algorithm fails to reconstruct a track.
There are reasons for this: first the ability of hardware to handle high rates and work 
effectively, second, the prohibitively high computer time needed to reconstruct 
an event when there are multiple hits in drift chambers caused by background, and third, 
at high rates it is possible to have two legitimate tracks and the tracking algorithm 
is not able to reconstruct either of these events. 
In order to minimize the track reconstruction time the Drift Chamber TDC window
is chosen to be 250 ns, it is required that each chamber have no more than 25 hits, and the maximum number 
of focal plane tracks should be less than 10.
\begin{figure}[ht]
\begin{center}
\epsfig{file=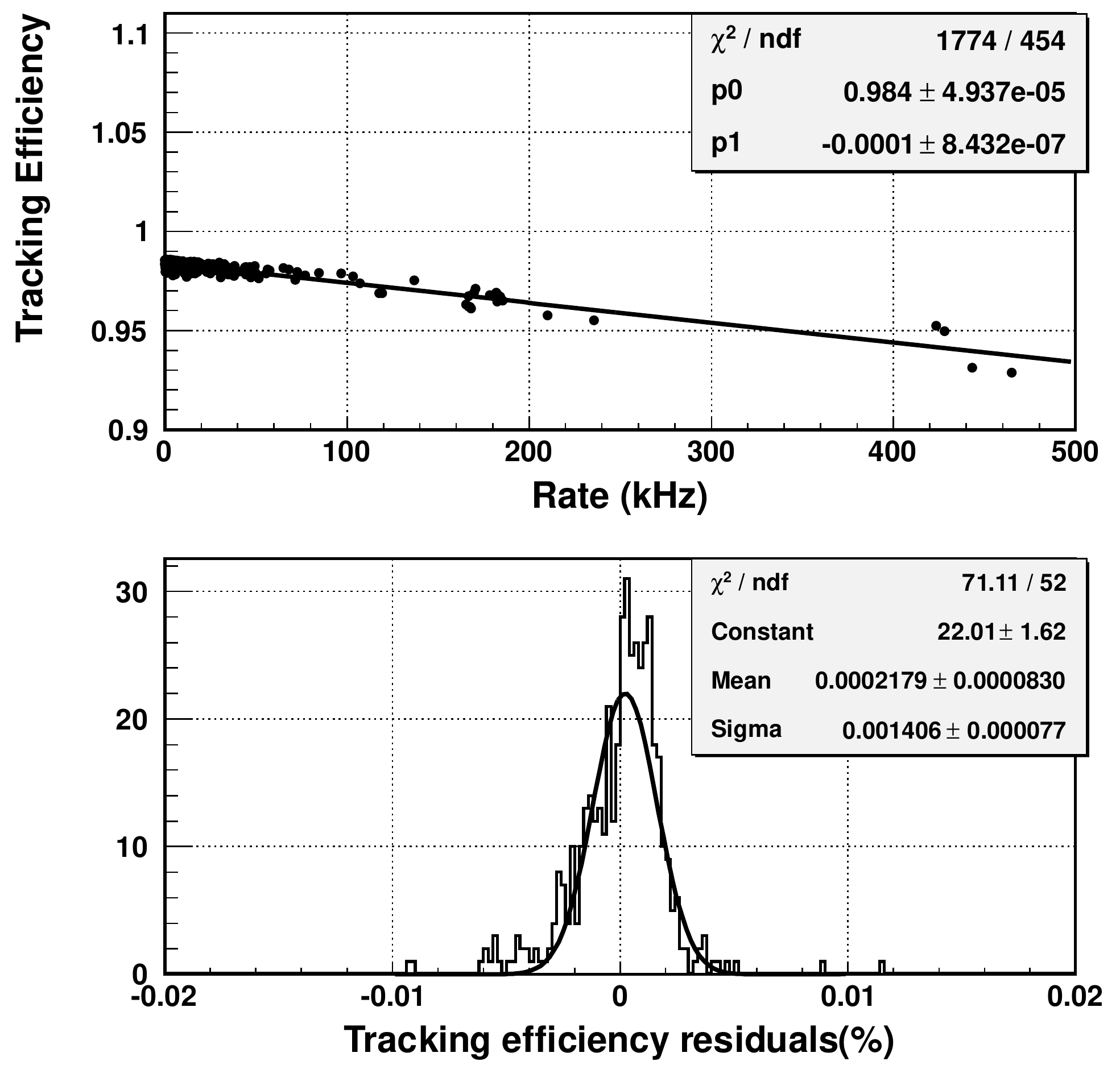, width=5.0in}
\end{center}
\caption{ Tracking efficiency as a function of the scintillator (3/4) rate.
\label{fig:track_eff}}
\end{figure}
In order to calculate the tracking efficiency, only the central part of the HMS acceptance
is chosen (a fiducial region). 
The fiducial area consists of paddles 4-13 in the X planes and 4-7 in the Y planes, as shown in Fig.~\ref{fig:swit_spot}.
This defines the fiducial area, and guarantees that only one particle passed through both drift  
chambers within the central area of the spectrometer acceptance and should have been tracked.

In order to calculate the tracking efficiency a clean sample of electrons 
is selected using electron identification cuts. 
The tracking efficiency is calculated as the number of events for which a track has been found, 
divided by the number of total events passing the electron identification cuts as, 
\vspace{-1.0cm}
\begin{center}
\begin{equation}
\epsilon_{(tracking)}={ Trigger \textrm{ \& } PID \textrm{ \& } Track \over Trigger \textrm{ \& } PID} \phantom{l},
\end{equation}
\end{center}
where Trigger indicates that there is a trigger, Track is that at least one track is found and PID represents the particle identification cut,
\begin{displaymath}
PID=hcer\_npe > 2.0 \hspace{1mm}\&\hspace{1mm} hcal\_et/hpcentral > 0.7.
\end{displaymath}

The tracking efficiency calculated for each run as a function of the scintillator (3/4) rate 
is shown in Fig.~\ref{fig:track_eff}. 
The tracking efficiency has been parametrized as a function of the scintillator (3/4) rate and
during the data analysis this parametrization has been used on a run-by-run basis. 
The systematic uncertainty is estimated as the spread of the measured tracking efficiency from the 
parametrization and is equal to 0.15\%. 
For a detailed description of tracking see Ref.~\cite{vlad}.

\subsubsection{Acceptance Calculation} \label{Sec:Acceptance}

Knowledge of the HMS acceptance is one of the dominant sources of uncertainties 
in determining the cross section.
For the HMS the acceptance is normally defined by a collimator and subsequent magnet apertures. 
As the magnet apertures partially define the acceptance, the magnetic model must be known in 
order to find the acceptance function. 

For a given angle and momentum the HMS can detect particles which have angles
and momentum around the given values.
The HMS acceptance is a function of the three target coordinates $X$, $Y$, $Z$ and three
spectrometer coordinates $\delta$, $X'$, $Y'$. 
For thin targets the cross section is independent of $X$, $Y$, $Z$ and therefore the acceptance is a function 
of only $\delta$, $X'$, $Y'$. 
Spectrometer angles $X'$, $Y'$, defined in Section~\ref{subsection:TargetCoordRec}, 
can be related to the polar angle $\theta$ by the following formula
\vspace{-1.0cm}
\begin{center}
\begin{equation}
\theta=\arccos(\cos(X')\cos(\theta_{HMS}-Y'))
\end{equation}
\end{center}
where $\theta_{HMS}$ is the central angle of the HMS.

\begin{figure}[ht]
\begin{center}
\epsfig{file=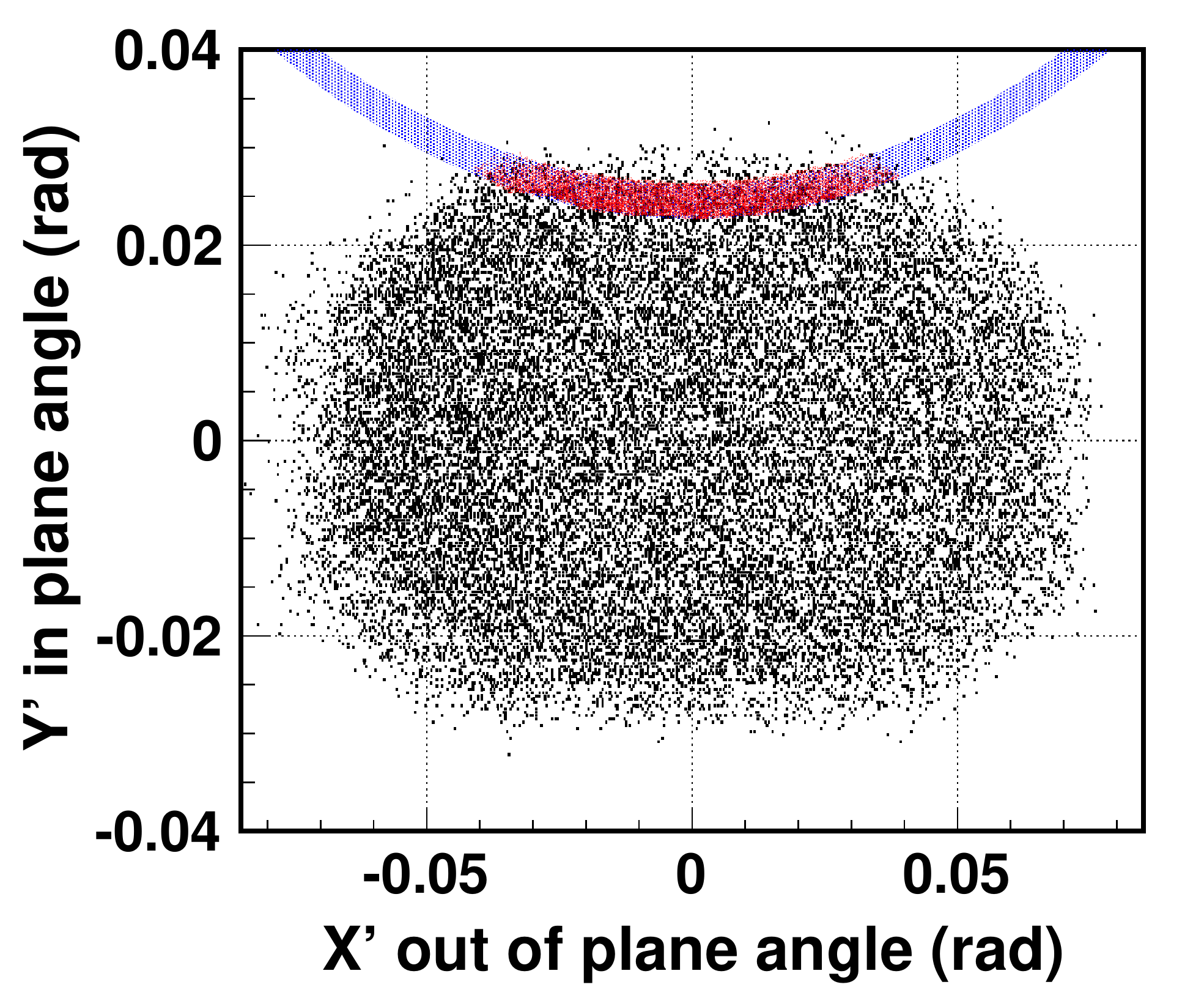, width=4.0in}
\end{center}
\caption{ Solid angle calculation for each theta bin. The blue region is the solid angle of a 
theta bin when there is no aperture, the red region is when the aperture of HMS is applied. 
The solid angle is equal to the ratio of events with aperture on and off multiplied by the total solid angle of $4 X'Y'$.}
\label{fig:solidangle}
\end{figure}

Since the inclusive cross section is independent of azimuthal angle $\phi$, the acceptance is a function of 
only two variables, $A(\delta,\theta)$ where $\delta$ is the momentum fraction $\theta$ and is the polar angle .
Before calculating the acceptance the solid angle of each $\theta$ bin is calculated.
During the cross section analysis events are binned in twenty $\theta$ bins from $-$0.035 to $+$0.035 radian.
For each bin the solid angle is calculated from a Monte Carlo. 
First, events are generated within the range $|X'|<0.1$ and $|Y'|<0.1$ and than the number of events that
passed through the acceptance of the HMS are compared to the number of initially generated events. 
In Fig.~\ref{fig:solidangle} the thickness of the blue region is the width of the $\theta$ bin (0.0035 radian). 
If the aperture is determined by the collimator only, the solid angle will be the blue region,
but since the HMS has other apertures, the real solid angle is determined in the following way.
The solid angle for that given $\theta$ bin is calculated as the ratio of number of events in red region 
to number of events in the blue region (events are sampled within $|X'|<0.08$ and $|Y'|<0.04$)
multiplied by total solid angle of $4 X'Y'$.

To calculate the acceptance function $A(E',\theta)$ events are generated in $X$, $Y$, $Z$, $\delta$, $X'$, $Y'$
and binned in $E',\theta$. 
In general it is convenient to use $W^{2}$ bins instead of $\delta(E')$ bins. 
There are several advantages in using $W^{2}$ bins over $\delta$ bins.
For example, in the kinematic range where the cross section varies significantly the density of
$W^{2}$ bins is higher than the density of $\delta$ bins. 
Also the radiative corrections are calculated at the centers of the $W^{2}$ bins, reducing the model cross section interpolation errors. 

In this analysis $W^{2}$ binning is not done in a direct way. Instead of binning the data in $W^{2}$ bins, 
data are binned in $E'$ bins in such a manner that the lowest and the highest $E'$ bins are within the 
known range of the HMS momentum acceptance of $|\delta|<8$ and at the same time correspond to a point in the $W^{2}$ grid. 
Here the $W^{2}$ grid points are given by this formula $W^{2}_{i}=0.05+dW^{2} \times i$, where $i=0,150$ and $dW^{2}=0.04$ GeV$^2$. 
This binning method is chosen over the direct $W^{2}$ binning based on the following consideration:
the acceptance does not always include a symmetric region in $\theta$ about the central value in $W^{2}$, 
for $W^{2}$ bins corresponding to high or low values of $\delta$ at the central angle only part of the $\theta$ 
acceptance lies within the spectrometer acceptance. 
The result is the maximum bin centering corrections are at the edge of the acceptance, where the acceptance is 
more poorly known than in the central part. 
Therefore, for $W^{2}$ bins at the edge of the momentum acceptance, the systematic error is larger than for $W^{2}$ bins in the central part. 
Also, taking into account the fact that there are less events available at the edge of the acceptance, the edge bins will have bigger statistical errors.

There are three main elements in the Monte Carlo simulation code: the event generator, the transport of 
the particles through the magnets, and the list of materials and apertures that cause multiple scattering or 
stop the particles. The initial coordinates are randomly generated along the target length, while the quantities
$\delta$, $\theta$ and $\phi$ are chosen randomly within their allowed limits. 
The particle is transported through the spectrometer to the detector hut using the computer program COSY 
Infinity~\cite{cosy}, which models the magnetic part of the spectrometer, the magnet positions, their internal 
dimensions and their magnetic field maps. 
If the particle successfully traversed the spectrometer and the detector stack, it is considered a success and 
contributes to $N_{success}$. 
The spectrometer acceptance is calculated by the following formula,
\vspace*{-1cm}
\begin{center}
\begin{equation}
Acceptance(\theta_{i}-\theta_{o},E'_{j})={N_{success}(\theta_{i}-\theta_{o},E'_{j}) \over N_{gen}(\theta_{i}-\theta_{o},E'_{j})}\phantom{l},
\end{equation}
\end{center}
where N$_{gen}(\theta_{i}-\theta_{o},E'_{j})$ and N$_{success}(\theta_{i}-\theta_{o},E'_{j})$ are the number of events generated 
and detected in a given $(\theta_{i}-\theta_{o},E'_{j})$ bin, respectively.
In the cross section analysis data are binned in $(\theta_{i}-\theta_{o},E'_{j})$  bins and each bin is
corrected by the corresponding $Acceptance(\theta_{i}-\theta_{o},E'_{j})$.
\begin{figure}[htb]
\begin{center}
\epsfig{file=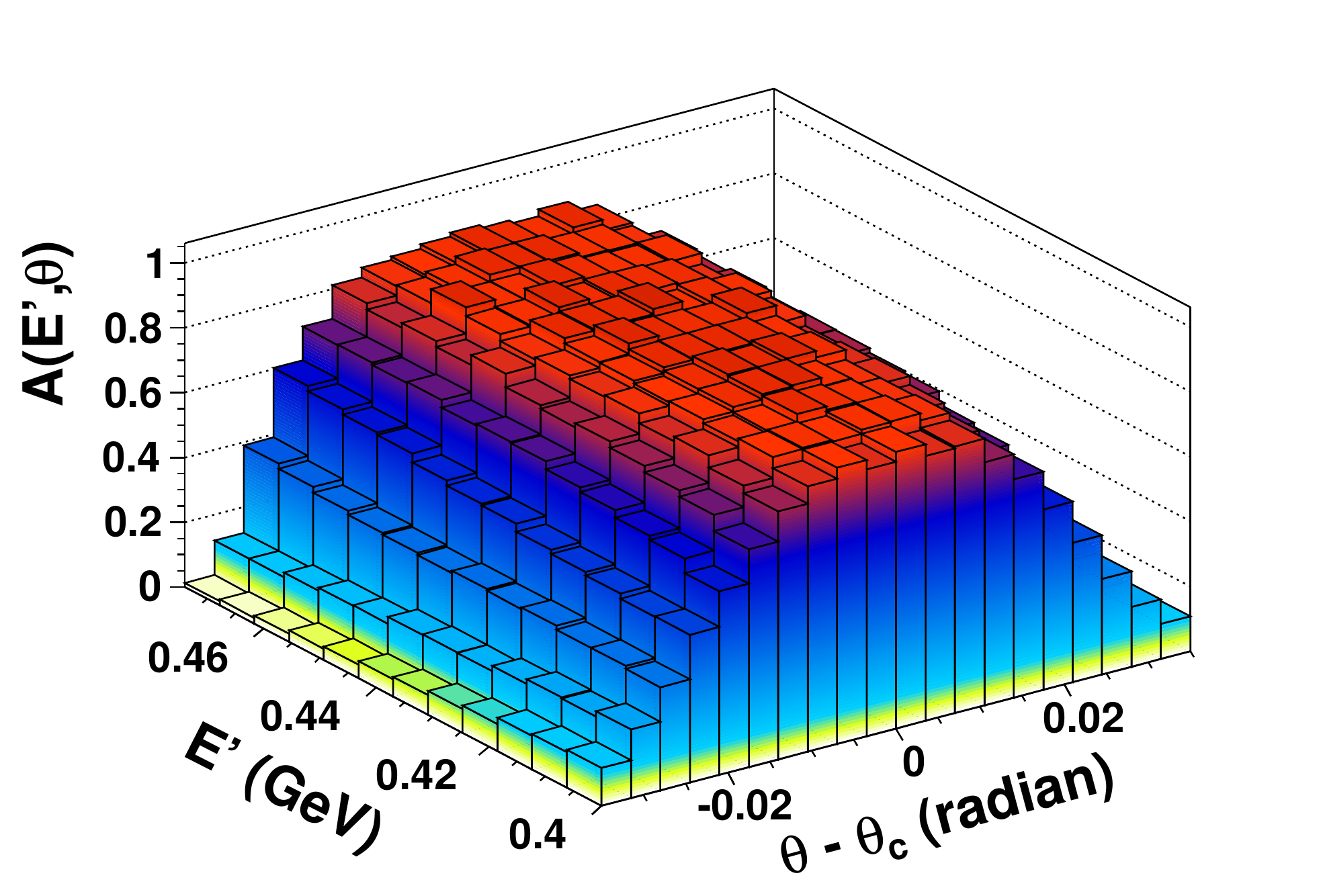, width=4.5in}
\end{center}
\caption{ HMS Acceptance function at $P_{HMS}=$ 0.44 GeV and $\theta=$ 75 degree, calculated by HMS Single-Arm Monte Carlo program.}
\label{fig:acceptance}
\end{figure}
In Fig.~\ref{fig:acceptance} the HMS acceptance is shown for $P_{HMS}=$ 0.44 GeV and $\theta=$75$^{o}$, 
obtained by generating five million Monte Carlo events. 
One can see that the momentum range used in this analysis ($|\delta|<8$\%) is well within the flat range, 
while the angular acceptance falls off quickly at the age of the angular acceptance.
This fall off is due to the shape of the collimator. 

\begin{figure}[htb]
\begin{center}
\epsfig{file=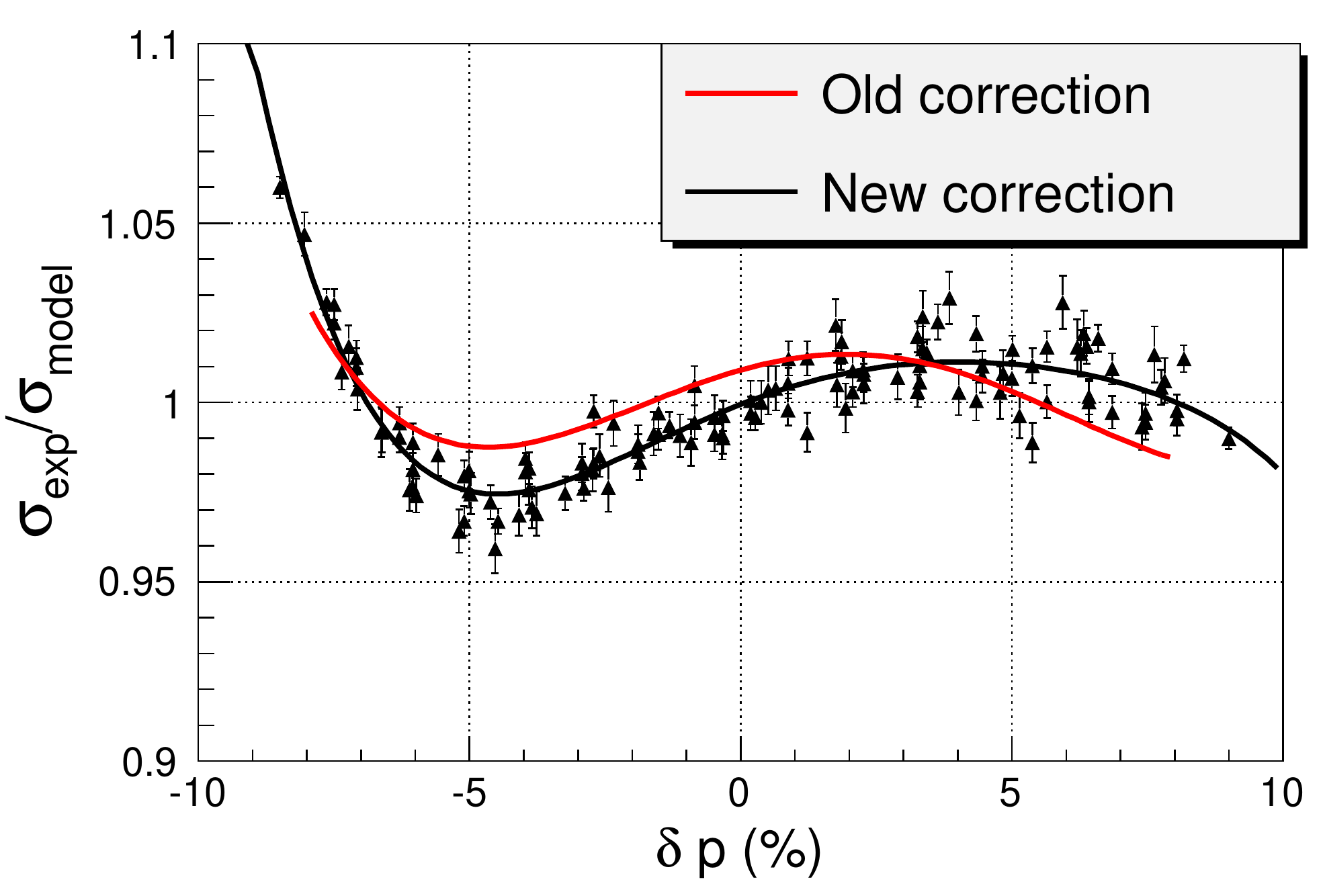, width=4.0in}
\end{center}
\caption{ Ratio of experimental to model cross sections versus $\delta p$. 
The red line is the same ratio obtained in a previous experiment, E99-118.
This effect is caused by optic mis-calibration.
}
\label{fig:dpcorr}
\end{figure}

During the cross section analysis a systematic dependence of the experimental cross section divided by the model cross 
section was observed and is shown in Fig.~\ref{fig:dpcorr}. 
After studying this effect under different kinematic conditions and offsets no target or kinematic ($E, E', \theta$) dependence was found. 
This effect is believed to be caused by mis-calibration of the optics of the HMS.
In order to correct for this effect a set of runs are selected at different HMS central momenta and angles. 
These runs are selected to have good agreement with the model cross section in order to rule out any
model dependence in the extraction of the correction function. 
The cross section ratios are parametrized as function of $\delta p$ and are used in cross section analysis in bin-by-bin basis.
The uncertainty associated with this parametrization is $\sim$ 0.6\%, estimated from the spread of data to the fit.

In addition, the position uncertainties on the target, collimator, magnets, and detector package contribute to both the 
\mbox{point-to-point} and normalized acceptance uncertainties. 
The total \mbox{point-to-point} uncertainty due to the position uncertainties on the target, collimator, magnets, and detector
package is $\sim$ 0.3\%. 
As a result, the total point-to-point (kinematic dependent) uncertainty on the acceptance correction 
was $\sim$ 0.7\%. 
The normalized (kinematic independent) uncertainty on the HMS acceptance is $\sim$ 0.6\%, 
which is determined by changing the positions slightly of the collimator, magnets, and detector package
respectively in the HMS Single-Arm Monte Carlo code and checking how this affected the extracted cross sections.

\subsubsection{Bin Centering Corrections}

The goal of the analysis is to extract electron-nucleus inclusive differential cross section for 
a range of E$'$ at a fixed scattering angles. 
During data acquisition each run is taken at a particular HMS angle and momentum. 
Since the HMS spectrometer has a relatively large angular acceptance of $\pm$ 2 degrees, 
the cross section can vary greatly across the angular acceptance. 
In order to calculate the cross section for each E$'$ bin at fixed $\theta$ value, a $\theta$
bin centering correction must be done. 
If the cross section varied linearly across a symmetric $\theta$ acceptance there 
will be no need for bin centering correction.

This correction removes the cross section dependence in the angular acceptance range using a known model cross section.
The bin centering correction depends on the model cross section used, and the model dependence in this correction 
can be a large systematic uncertainty in the analysis. 
The correction is applied by rescaling each $(E',\theta)$ bin by the ratio of its cross section at the center of the bin 
to the central cross section $(E',\theta_{o})$. 
The bin centering correction for $(E'_{i},\theta_{o})$ bin is calculated with this formula:
\vspace*{-1cm}
\begin{center}
\begin{equation}
BC(E, E', \theta) ={\sigma^{model}_{rad}(E, E'_{i}, \theta^{central})} \Bigg/ {\sigma^{model}_{rad}(E, E'_{i}, \theta)}, 
\label{eq:bccorr} \phantom{l}
\end{equation}
\end{center}
where $E'_{i}$ is the center of the $E'$ bin, $\sigma^{model}_{rad}$ is the total radiated cross section.

\subsubsection{Radiative Corrections} \label{sec:RadCorr}

\begin{figure}[htb]
\begin{center}
\epsfig{file=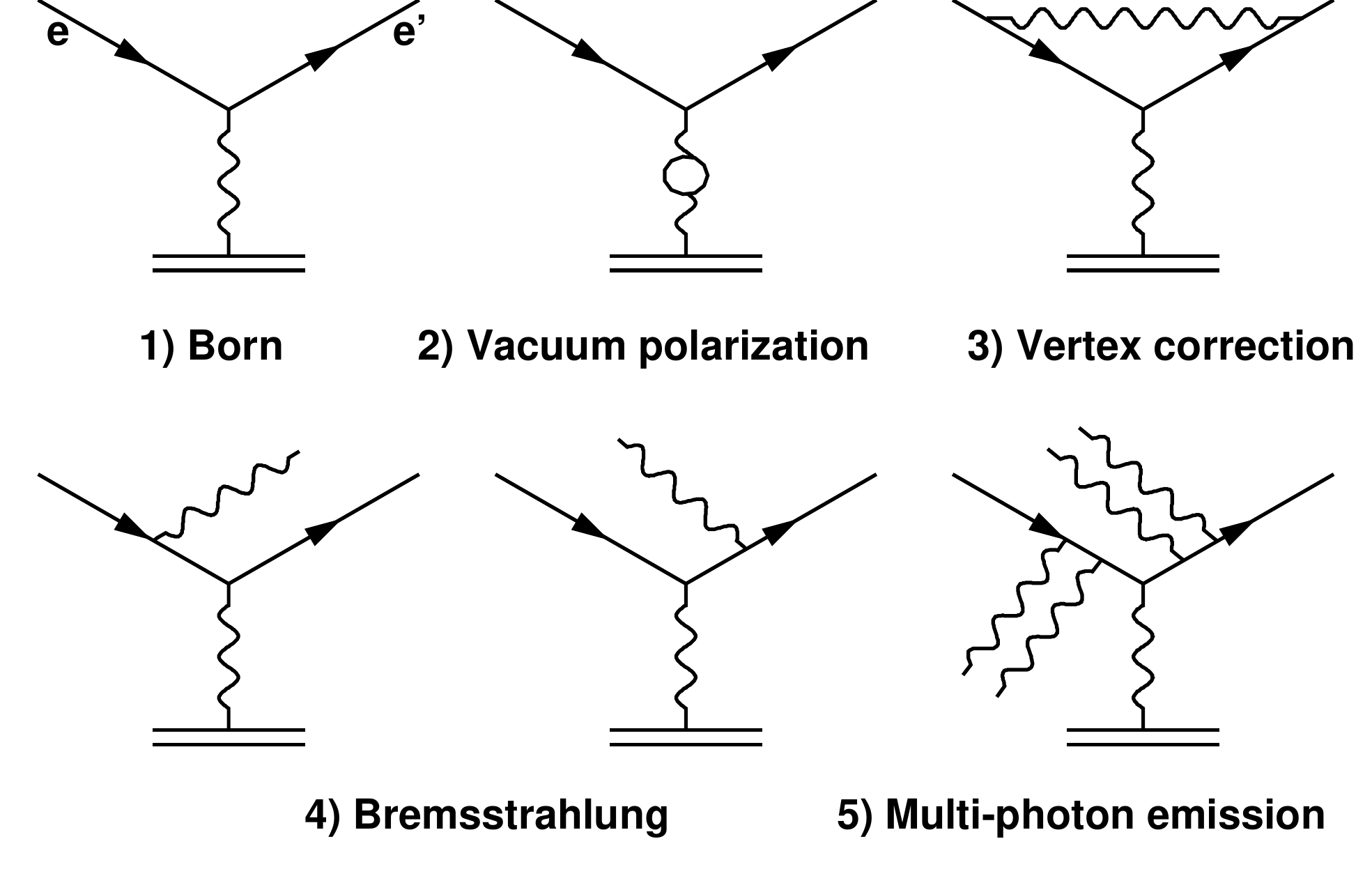, width=5.0in}
\end{center}
\caption{ 1) is the lowest-order diagram for charged-lepton-nucleon scattering.
2)-4) are the lowest-order electromagnetic radiative corrections. }
\label{fig:internal_rad}
\end{figure}

The differential cross sections measured in the resonance region for electron-nucleon 
scattering may have large contributions from processes other than the one 
photon exchange approximation.
This is also called Born approximation and is shown in Fig.~\ref{fig:internal_rad}.
The Feynman diagrams for higher order electromagnetic processes in $\alpha$ 
are shown in ( 2, 3, 4, 5, ) which include 
vacuum polarization (creation and annihilation of particle-antiparticle pairs), 
vertex processes (emission and reabsorption of virtual photons), 
and bremsstrahlung (emission of real photons in the field of the nucleon during interaction). 

The cross section measured in the kinematic range of this experiment may have up to 
a 30\% contribution from those processes.
In order to determine the differential cross section for the one$-$photon exchange process, 
all the other contributions from the higher order processes in $\alpha$ have to be estimated
and corrected for in the measured cross section. 

The impact of the radiative process to experimental data has the following three effects:

\begin{itemize}
\item{ The incoming particle's momentum may be reduced, thereby changing the kinematic configuration for the event, 
       and also the probability for the interaction (through the energy dependence of the cross section).}
\item{ The outgoing particle's momentum may be reduced, moving events from one kinematical configuration to another. }
\item{ The overall cross section for the process in a particular kinematical configuration will change. }
\end{itemize}

\begin{figure}[htb]
\begin{center}
\epsfig{file=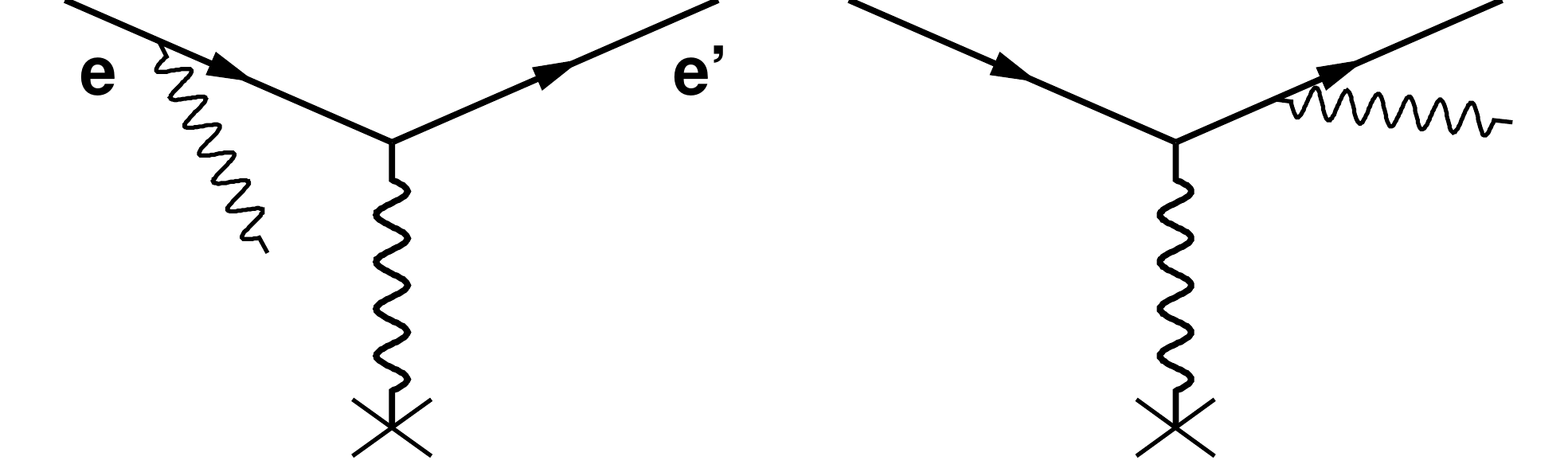, width=4.0in}
\end{center}
\caption{ The Feynman diagrams for external radiative processes.}
\label{fig:external_rad}
\end{figure}

These effects are taken into account by applying radiative corrections to the measured cross section. 
Electromagnetic radiative effects in electron scattering are divided into two categories: 
external and internal.
The external radiative processes, shown in Fig.~\ref{fig:external_rad}, can take place before and 
after the scattering in the material the electrons passes through (the largest contribution happens in the target material). 
Both cases have to be corrected in order to calculate the Born cross section. 
Corrections for ionization energy losses are also applied, but have a much smaller contribution than the radiative processes.

Internal effects occur at the scattering vertex and are calculable in Quantum Electrodynamics.
The internal effects include internal bremsstrahlung (emission of photons in the field of 
the nucleon during the scattering process), vacuum polarization, vertex processes, 
and multiple photon exchange (important at low $Q^2$).
In the first order in $\alpha$ there is also contribution from the diagrams of vacuum polarization by 
hadrons, which is not shown here but is taken into account in Ref.~\cite{Bardin}.

The program used to calculate radiative corrections for this experiment, including
both the internal and external effects and the elastic tail, are based on the
program developed at SLAC which is described in Ref.~\cite{Dasu}.

The measured cross section is the combination of radiated elastic, quasi-elastic and inelastic cross sections,
\begin{figure}[!htb]
\begin{center}
\epsfig{file=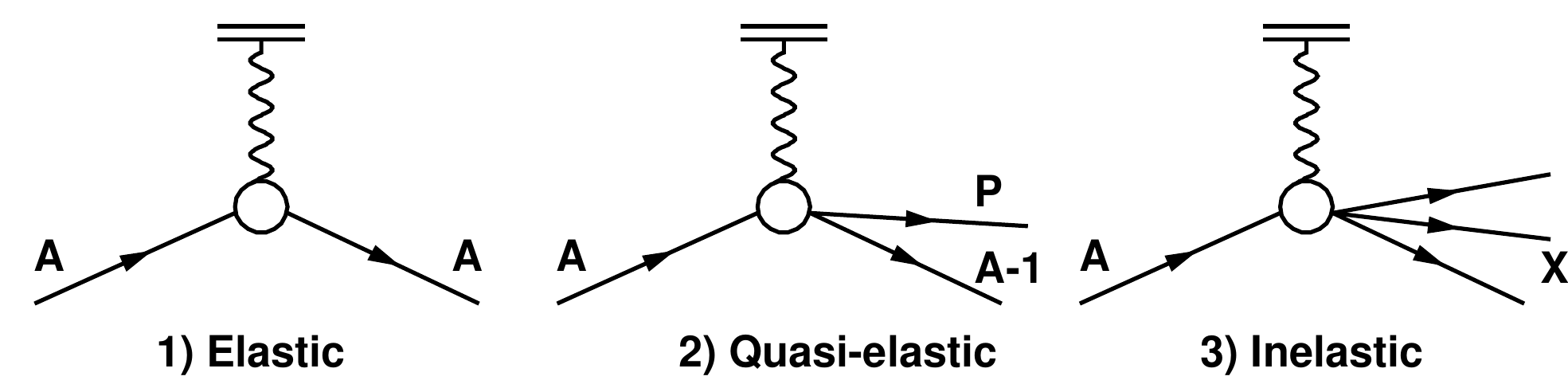, width=5.5in}
\end{center}
\caption{ Feynman diagrams of processes present in the kinematic range of this experiment. }
\label{fig:LepHadrScatt}
\end{figure}
are depicted in Fig.~\ref{fig:LepHadrScatt}, and the total measured cross section is the sum of all three, 
\vspace{-1.0cm}
\begin{center}
\begin{equation}
\sigma_{meas}  = \sigma_{el}^{rad} + \sigma_{qe}^{rad} + \sigma_{inel}^{rad} \phantom{l}, 
\end{equation}
\end{center} 
and in inclusive experiments the inelastic and quasi-elastic cross sections are not separated. 

The total radiative cross section (quasi-elastic $+$ inelastic) is proportional to the Born cross section, 
and can be calculated by the following formula:
\vspace{-1.0cm}
\begin{center}
\begin{equation}
\displaystyle\sigma_{Born}^{inel+qe}  = \displaystyle \left[ \frac{\sigma_{meas}-\sigma_{el}^{rad}}{\sigma_{rad}^{mod}-\sigma_{el}^{rad}}\right] \times \sigma_{Born}^{mod}
\label{eq:CSBornMesured}
\end{equation}
\end{center} 
where $\sigma_{rad}^{mod}$ is the total radiated model cross section (internal + external ), $\sigma_{el}^{rad}$ is the elastic radiated model 
cross section (internal + external ).
The radiated model cross section is the convolution of internal and external radiative cross sections:
\vspace{-1.0cm}
\begin{center}
\begin{equation}
\displaystyle\sigma_{rad}^{mod}=internal \otimes external \otimes Born.
\end{equation}
\end{center} 
For this experiment, the external radiative corrections are computed using a complete calculation
of Mo-Tsai~\cite{Mo:1968cg} with a few approximations. 
This approach, “MTEQUI”, uses the equivalent radiator approximation~\cite{Dasu}. 
In the equivalent radiator method, the effect of internal Bremsstrahlung is calculated using two 
hypothetical radiators of equal radiation length, one placed before and one after the scattering. 
It is important to note that the energy-peaking approximation is not used for the computation of
external contributions. 
The internal contribution in MTEQUI method is evaluated by setting the radiation length of the 
material before and after the scattering point to zero, and ignoring the target length 
integral (see Eqn. C1 in Ref.~\cite{Mo:1968cg}). 

The internal radiative corrections $\sigma_{int}^{rad}$ have been calculated using the formalism of Bardin~\cite{Bardin}.
The calculations were done for all diagrams in Fig.~\ref{fig:internal_rad} without any approximation. 

In order to reduce the effects of any approximations in treating the external radiative effects we use the following 
expression to calculate $\sigma_{Born}$, 
\vspace{-1.0cm}
\begin{center}
\begin{equation}
\sigma_{rad}^{model}  = \displaystyle\sum_{ii=el,qe,inel}\sigma_{ii,int}^{rad}(Bardin)\times\left( \displaystyle\frac{\displaystyle\sum_{ii=el,qe,inel}(\sigma_{ii,int}^{rad}+\sigma_{ii,ext}^{rad})}{\displaystyle\sum_{ii=el,qe,inel}\sigma_{ii,int}^{rad}}\right)_{Mo,Tsai}.
\end{equation}
\end{center} 
It was found that using Mo and Tsai's MTEQUI method to treat external radiative correction 
is consistent for targets with different radiation lengths, see Ref.~\cite{Dasu}.

\begin{figure}[p]
\begin{center}
\epsfig{file=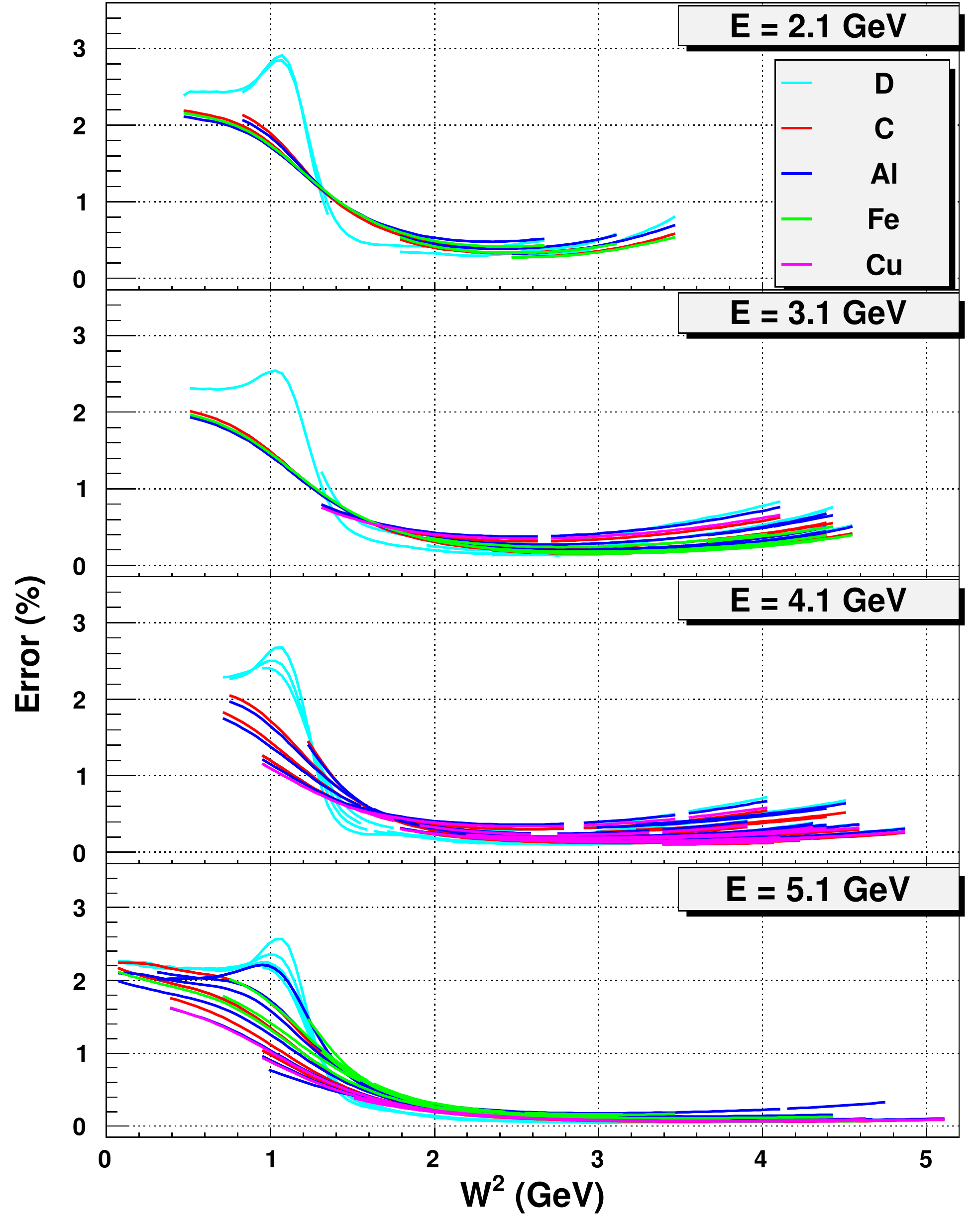, width=5.5in}
\end{center}
\caption{ Systematic errors due to contribution from $\alpha^2$ processes are shown in four different beam energies.
The error is set to be 3\% of the size of the quasi-elastic tail, relative to the inelastic~\cite{BostedPCom}. 
The error has a slight $\epsilon$ (beam energy) dependence following Ref.~\cite{Dasu}.
}
\label{fig:RadCorrSystAlpha2}
\end{figure}

The uncertainty on the cross sections due to the radiative correction is estimated
at 1\% for the $\epsilon$ dependence, and the normalized uncertainty is also $\sim$ 1\%,
according to the radiative correction studies done at SLAC~\cite{Dasu}. 
The $\epsilon$ dependent uncertainty and normalization uncertainty caused by the 
two-photon corrections to the one photon exchange diagram are estimated to be 
$\sim$ 0.3\% respectively, according to the results of an early SLAC experiment~\cite{Fancher:1976ea}.
That experiment measured the difference of electron/positron cross sections at Q$^2$=1.2-3.3 GeV$^2$ and 
W$^{2}$=1.3-17.0 GeV$^{2}$ range (comparable to the kinematic range of this experiment) and 
found $\sigma_{e^{+}}/\sigma_{e^{-}}$=1.0027$\pm$ 0.0035 with no significant dependence on Q$^2$ or W$^{2}$.
Since no significant difference was found between electron/positron cross sections in that experiment, one can conclude that 
two photon exchange contributions are small in the kinematic range of this experiment. 

In the kinematic range of this experiment the radiative corrections are a maximum of 30\%.
Systematic errors due to contributions from second order in $\alpha$ processes are set at 3\% of the size of the quasi-elastic tail, 
relative to the inelastic~\cite{BostedPCom}, see Fig.~\ref{fig:RadCorrSystAlpha2}. 
The error is less than 1\% for $W^2>1.4$ and has slight $\epsilon$ (beam energy) dependence. 

The Born model cross section used for this analysis is discussed in Sec.~\ref{sec:ModelCS}.
The radiative correction is applied to measured cross sections on bin-by-bin basis to get the corrected cross sections.

The recoil polarization measurements of the form factor ratio $\mu_pG_E^p/G_M^p$\cite{Jones:1999rz,Gayou:2001qd}  contradict the Rosenbluth measurements (See Ref.~\cite{Bosted:1994tm} for a compilation and references) and it has been suggested  that the earlier experiments might have not have fully understood their  systematic errors or had normalization problems. The Rosenbluth measurements have been reexamined~\cite{Arrington:2003df} and this global reanalysis could find no systematic or normalization problems that could account for the discrepancy. The author of Ref.~\cite{Arrington:2003df} concluded that a modest linear $\epsilon$-dependence correction (of origin yet unknown) to the cross section measurements might explain the difference.  Several investigators~\cite{Blunden:2003sp,Rekalo:2003xa,Chen:2004tw} have explored the possibility of two-photon exchange corrections (which would be less important in the direct ratio measurement of recoil polarization) to explain the discrepancy. While only incomplete calculations exist, the results of Ref.~\cite{Blunden:2003sp,Chen:2004tw} account for part of the difference. 

As this experiment exploits the Rosenbluth technique the question naturally arises whether two-photon effects might play role. Unfortunately no calculations have been done specifically to answer this question.  Nonetheless we have attempted to gauge what effect an $\epsilon$ dependence might have in our results. To do so we artificially introduced a 2\% $\epsilon$ dependence in the radiative corrections which resulted in $\pm$0.028 uncertainty in $R$, see Sec.~\ref{sec:RosSepR}. Hence we can conclude that for our specific conditions the neglect of two photon effects (as manifested in a $\epsilon$-dependence) has only modest consequences.

\subsubsection{Coulomb Correction} \label{sec:CoulumbCorr}

In the field of high Z nuclear targets the incoming and outgoing electron wave functions are affected by the Coulomb field.
From the classical point of view the incoming electron accelerates in the field of the positive 
charged nuclei and the scattered outgoing electron decelerates.
This effect causes an increase in the momentum of incoming beam electron 
and a decrease in the momentum of the scattered electron relative to the vertex values.
At high enough beam energies this effect is negligible while at the beam energies of this experiment 
it is not and has to be taken into account.
The change of the vertex kinematic variables can have a big impact on the measured cross sections. 
The change of the energy of the incoming electron will change the cross section of the process of interest 
while the change of the scattered electrons energy change will cause it to populate a different kinematic bin
and change the measured cross section. 

Coulomb corrections have the result that the plane wave Born approximation loses its validity and hence 
requires a correction. 
For this experiment the Effective Momentum Approximation (EMA) is used as described in Ref.~\cite{Aste:2005wc}
and discussed below.

Assuming a spherical charge distribution in the nucleus, the electrostatic
potential inside the charged sphere can be defined as followed:
\vspace{-1.0cm}
\begin{center}
\begin{equation}
V(r)=-\frac{3\alpha(Z-1)}{2R}+\frac{\alpha(Z-1)}{2R}\frac{r}{R}
\end{equation}
\end{center}
where the radius of A is given by this formula:
\vspace{-1.0cm}
\begin{center}
\begin{equation}
R=1.1A^{1/3}+0.86A^{-1/3}.
\end{equation}
\end{center}
Because most the nucleons of heavy nuclei are located in the periphery of the nucleus, 
taking the electrostatic potential at the center of the nucleus will be an overestimate of the Coulomb effect. 
In the effective momentum approximation (EMA)~\cite{Aste:2005wc}, the effective potential used is 
$V_{eff}\approx(0.75-0.8)V(r=0)$. This value for $V_{eff}$ agrees with the extracted effective 
potentials from positron and electron inclusive scattering experiments, see Ref.~\cite{Gueye:1999mm}.
In order to take account the $V_{eff}$, Q$^2$ is replaced by Q$^2_{eff}$ where 
\vspace{-1.0cm}
\begin{center}
\begin{equation}
Q^{2}_{eff}=4(E+V_{eff})(E'+V_{eff})\sin^{2}\left(\frac{\theta}{2}\right)
\end{equation}
\end{center}

\begin{figure}[p]
\begin{center}
\epsfig{file=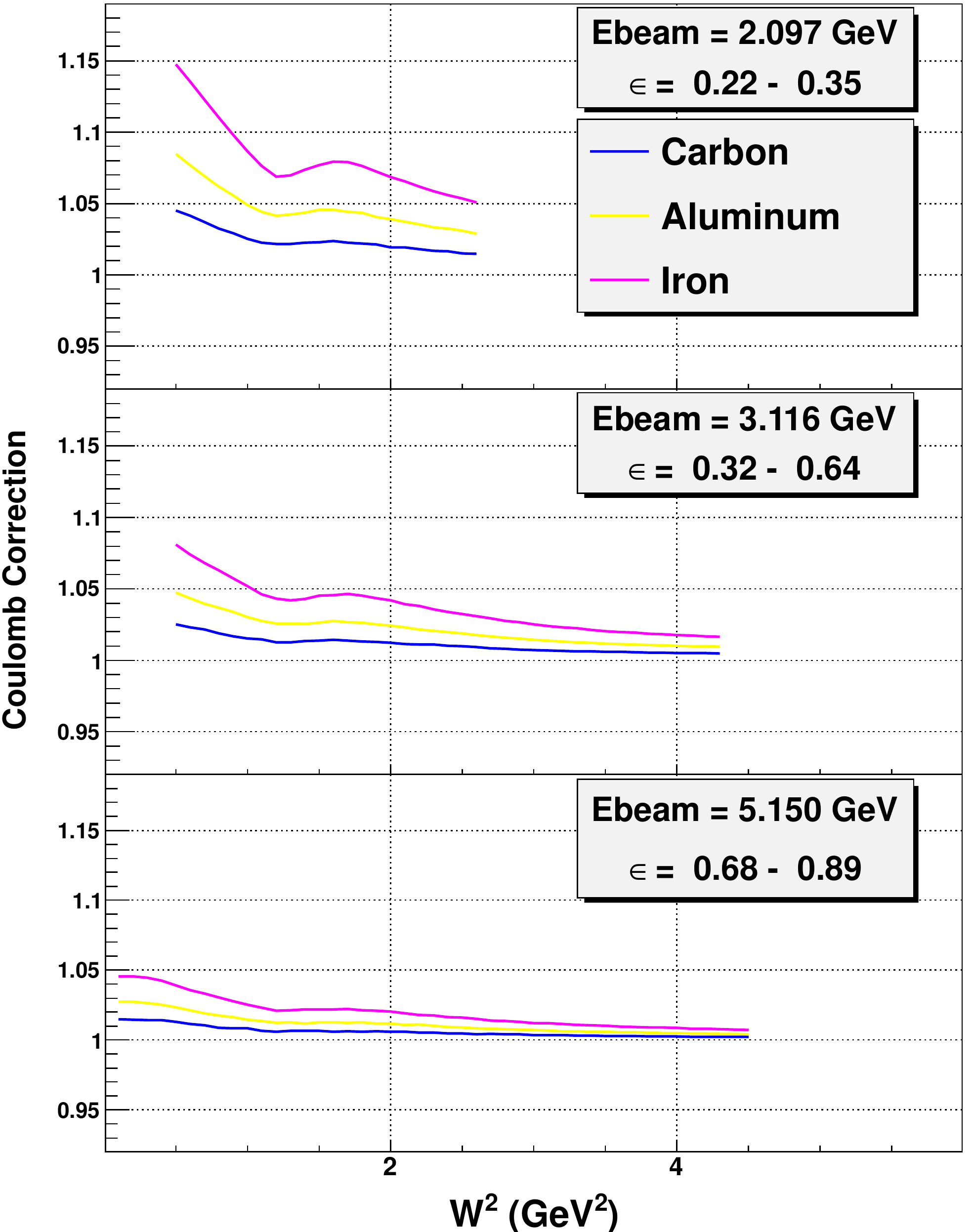, width=5.5in}
\end{center}
\caption{ The Coulomb correction factor as a function of $W^2$, 
for low to high $\epsilon$ range and for carbon, aluminum and iron targets. 
For heavy nuclei, the Coulomb correction factor is significant, reaching a maximum of $\sim$ 15\% at low $W^2$, near the quasi-elastic peak. 
The Coulomb correction is smaller for high beam energy (high $\epsilon$ range) than for low energy (low $\epsilon$ range).
}
\label{fig:CoulombCorr}
\end{figure}

There are two methods to do Coulomb correction: one assumes that the Mott cross section is 
unchanged while the nuclear spectral function is subjected to a transformation when Q$^2$ 
is replaced by Q$^2_{eff}$ defined above. 
The second method changes both the Mott cross section ($Q^2 \rightarrow Q^{2}_{eff}$ and $E' \rightarrow E'+V_{eff}$) 
and the nuclear spectral function. 
Also, for the second method a focusing factor for the incoming electron is given by the following formula 
\vspace{-1.0cm}
\begin{center}
\begin{equation}
F_{foc}=\frac{E_{in}+V_{eff}}{E_{in}}
\end{equation}
\end{center} 
is applied. 
In this experiment the average potential of $V_{eff}=0.775 \times V(r=0,Z,A)$ is used. 
The Coulomb correction is applied to extracted cross section in the following way
\vspace{-1.0cm}
\begin{center}
\begin{equation}
\sigma_{meas}^{cc}(E,E')=\sigma_{meas}(E,E') \times \displaystyle\frac{\sigma_{model}(E,E')}{\sigma^{CC}_{model}(E+V_{eff},E'+V_{eff})}\times\frac{1}{F_{foc}^{2}} \label{eq:CoulombCorr}
\end{equation}
\end{center} 

The Coulomb correction factors versus $W^2$ are shown in Fig.~\ref{fig:CoulombCorr} for beam energies 2.1, 3.1, 5.1 GeV. 
For heavy nuclei, the Coulomb correction factor is significant, reaching a maximum of $\sim$ 15\% at low $W^2$, near the quasi-elastic peak.
Different beam energies indicate a different range in $\epsilon$. 
The Coulomb correction is bigger for low epsilon (low beam energy) data than for high epsilon data (high beam energy). 
It is necessary to apply the Coulomb correction to properly extract the nuclear R=$\sigma_{L}/\sigma_{T}$. 

\subsection{Cross Section Model} \label{sec:ModelCS}

The purpose of this experiment is to measure the Born cross section for inclusive electron nucleus scattering in resonance 
region. 
Since the exact Born cross section for inelastic electron-nuclei scattering is impossible to measure 
due to superposition of different reactions (see Fig.~\ref{fig:LepHadrScatt}) 
present in the kinematic range of this experiment, a model cross section is necessary.
Also a cross section model is necessary for the bin centering correction, radiative corrections, and the Coulomb corrections.

\subsubsection{Quasi-elastic model}

In earlier sections the scaling of the deep inelastic cross section has been discussed. 
This refers to the observation that the cross section for scattering from a nucleon, 
normally of function of two kinematic variables, reduces to a function of a single variable, 
$x=\frac{Q^2}{2m\nu}$ and reflects that the scattering takes place from a structureless, point 
like object that is essentially free within the nucleon.

There exists a analogous scaling behavior in scattering from quasifree nucleons in the nucleus~\cite{Day:1990mf}. 
In the PWIA quasi-elastic scattering, where the nucleons are moving independently in the mean field of the nucleus, 
the cross section as a function of final electron energy $E'$ is a convolution of the spectral function $S(k,E)$ 
with the elementary electron-nucleon cross section $\sigma_{ei}$. 
The spectral function $S(k,E)$ is the joint probability to find a nucleon with momentum $k$ and separation energy 
$E$ in the nucleus. Schematically then, 

$$\frac{d^2\sigma}{d\Omega dE'} \propto \int d\vec{k} \int dE \sigma_{ei}  S_i(k,E) \delta()$$ where a summation 
over all the protons and nucleons is implied and the delta function argument conserves energy and momentum.

At large momentum transfer the expression can be rewritten (with a few important assumptions, see Ref.~\cite{Day:1990mf}) as
 $$\frac{d \sigma^2}{d \Omega dE'} = \sum_{ei} \sigma_{ei} \cdot K \cdot F(y)$$ with K a kinematic factor and $F(y)$  
is the longitudinal momentum distribution. $y$ is the scaling variable and is determined from energy conservation with $\nu = E - E'$,
$$\nu + M_A =[(M_{A-1} + E_{A-1})^2 + \vec{k}^2]^{1/2} + [M^2+(\vec{k}+\vec{q})^2]^{1/2}.$$  $y$ is the minimum momentum of 
the struck nucleon satisfying the previous expression - the longitudinal momentum of the struck nucleon.
 
Hence the quasi-elastic cross section from a particular nucleus is a function $F(y)$ of a single variable, $y$, 
itself a function of $\vec{q}$ and $\nu$, and is independent of them separately - the cross sections {\sl scale}. 
This is called scaling of the first kind.
 
The difference between $x$ and $y$ scaling is that in $y$ scaling the struck object has structure and must be accounted 
for through the division of the cross section by the elementary elastic electron-nucleon cross section (which is a 
strong function of momentum transfer). While useful for the modeling of the cross section, $y$ scaling can provide 
access to the nucleon momentum distribution $n(k)$ through $F(y)$, the scaling function

 $$F(y) = 2 \pi\int_{-y}^\infty kdk \int_{E_{\rm min}}^{E_{\rm max}} dE S(k,E)$$ and
 if  $E_{\rm max} = \infty$ then
 $$F(y) = 2\pi \int_{-y}^\infty kdkn(k) $$ with
 $F(y)$  the longitudinal momentum distribution. Hence quasi-elastic scattering would provide a direct measure of $n(k)$. 
$$n(k) = -\frac{1}{2\pi y}\frac{dF(y)}{dy}.$$


In a series of papers by Donnelly and Sick~\cite{Donnelly:1998xg,Donnelly:1999sw,Maieron:2001it,Amaro:2004bs} and collaborators it was found that scaling of the quasi-elastic cross section could be extended to account for the difference in atomic number A.  This is scaling of the second kind. 
The scaling variable $\psi$ is given by the following formula 
\vspace{-1.0cm}
\begin{center}
\begin{equation}
\psi=\displaystyle\frac{1}{\sqrt{\xi_{F}}}\frac{\lambda-\tau}{\sqrt{(1+\lambda)\tau+\kappa\sqrt{\tau(\tau+1)}}}
\end{equation} 
\label{eq:scale_psi}
\end{center} 
where $\lambda,\kappa,\tau$ are dimensionless energy, 3-momentum and 4-momentum transfers respectively (these variables 
are normalized by a factor ${1}/{m_{N}}$, where $m_{N}$ is the nucleon mass.).
The new scaling variable $\psi$ is related to $y$ but lacks one important property, it does not take 
account the small energy shift $E_{shift}$ which is included in separation energy $E_{s}$. 
An improved phenomenological dimensionless scaling variable is employed in treatments of 
superscaling~\cite{Cenni:1996zh}, which included the empirical shift E$_{shift}$. 
One should note that scaling function is very sensitive to $k_{F}$ and yet it is possible to find a 
value for it to line up all data on one curve. 
The dependence from E$_{shift}$ is not that pronounced but it needs to be taken into account to put the quasi-elastic peak in the
place where the scaling variable $\psi$ is zero.

In order to fix this a new variable $\psi'$ is introduced, which has the same form as $\psi$
but the dimensionless variables $\lambda$ and $\tau$ are shifted.
The new scaling variable is given by the following formula
\vspace{-1.0cm}
\begin{center}
\begin{equation}
\psi'=\displaystyle\frac{1}{\sqrt{\xi_{F}}}\frac{\lambda'-\tau'}{\sqrt{(1+\lambda')\tau'+\kappa'\sqrt{\tau'(\tau'+1)}}}
\end{equation} 
\label{eq:scale_psi1}
\end{center} 
where $\lambda_{shift}={E_{shift}}/{2m_{N}}$, $\lambda'=\lambda-\lambda_{shift}$, $\tau'=\sqrt{(k^2-\lambda'^2)}$.
The scaling function in $\psi'$ is given by the following formula:  
\vspace{-1.0cm}
\begin{center}
\begin{equation}
F(\psi')=\frac{1.5576 }{(1+1.772^2(\psi'+0.3014)^2)(1+e^{(-2.4291\psi')})k_{F}}, 
\end{equation}
\end{center} 
see Ref.~\cite{Amaro:2004bs} for details. The values of $E_{shift}$ and $k_{F}$ are given in the Table~\ref{tab:EshiftandKF}.
The structure functions $F_{1}$ and $F_{2}$ are calculated using the following formulas
\vspace{-1.0cm}
\begin{center}
\begin{equation}
\begin{array}{l}
F_{1} = M_{p}F(\psi') G_{T}/2\\
F_{2} = \nu F(\psi') (\nu_{L} G_{L} + \nu_{T} G_{T})
\end{array}
\end{equation}
\end{center} 
In the above equation the $G_{L}$ and $G_{T}$ are related to the nucleon elastic form-factors 
and their values are taken from the Bosted fit for the nucleon form factors~\cite{Bosted:1994tm}. 
Pauli suppression is taken into account according to Eq. B54 of Ref.~\cite{Tsai:1973py}.
\begin{table}
\begin{center} 
\begin{tabular}{|c|c|c|}
\hline
A & E$_{shift}$ (GeV) & $k_{F} (GeV)$  \\
\hline
Carbon    & 0.020  &  0.228 \\
Aluminum  & 0.018  &  0.236 \\
Iron      & 0.018  &  0.241 \\
Copper    & 0.018  &  0.245  \\
\hline 
\end{tabular}
\end{center}
\caption{Values of E$_{shift}$ and $k_{F}$ for nuclear targets used in this experiment. } \label{tab:EshiftandKF}
\end{table}   
From Table~\ref{tab:EshiftandKF} one can see that the energy shift does not vary too much for different nuclei. 
The values of $k_{F}$ don't vary too much after carbon.
\subsubsection{Inelastic Model}

The kinematic range of this experiment covers \mbox{$W^2=$ 0.0-4.5 GeV$^2$} and \mbox{Q$^2=$ 0.5-4.5 GeV$^2$}. 
The inelastic model cross section used in this experiment is based on the structure functions 
of proton and deuterium, as well as on the EMC~\cite{Aubert:1983xm} effect measured in the DIS region.
The proton structure functions are extracted from an inclusive inelastic electron-proton cross sections fit 
described in Ref.~\cite{Eric:proton}. 
The fit is constrained by photoproduction data at $Q^2$=0 and makes a smooth transition to DIS at higher W$^2$
values.
The other important aspect of this fit is that it is based on $R=\sigma_{L}/\sigma_{T}$ obtained by
Rosenbluth separation. 
Since it is impossible to extract the on-mass shell neutron structure functions directly by measuring them 
using a neutron target, a model based on the deuteron structure function is used.
The deuteron fit, described in Ref.~\cite{Bosted:2007xd}, relies on a fit of the ratio $R_{p}=\sigma_{L}/\sigma_{T}$ 
and the assumption $R_{p}=R_{n}$.
The fit includes photoproduction and low $Q^2$ data points and data from several other experiments.
Since there are not enough data at low $Q^2$ and no Rosenbluth separation has been done 
only the transverse portion of the cross section is fitted. 
Fermi motion is taken into account in a Plane Wave Impulse Approximation (PWIA). 
The region (dip) between the quasi-elastic peak and the $\Delta$(1232) resonance is systematically under-predicted 
at low $Q^2$, possibly because Meson Exchange Currents (MEC) and Final State Interactions (FSI) 
have been ignored. 
An additional empirical function was used to fill in the missing strength in the dip region.
$R_{D}$ for deuteron is evaluated by doing Fermi-smearing to proton $\sigma_{L}$ and $\sigma_{T}$ described 
in Ref.~\cite{Eric:proton}. 
The Fermi-motion of the nucleons in the deuterium is taken into account using a PWIA calculation and the 
Paris~\cite{Lacombe:1980dr} deuterium wave function.
After performing Fermi smearing of $F_{1}^{D}$ and $F_{1}^{P}$ the neutron $F_{1}^{N}$ 
is calculated as 
\vspace{-1.0cm}
\begin{center}
\begin{equation}
F_{1}^N=2F_{1}^{D}-F_{1}^{P}
\end{equation}
\end{center} 
where $F_{1}^{D}$ is deuterium structure function per nucleon.
The structure functions of the nuclei (A,Z) is calculated by the following formula, 
\vspace{-1.0cm}
\begin{center}
\begin{equation}
\begin{array}{l}
F_{1}^{A}=2ZF_{1}^{D}+\displaystyle\left( A - 2Z\right) F_{1}^{N}\\
F_{2}^{A}=F_{1}^{A}{\left( 1 + R_{P}\right)}/{\left( 1 + \nu^{2}/Q^{2}\right)/\nu }. \label{eq:F1F2}
\end{array}
\end{equation}
\end{center}
The structure functions given in the Eq.~\ref{eq:F1F2} are further corrected for the EMC effect.
The EMC effect has been studied by comparing $F_{2}$ measured on bound nucleons in nucleus $A$ and deuterium 
($x=0.0085-0.09$)~\cite{Amaudruz:1991cca}, ($x > 0.125$)~\cite{Gomez:1993ri}.
\vspace{-1.0cm}
\begin{center}
\begin{equation}
EMC(x,A)=(\sigma^{A}/\sigma^{D})_{is}=C(x)A^{\alpha(x)}
\end{equation}
\end{center} 
where $\log C(x) = a + b\log(x) + c (\log x)^2$ and $\alpha(x)$ is an eight order polynomial function of $x=Q^{2}/({2M\nu})$.
The ``is'' subscript means the isoscalarity correction (taking into account the fact that Z$\neq$A/2 for all nuclei). 
For $x$ lower than 0.0085 the EMC correction is taken to be that for $x=0.0085$ (no reliable data exists below this value). 
For $x$ higher than 0.7 the EMC correction is taken to be that for $x=$0.7, and the rest is taken into account when Fermi smearing is done.
An empirical fit form is added to the dip region between the quasi-elastic peak and the $\Delta$(1232) resonance.
The function has the same form as in the deuteron fit.

\subsection{Global Fit and Model Iteration} \label{sec:GlobalFit}

In order to extract the Born cross section from experimental data a model cross section is necessary. 
This can be seen from Eq.~\ref{eq:CSBornMesured} where the extracted cross section is proportional to the 
model cross section. 
Also, the model cross section is used to calculate radiative corrections (Eq.~\ref{eq:CSBornMesured}), 
Coulomb corrections (Eq.~\ref{eq:CoulombCorr}) and apply bin centering corrections (Eq.~\ref{eq:bccorr}).
Since the model cross section is used to apply all the corrections mentioned above, the extracted cross section 
clearly depends on knowledge of the model cross section. 
In order to minimize the model dependence, an iteration procedure is used. The procedure is described below, step-by-step.

\begin{figure}[p]
\begin{center}
\epsfig{file=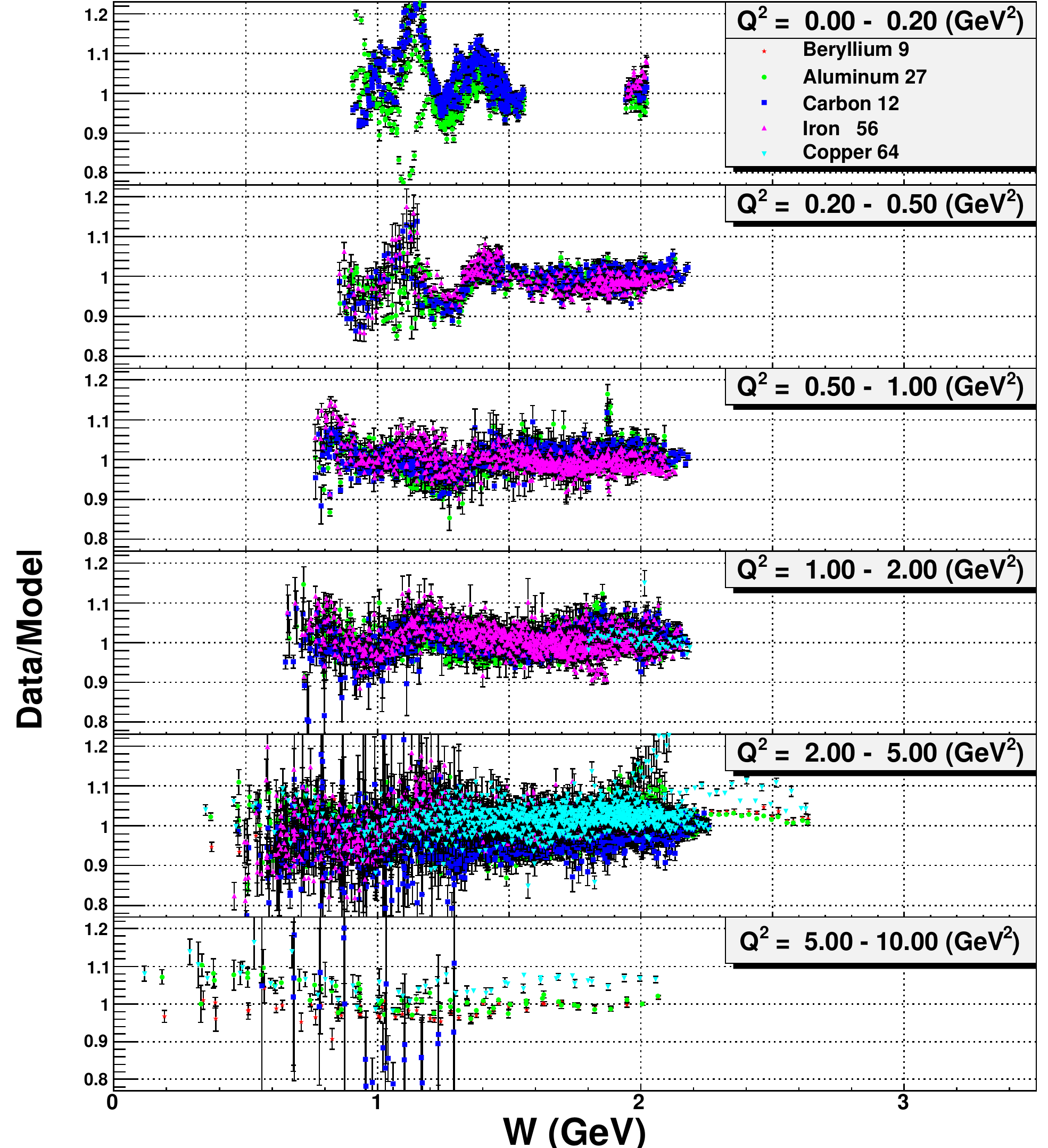, width=6.0in}
\end{center}
\caption{ The agreement between experimental data and model cross section is shown for six $Q^2$ bins after doing the first iteration of the global fit.}
\label{fig:global_fit}
\end{figure}

\begin{enumerate}
\item The Born cross section is extracted using the model cross section described in Sec.~\ref{sec:ModelCS}. 
The same model is also used to calculate radiative corrections and bin centering correction.

\item The extracted cross section is combined with the cross section results of other experiments~\cite{NadiaThesis, Benhar:2006er}, some of them done dating from 1970.
These data are fitted with the model cross section described in Sec.~\ref{sec:ModelCS} with some 
additional corrections. These corrections are listed below. 
The quasi-elastic structure function $F_{1}^{qe}$ is multiplied by polynomial function of fourth degree in $y$.  
The parameters of the polynomial are constrained in such way that the value of the polynomial function is always positive
in the kinematic range of quasi-elastic scattering.
The inelastic structure function $F_{1}^{in}$ is multiplied by a function of $W^{2}$ and $Q^{2}$~\cite{BostedVaheFit}.
When $Q^{2} \rightarrow \infty$ the value of this function goes to 1 and does not have any impact
on $F_{1}^{in}$. 
A linear A (nuclear) dependence is introduced to $R_{A}$.
The result of the first iteration is shown in Fig.~\ref{fig:global_fit}. 
\item The corrected model is used to calculate radiative corrections again and redo the cross section analysis. 
Steps 1 and 2 are repeated until the fitted model cross section converges.
\item Using the corrected model the $R_{A}-R_{D}$ is extracted (by means of a Rosenbluth separation) for each $W^{2}$ bin. 
Since the $Q^{2}$ dependence of $R_{A}-R_{D}$ is very weak two $Q^{2}$ bins, 0.5$-$2.5 GeV$^2$ and 2.5$-$4.5 GeV$^2$, 
are used to parametrize the $R_{A}-R_{D}$ versus $W^{2}$. 
The parametrizations of $R_{A}-R_{D}$ are used in the global fit program to run steps 1,2,3,4 repetitively until
no improvement in model cross section can be found. 
\end{enumerate} 
\subsection{Extraction of \texorpdfstring{$R$}{R} and \texorpdfstring{$F_{2}$}{F2}}

The electron-nuclear inclusive scattering cross section can be written in terms of photoabsorption 
cross sections for transverse (helicity $\pm$ 1) photons and longitudinal (helicity 0) photons, 
see Eq.~\ref{eq:cs_photoabs}.
\vspace{-1cm}
\begin{center}
\begin{equation}
{1 \over \Gamma }{d^2\sigma \over d\Omega dE'} = \sigma_{T}+\epsilon\sigma_{L}  \phantom{l}.
\label{eq:cs_photoabs}
\end{equation} 
\end{center}
In Eq.~\ref{eq:cs_photoabs}, where $\Gamma$ is the flux of transverse virtual photons, the right side 
is a function of $W^2$, $Q^2$ and $\epsilon$ (beam energy).
In order to determine $\sigma_{T}$ and $\sigma_{L}$ from this equation, 
measurements are done at the same ($W^2, Q^2$) but different $\epsilon$ (beam energy).
After calculating $\sigma_{T}$ and $\sigma_{L}$ the structure functions $R$ and $F_{2}$ can be determined 
by the following formulas:
\vspace{-1.0cm}
\begin{center}
\begin{equation}
R = {\sigma_{L} \over \sigma_{T}} \hspace{0.5cm} and 
\label{eq:Rsepar}
\end{equation}
\end{center}
\vspace{-1.0cm}
\begin{center}
\begin{equation}
F_{2} = { \nu \frac{K}{4\pi^2\alpha} \left( \sigma_{T}+\sigma_{L}\right) }\frac{Q^2}{Q^2+\nu^2},
\label{eq:f2nucl}
\end{equation}
\end{center}
where $K$ is called ``equivalent photon energy''.
\begin{figure}[t]
\begin{center}
\epsfig{file=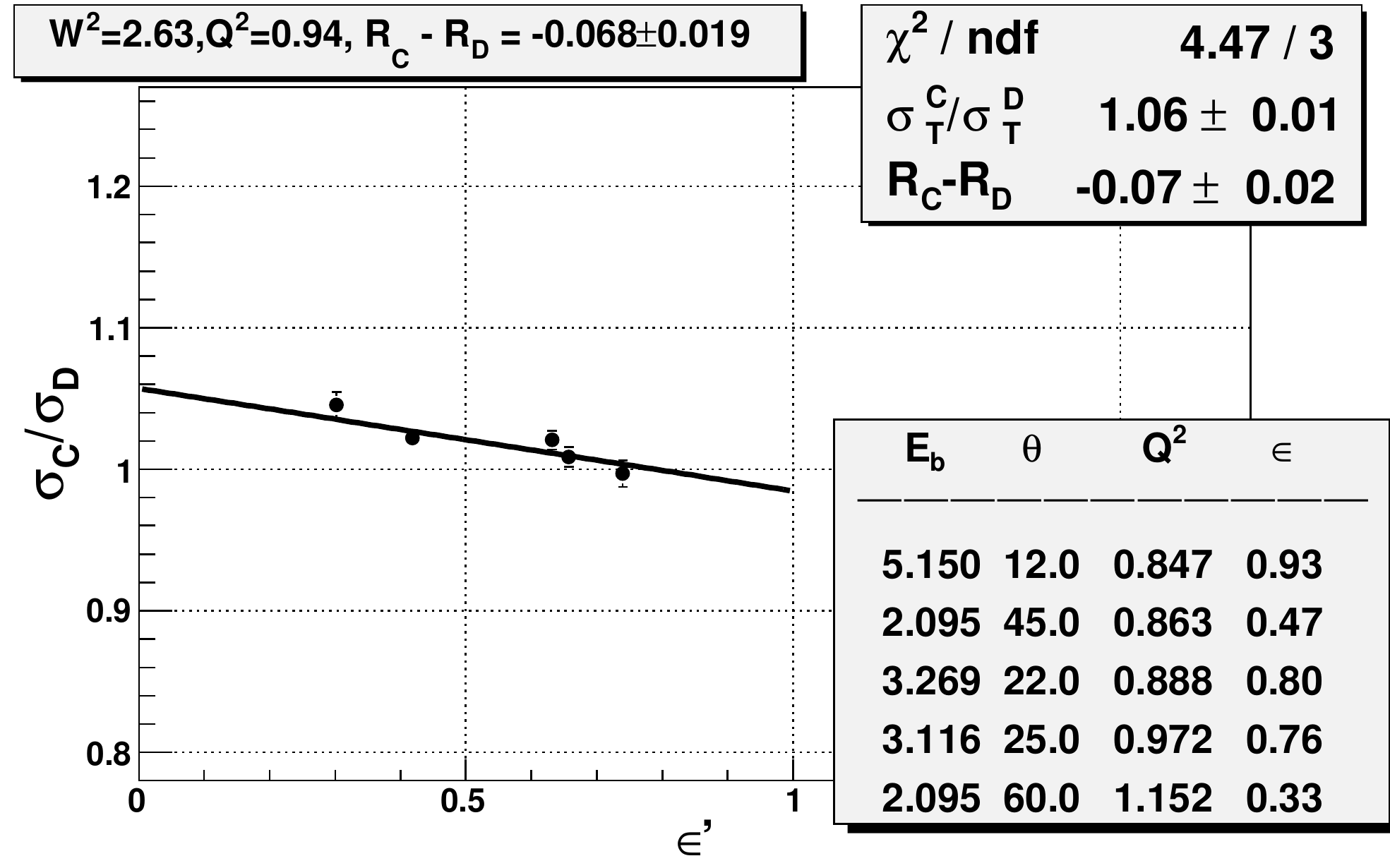, width=5.0in}
\caption{ An example of Rosenbluth separation for carbon target. The slope of the curve is $R_{A}-R_{D}$, the intercept is
$\sigma_{T}^{A}/\sigma_{T}^{D}$.}
\label{fig:ros_sep}
\end{center}
\end{figure}  
In this experiment the Rosenbluth separation is done using the nuclear cross section ratio to the deuterium cross section~\cite{Dasu}.
The ratio of the cross sections is given by
\vspace{-1cm}
\begin{center}
\begin{equation}
\displaystyle\frac{\sigma_{A}}{\sigma_{D}}=\displaystyle\frac{\sigma_{A}^{T}}{\sigma_{D}^{T}}\left[ 1 + \epsilon' \left( R_{A} - R_{D}\right) \right], 
\label{eq:Rsepar_Ratio}
\end{equation}
\end{center}
where $R_{D}$ is the deuterium $R$ and $\epsilon'=\epsilon/(1+\epsilon R_{D})$. 
This method has it's advantage compared to Rosenbluth separation method which uses Eq.~\ref{eq:cs_photoabs}. 
If Rosenbluth separation is done using Eq.~\ref{eq:cs_photoabs}, systematic uncertainties may have big impact.
Since $\epsilon R_{D}$ is small, it's impact on $\epsilon'$ is small, for a 20\% uncertainty when $R_{D}=0.2$, the uncertainty on  
$\epsilon'$ is only 4\%. 
Taking the ratios of cross sections $\sigma_A/\sigma_{D}$ cancels most systematic uncertainties:
acceptance, beam charge, offsets in beam energy, HMS angle, HMS momentum, model dependence to some extent, radiative corrections.  

An example of Rosenbluth separation is shown in Fig.~\ref{fig:ros_sep}. 
Cross sections are not measured at exactly same $W^2, Q^2$ values, instead, for each $W^2$ there are 
several $Q^2$s, which can be combined to have one average $Q^2_{mean}$ value. 
In this analysis the values of $Q^2$s that are combined to perform Rosenbluth separation 
are chosen to be within a $\Delta Q^2=$0.5 $GeV^2$ range ($Q^2_{mean}$ is calculated as the mean of these $Q^2$s). 
After calculating $Q^2_{mean}$, all cross sections are centered to the same value of $W^2, Q^2_{mean}$ 
using the model cross section, via 
\vspace{-1cm}
\begin{center}
\begin{equation}
\sigma(W^2,Q^2_{mean})=\sigma_{exp}(W^2,Q^2)\displaystyle\frac{\sigma_{model}(W^2,Q^2_{mean})}{\sigma_{model}(W^2,Q^2)}.
\label{eq:Rsep_centering}
\end{equation}
\end{center}
If the model cross section correction is greater than 50\% that particular point is excluded from the fit.

In order to have enough range in $\epsilon$, each Rosenbluth separation is only done if the $\Delta\epsilon>0.3$ condition is satisfied.
Each data point is used only once to avoid correlated uncertainties.
For this experiment 280 Rosenbluth separations are done for carbon and aluminum, 120 for iron, 110 for copper.
A program is written to search the cross section data for Rosenbluth separation candidate cross section values and perform the 
Rosenbluth separation. 
The program is designed in such a way that adding data from another experiment to the data set of this experiment requires no modification. 
An algorithm was developed to exclude a given cross section value being used more than once, and to find $Q^2$ data 
within $\Delta Q^2=$0.5 range while at same time have the minimum spread around the calculated mean $Q^2_{mean}$ value.
This minimizes the model dependence introduced by Eq.~\ref{eq:Rsep_centering}. 

%
%
Though $R$ can be extracted in a model dependent way by using the expression 
\vspace{-1cm}
\begin{center}
\begin{equation}
{ d^{2}\sigma \over d\Omega dE' } = \sigma_{Mott}{2MxF_{2} \over Q^{2}\epsilon}\Bigg({1+\epsilon R \over 1+R}\Bigg)  \phantom{l},
\label{eq:RmodelExtract}
\end{equation}
\end{center}
the experimental cross section and a model of the structure function $F_{2}$, 
it was not done for this analysis.

\subsection{Systematic Uncertainties}
The total systematic uncertainty in the cross section extraction is taken as the 
sum in quadrature of all systematic uncertainties of the quantities that contribute to the cross section. 

Systematic uncertainty can be divided into two groups: point-to-point uncertainties and normalization uncertainties. 
Point-to-point uncertainties are caused by changes in experimental conditions during 
data acquisition, and therefore their effect is uncorrelated between different data points. 
These include uncertainties arising from a variation in the efficiencies of detectors 
and data acquisition systems, changes from one spectrometer setting to another.
Point-to-point uncertainties can be removed if the same measurement is done more than once, while in contrast, 
normalization uncertainties can't be avoided by doing more measurements.
An example of normalization uncertainty can be a systematic shift in the offset of a current measurement device, 
target thickness measurement, acceptance, etc.

\begin{table}[ht]
\begin{center}
\begin{tabular}{|c|c|c|c|c|}
\hline
Section             & Quantity           \T \B               & Uncertainty   & $\delta_\sigma\%$ \\
\hline
           	    & Beam Energy		\T \B        &  0.05\%         & 0.25\% \\
\ref{sec:bcm}	    & Beam Charge                 \T \B      &  0.3 $\mu$A        & 0.37-0.75\%\\
	            & Scattered $e'$ Energy	\T \B        &  0.06\%         & 0.025\%\\
	            & Scattered $e'$ Angle	\T \B        &  $\sim$ 0.2 mrad   & 0.3\%  \\
\ref{sec:DeadTime}  & Elect. Dead Time Correction   \T \B    &  27.0\%        & 0.04\% \\
\ref{sec:DeadTime}  & Comp. Dead Time Correction    \T \B    &  0.2\%          & 0.2\%\\
\ref{Sec:Acceptance}& Acceptance                     \T \B   &  0.6\%           & 0.6\%\\
                    & Model Dependence               \T \B   &  0.6\%           & 0.6\% \\
\ref{sec:RadCorr}   & Radiative Correction           \T \B   &  1.0\%           & 1.0\% \\
\hline
\end{tabular}
\end{center}
\caption{Normalization systematic uncertainties in the experimental parameters (column 3) and the 
corresponding systematic uncertainties in the differential cross section (column 4).}
\label{tab:systematics}
\end{table}

\begin{table}[ht]
\begin{center}
\begin{tabular}{|c|c|c|c|c|}
\hline
Section             & Quantity           \T \B               & Uncertainty   & $\delta_\sigma\%$ \\
\hline
           	    & Beam Energy		\T \B        &  0.05\%            & 0.25\% \\
\ref{sec:bcm}	    & Beam Charge                 \T \B      &  0.3 $\mu$A        & 0.3$\%$\\
	            & Scattered $e'$ Energy	\T \B        &  0.06\%            & 0.025\%\\
	            & Scattered $e'$ Angle	\T \B        &  $\sim$ 0.2 mrad   & 0.3\%  \\
\ref{sec:DeadTime}  & Elect. Dead Time Correction   \T \B    &  27.0\%            & 0.008\% \\
\ref{sec:DeadTime}  & Comp. Dead Time Correction    \T \B    &  0.2\%             & 0.2\%\\
\ref{sec:trigeff}   & Trigger Efficiency             \T \B   &  0.007\%           & 0.007\%\\
\ref{sec:TrackEff}  & Tracking Efficiency            \T \B   &  0.15\%            & 0.15\% \\
\ref{sec:PIDcuts}   & \u{C}erenkov Efficiency        \T \B   &  0.15\%            & 0.15\%\\
\ref{sec:PIDcuts}   & Calorimeter Efficiency        \T \B    &  0.05-0.2\%        & 0.05-0.2\%\\
\ref{sec:CSB}       & Charge Symmetric Background    \T \B   &  0.1-0.4\%         & 0.1-0.4\%\\
\ref{Sec:Acceptance}& Acceptance                     \T \B   &  0.7\%             & 0.7\%\\
\ref{sec:RadCorr}   & Radiative Correction           \T \B   &  3.0\% of QE       & $\leq$ 1.0\% \\
\hline
\end{tabular}
\end{center}
\caption{Point-to-point systematic uncertainties in the experimental parameters (column 3) and the 
corresponding systematic uncertainties in the differential cross section (column 4).}
\label{tab:systematicsptp}
\end{table}

Kinematic uncertainties are caused by uncertainty in beam energy, spectrometer momentum, and spectrometer angle. 
These uncertainties are estimated from a study of elastic electron-proton scattering data. 
A procedure described in Ref.~\cite{vlad} provides a way to estimate the uncertainties in beam energy, 
spectrometer momentum, and spectrometer angle, and their values are shown in the third column of Table~\ref{tab:systematics}.
A model cross section is used to estimate their impact on measured cross section and is shown in the fourth 
column of Table~\ref{tab:systematics}.

The beam charge measurement is discussed in Sec.~\ref{sec:bcm}. 
The point-to-point uncertainty is estimated to be 0.3\%. 
This is obtained by studying the residuals of the measured currents during the calibration procedure. 
An additional scale uncertainty of 0.3\% is assumed for the charge measured, due to the UNSER
calibration. 
This is estimated by examining runs taken with the same kinematics settings but different beam 
currents on a carbon target ( with 10 $\mu A$ steps starting from 20 $\mu A$ up to 100 $\mu A$).

The scale uncertainty of the HMS acceptance correction is 0.6\%. 
This is estimated by changing positions of the target, collimator, magnets and then calculating the acceptance.
Cross sections are extracted using this shifted acceptance. 
The spread of distribution of the ratio of shifted cross sections to cross section with a known 
positions of the target, collimator, magnets is assigned as the systematic uncertainty for the acceptance correction.
The momentum offset correction discussed in Sec.~\ref{Sec:Acceptance} contributes about 0.7\% point-to-point 
uncertainty.

No normalization uncertainty is assigned to the tracking efficiency. 
At a fixed rate the tracking efficiency should scale linearly with the DC time window.
This was studied by varying the DC TDC time window and plotting the tracking efficiency versus time window width. 
No nonlinearity is observed according to Ref.~\cite{THorn}. 
Also a point-to-point uncertainty of 0.15\% is assigned to the tracking efficiency 
based on the spread of the points in tracking efficiency versus SCIN rate plot, as in Fig.~\ref{fig:track_eff}.

The trigger efficiency is better than 99.9\%, and no scale uncertainty is assigned to it. 
The point-to-point uncertainty is 0.007\%, see Fig.~\ref{fig:trigeff}, and is negligible.

Electronic and computer dead times are discussed in Sec.~\ref{sec:DeadTime}. 
A scale uncertainty of 0.04\% (20 kHz average rate) is found for the electronic dead time, mainly from the deviation
of the measured value of $\tau$ (from the plot of electronic dead time versus pretrigger rate, electronic 
dead time is found to be $\sim$ 80 ns ) from the expected value of 60 ns. 
The point-to-point uncertainty (spread of electronic dead time from expected value defined by $1-60ns \times rate$) 
is 0.008\% and also is negligible. 
The uncertainty in computer dead time is estimated by taking runs at the same kinematics but different prescale factors. 
Results are found to agree within 0.2\%, see Ref.~\cite{vlad}. 
A point-to-point uncertainty of 0.2\% is assigned.

Pion contamination can be up to 3\% at some kinematics. 
It should be the same for positron runs, so no explicit subtraction is done. 
It is automatically subtracted when the charge symmetric background is subtracted. 
The uncertainty is absorbed into the charge symmetric background uncertainty.

The charge symmetric background is discussed in Sec.~\ref{sec:CSB}. 
At scattered electron angles smaller than 50$^o$ (low E$'$)  point-to-point uncertainty is estimated 
to be 0.1\%, for angles greater than 50$^o$ (high E$'$) the point-to-point uncertainty is estimated to be 0.4\%, 
(see Fig.~\ref{fig:CSB_SYSTEM}).

The effect of the model on the bin centering corrections is studied by varying the 
shape of the model. 
This is done by supplying an artificial Q$^2$ dependence as input to the individual 
DIS and QE cross sections.
The point-to-point of 0.5\% uncertainty due to model dependence is assigned as a result of the study.

In the kinematic range of this experiment radiative correction are maximum 30\%.
The systematic error for higher order $\alpha$ contributions are set to be 3\% of the size of the quasi-elastic tail, 
relative to the inelastic~\cite{BostedPCom}. 
The uncertainty on the cross sections due to the radiative correction is estimated
at 1\% for $\epsilon$ dependence, and the normalized uncertainty is also $\sim$1\%,
according to the radiative correction studies done at SLAC~\cite{Dasu}. 

\chap4{Results and Discussions}
In the last chapters we have discussed the calibration of the detectors and 
the corrections applied to the experimental data to extract the cross sections. 
In this chapter we will present the results of the analysis.
First, the extracted differential cross sections for the aluminum target will be presented. 
Next, the results of nuclear dependence of $R=\sigma_{L}/\sigma_{T}$ will be shown and 
nuclear dependence of $R_{A}-R_{D}$ and $R_{A}-R_{C}$ will be discussed.
The Rosenbluth separated structure functions $R=\sigma_{L}/\sigma_{T}$ and $F_{1}$, 
$F_{2}$ and $F_{L}$ will be compared with a model obtained from an empirical 
fit performed in the nuclear resonance region.
Next, the extracted $F_{2}$ structure function using model cross sections will be discussed 
in terms of quark-hadron duality.
Finally, Rosenbluth separated structure functions will be used to study 
quark-hadron duality in nuclei. Before beginning this presentation of the results I will take 
a short detour to discuss how elastic electron-proton scattering is used to gauge our estimate of
systematic errors.

\subsection{Elastic Electron-Proton Cross Section}
Elastic electron-proton scattering can be used as a cross check of the systematic 
uncertainties present in the setup of an experiment.
Since the elastic electron-proton cross section has been measured in the past with great precision, 
any deviation from that measurement would indicate a problem present in the collected data or 
in calibrations. 
In addition, elastic electron-proton scattering can be used to determine if normalization 
uncertainties like charge, spectrometer acceptance and detector inefficiencies and dead times 
are taken into account correctly.
\begin{figure}[ht]
\begin{center}
\epsfig{file=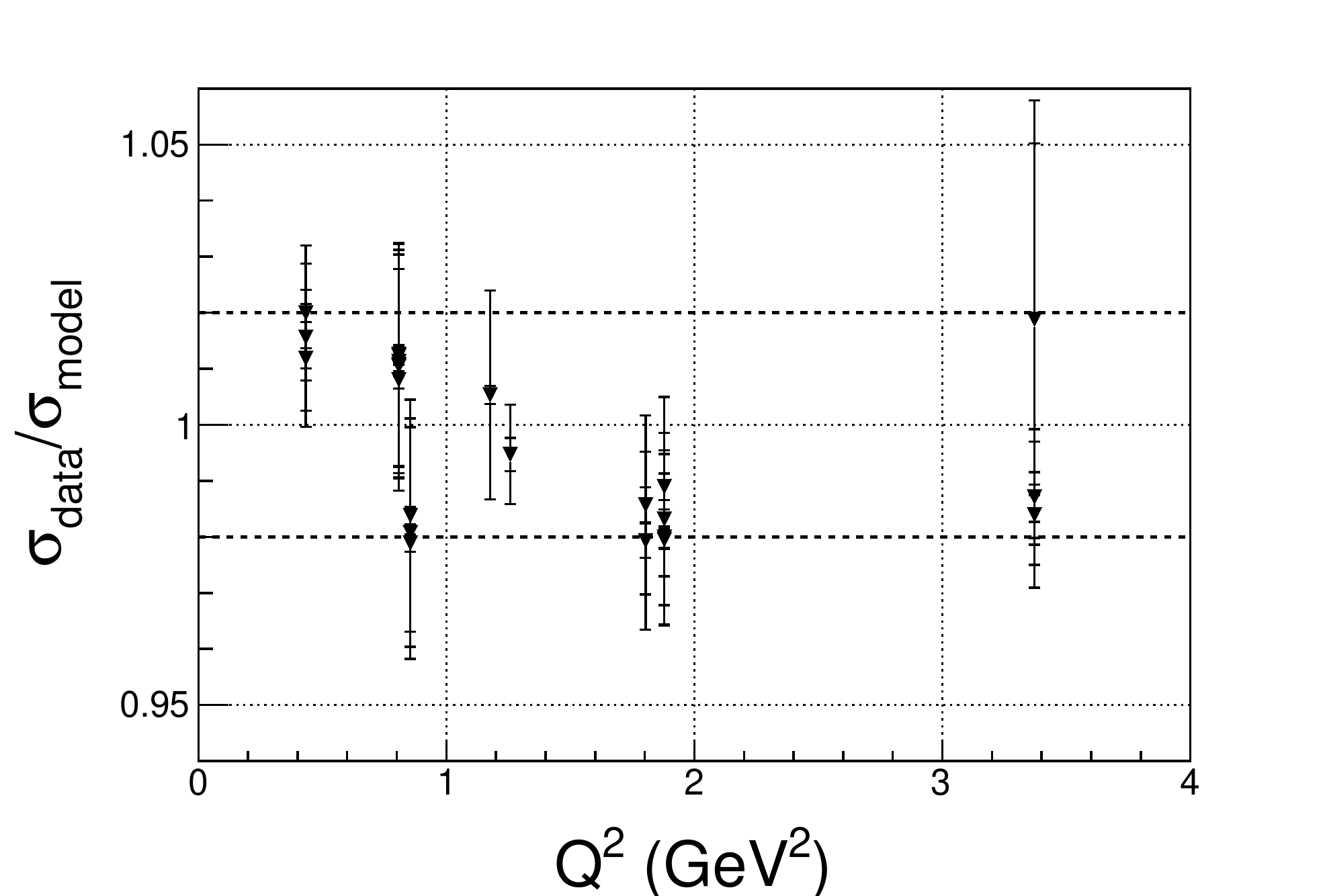,width=5.0in}
\end{center}
\caption { Ratio of extracted elastic cross sections to a fit of the world's
elastic data~\cite{Arrington:2003df}. Point-to-point systematic 
errors are also shown and are the dominant uncertainty.
The dashed lines indicate $\pm$ 2\% range.}
\label{fig:cselastic}
\end{figure}

In this experiment elastic data are taken at twelve different kinematic settings. 
The elastic data were analyzed using the same technique as the data in this experiment 
with a few exceptions. 
In addition to the cuts used in resonance cross section calculation, a cut $0.8 < W^{2} < 1.15$ GeV$^2$ 
is applied to avoid the pion electro-production region. 
In order to compensate for radiated elastic events with $W^{2}>1.15$ GeV$^2$ (low $W^{2}$ bins radiate
to higher $W^{2}$ bins) a correction factor is calculated and applied to each $\theta$ bin. 
For each run the elastic cross sections is averaged over $\theta$ bins using an elastic cross section model 
and the results are shown at a given central $Q^{2}$ value in Fig.~\ref{fig:cselastic}.
These elastic cross sections agree with the world's elastic cross section data within better than 2\%. 
Since the measurement of elastic cross sections are more sensitive to kinematic offsets  
than inelastic resonance cross section, it is expected that systematic uncertainties 
in the resonance region are, in the worst case, the same as in the elastic case. 
For more information about elastic electron-proton cross section extraction method see Ref.~\cite{Christy:2004rc}.
\subsection{Differential Cross Sections}
The method of extracting the cross sections was discussed in the Section~\ref{Sec:CSEC}. 
The differential cross sections were extracted for four different targets: carbon,
aluminum, iron and copper. 
Carbon and aluminum data were taken in the range $0.0 < W^{2} < 4.5 $ GeV$^{2}$ and 
$0.5 < Q^{2} < 4.5 $ GeV$^{2}$. 
For iron data were taken in the range $0.0 < W^{2} < 4.5 $ GeV$^{2}$ and 
$0.5 < Q^{2} < 2.5 $ GeV$^{2}$, and for copper $0.0 < W^{2} < 4.5 $ GeV$^{2}$ and 
$2.5 < Q^{2} < 4.5 $ GeV$^{2}$. 

In the analysis process, the data were stored in equally spaced $W^{2}=0.04$ GeV$^{2}$ bins. 
After iterating the model described in the Section~\ref{sec:GlobalFit}, 
double differential electro-production cross sections were extracted. 

\begin{figure}[htp]
\begin{center}
\epsfig{file=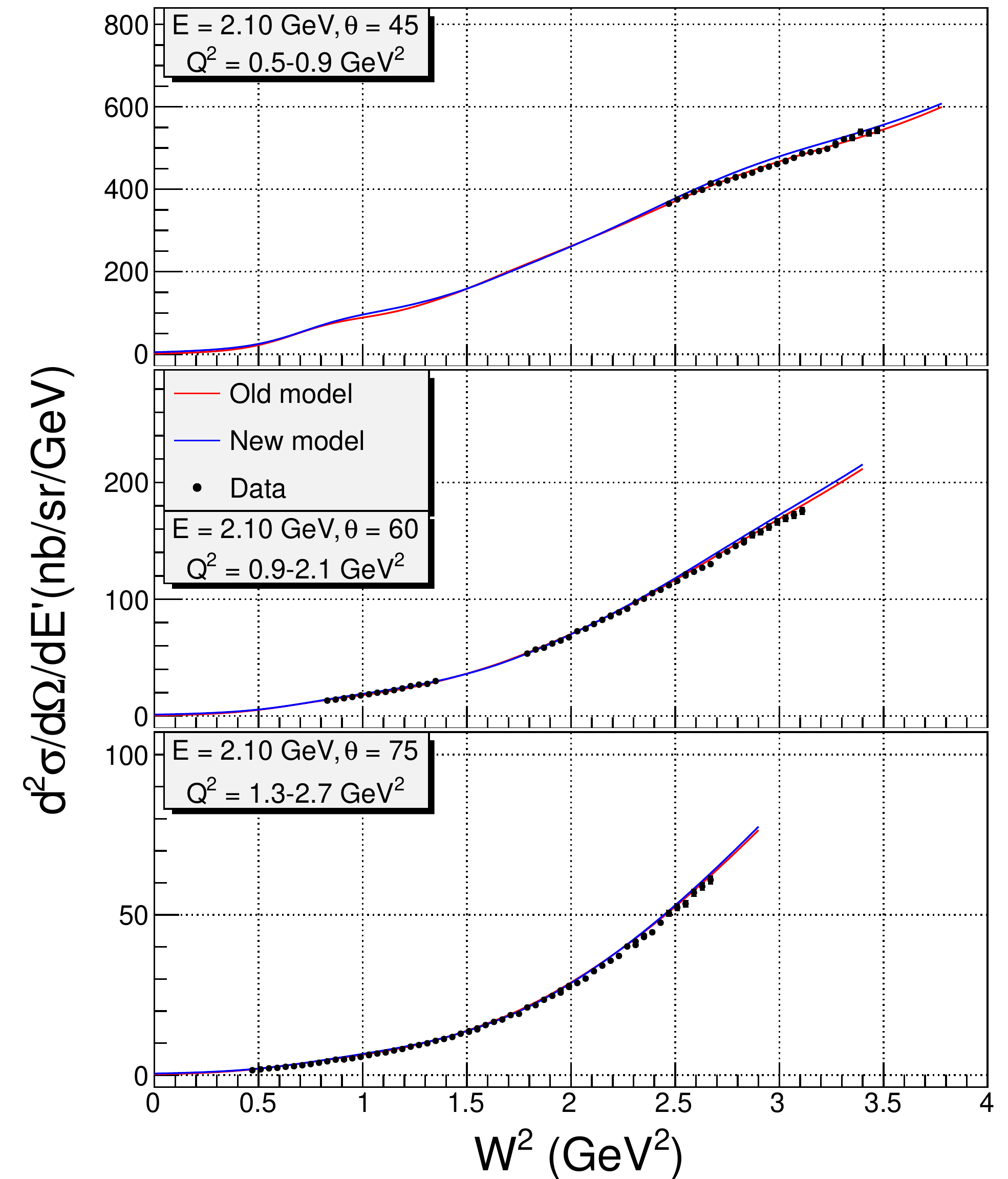,width=5.0in}
\end{center}
\caption { Extracted differential cross section for aluminum compared to the model cross section. }
\label{fig:csAl2097}
\end{figure}

Figures~\ref{fig:csAl2097} to \ref{fig:csAl5151_1} show examples cross section spectra for aluminum (cross sections for C, Fe, and Cu are in Appendix~\ref{sec:CrossSections}). 
The blue solid curve in these figures is the fit to the extracted cross sections of this experiment. 
The red curve is the fit based on proton and deuteron fits described in the Section~\ref{sec:GlobalFit} 
and was used as the starting input cross section model in this analysis.
Before starting the iteration procedure, 77\% of the cross section points agreed within 5\% of the corresponding model value, 
after the iteration procedure 90\% agreed. 
The iterative procedure is stopped when convergence is achieved. 
Convergence is confirmed by taking the difference with the extracted cross sections 
between successive iterations and making sure that the difference is less than 0.5\%. 
In total three iterations were necessary to achieve convergence.
Only the statistical uncertainty of the data is shown in the figures, 
and it is typically smaller than the symbol size.

\begin{figure}[p]
\begin{center}
\epsfig{file=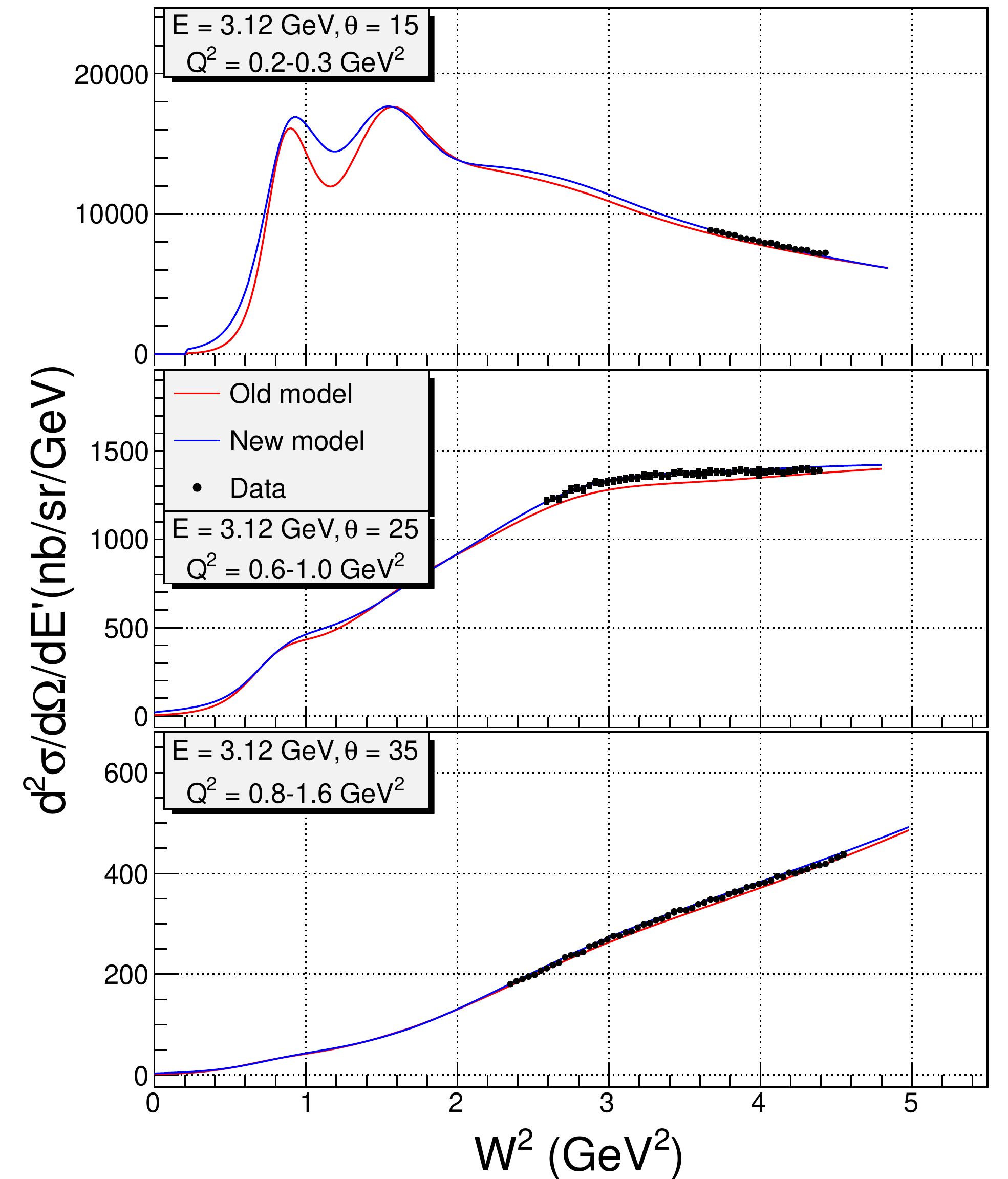,width=5.5in}
\end{center}
\caption { Extracted differential cross section for aluminum compared to the model cross section.}
\label{fig:csAl3116_0}
\end{figure}

\begin{figure}[p]
\begin{center}
\epsfig{file=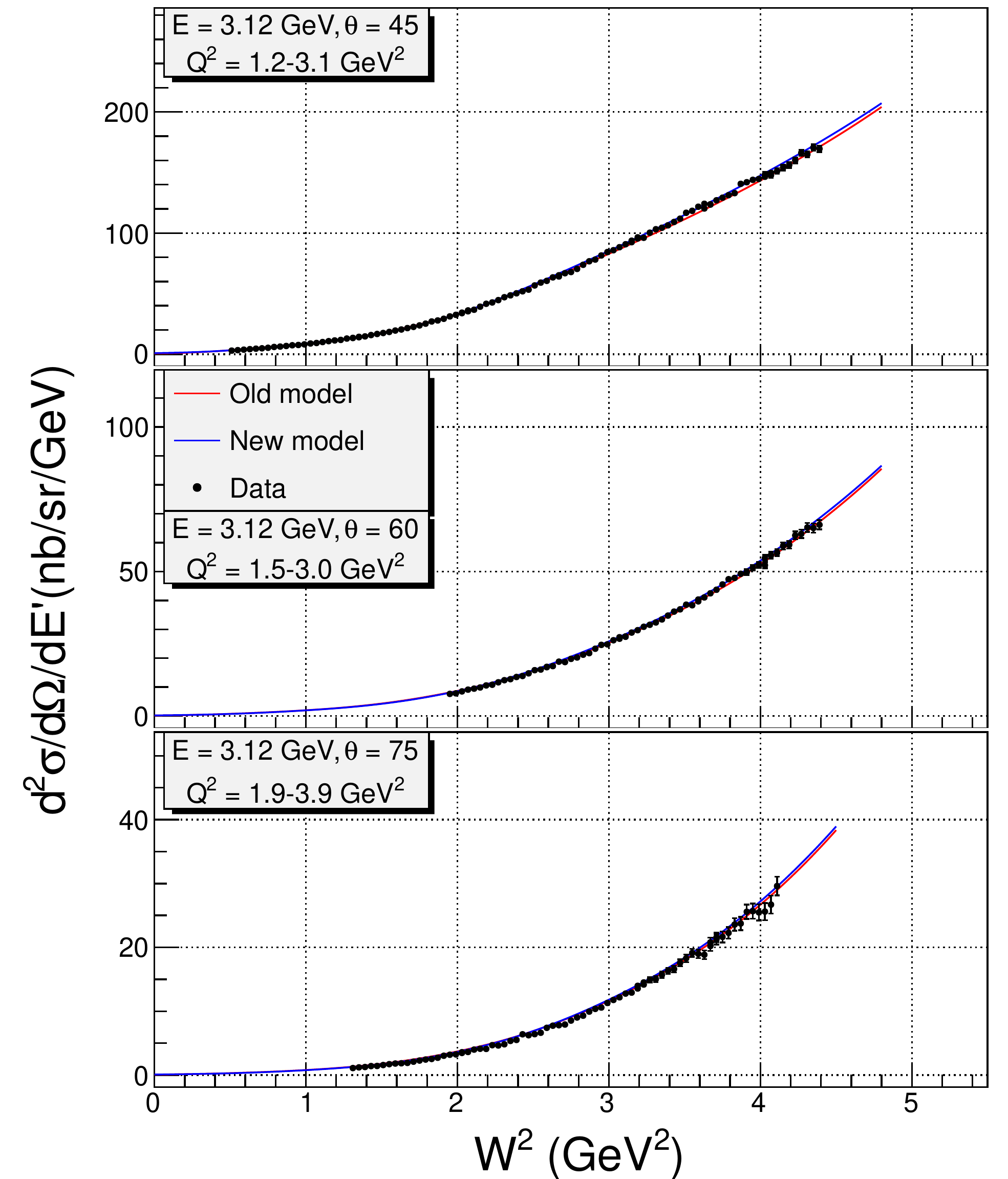,width=5.5in}
\end{center}
\caption { Extracted differential cross section for aluminum compared to the model cross section. }
\label{fig:csAl3116_1}
\end{figure}

\begin{figure}[p]
\begin{center}
\epsfig{file=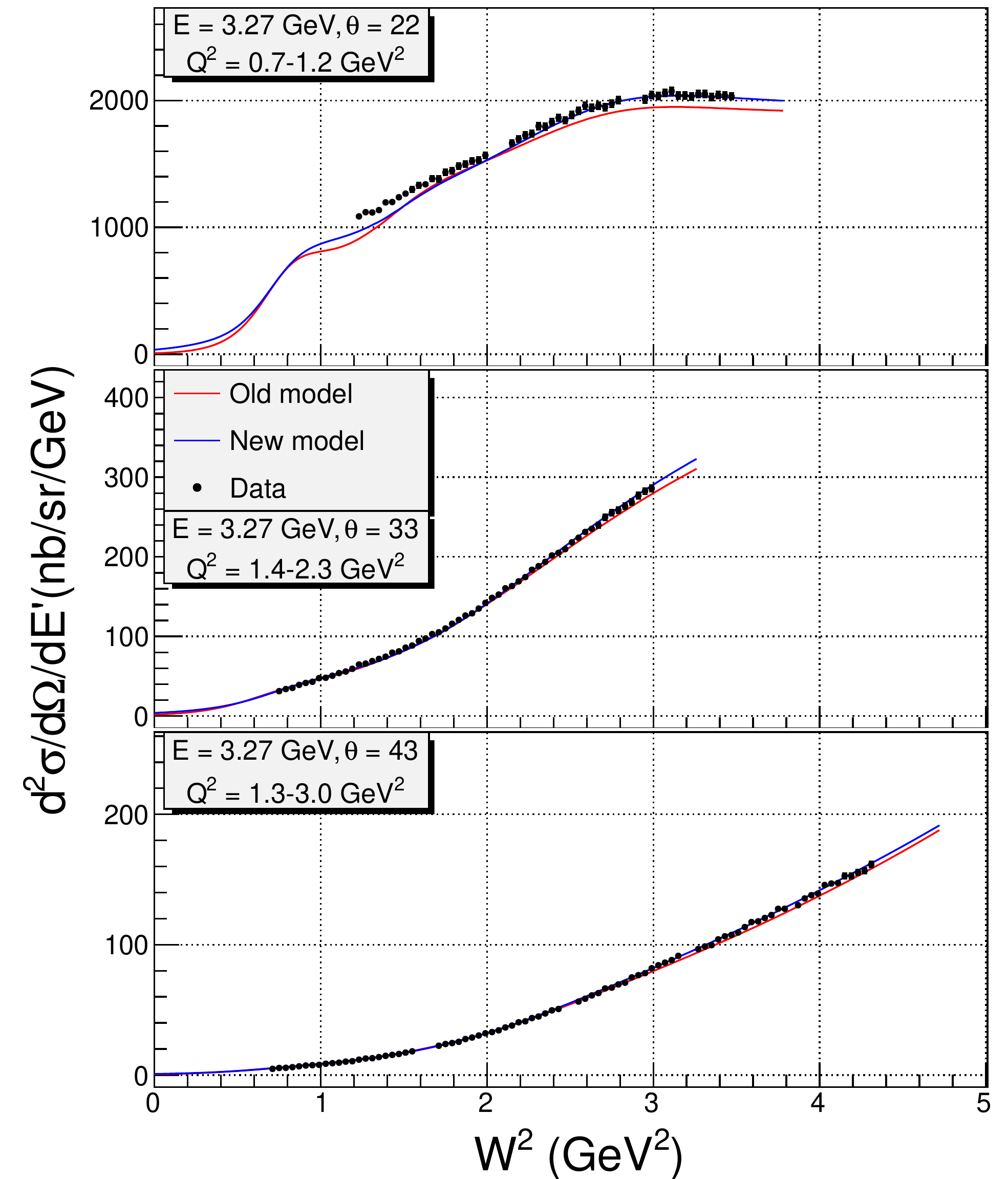,width=5.5in}
\end{center}
\caption { Extracted differential cross section for aluminum compared to the model cross section. }
\label{fig:csAl3270_0}
\end{figure}

\begin{figure}[p]
\begin{center}
\epsfig{file=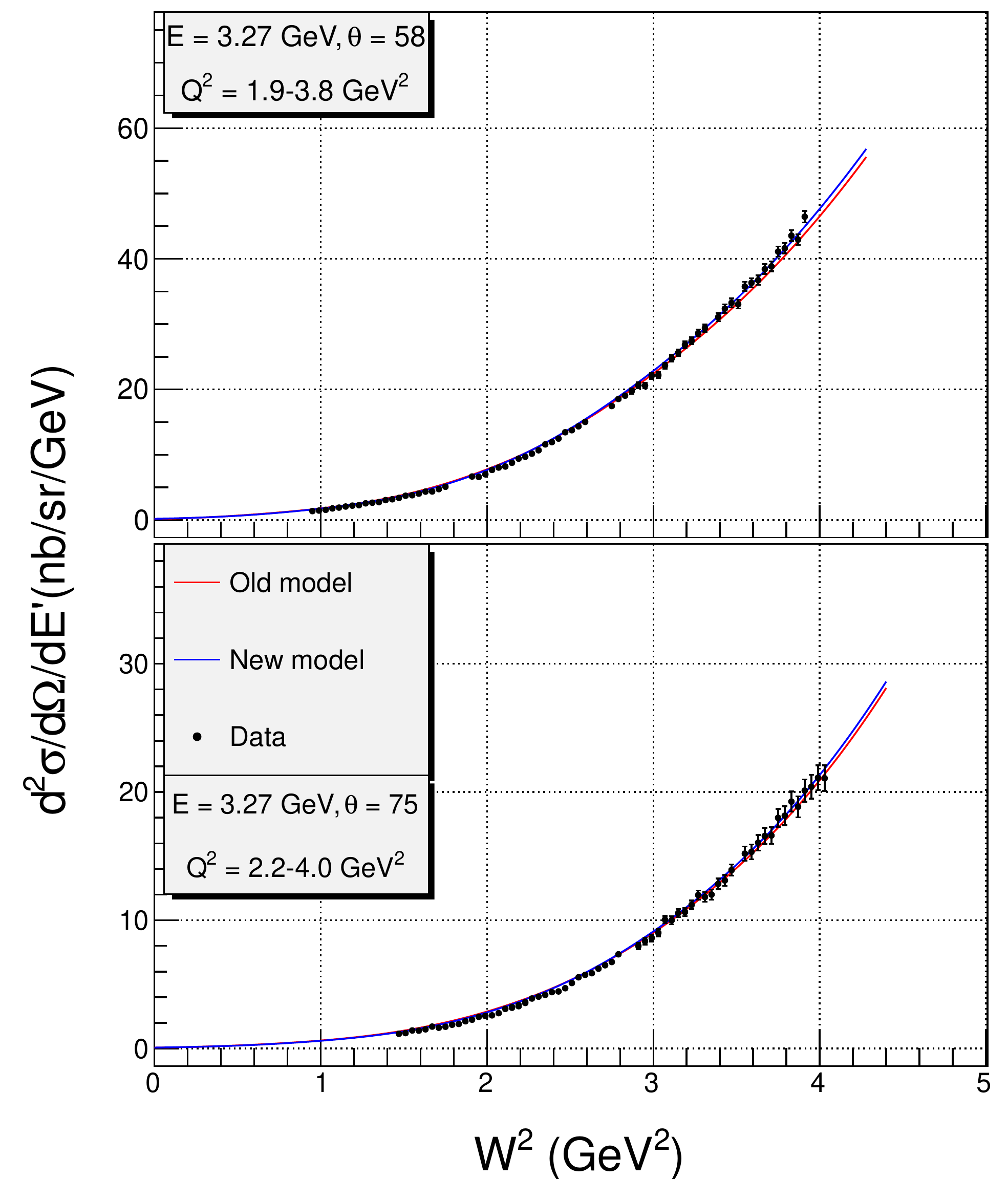,width=5.5in}
\end{center}
\caption { Extracted differential cross section for aluminum compared to the model cross section. }
\label{fig:csAl3270_1}
\end{figure}

\begin{figure}[p]
\begin{center}
\epsfig{file=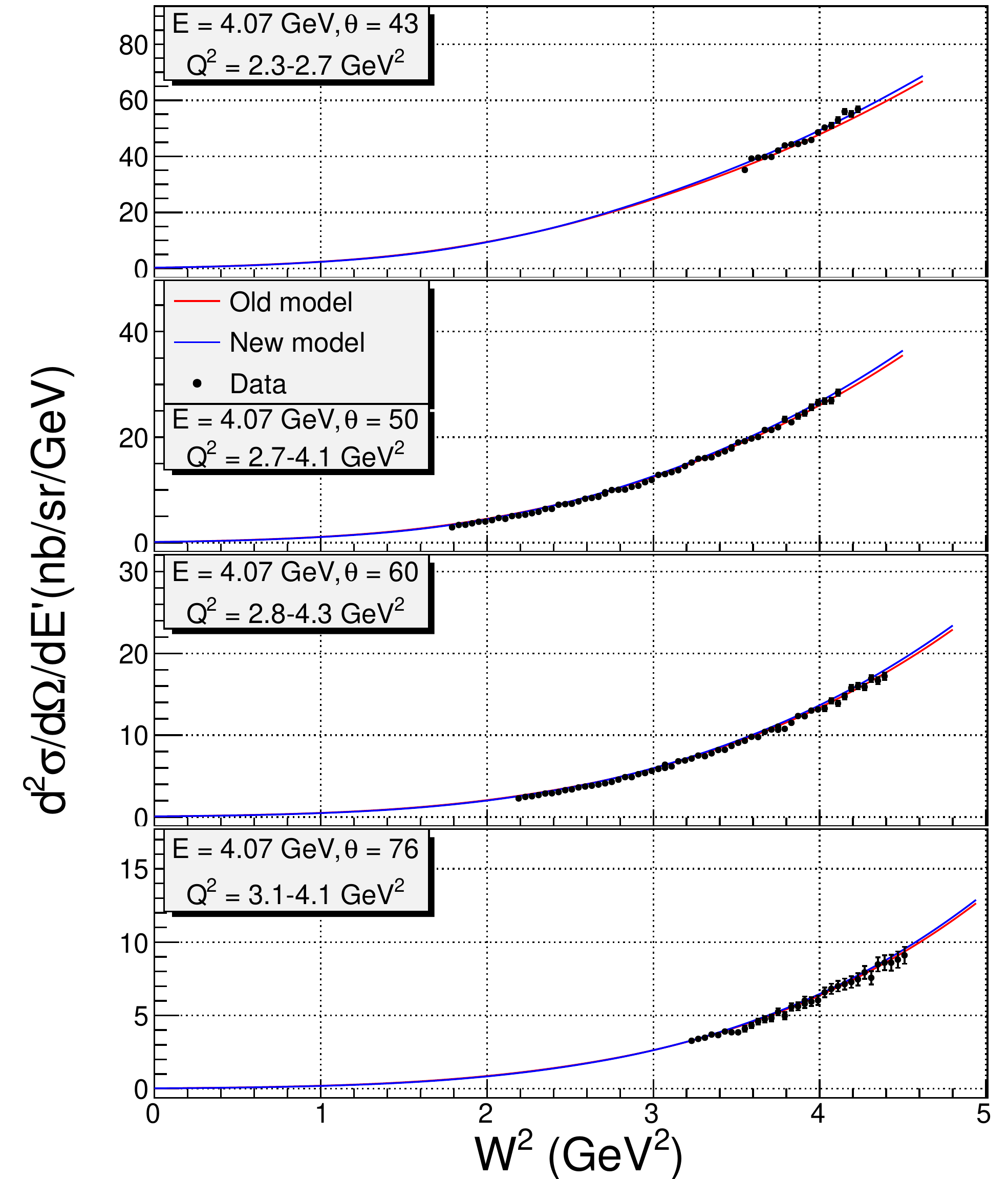,width=5.5in}
\end{center}
\caption {  Extracted differential cross section for aluminum compared to the model cross section.}
\label{fig:csAl4074_0}
\end{figure}

\begin{figure}[p]
\begin{center}
\epsfig{file=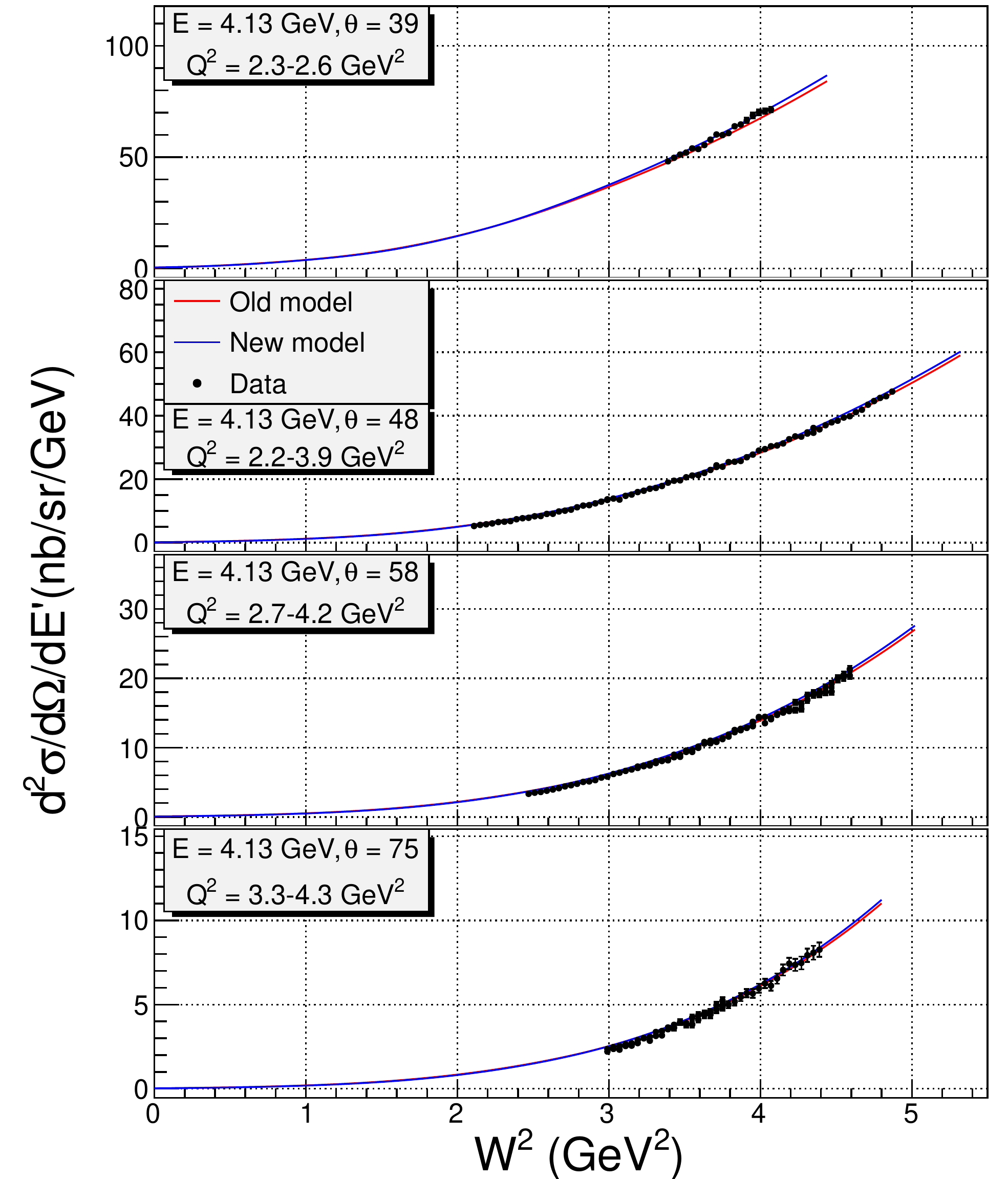,width=5.5in}
\end{center}
\caption {  Extracted differential cross section for aluminum compared to the model cross section.}
\label{fig:csAl4134_0}
\end{figure}

\begin{figure}[p]
\begin{center}
\epsfig{file=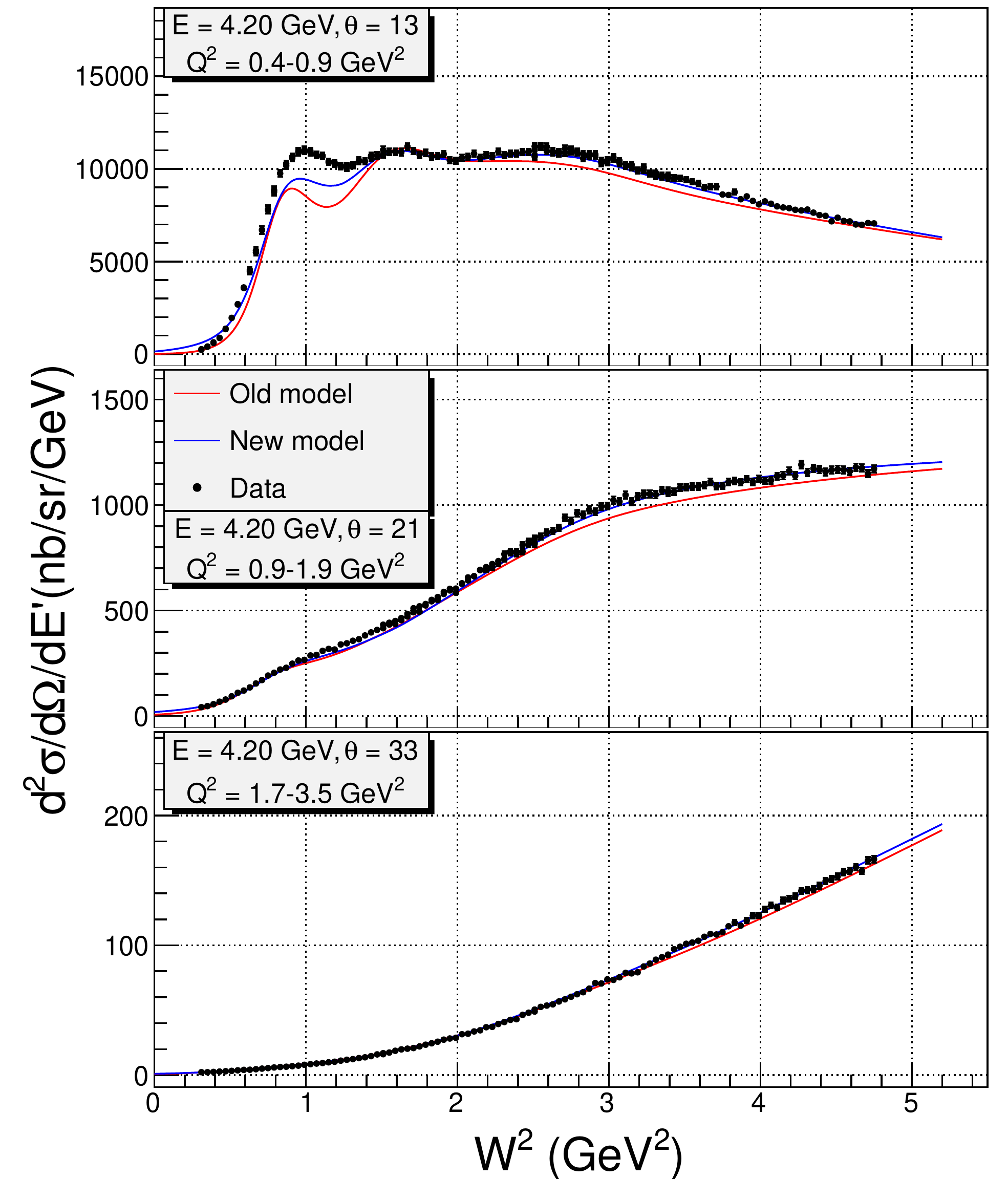,width=5.5in}
\end{center}
\caption { Extracted differential cross section for aluminum compared to the model cross section. }
\label{fig:csAl4199_0}
\end{figure}

\begin{figure}[p]
\begin{center}
\epsfig{file=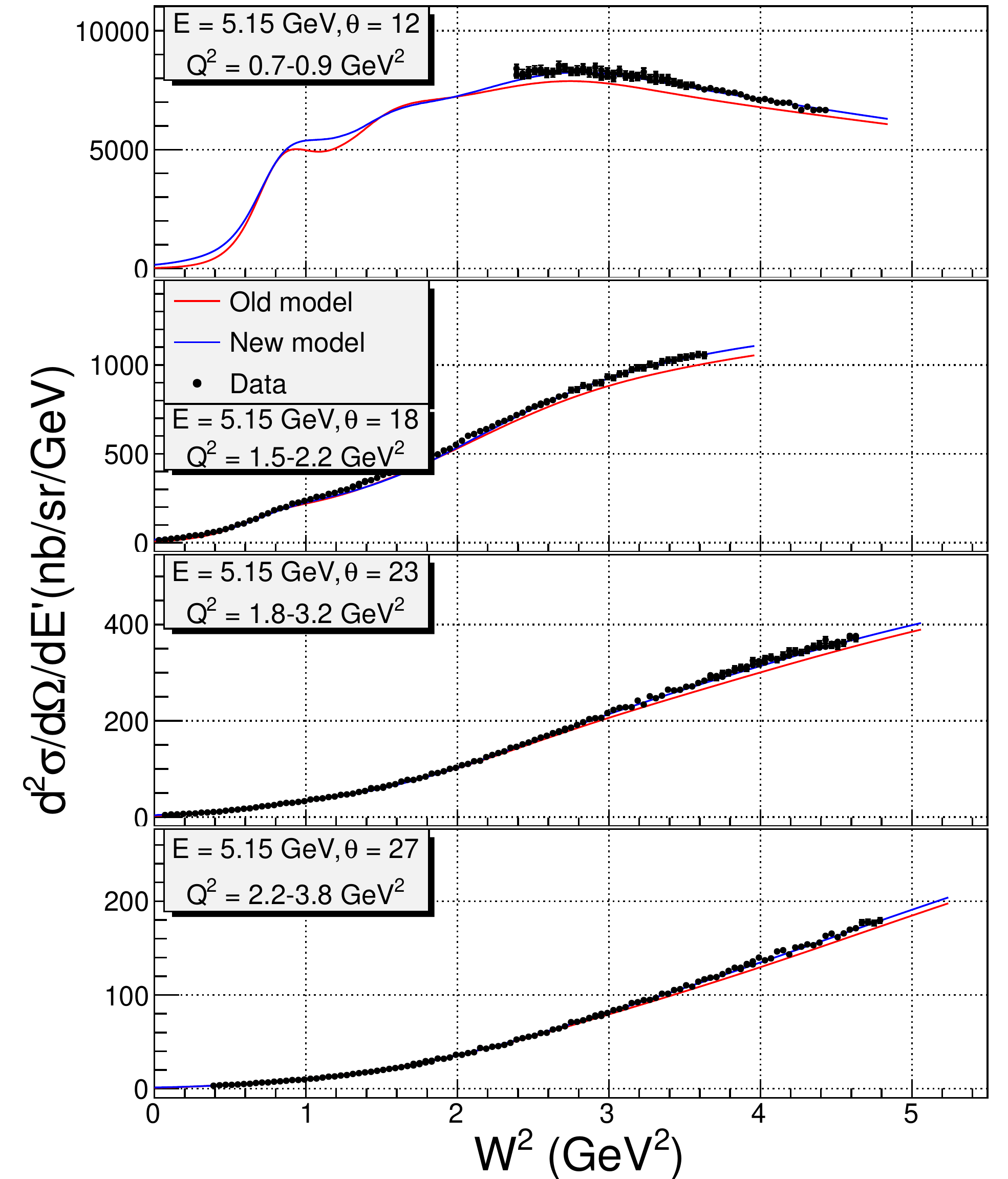,width=5.5in}
\end{center}
\caption { Extracted differential cross section for aluminum compared to the model cross section. }
\label{fig:csAl5151_0}
\end{figure}

\begin{figure}[p]
\begin{center}
\epsfig{file=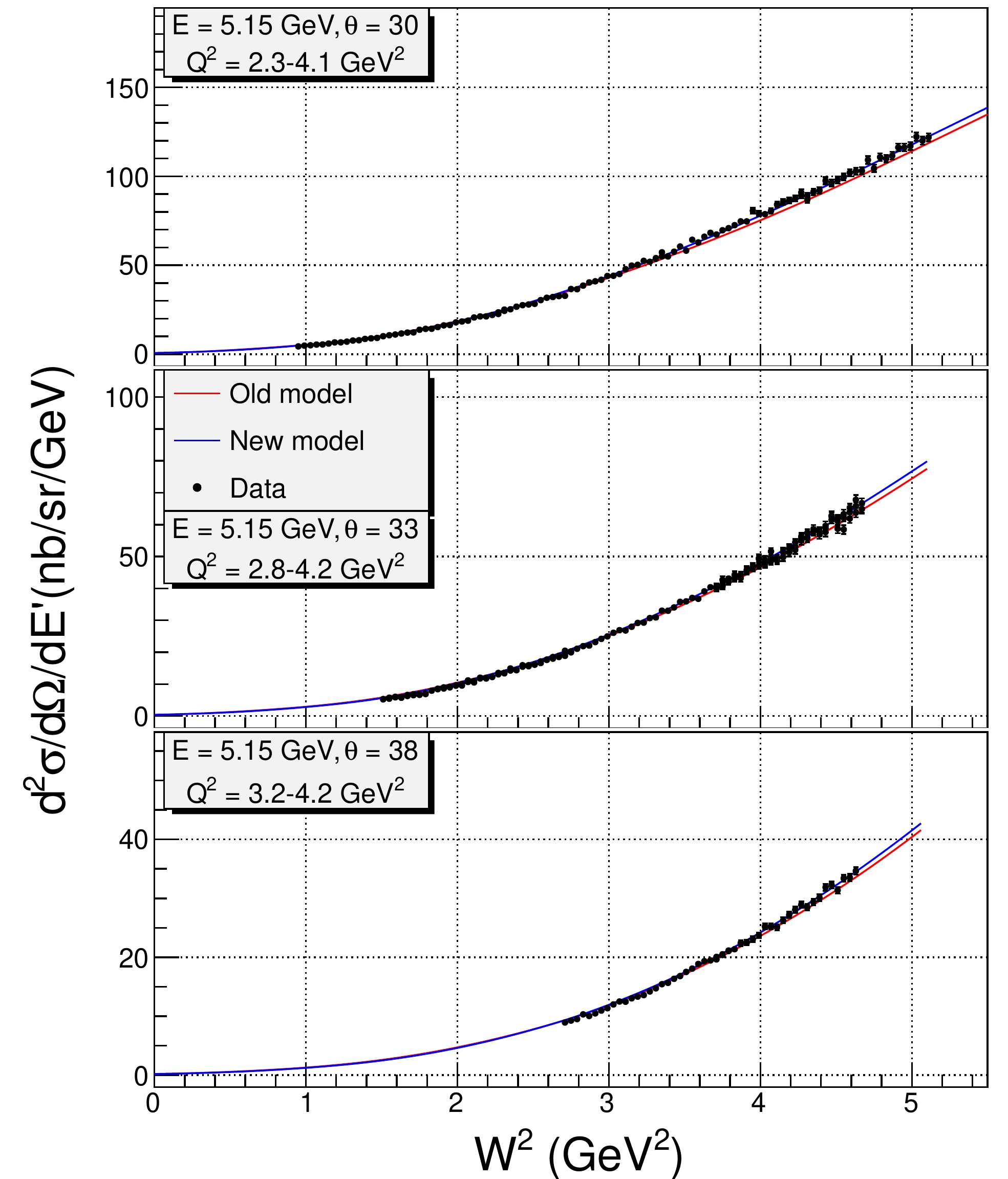,width=5.5in}
\end{center}
\caption { Extracted differential cross section for aluminum compared to the model cross section. }
\label{fig:csAl5151_1}
\end{figure}

\newpage
\subsection{Cross Section Ratios and Extraction of \texorpdfstring{$R_{A}-R_{D}$}{RA-RD} } \label{sec:RARD}
During the running period of this experiment the experiment E06-009~\cite{E06009}  
took data on a deuterium target at the same kinematic points as the current experiment. 
In order to study the nuclear dependence of $R$, the deuteron cross sections were also 
extracted by the author in addition to the cross sections from heavier nuclei of this experiment. 

\begin{figure}[p]
\begin{center}
\epsfig{file=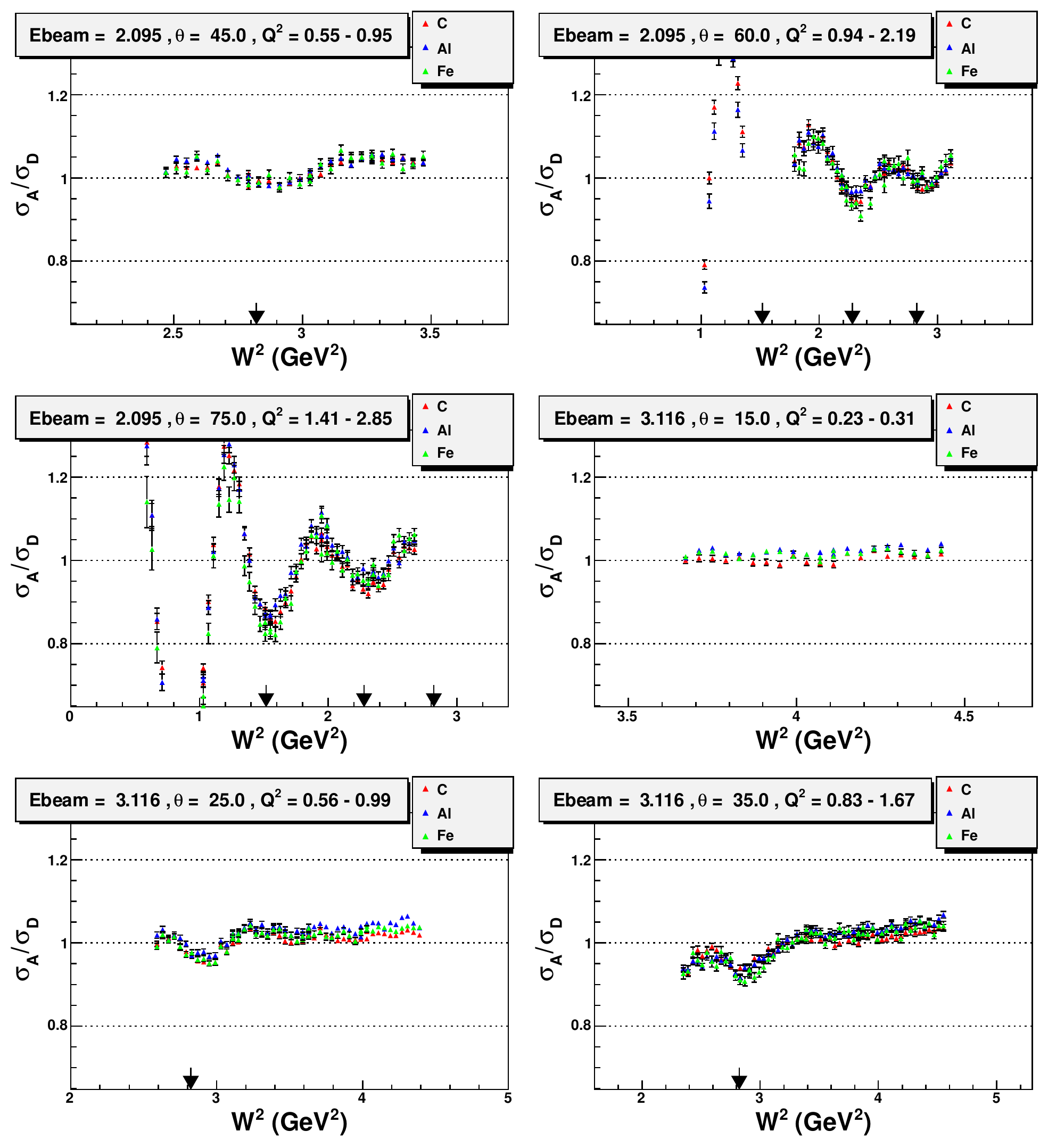,width=6.0in}
\end{center}
\caption { Cross section ratios of nuclear targets to deuteron versus $W^2$ is shown. 
Colors indicate different nuclear targets. 
The three dip regions (where enough $W^2$ range exists) correspond to the nucleon resonances 
$P_{33}$ (\mbox{$W^2$ = 1.52 GeV$^2$}), $S_{11}$ (\mbox{$W^2$ = 2.28 GeV$^2$}), $F_{15}$ (\mbox{$W^2$ = 2.82 GeV$^2$)} and 
are indicated by arrows. 
Since the Fermi momentum of deuteron is the smallest of all nuclei, the resonance structure is not washed out for it,  
and the cross sections for deuteron at the positions of resonances are larger than that of 
other nuclei. Therefore, resonances are seen as dips in the cross section ratios. 
}
\label{fig:csRatio1}
\end{figure}

\begin{figure}[p]
\begin{center}
\epsfig{file=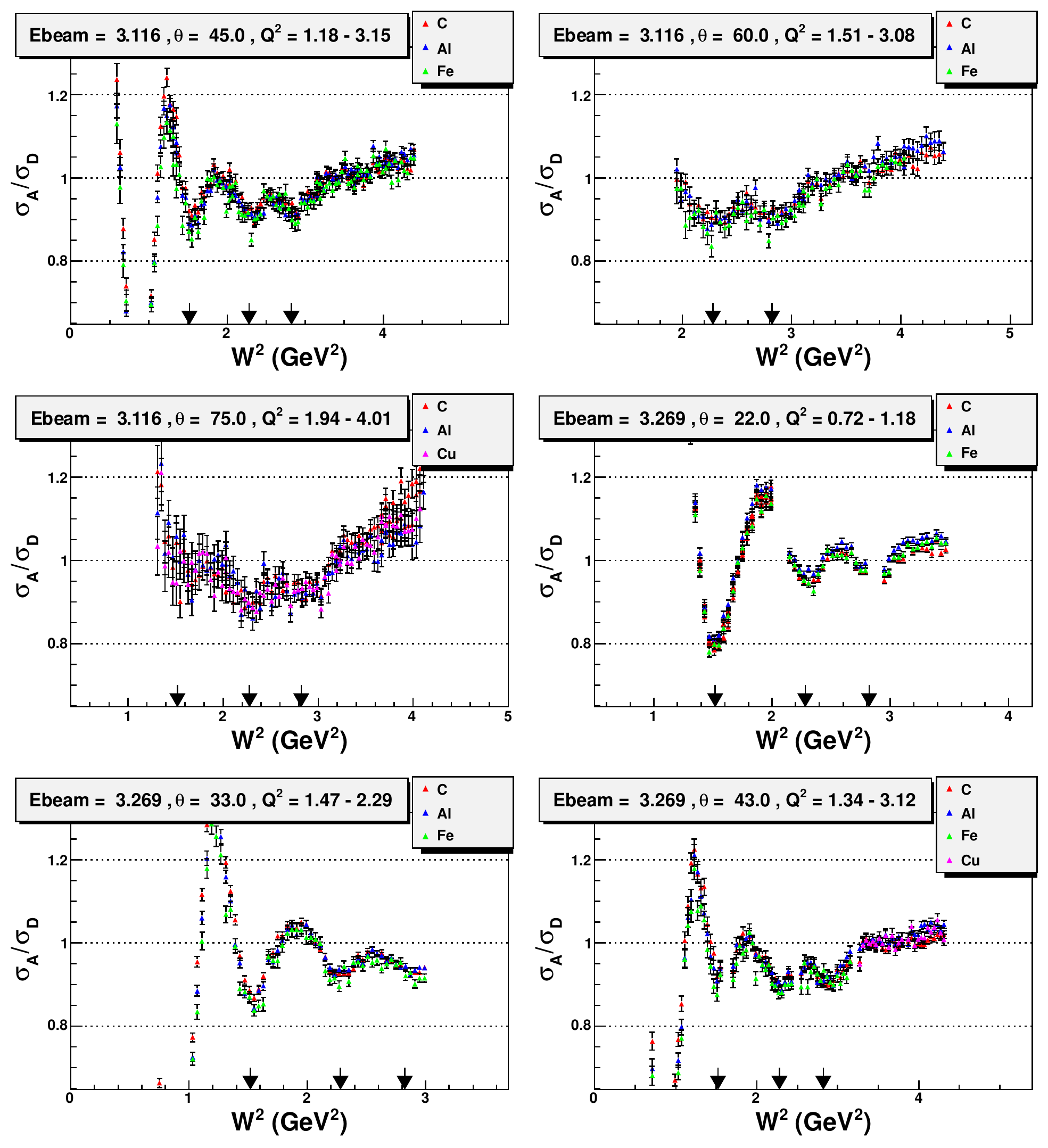,width=6.0in}
\end{center}
\caption { As in Fig.~\ref{fig:csRatio1}.
}
\label{fig:csRatio2}
\end{figure}

\begin{figure}[p]
\begin{center}
\epsfig{file=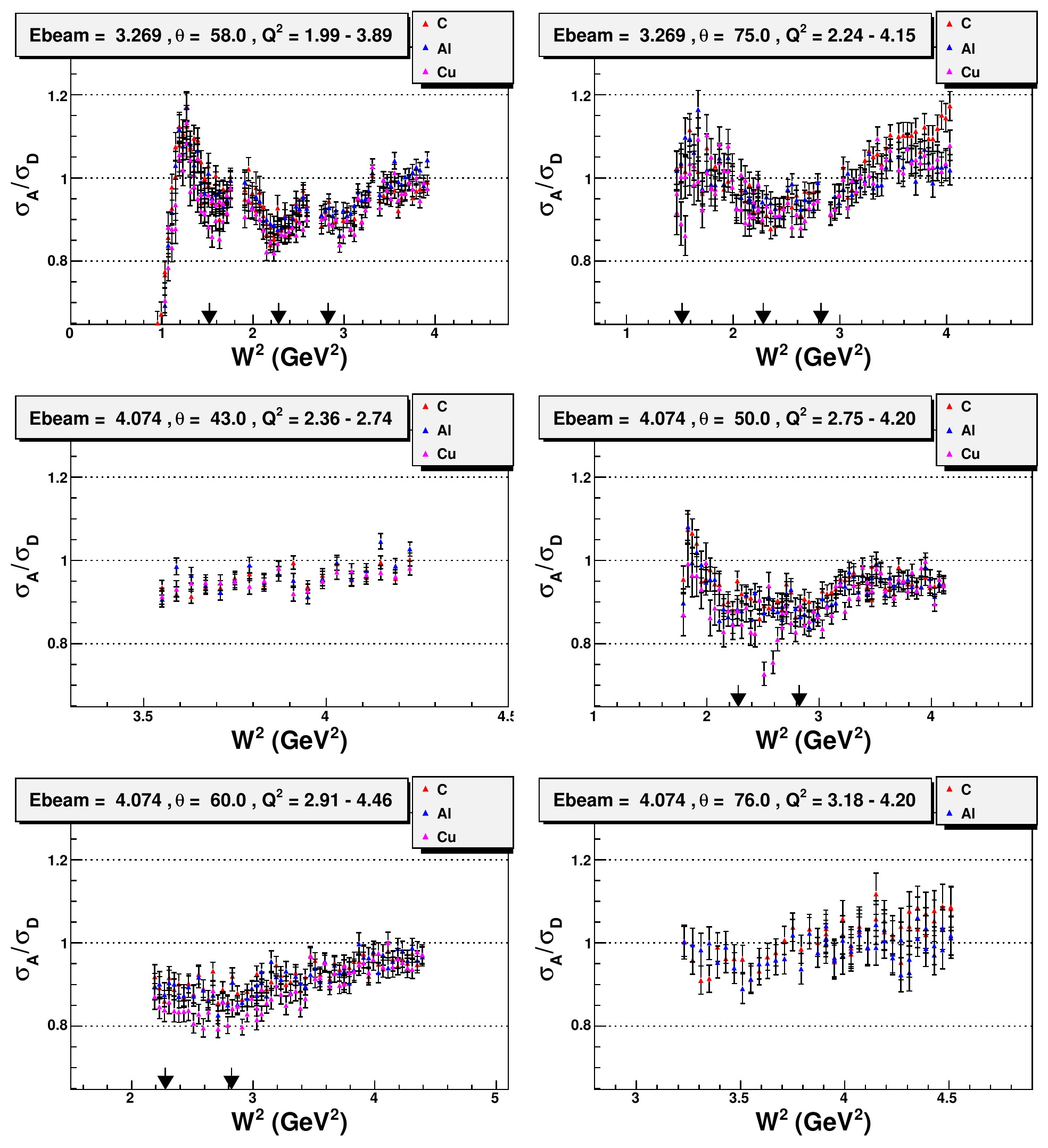,width=6.0in}
\end{center}
\caption { As in Fig.~\ref{fig:csRatio1}.
}
\label{fig:csRatio3}
\end{figure}

\begin{figure}[p]
\begin{center}
\epsfig{file=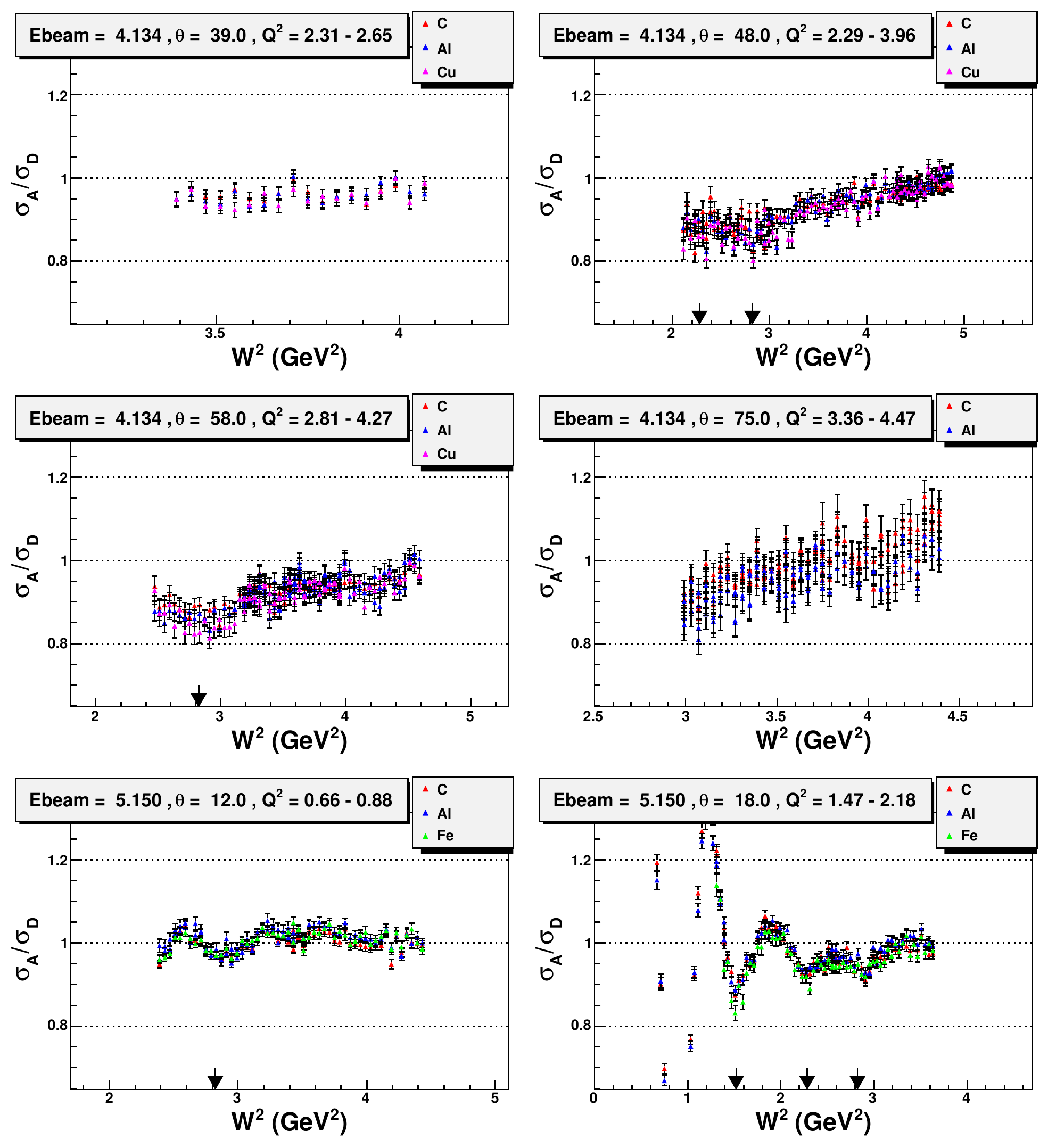,width=6.0in}
\end{center}
\caption {  As in Fig.~\ref{fig:csRatio1}.
}
\label{fig:csRatio4}
\end{figure}

\begin{figure}[!p]
\begin{center}
\epsfig{file=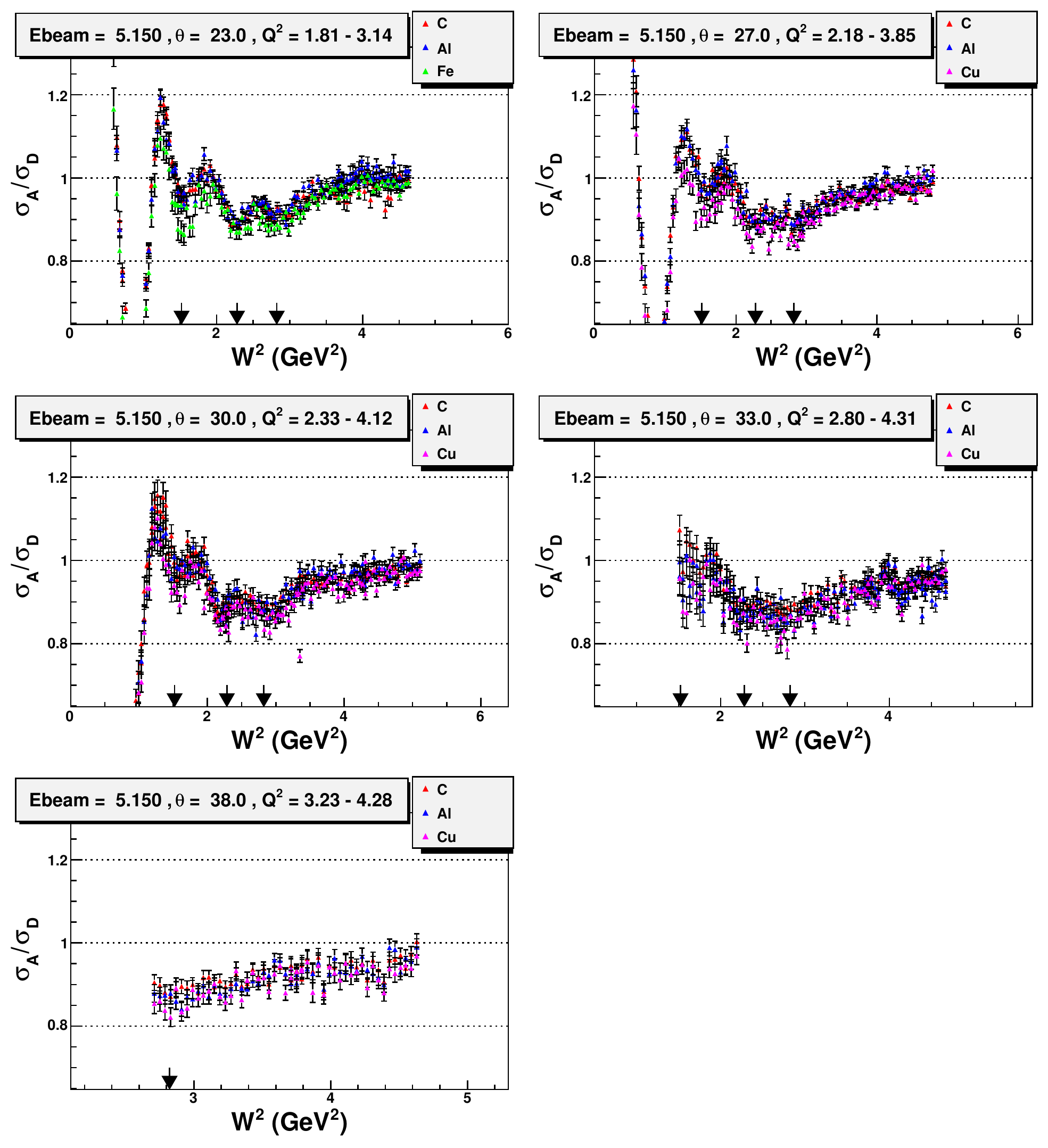,width=6.0in}
\end{center}
\caption { As in Fig.~\ref{fig:csRatio1}.
}
\label{fig:csRatio5}
\end{figure}

The ratio of cross sections $\sigma_{A}/\sigma_{D}$ are shown in Figs.~\ref{fig:csRatio1} to Fig.~\ref{fig:csRatio5}.
The cross section ratios of nuclear targets to deuterium are indicated by different colors. 
The three dip regions (where enough $W^2$ range exists) correspond to the nucleon resonances 
$P_{33}$ ($W^2$ = 1.52 GeV$^2$), $S_{11}$ ($W^2$ = 2.28 GeV$^2$), and $F_{15}$ (\mbox{$W^2$ = 2.82 GeV$^2$)}. 
Since the Fermi momentum of deuteron is the smallest of all nuclei, the resonance structure is not washed out,  
and the cross sections for deuteron at the positions of resonances are larger than that of 
other nuclei. Therefore, resonances are seen as dips in the cross section ratios. 
Coulomb corrections were applied to the cross section ratios shown in these figures as discussed in 
Section~\ref{sec:CoulumbCorr} and the strength of the correction is shown in Fig.~\ref{fig:CoulombCorr}. 
Since these are ratios of cross sections and not of separated structure functions,
therefore it is necessary to assume that there is no nuclear dependence of $R$ in order to interpret these
data. This follows from Eq.~\ref{eq:RA-RD_app} below.

The nuclear to deuteron cross section ratio can be written in terms of structure function 
$R$ as shown in the following equation 
\vspace{-1.0cm}
\begin{center}
\begin{equation} \label{eq:RA-RD_app}
\displaystyle\frac{\sigma_{A}}{\sigma_{D}}=\displaystyle\frac{\sigma_{T}^{A}}{\sigma_{T}^{D}}\left[\frac{1+\epsilon R_{A}}{1+\epsilon R_{D}}\right] \approx \frac{\sigma_{T}^{A}}{\sigma_{T}^{D}} \displaystyle \left[ 1 + \epsilon' \left( R_{A} - R_{D}\right)  \right]
\end{equation}
\end{center}
where $\epsilon'=\epsilon/(1+\epsilon R_{D})$. 
At fixed values of $W^2$ and $Q^{2}$ the right side of this equation is function of $\epsilon'$ only  
and represents the equation of a straight line with constant 
$\sigma_{T}^{A}/\sigma_{T}^{D}$ and slope $(\sigma_{T}^{A}/\sigma_{T}^{D})(R_{A}-R_{D})$. 

Since $\epsilon R_{D}$ is small, the uncertainty of $\epsilon'$ due to $R_{D}$ is negligible. 
The $R_{A}-R_{D}$ results are consistent with zero for $W^{2}>2$ GeV$^{2}$, in agreement
with DIS data where no significant A dependence of $R$ was found. 
These results indicate that contributions to $R$ from nuclear effects are small. 
The $Q^{2}$ range of this data is not low enough to investigate the predictions of Miller~\cite{Miller} 
who suggests a significant enhancement of longitudinal cross section at $x$ = 0.4 and $Q^{2}=0.3$ GeV$^2$ for 
heavy nuclei.
This will be possible when the results of E02-109~\cite{E02-109} experiment are available. 
Combining these data with the data of E02-109 will allow to perform a precise Rosenbluth 
separation in wide range of $Q^2$ for carbon, aluminum and iron.

\begin{figure}[p]
\begin{center}
\epsfig{file=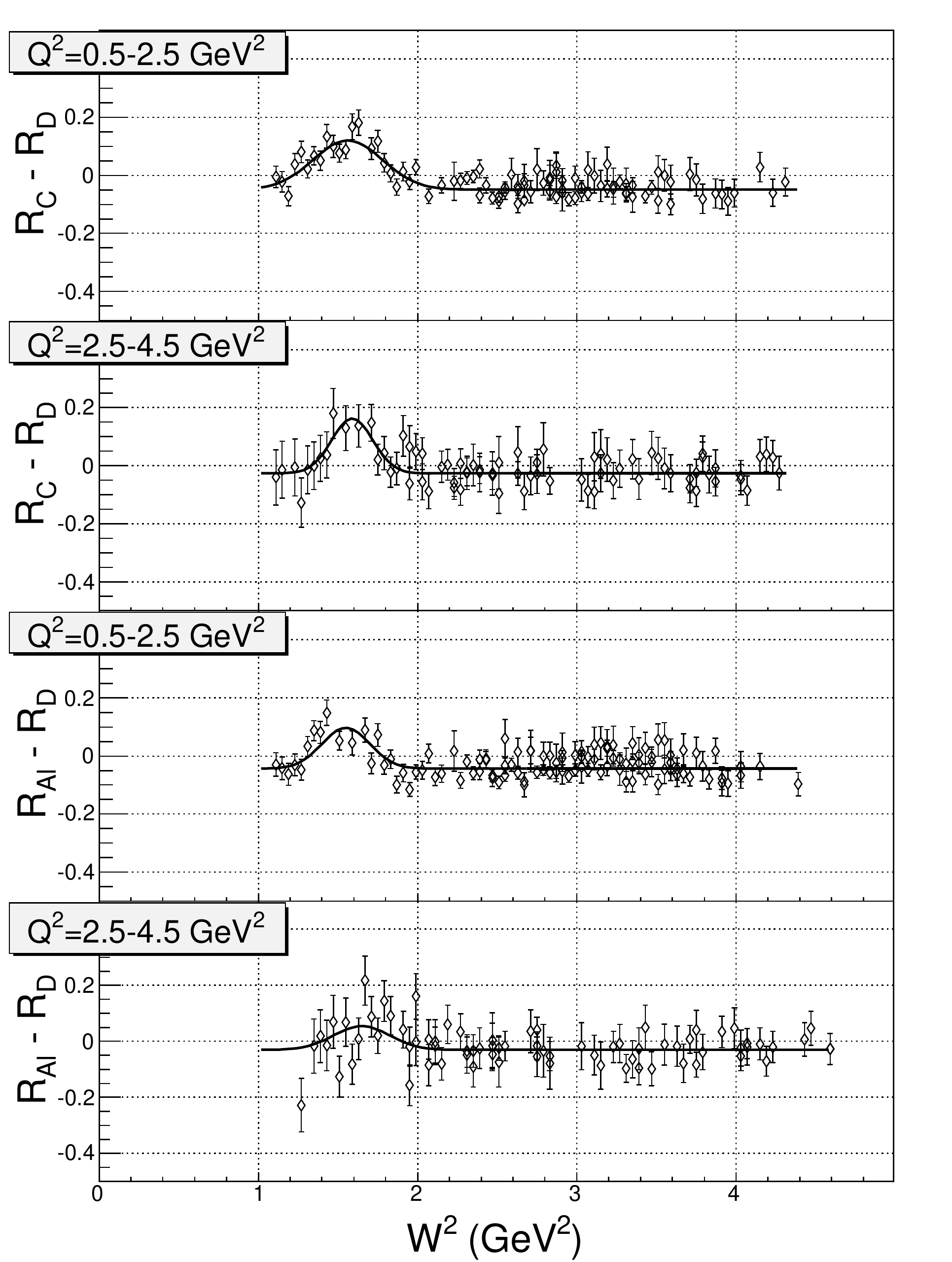,width=5.5in}
\end{center}
\caption { The results for $R_{A}-R_{D}$ are plotted versus $W^{2}$ for two $Q^{2}$ bins. 
Top two plots are for carbon, and bottom two are for aluminum.}
\label{fig:RA-RD}
\end{figure}

\begin{figure}[ht]
\begin{center}
\epsfig{file=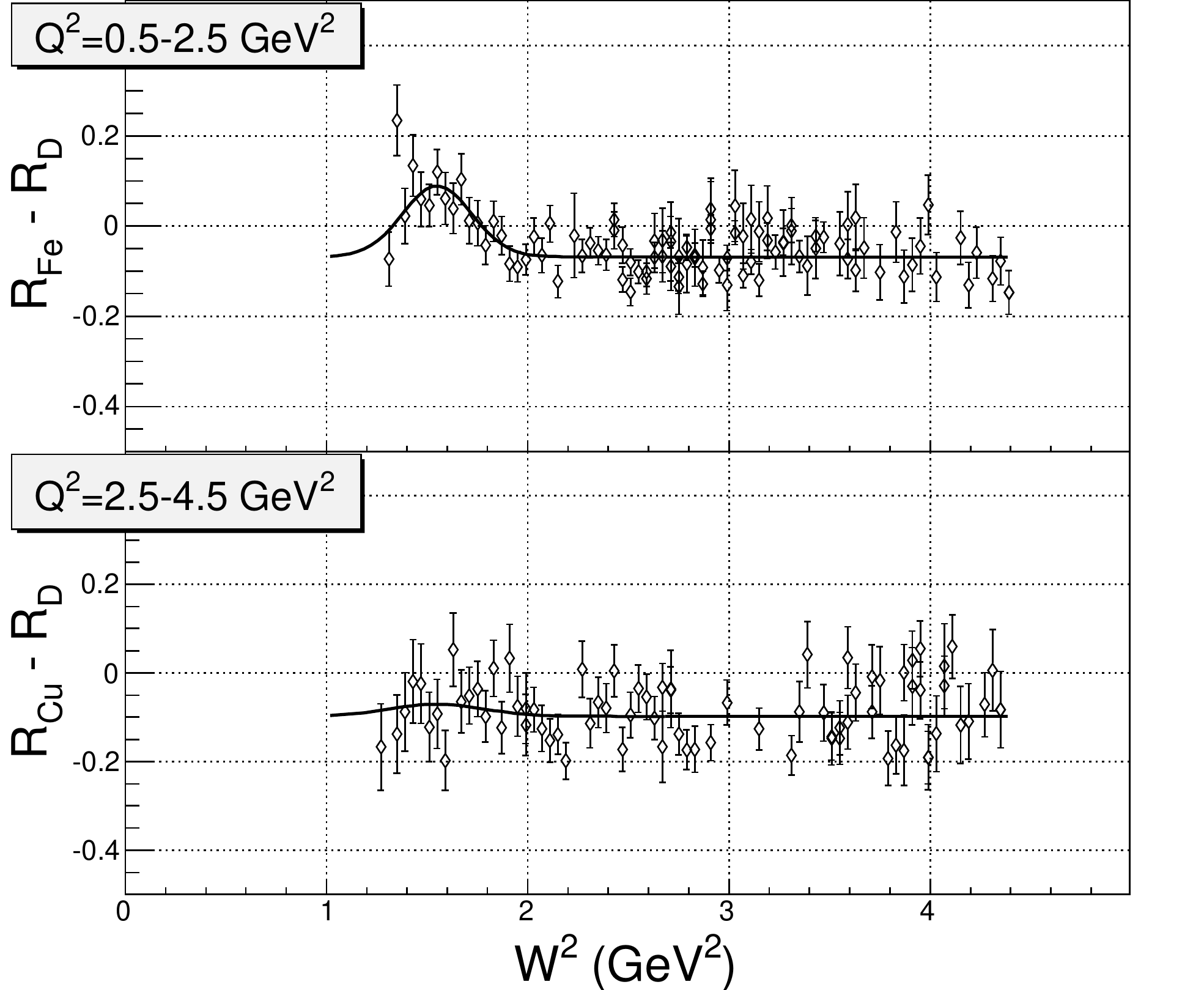,width=5.5in}
\end{center}
\caption { The results for $R_{A}-R_{D}$ are plotted versus $W^{2}$ for two $Q^{2}$ bins. 
Top plot is for iron, and bottom one is for copper.}
\label{fig:RA-RDFeCu}
\end{figure}

The observed enhancement of $R_{A}-R_{D}$ near $W^{2} = $ 1.5 GeV$^2$ (Figs.~\ref{fig:RA-RD} and ~\ref{fig:RA-RDFeCu}) indicates some nuclear 
dependence, which vanishes with increase of $Q^{2}$ and with nuclear number $A$. 
A final conclusion about the origins of the enhancement of $R_{A}-R_{D}$ can not be drawn until the analysis of experiment 
E06-009~\cite{E06009} is complete and precision deuteron data become available which will allow a better 
representation of $\sigma_{D}$ in Eq.~\ref{eq:RA-RD_app}. 
These data will allow improvements to the deuteron model and precise cross sections in the same kinematic 
region as the current experiment.

\begin{figure}[p]
\begin{center}
\epsfig{file=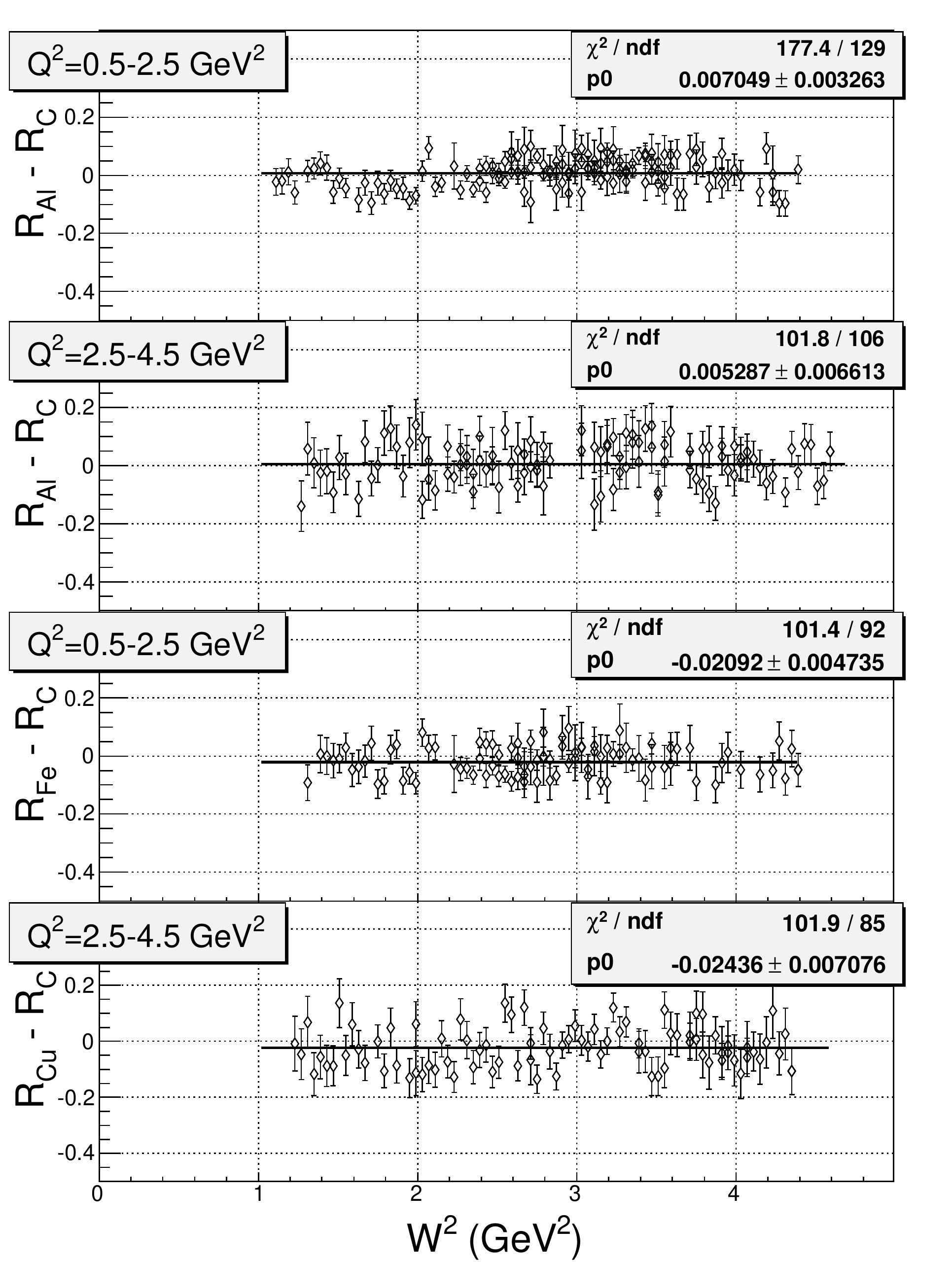,width=5.5in}
\end{center}
\caption { The results for $R_{A}-R_{C}$ are plotted versus $W^{2}$ for two $Q^{2}$ bins. 
Top two plots are for aluminum, then iron and copper.}
\label{fig:RA-RC}
\end{figure}

In addition to extracting $R_{A}-R_{D}$, $R_{A}-R_{C}$ was also extracted, shown in Fig.~\ref{fig:RA-RC}.
The model for $R_{C}$ was obtained from the current fit to the data as described in Section~\ref{sec:GlobalFit}. 
No nuclear dependence of $R_{A}-R_{C}$ was found over the range of $W^{2}$ and in the four $Q^{2}$ bins 
for aluminum, iron and copper.

\subsection{Rosenbluth Separated R} \label{sec:RosSepR}
\begin{figure}[p]
\begin{center}
\epsfig{file=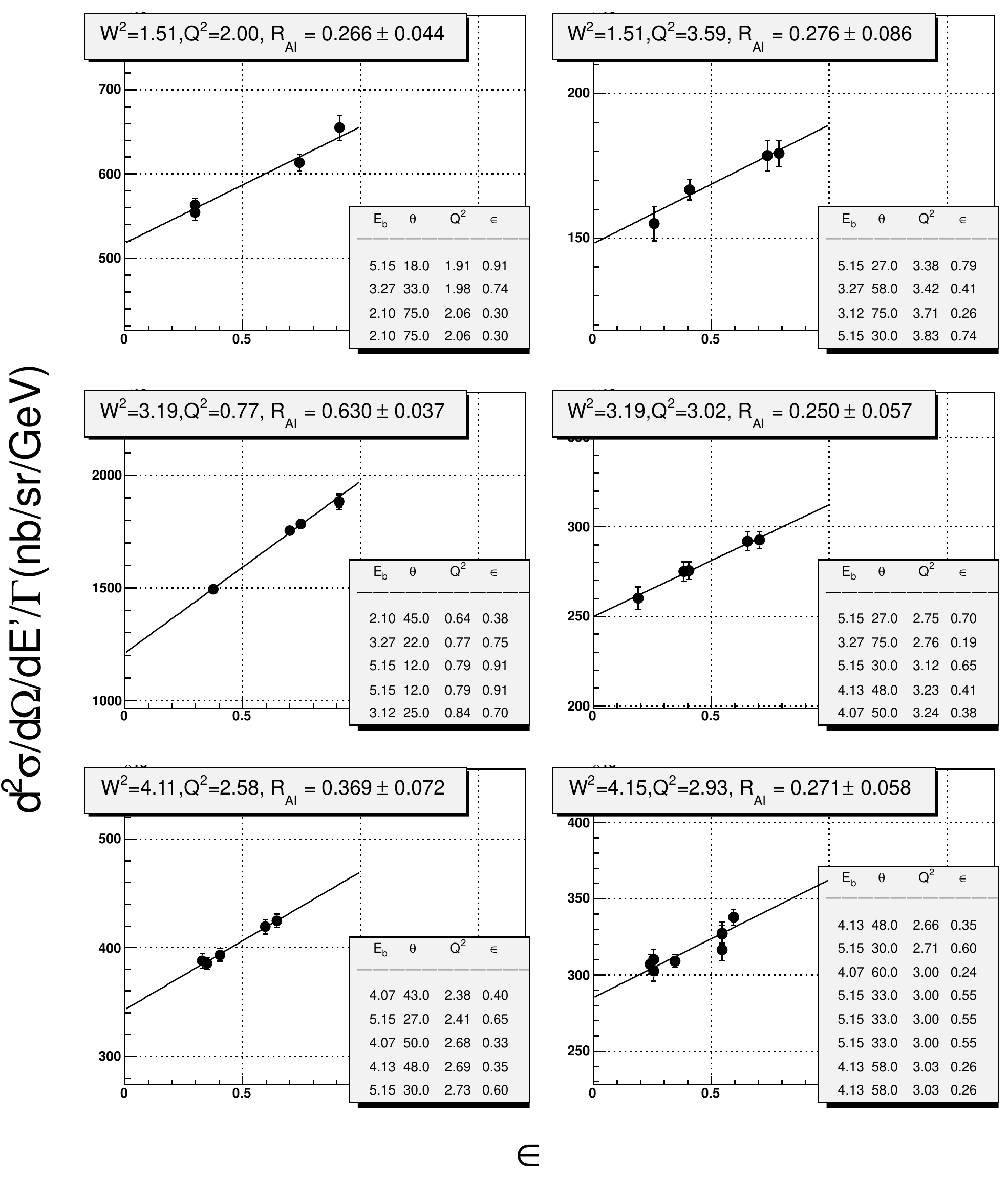,width=5.8in}
\end{center}
\caption { Some examples of Rosenbluth separated $R$ for aluminum. 
The horizontal axis, $\epsilon$, is the relative polarization of longitudinal photons. 
The vertical axis is the reduced cross section defined in the left side of Eq.~\ref{eq:RosSepR}.
The intercept of the linear fit with the $y$ axis defines $\sigma_{T}$, while the 
slope defines $\sigma_{L}$. $R$ is the ratio $\sigma_{L}/\sigma_{T}$. 
For each Rosenbluth separation the beam energy, electron scattering angle, $Q^2$ and $\epsilon$ are given in the legend.  
}
\label{fig:R_separ}
\end{figure}

Once the cross section model has converged, the final cross section can be extracted and used 
to perform Rosenbluth separation using the following formula 
\vspace{-1cm}
\begin{center}
\begin{equation}
\displaystyle\frac{d^{2}\sigma_{A}}{d\Omega dE'}\bigg/\Gamma=\displaystyle\left( \sigma_{T} + \epsilon \sigma_{L}\right) \label{eq:RosSepR}
\end{equation}
\end{center}
where $\Gamma$ is the flux of virtual photons and is given by Eq.~\ref{eq:GammaFlux}. 
In order to perform a Rosenbluth separation using the Eq.~\ref{eq:RosSepR} 
it is necessary to measure cross sections at the same values of $W^{2}$ and $Q^{2}$ 
but at different values of $\epsilon$. 
In this analysis the cross sections are calculated at fixed $W^{2}$ bins, as described in Section~\ref{Sec:Acceptance}, 
where for each $W^{2}$ there are several values of $Q^{2}$ corresponding to different 
$\epsilon$'s. 
In order to reduce systematic uncertainties of $R$, each Rosenbluth separation was performed 
when there were more then three points with an $\epsilon$ range larger than 0.3, 
one of the $\epsilon$ points must be greater or smaller than 0.5 and all points were 
within $\Delta Q^{2} < 0.5$ GeV$^2$ range. 
The average $\epsilon$ range for all Rosenbluth separations is 0.5$\pm$0.1. 
The model cross section obtained after the iteration procedure was used 
to interpolate all cross sections within the $\Delta Q^{2} < 0.5$ range into an average value of $Q^{2}$. 
In order to avoid having correlated errors no data point was used twice in the Rosenbluth separation.
The value of $\chi^{2}$ per degree of freedom is about 1 indicating that point-to-point systematic 
uncertainties are taken into account correctly. 
A few Rosenbluth separations are shown in Fig.~\ref{fig:R_separ}. 

Rosenbluth separations were performed for carbon, aluminum iron and copper. 
The Coulomb field of these nuclei are not negligible and
a correction is necessary to take into account its effect.
This is very important for Rosenbluth separation since the Coulomb correction is 
$\epsilon$ (beam energy) dependent and can change the value of $R$ dramatically. 
\begin{figure}[p]
\begin{center}
\epsfig{file=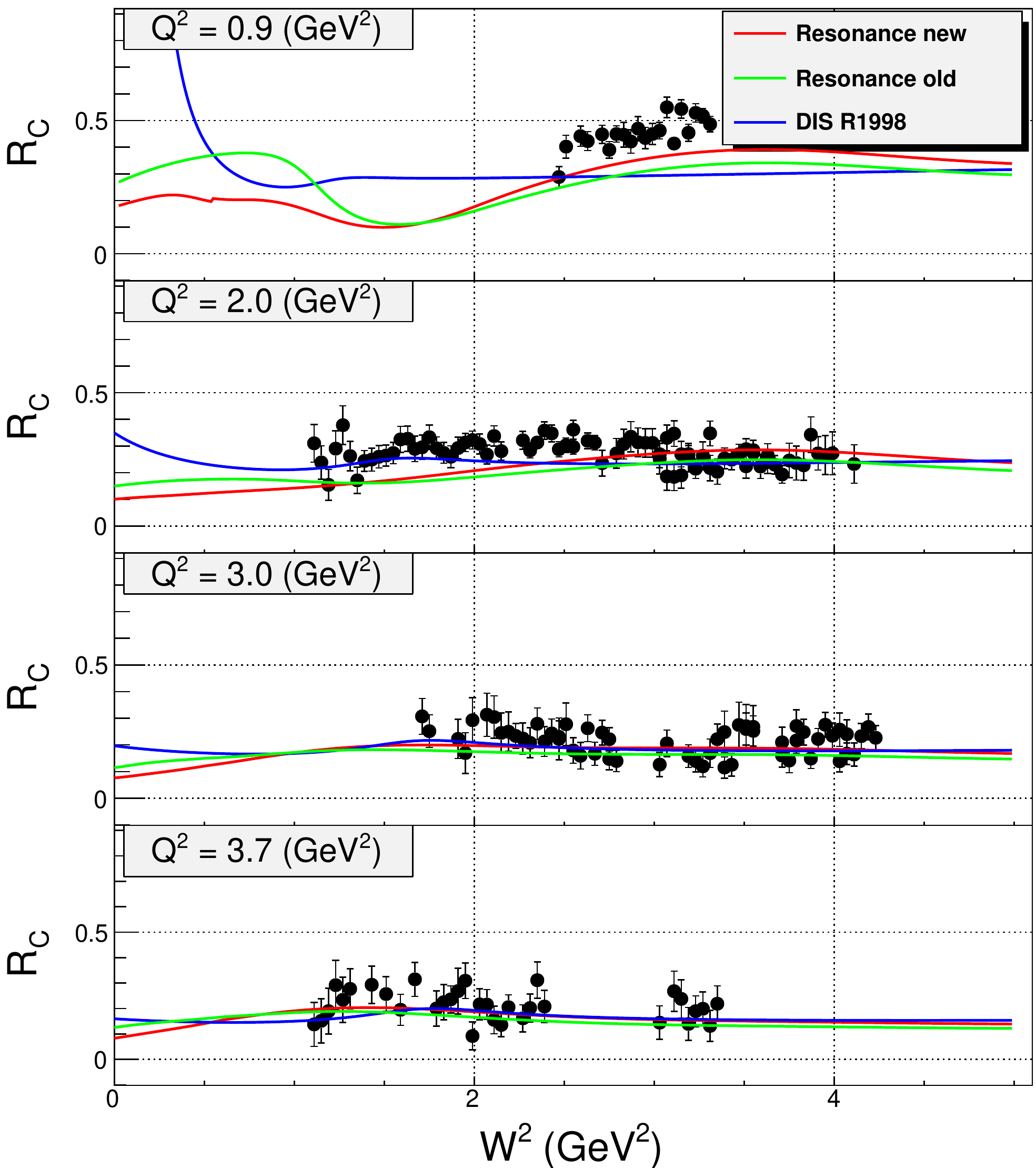,width=5.8in}
\end{center}
\caption { Rosenbluth separated $R$ of carbon as a function of $W^{2}$ at four different $Q^{2}$ ranges. 
In order to take into account the $Q^{2}$ dependence of $R$ for each range, 
the model obtained from the global fit is used to move $R$s to an average $Q^2$ shown in the plot.}
\label{fig:R_carbon}
\end{figure}

\begin{figure}[p]
\begin{center}
\epsfig{file=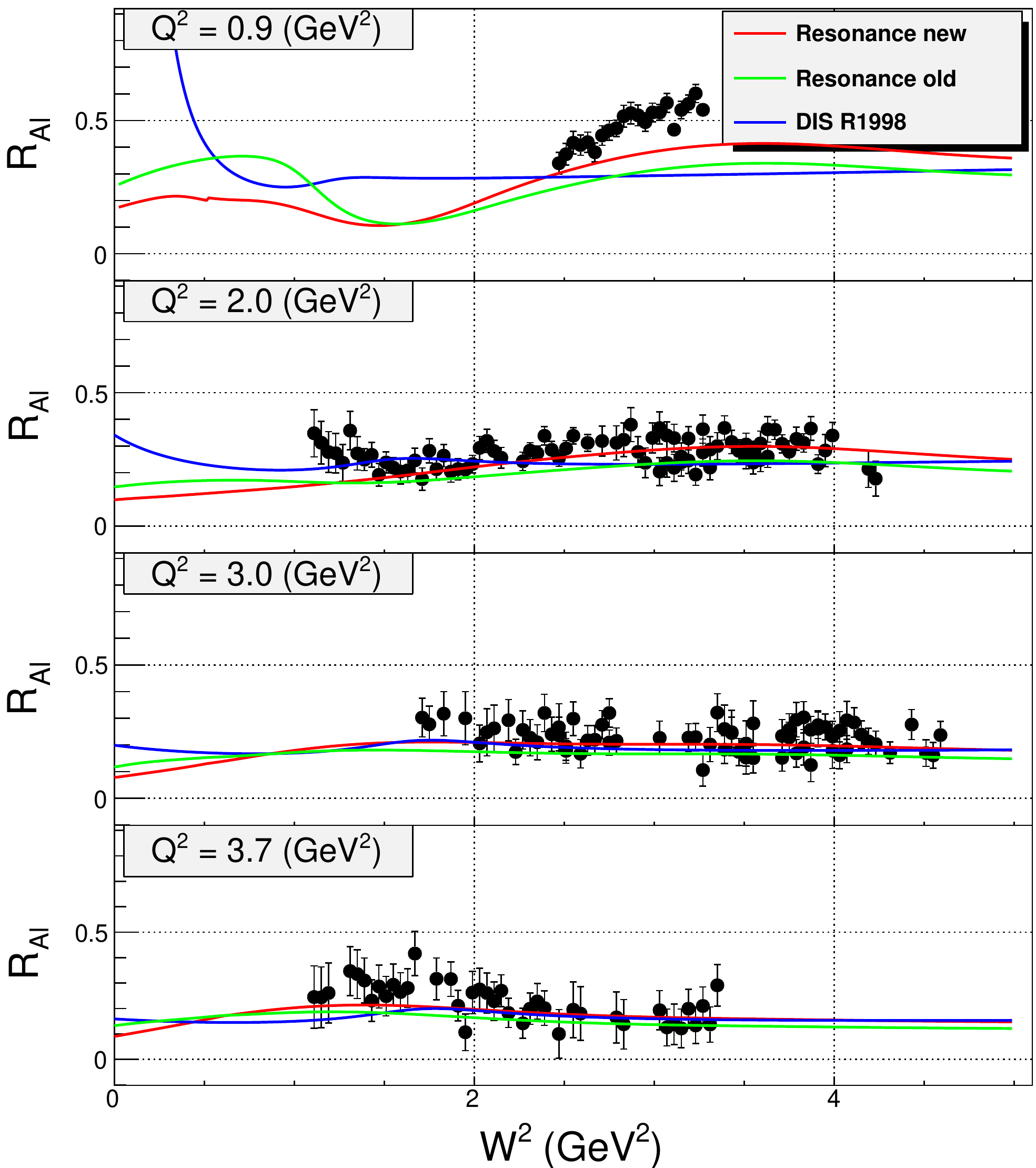,width=5.8in}
\end{center}
\caption { Rosenbluth separated $R$ of aluminum as a function of $W^{2}$ at four different $Q^{2}$ ranges. 
In order to take into account the $Q^{2}$ dependence of $R$ for each range, 
the model obtained from the global fit is used to move $R$s to an average $Q^2$ shown in the plot. }
\label{fig:R_aluminum}
\end{figure}

\begin{figure}[p]
\begin{center}
\epsfig{file=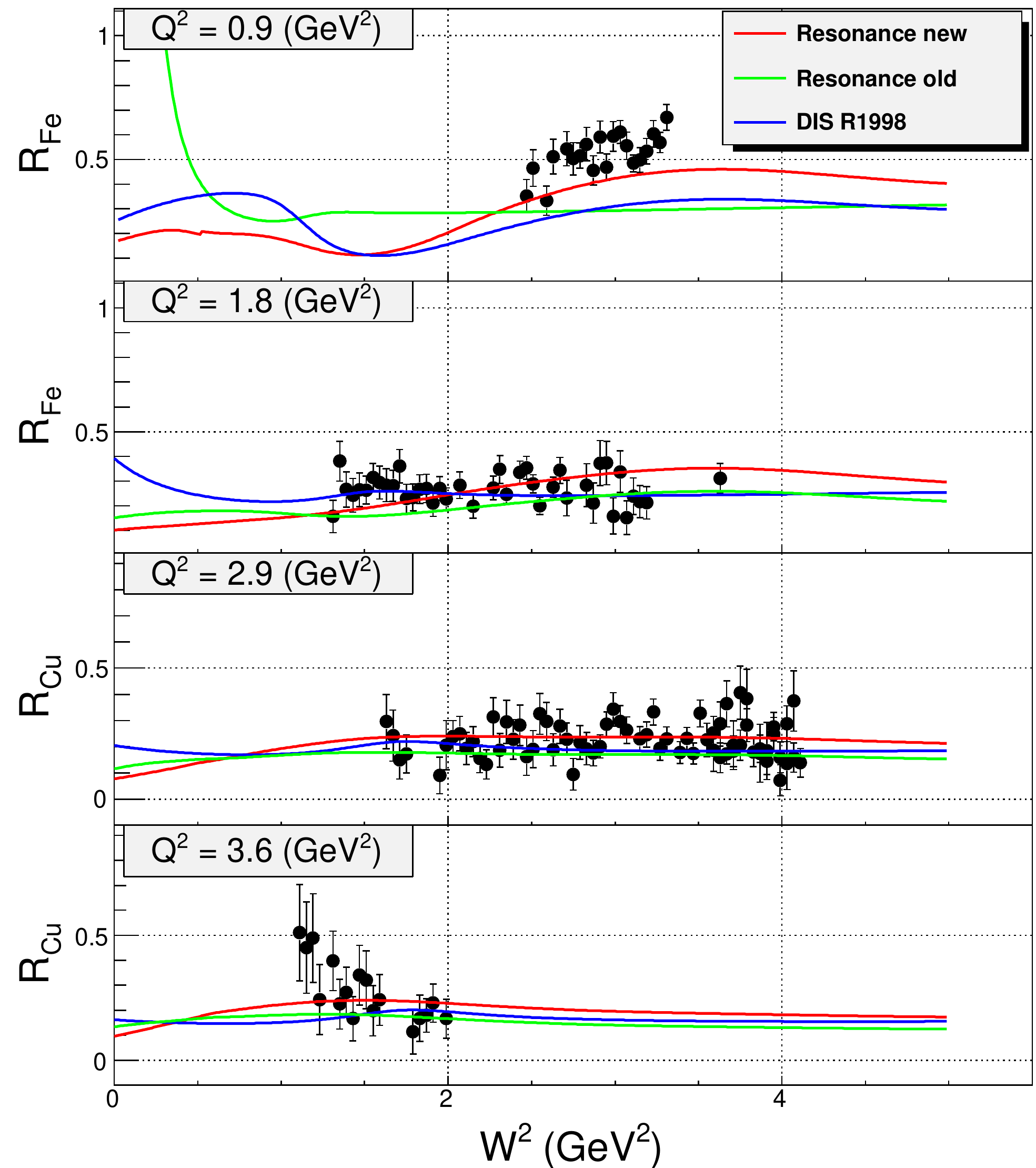,width=5.8in}
\end{center}
\caption { Rosenbluth separated $R$ of iron and copper as a function of $W^{2}$ at four different $Q^{2}$ ranges. 
In order to take into account the $Q^{2}$ dependence of $R$ for each range, 
the model obtained from the global fit is used to move $R$s to an average $Q^2$ shown in the plot. }
\label{fig:R_FeCu}
\end{figure}

Another source of systematic uncertainty is the $\epsilon$ dependence of radiative corrections.
Although the radiative corrections for inclusive electron-nuclei are well understood, at the level of $\sim$ 1\%, 
it is still possible to have a systematic error in $R$ if the correction is large. 
In this experiment the maximum radiative correction was about 30\%. 
In order to study the effect of radiative corrections, an artificial $\epsilon$ dependence is introduced to 
cross sections during the Rosenbluth separation and then the separated $R$s are compared to the $R$s 
separated without the artificial $\epsilon$ dependence. 
The size of this artificial $\epsilon$ dependence is set to be within $\pm$ 1\% range. 
A systematic uncertainty of $\Delta R$ = 0.028 is found and is assigned to all values of $R$ attributed 
to the $\epsilon$ dependence of the radiative corrections.

The Rosenbluth separated $R$ for carbon, aluminum, iron and copper are shown as a function of 
$W^{2}$ over the resonance region in Figs.~\ref{fig:R_carbon}, \ref{fig:R_aluminum} and 
\ref{fig:R_FeCu} respectively.
Each figure has four plots indicating four $Q^{2}$ ranges from top to bottom respectively: 
0.5$-$1.0 GeV$^2$, 1.0$-$2.5 GeV$^2$, 2.5$-$3.5 GeV$^2$ and 3.5$-$4.5 GeV$^2$. 
For each $Q^2$ range the $R$ is interpolated, using models of $R$, to the average $Q^2$ 
value shown on top of each plot. 
The lines shown on all plots correspond to different models for $R$. 
The red line (Resonance new) indicates the model obtained after iterating the cross section 
model. 
The blue line is the initial model used to extract the cross sections. 
The green line is the DIS $R$ fit ``R1998''~\cite{Whitlow:1990gk} and is 
based on a fit to world $R$ measurements. 
In general all three models agree with the data, except for the lowest $Q^{2}$ case.
At lowest $Q^{2}=$ 0.9 GeV$^2$ the agreement is better with the ``Resonance new'' model, which 
is natural, since this model is based on the first cross section measurements done in the 
nucleon resonance region, but it still needs improvement. 
This improvement will be done when the low $Q^{2}$ data of \mbox{E02-109}~\cite{E02-109} is available 
to include in the iterative procedure discussed in Sec.~\ref{sec:GlobalFit}.
At $Q^{2}=$ 2.0 GeV$^2$ for carbon and aluminum the agreement between data and DIS model for 
$W^2 <$ 3 is lacking.
This, again, may be caused by not having the low $Q^{2}$ data of the \mbox{E02-109} 
in the iterative procedure. 
For iron at $Q^{2}=$ 2.0 GeV$^2$ the data have larger uncertainties and it is impossible to make a definite 
observation about the agreement of data and models.
This is further complicated by the fact that the difference between the different models is not significant.
At $Q^{2}=$ 3.0 and 3.7 GeV$^2$ for carbon, aluminum and copper the agreement of the data and models is 
good within the statistical errors and the difference between the models decreases with increasing $Q^{2}$.

\subsection{\texorpdfstring{$F_{2}$}{F2} Structure Function and Duality in Nuclei}
The extraction of the $F_{2}$ structure function from cross section data can 
only be accomplished with some input for the ratio $R$ of the longitudinal to 
transverse cross sections.
In order to perform a model-independent extraction of the unpolarized structure function $F_{2}$ 
from  inclusive cross section data, $R$ is obtained by performing a Rosenbluth separation, 
described earlier. 
The Rosenbluth separated $R_{A}$ still contains some model dependence since 
the $R_{A}=R_{D}+F_{fit}$ where $R_{D}$ is known from a model and $F_{fit}$ is 
found by fitting $R_{A}-R_{D}$ at two $Q^{2}$ bins as shown in Figures~\ref{fig:RA-RD} and~\ref{fig:RA-RDFeCu}.
However, it should be noted that the model of $R_{D}$ is based on a fit of
Rosenbluth separated data~\cite{liang}. 

\begin{figure}[p]
\begin{center}
\epsfig{file=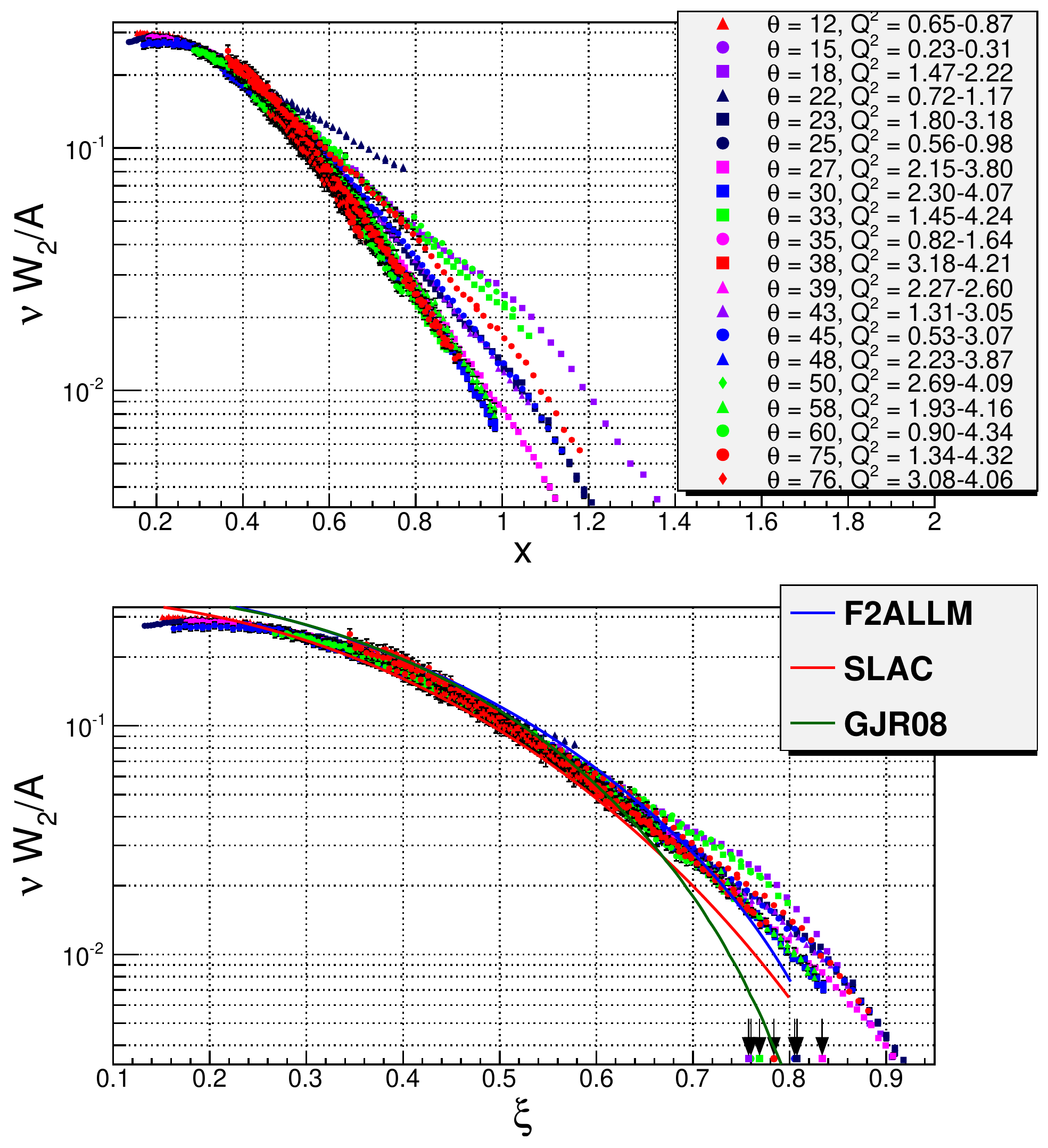,width=5.5in}
\end{center}
\caption { Top plot: Structure function per nucleon versus Bjorken $x$ for carbon (per nucleon) from the present measurement. 
Bottom plot: The $\nu W_{2}=F_{2}$ structure function for carbon (per nucleon) as a function of Nachtmann $\xi$.
The $Q^{2}$ ranges are given at each $\theta$ angle. 
The curves are the proton $F_{2}$ parametrizations at $Q^{2}$ = 10 GeV$^{2}$, corrected for the nuclear EMC effect.
Arrows on the right side ($\xi \sim$ 0.8) of the bottom plot indicate positions of quasi-elastic peaks. 
}
\label{fig:F2C_scaleing}
\end{figure}

\begin{figure}[p]
\begin{center}
\epsfig{file=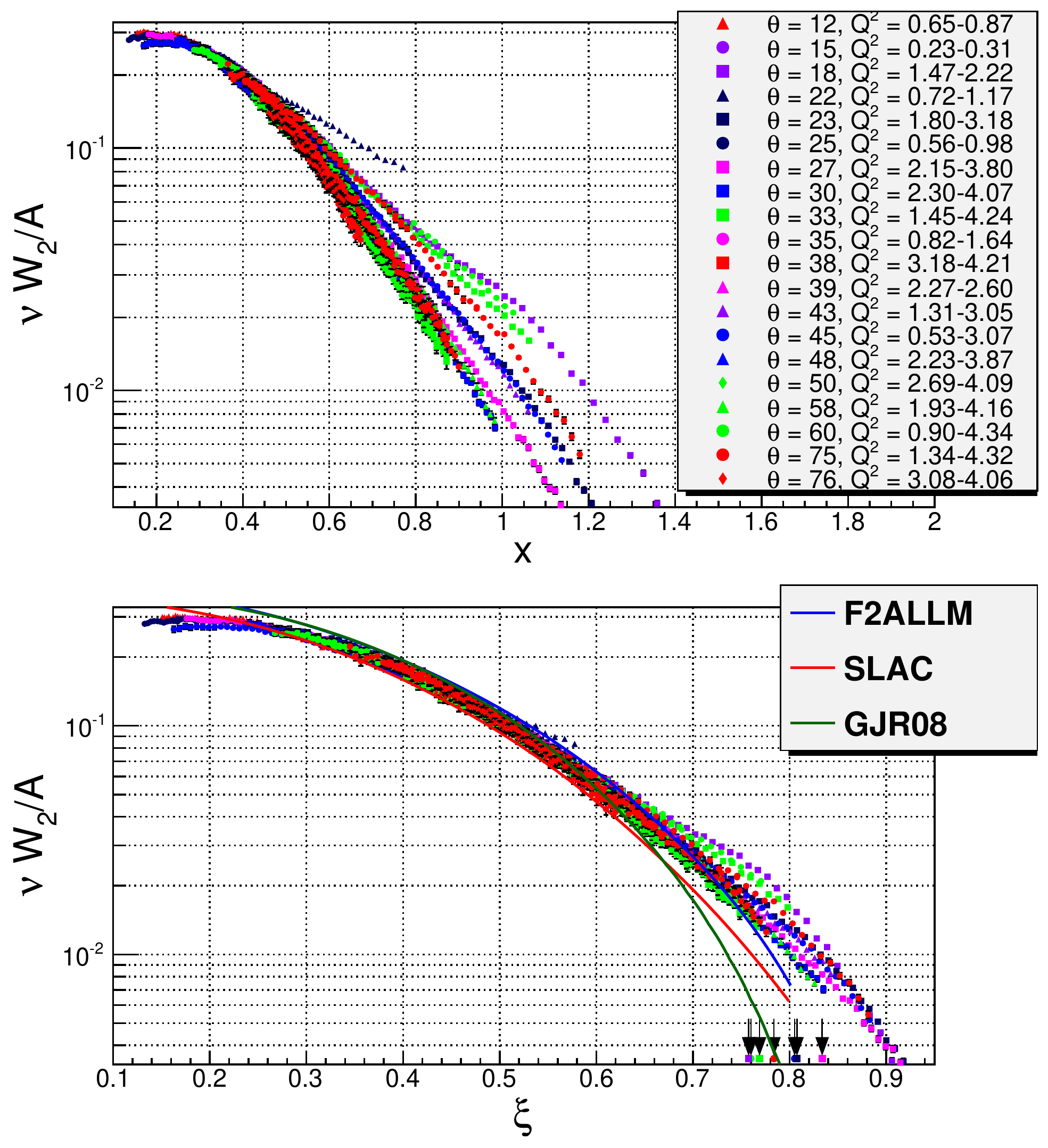,width=5.5in}
\end{center}
\caption { Top plot: Structure function per nucleon versus Bjorken $x$ for aluminum (per nucleon) from the present measurement. 
Bottom plot: The $\nu W_{2}=F_{2}$ structure function for aluminum (per nucleon) as a function of Nachtmann $\xi$.
The $Q^{2}$ ranges are given at each $\theta$ angle.
The curves are the proton $F_{2}$ parametrizations at $Q^{2}$ = 10 GeV$^{2}$, corrected for the nuclear EMC effect. 
Arrows on the right side ($\xi \sim$ 0.8) of the bottom plot indicate positions of quasi-elastic peaks. 
}
\label{fig:F2Al_scaleing}
\end{figure}

\begin{figure}[p]
\begin{center}
\epsfig{file=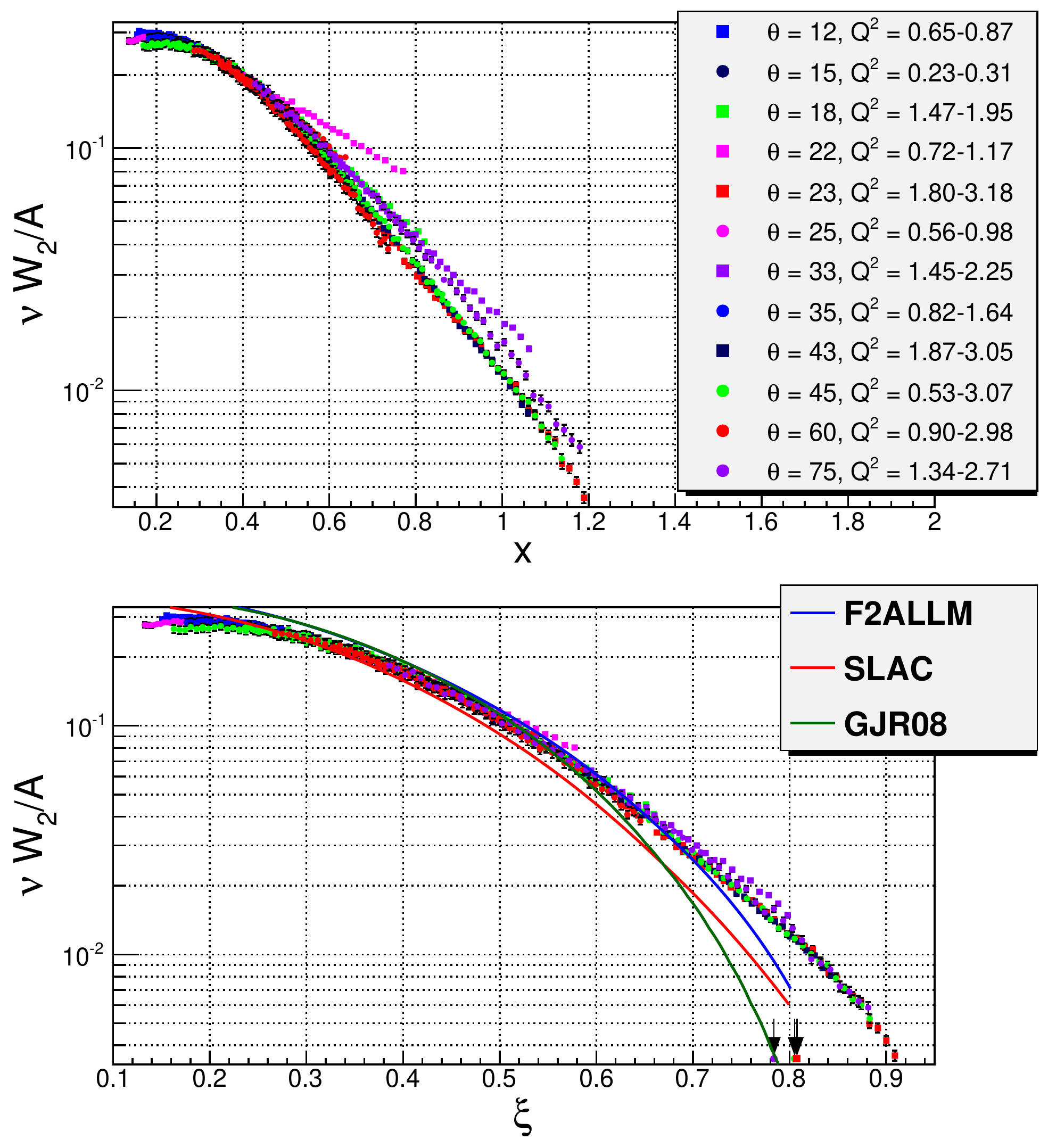,width=5.5in}
\end{center}
\caption { Top plot: Structure function per nucleon versus Bjorken $x$ for iron (per nucleon) from the present measurement. 
Bottom plot: The $\nu W_{2}=F_{2}$ structure function for iron (per nucleon) as a function of Nachtmann $\xi$.
The $Q^{2}$ ranges are given at each $\theta$ angle.
The curves are the proton $F_{2}$ parametrizations at $Q^{2}$ = 10 GeV$^{2}$, corrected for the nuclear EMC effect.
Arrows on the right side ($\xi \sim$ 0.8) of the bottom plot indicate positions of quasi-elastic peak. 
}
\label{fig:F2Fe_scaleing}
\end{figure}

\begin{figure}[p]
\begin{center}
\epsfig{file=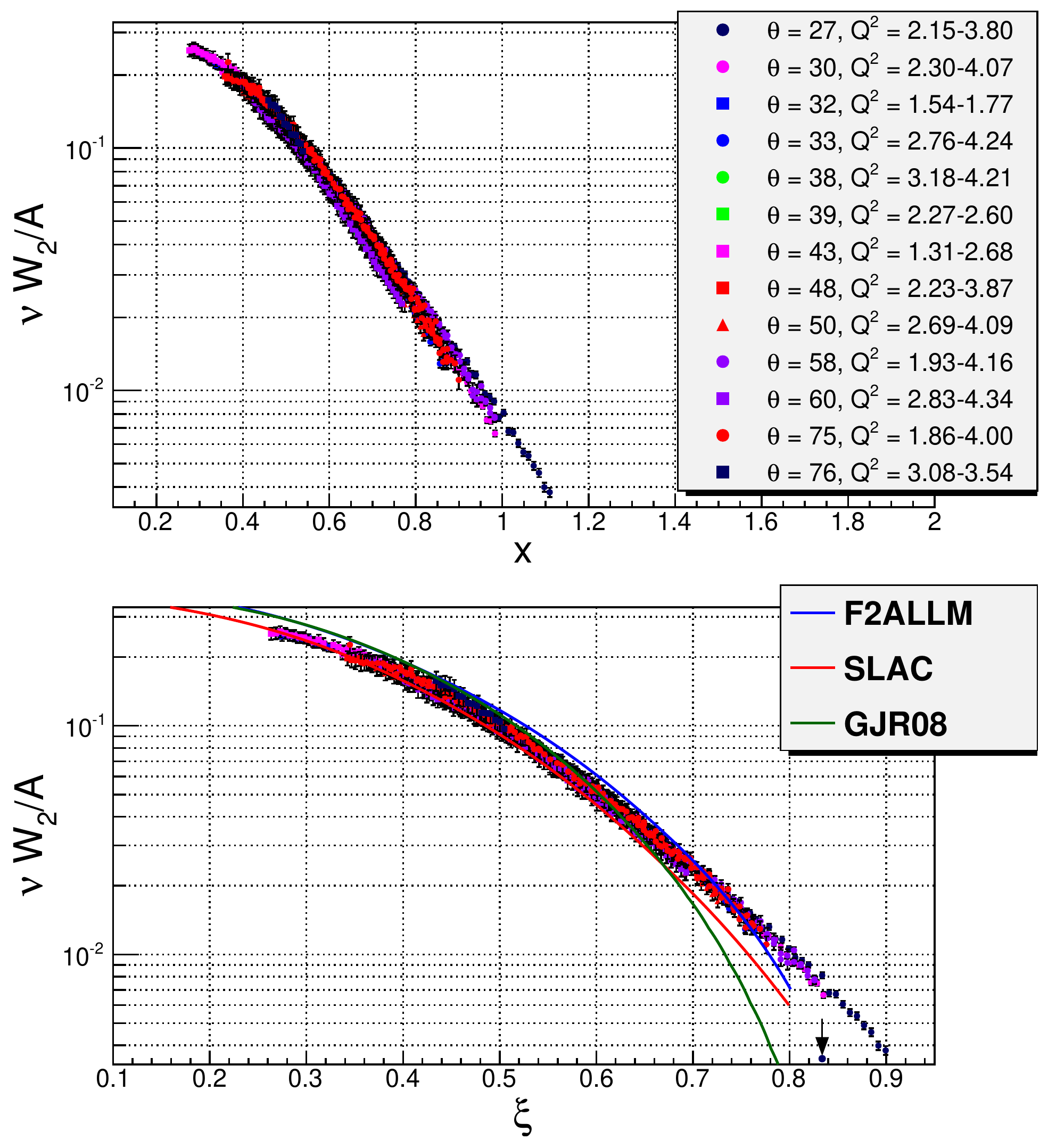,width=5.5in}
\end{center}
\caption { Top plot: Structure function per nucleon versus Bjorken $x$ for copper (per nucleon) from the present measurement. 
Bottom plot: The $\nu W_{2}=F_{2}$ structure function for copper (per nucleon) as a function of Nachtmann $\xi$.
The $Q^{2}$ ranges are given at each $\theta$ angle. 
The curves are the proton $F_{2}$ parametrizations at $Q^{2}$ = 10 GeV$^{2}$, corrected for the nuclear EMC effect.
The arrow on the right side ($\xi \sim$ 0.85) of the bottom plot indicate the position of quasi-elastic peak. 
}
\label{fig:F2Cu_scaleing}
\end{figure}

After extracting $R_{A}$ one can calculate $F_{2}^{A}$ using 
\begin{equation} \label{eq:F2Model}
F_{2}^{A}=\displaystyle\frac{\sigma}{\sigma_{Mott}}\nu\epsilon\displaystyle\frac{1+R}{1+\epsilon R}.
\end{equation}
The results of extracted $F_{2}^{A}$ are shown in Figs.~\ref{fig:F2C_scaleing}, \ref{fig:F2Al_scaleing}, 
\ref{fig:F2Fe_scaleing} and \ref{fig:F2Cu_scaleing} for carbon, aluminum, iron and copper respectively. 
In the top plot the $F_{2}^{A}$ structure function is shown versus the Bjorken $x$ variable, in the 
second plot, it is shown versus the Nachtmann variable $\xi$ defined in Eq.~\ref{eq:eqnach}.
The curves shown on the bottom plots of Figs.~\ref{fig:F2C_scaleing}, \ref{fig:F2Al_scaleing}, 
\ref{fig:F2Fe_scaleing} and \ref{fig:F2Cu_scaleing} are the proton $F_{2}$ based on different parametrizations 
and are corrected for the nuclear EMC effect. 
The parametrization ALLM97~\cite{Abramowicz:1997ms} is $F_2$ of the proton in the DIS region. 
It rests on a Regge motivated approach, similar to that used earlier by
Donnachie and Landshoff~\cite{Donnachie:1993it}, extended into a large $Q^2$ regime in a way that is 
compatible with QCD expectations. 
The parametrization of SLAC data is based on the old SLAC structure functions and authored by Bodek and Atwood. 
The parametrization GJR08~\cite{Gluck:2007ck} stems from the dynamical parton distributions 
of the nucleon. 

In the top plot of Fig.~\ref{fig:F2C_scaleing} (carbon) and Fig.~\ref{fig:F2Al_scaleing} (aluminum), 
scaling in Bjorken $x$ is not observed for the most part. 
This is expected since the $F_{2}$ set includes values at momentum transfers ranging 
from $Q^{2}$=0.5 GeV$^{2}$ at low $x$ to $Q^{2}$=4.5 GeV$^{2}$ at the higher $x$ values.
In addition the quasi-elastic peak and resonances are not completely washed out by the Fermi motion of the 
nucleons inside the nucleus. 
The scaling is violated the most at $x=$ 1, which corresponds to the quasi-elastic peak. 
For aluminum and carbon the Fermi momentum is lower than for heavier targets and the quasi-elastic 
peak is not broadened enough to show scaling. 
In the bottom plot of Fig.~\ref{fig:F2C_scaleing} and Fig.~\ref{fig:F2Al_scaleing} 
$F_{2}$ is plotted versus the Nachtmann variable $\xi$, which takes account the finite target mass, 
and seems to become independent of $Q^{2}$. 
Note that the $F_{2}$ of iron, shown in Fig.~\ref{fig:F2Fe_scaleing}, 
has a $Q^{2}$ range of 0.5-3.0 GeV$^{2}$ in 
contrast to both carbon and aluminum where the is 0.5-4.5 GeV$^{2}$. 
Even though Fermi momentum of iron is larger than that of carbon and aluminum 
the $F_{2}$ scaling in Bjorken $x$ is still not observed (top plot).
In the bottom plot, where $F_{2}$ is plotted versus Nachtmann $\xi$, 
the situation is completely different and scaling is observed. 
The quasi-elastic peak is not visible. 
This is an important observation, since it shows that Fermi momentum of iron is 
enough to remove $Q^{2}$ dependence of $F_{2}$ structure function near the quasi-elastic peak 
even at $Q^{2}$ values around 1 GeV$^{2}$.
In Fig.~\ref{fig:F2Cu_scaleing} $F_{2}$ is shown for copper in 
the $Q^2$ range between 1.3-4.3 GeV$^{2}$ higher than that for other targets. 
As it can be seen from the figure that $F_{2}$ scales versus both Bjorken $x$ and Nachtmann $\xi$.
For copper it can be concluded that smearing effect of Fermi motion on the resonances 
combined with high enough $Q^{2}$ ($>$ 1.5 GeV$^{2}$), scaling of $F_{2}$ is achieved even without 
applying target mass corrections. 

For nuclei heavier than iron even the quasi-elastic peak is washed out 
by the Fermi smearing at higher $Q^{2}$, and scaling is seen at all values of $x$ and $\xi$. 
Here the resonance region is essentially indistinguishable from the DIS scaling regime.
The observed scaling is consistent with the duality arguments discussed in Sec.~\ref{eq:DualBGsumrule}. 
There local averaging of the structure function over $\xi$ for the nucleon resonances seen at low Q$^{2}$ 
and quasi-elastic peak, produces structure function consistent with the high $Q^{2}$ 
scaling limit of structure function.
Here we see indications that the Fermi motion of the nucleons in the nucleus are performing local
averaging and there is no need to use the finite energy sum rule in order 
to quantify the similarity of scaling functions in resonance and deep inelastic regimes.

Qualitatively, the nuclear effects in the resonance region appear to be similar to those in
the deep inelastic region. 
This is surprising since the nuclear dependence of the
scaling structure functions is not expected to be the same as the nuclear dependence of
resonance production.
There is a priori no reason why these modifications would be the
same as those for structure functions measured in deep inelastic scattering. 
On the other hand, this may be viewed as another consequence of quark-hadron duality.


\begin{figure}[ht]
\begin{center}
\epsfig{file=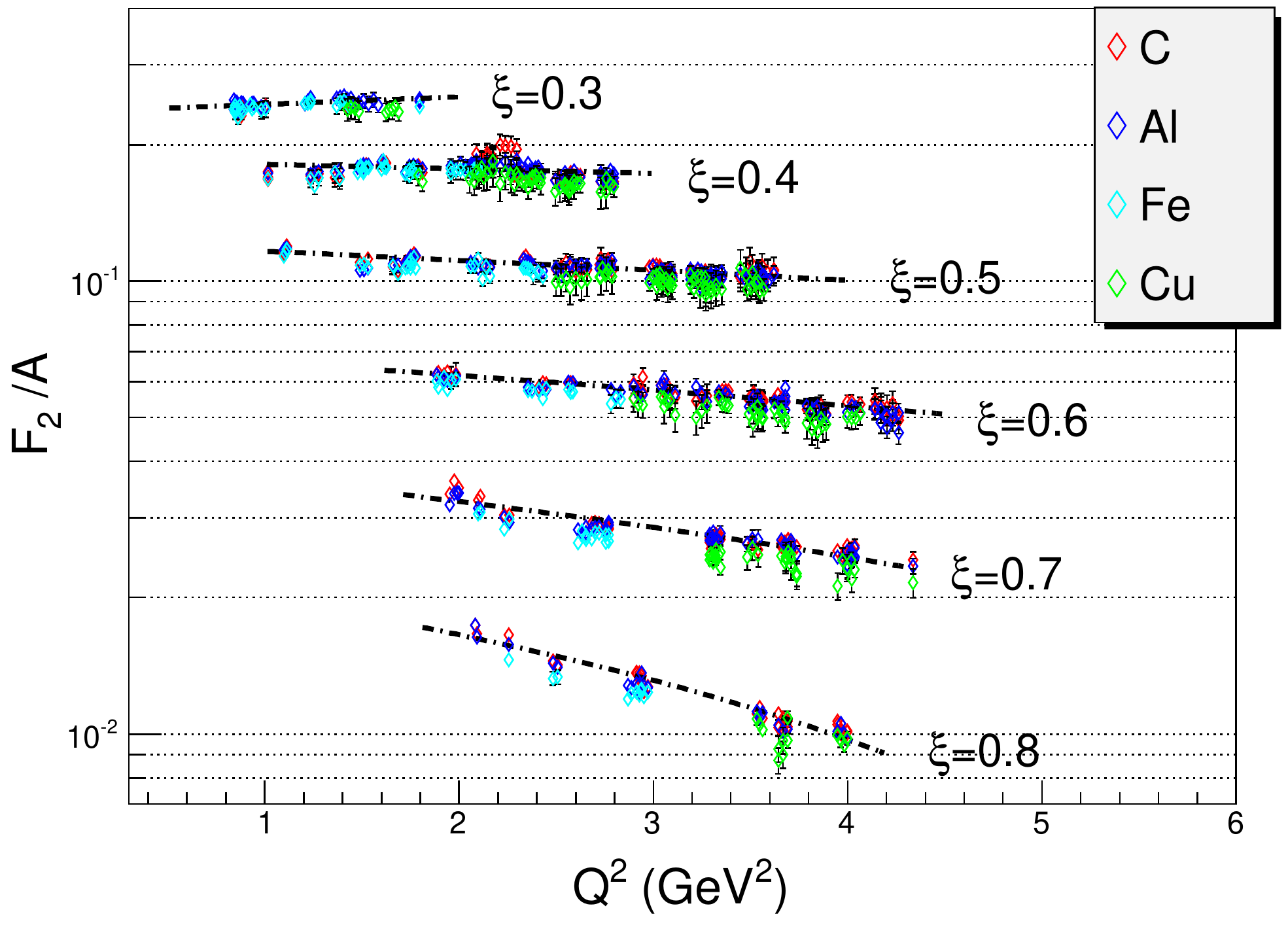,width=5.5in}
\end{center}
\caption { Structure function per nucleon for C, Al, Fe and Cu as a function of $Q^{2}$ at fixed values of
Nachtmann variable $\xi$. The dashed line serves only to guide the eye.}
\label{fig:F2A_vsXi}
\end{figure}
The extracted $F_{2}$ structure functions (per nucleon) are similar for all nuclear targets 
as it can be seen in Fig.~\ref{fig:F2A_vsXi}.
Figure~\ref{fig:F2A_vsXi} shows the structure function for carbon, aluminum, iron 
and copper at $\xi=$0.3, 0.4, 0.5, 0.6, 0.7 and 0.8.
The dashed lines indicate fixed values of Nachtmann $\xi$.
The kinematic coverage for the iron and copper targets is less than for carbon and aluminum. 
For iron $Q^{2}$ ranges from 0.5-2.5 GeV$^2$, for copper from 2.5-4.5 GeV$^2$.
At values of $\xi$ corresponding to the top of the quasi-elastic peak, 
the structure function decreases slightly with $A$, as the increased Fermi momentum broadens 
and lowers the peak. 
At lower values of $\xi$, far away from the quasi-elastic peak, the structure function per 
nucleon is nearly identical for all of the nuclei. 

In Fig.~\ref{fig:WorldF2DIS} world data on the structure function $F_{2}$ of the proton 
versus $Q^2$ is shown at several values of $x$. 
In Fig.~\ref{fig:F2A_vsXi} a similar plot is depicted for nuclear $F_{2}^{A}$ structure function 
in a limited $Q^2$ range. 
Overall the two plots follow the same pattern, at small $\xi$ ($x$) $F_{2}^{A}$ ($F_{2}^{p}$) rises while at high $\xi$ ($x$) 
decreases. 
As it is described in Sec.~\ref{sec:DISQPM}, this behavior is consistent with the QCD predictions indicating scaling violations. 
In the $F_{2}^{A}$ structure function at higher $\xi$ and in a much smaller range of $Q^2$ (compared to $F_{2}$ of proton) observed scaling violation is 
significant and can not be considered as logarithmic in $Q^2$. 
Higher values of $\xi$ are near the quasi-elastic peak meaning higher twist contributions are large  
and the observed violation should have $1/(Q^{2})^{n}$ (power correction) dependence.

\subsection{Structure Functions and Duality Studies}
The Rosenbluth separation allows the extraction of $\sigma_{L}$ and $\sigma_{T}$ independently. 
Then the structure functions $F_2$, $F_1$, and $F_L$ can be calculated using 
the knowledge of $\sigma_{L}$ and $\sigma_{T}$ by the following formulas
\begin{equation}
F_{1}= \displaystyle\frac{K}{4\pi^{2}\alpha} M \sigma_{T}
\label{eq:F1RSep1}
\end{equation}
\begin{equation}
F_{2}= \displaystyle\frac{K}{4\pi^{2}\alpha} \left( 1 + \frac{Q^{2}}{\nu^{2}} \right) \left[ \sigma_{T} + \sigma_{L}\right]
\label{eq:F2RSep1}
\end{equation}
where $K$ is a kinematic variable.
From the definition of $F_{1}$ one can see that it is related only to the transverse virtual photon
coupling, while $F_2$ is a combination of both transverse and longitudinal couplings. 
It is useful therefore to define a purely longitudinal structure function $F_{L}$
\begin{equation}
F_{L}= \left( 1 + \frac{Q^{2}}{\nu^{2}} \right)F_{2}-2xF_{1}.
\label{eq:FLRSep1}
\end{equation}

One of the goals of this experiment was to study quark-hadron duality for the 
transverse structure function $F_{1}$, the longitudinal structure function $F_{L}$ and for 
$F_{2}$. 
In Section~\ref{sec:RosSepR}, where $R$ is extracted in the resonance region and 
compared to $R$ in DIS region, no significant difference was found in these 
regimes, except at the lowest $Q^{2}$ (0.9 GeV$^{2}$). 
The similarity of $R$ in the two kinematically distinct regions for high enough $Q^{2}$ is expected 
from quark-hadron duality and for the proton has been studied extensively. 
The results of this experiment allow a similar comparison to be made in the resonance region 
for heavy nuclei. 
Studying quark-hadron duality in nuclear targets can provide information 
on what role the nuclear effects have.

\begin{figure}[p]
\begin{center}
\epsfig{file=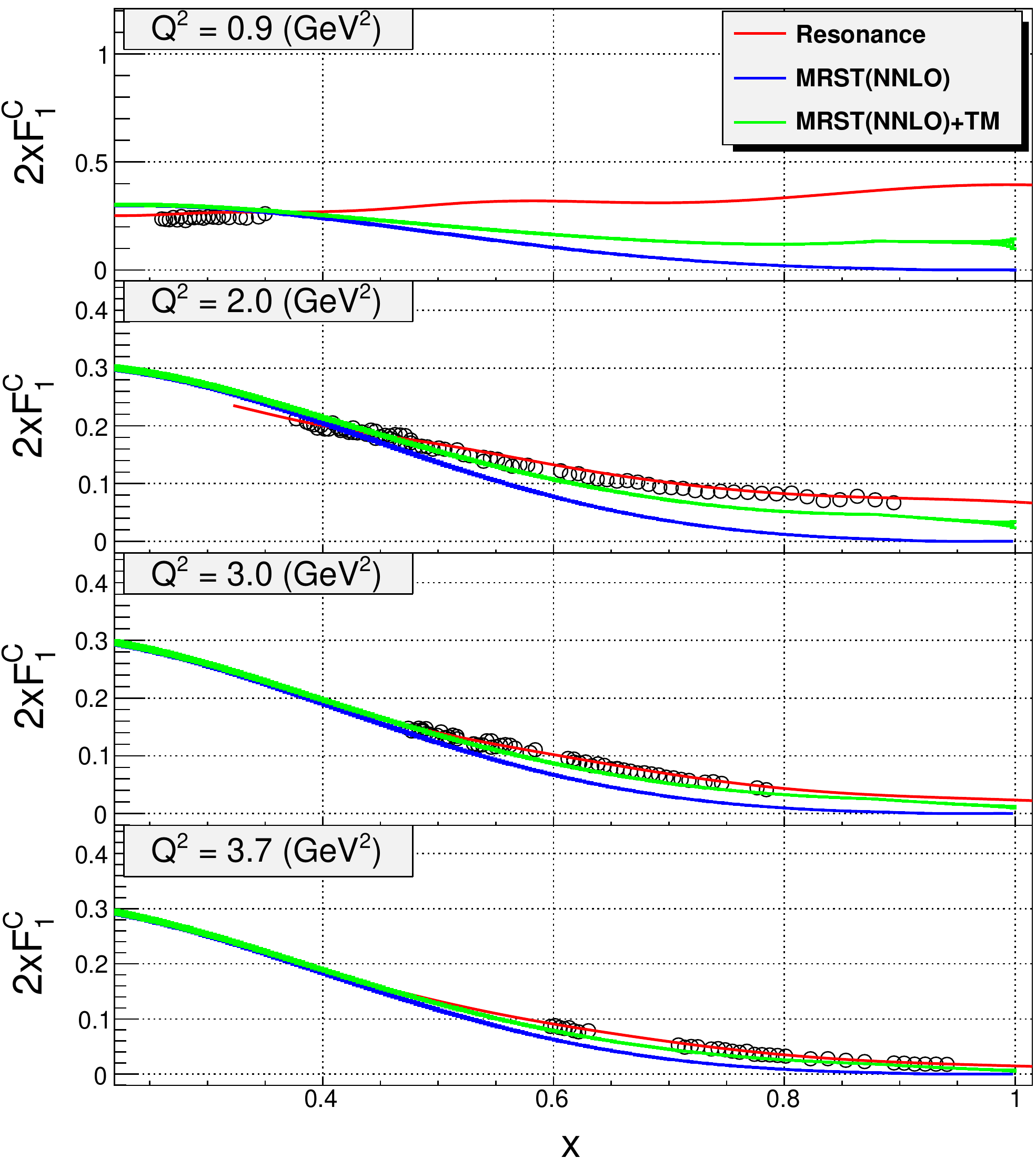,width=5.8in}
\end{center}
\caption { The purely transverse structure function $2xF_{1}$ of carbon per nucleon, measured in the resonance
region as a function of Bjorken $x$ is compared with MRST(NNLO) with and without target mass correction.
Both curves are corrected for the nuclear EMC effect and isospin weighting.
Errors are smaller than the symbol size.}
\label{fig:F1CRosSep}
\end{figure}

\begin{figure}[p]
\begin{center}
\epsfig{file=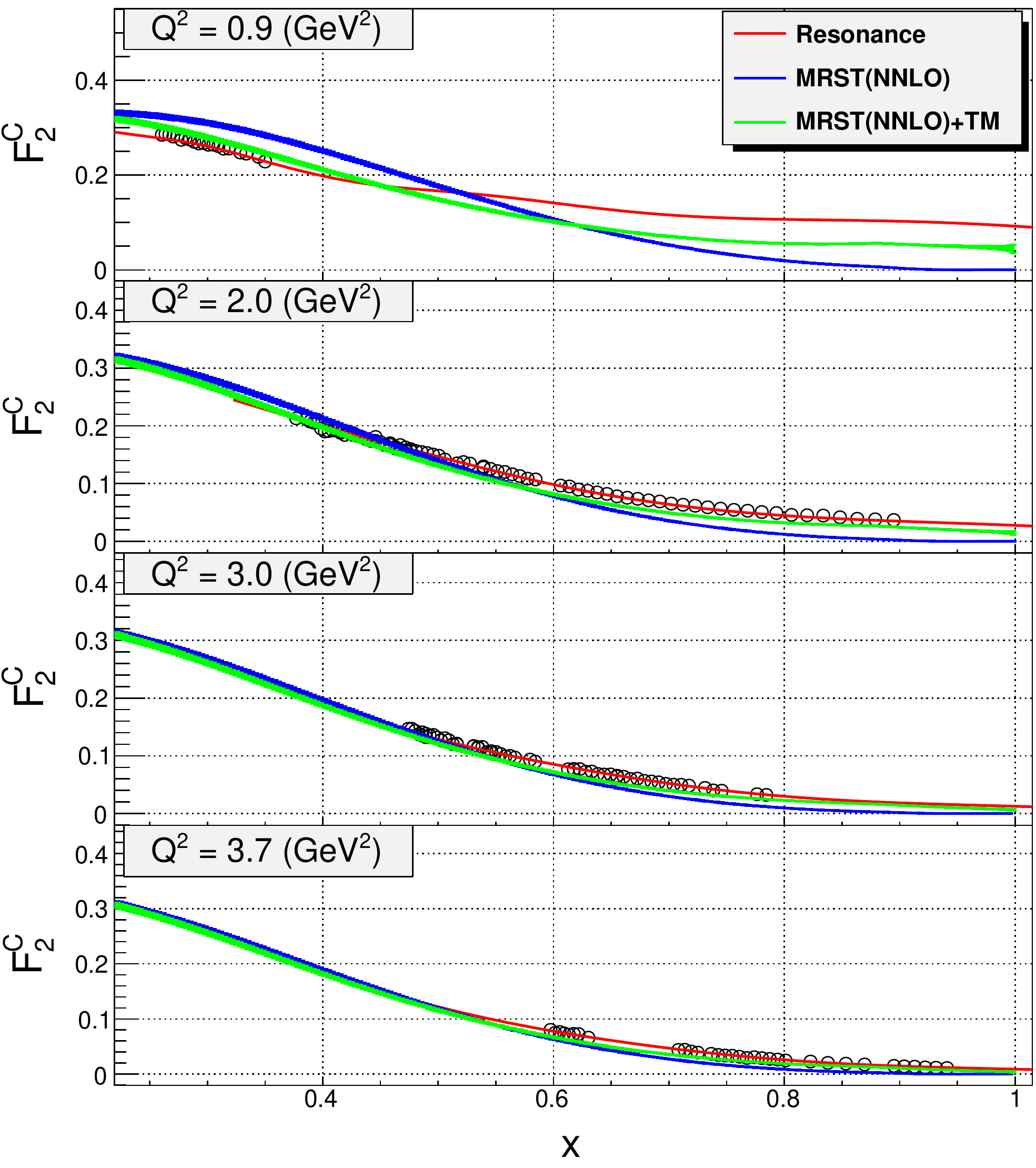,width=5.8in}
\end{center}
\caption {$F_{2}$ structure function of carbon per nucleon, measured in the resonance
region as a function of Bjorken $x$ is compared with MRST(NNLO) with and without target mass correction.
Both curves are corrected for the nuclear EMC effect and isospin weighting.
Errors are smaller than the symbol size.
}
\label{fig:F2CRosSep}
\end{figure}

\begin{figure}[p]
\begin{center}
\epsfig{file=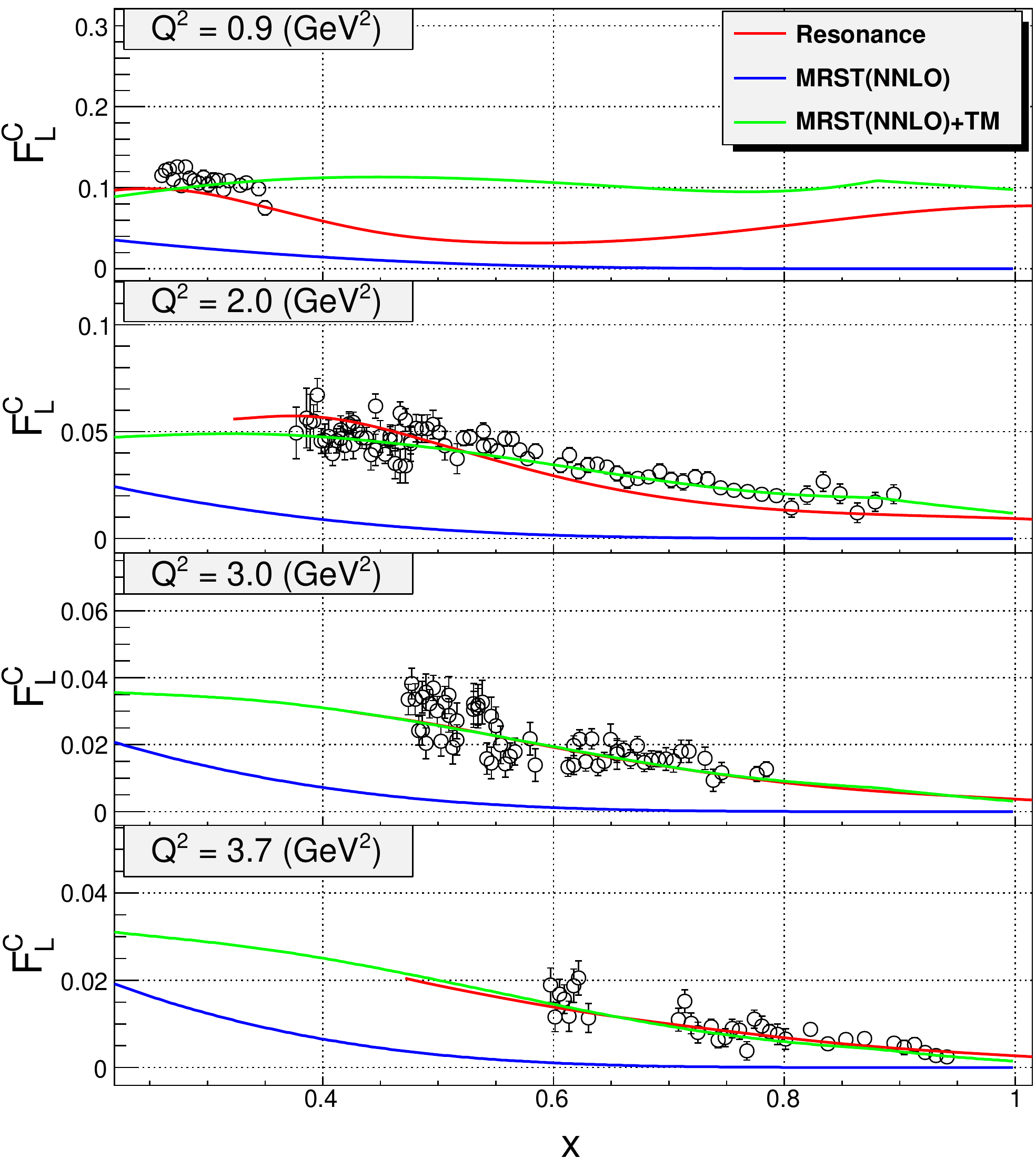,width=5.8in}
\end{center}
\caption { The purely longitudinal structure function $F_{L}$ of carbon per nucleon, measured in the resonance
region as a function of Bjorken $x$ is compared with MRST(NNLO) with and without target mass correction.
Both curves are corrected for the nuclear EMC effect.
}
\label{fig:FLCRosSep}
\end{figure}

\begin{figure}[p]
\begin{center}
\epsfig{file=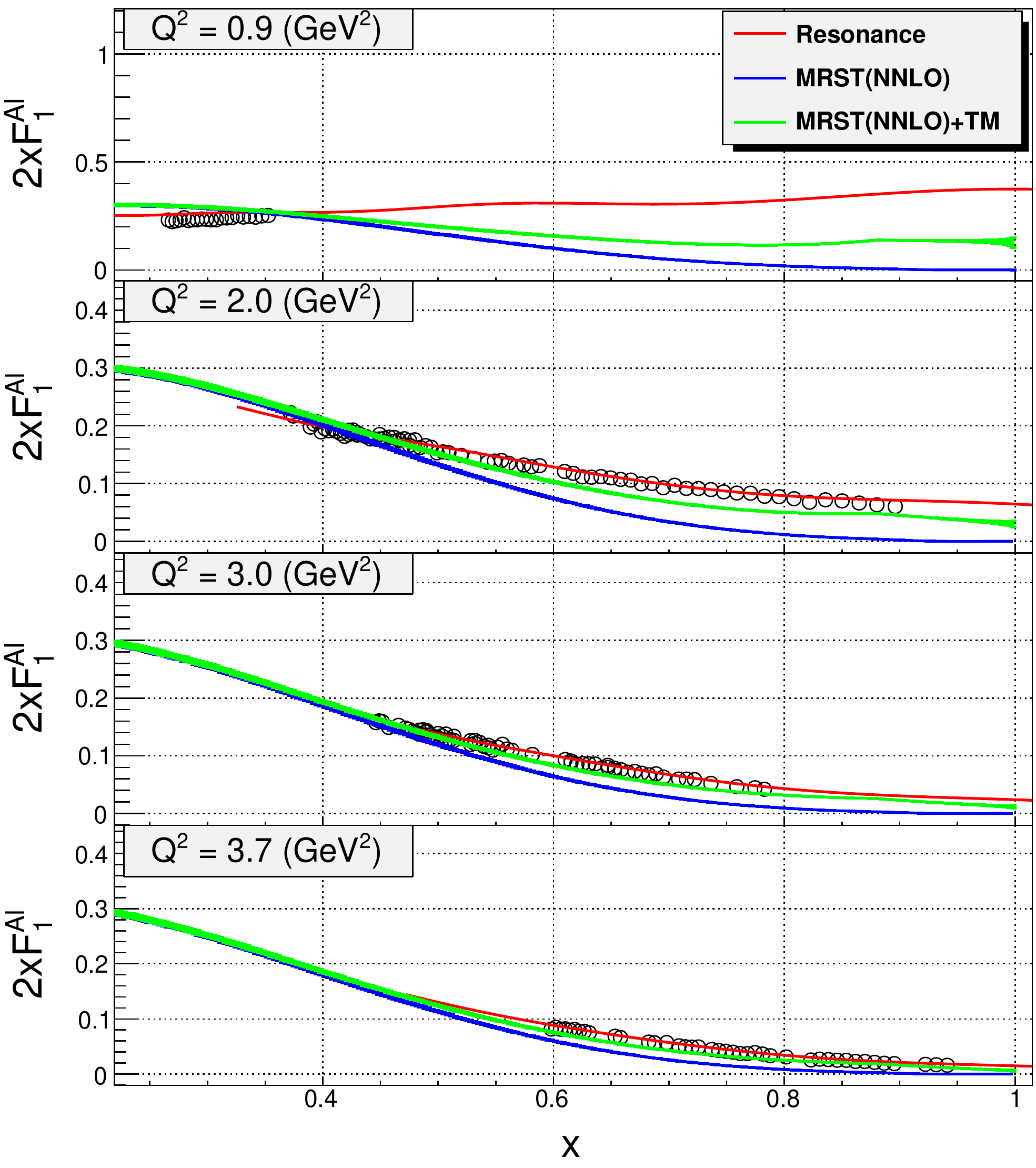,width=5.8in}
\end{center}
\caption { The purely transverse structure function $2xF_{1}$ of carbon per nucleon, measured in the resonance
region as a function of Bjorken $x$ is compared with MRST(NNLO) with and without target mass correction.
Both curves are corrected for the nuclear EMC effect and isospin weighting.
Errors are smaller than the symbol size. }
\label{fig:F1AlRosSep}
\end{figure}

\begin{figure}[p]
\begin{center}
\epsfig{file=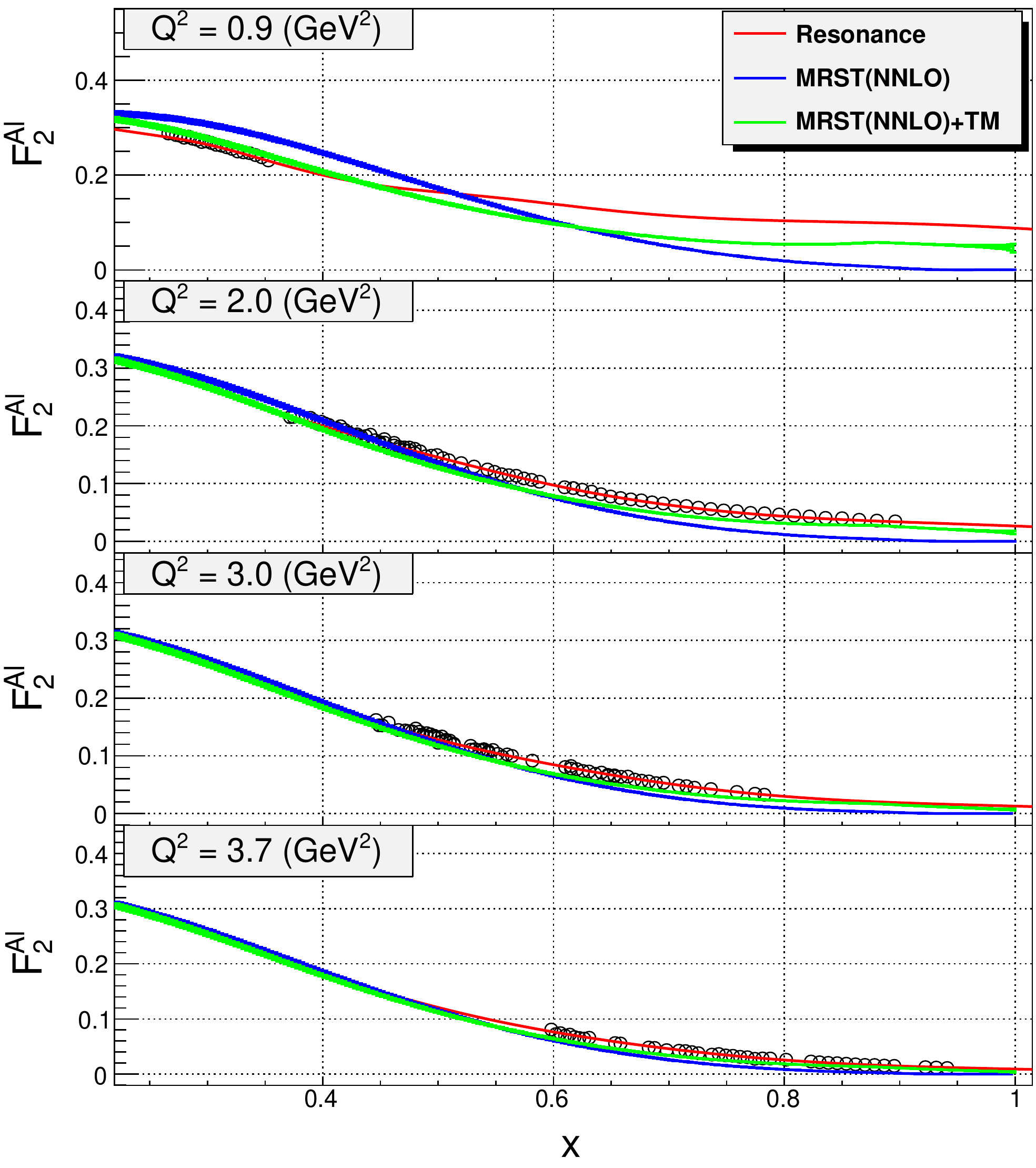,width=5.8in}
\end{center}
\caption { $F_{2}$ structure function of aluminum per nucleon, measured in the resonance
region as a function of Bjorken $x$ is compared with MRST(NNLO) with and without target mass correction.
Both curves are corrected for the nuclear EMC effect and isospin weighting.
Errors are smaller than the symbol size.}
\label{fig:F2AlRosSep}
\end{figure}

\begin{figure}[p]
\begin{center}
\epsfig{file=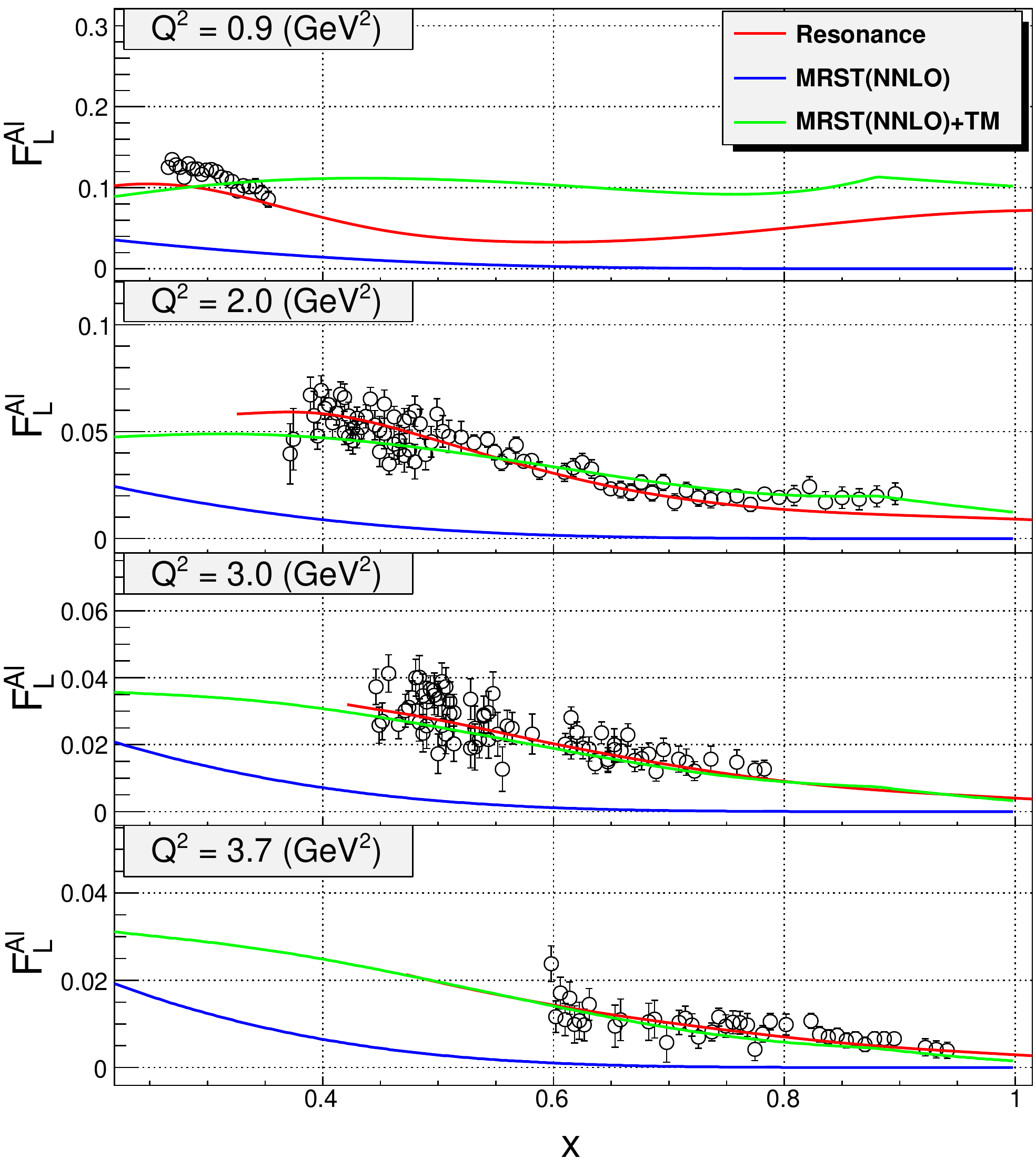,width=5.8in}
\end{center}
\caption { The purely longitudinal structure function $F_{L}$ of aluminum per nucleon, measured in the resonance
region as a function of Bjorken $x$ is compared with MRST(NNLO) with and without target mass correction.
Both curves are corrected for the nuclear EMC effect.
}
\label{fig:FLAlRosSep}
\end{figure}

\begin{figure}[p]
\begin{center}
\epsfig{file=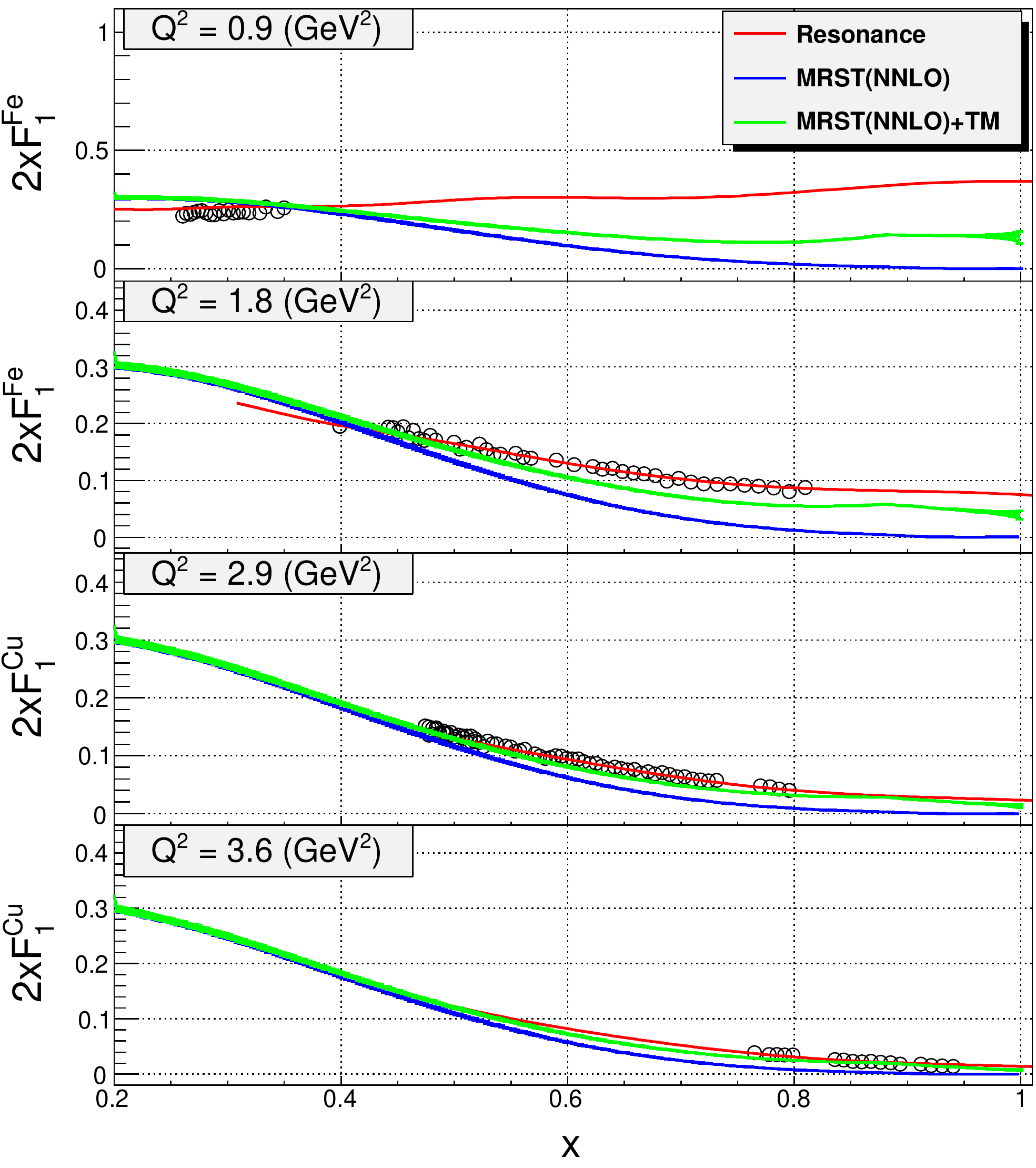,width=5.8in}
\end{center}
\caption { The purely transverse structure function $2xF_{1}$ of iron and copper per nucleon, measured in the resonance
region as a function of Bjorken $x$ is compared with MRST(NNLO) with and without target mass correction.
Both curves are corrected for the nuclear EMC effect and isospin weighting.
Errors are smaller than the symbol size. }
\label{fig:F1FeCuRosSep}
\end{figure}

\begin{figure}[p]
\begin{center}
\epsfig{file=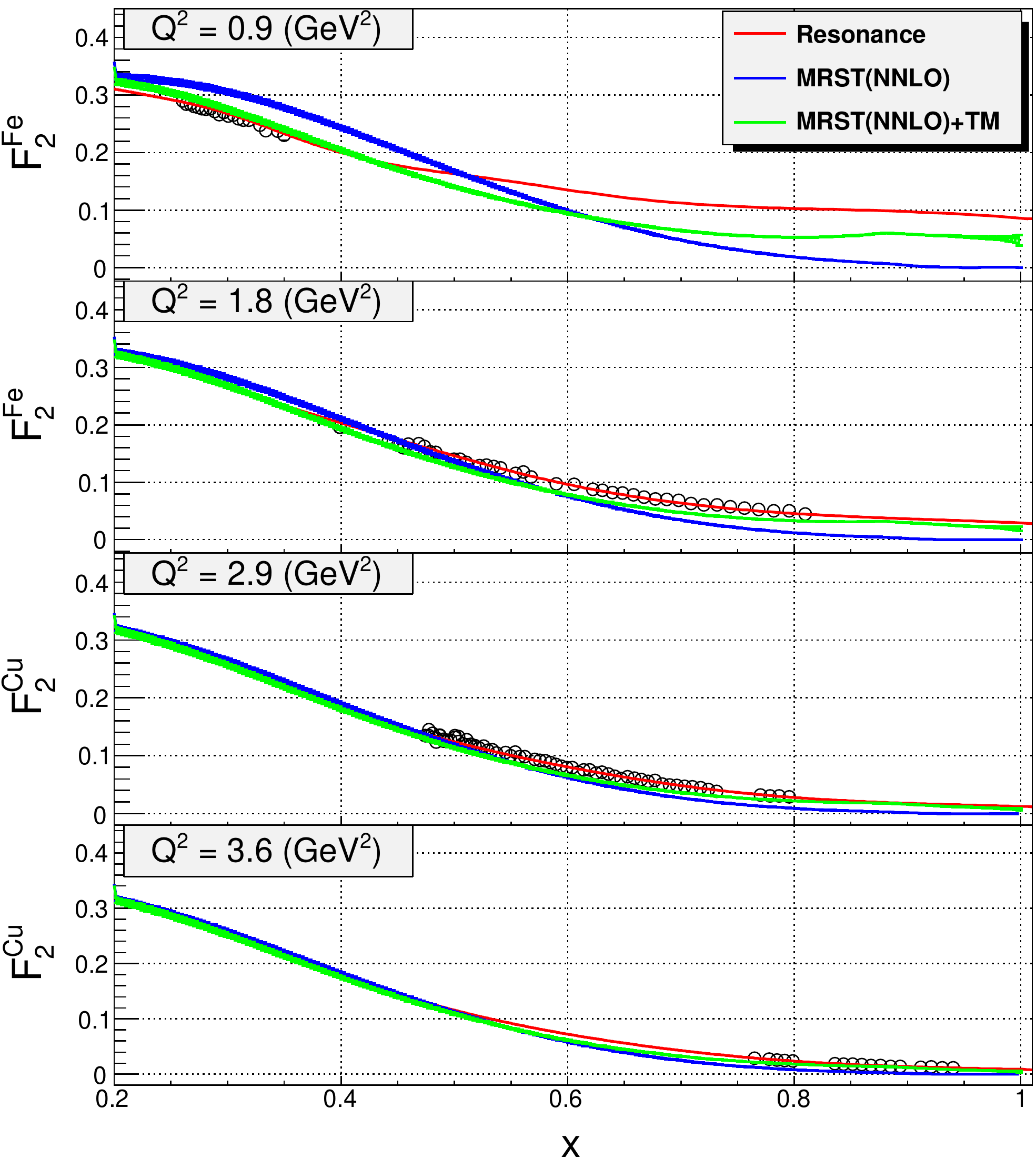,width=5.8in}
\end{center}
\caption { $F_{2}$ structure function of iron and copper per nucleon, measured in the resonance
region as a function of Bjorken $x$ is compared with MRST(NNLO) with and without target mass correction.
Both curves are corrected for the nuclear EMC effect and isospin weighting.
Errors are smaller than the symbol size. }
\label{fig:F2FeCuRosSep}
\end{figure}

\begin{figure}[p]
\begin{center}
\epsfig{file=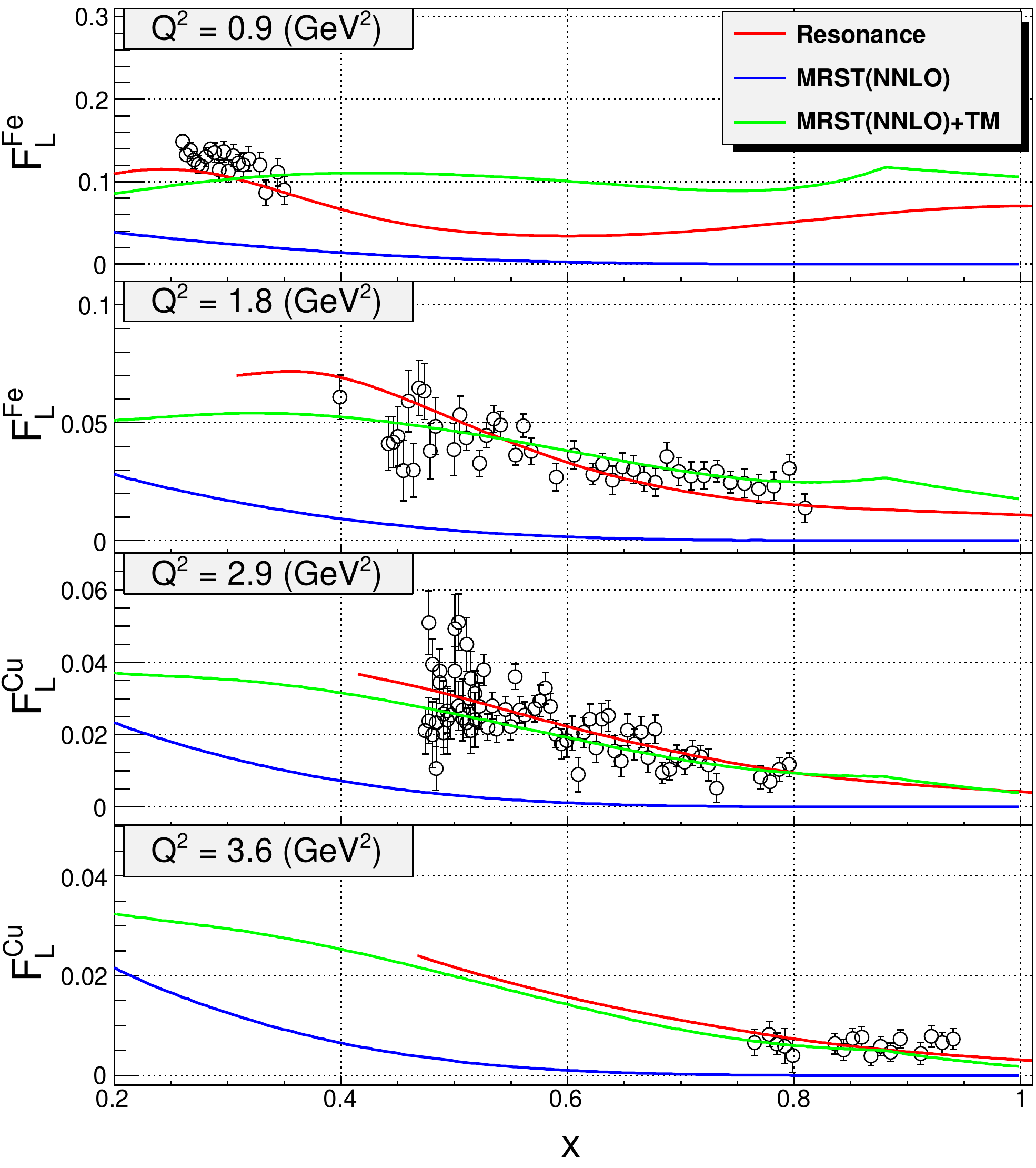,width=5.8in}
\end{center}
\caption { The purely longitudinal structure function $F_{L}$ of iron and copper per nucleon, measured in the resonance
region as a function of Bjorken $x$ is compared with MRST(NNLO) with and without target mass correction.
Both curves are corrected for the nuclear EMC effect.
}
\label{fig:FLFeCuRosSep}
\end{figure}

The results of structure functions $F_{1}$, $F_{L}$ and $F_{2}$ are presented below 
and compared to models of the structure functions in DIS region.
The $2xF_{1}$ and $F_L$ are plotted as a function of $x$ at different $Q^2$ to provide
a direct comparisons of the resonance transverse and longitudinal
structure functions to those in the DIS region, and are seen in Figs.~\ref{fig:F1CRosSep}$-$\ref{fig:FLFeCuRosSep}.
In all plots the resonance structure is not visible because of averaging effect of 
Fermi motion of nucleons inside nuclei. 
The red curve shown in the plots is calculated from the fit to current model, 
the blue curve is calculated from MRST (NNLO) fit~\cite{Martin:2004ir}, without target 
mass corrections.
The green curve is calculated from the MRST (NNLO) fit~\cite{Martin:2004ir} with target mass corrections.
In MRST (NNLO) fit~\cite{Martin:2004ir} the parton distributions are determined by a 
global analysis of a wide range of deep inelastic and related hard scattering data. 
The Bjorken $x$ dependence of the distributions are parametrized at some 
low momentum scale, and a fixed order (either LO or NLO or NNLO) DGLAP~\cite{Gribov:1972ri,Dokshitzer:1977sg,Altarelli:1977zs} evolution performed to
specify the distributions at the higher momentum scales where data exist. 
The target mass correction is done (calculated based on prescription of 
Georgi and Politzer~\cite{Georgi:1976vf}) using a program provided by M.E. Christy~\cite{EricChr}. 
The MRST and MRST (NNLO) are also corrected for EMC effect, and isospin weighting (see below) 
and nucleon structure functions from Ref.~\cite{Arneodo:1995cq}. 
The isospin correction is given as 
\begin{equation} \label{eq:F2nF2pNMC}
C_{is}= \left( Z + ( A - Z) (F_{2}^{n}/F_{2}^{p}) \right)\bigg/A
\end{equation}
and multiplies the structure functions of the proton.

The comparison of the DIS data at high $Q^{2}$ to resonance data at low $Q^{2}$ 
values follows the original idea of Bloom and Gilman, discussed in Sec.~\ref{sec:BGDuality}.
Bloom and Gilman compared low $Q^2$ resonance data to high $Q^2$ DIS data and argued that the 
observed similarity between the two structure functions indicates that resonances are not a separate entity 
but are an intrinsic part of the scaling behavior of $\nu W_{2}(\nu,Q^{2})$ structure function~\cite{Bloom:1971ye}.
The shortcoming of this approach is that the comparison is made at different $Q^2$ values, although at 
the same $\xi$, which is equivalent to making the comparison at different Bjorken $x$ and sensitive 
therefore to different parton distributions. 
In order to be more precise the comparison of DIS data should be made at the same $Q^2$ where the resonance 
data is taken. 
This is achieved by utilizing parton distributions determined by a global analysis of a wide 
range of deep inelastic and related hard scattering data, such as MRST (NNLO) fit~\cite{Martin:2004ir} 
(the green curve).
Comparison of resonance region data with PDF-based global fits, DGLAP~\cite{Gribov:1972ri,Dokshitzer:1977sg,Altarelli:1977zs} $Q^2$ evolution 
and target mass correction allows the resonance-scaling comparison to be made at the same values of 
$x$ and $Q^2$.

The transverse structure function $2xF_{1}$ versus Bjorken $x$ for carbon, aluminum, 
iron and copper is shown in Figs.~\ref{fig:F1CRosSep}, ~\ref{fig:F1AlRosSep} and ~\ref{fig:F1FeCuRosSep}; 
the longitudinal structure function $F_{L}$ versus Bjorken $x$ is shown in 
Figs.~\ref{fig:FLCRosSep}, ~\ref{fig:FLAlRosSep} and ~\ref{fig:FLFeCuRosSep}.
The structure function $F_{2}$ is shown in Figs.~\ref{fig:F2CRosSep}, ~\ref{fig:F2AlRosSep}, ~\ref{fig:F2FeCuRosSep}.
The structure functions are compared at four different $Q^{2}$ ranges for carbon and aluminum. 
For iron only the two lowest $Q^{2}$ ranges are shown, for copper only the two highest since 
there is no data at high $Q^{2}$ for iron and at low $Q^{2}$ data for copper. 
The agreement of MRST (NNLO) parametrization with the data is the poorest as can be seen 
in all the plots, while  MRST (NNLO) with target mass corrections does a better job at describing the data. 
This makes it clear that target mass effects are required to describe the data.

These results confirm the validity of quark-hadron duality in the resonance region in the separated 
transverse $F_{1}$ and longitudinal structure functions $F_{L}$ of the nuclei for $Q^{2}$ larger 
than 2 GeV$^2$ and $x > $ 0.6. 
The duality is similarly observed in the $F_{2}$ structure functions for $Q^{2}$ larger than 
2 GeV$^2$ and $x > $ 0.6. 
For $Q^{2}$ larger than 2~GeV$^2$ and $x < $ 0.6 duality is violated up to maximum 20\% in all structure functions. 
This can be caused by nuclear effects. 

Qualitatively, the nuclear effects in the resonance region appear to be similar to those in
the deep inelastic region. 
This similarity may be viewed as another consequence of quark-hadron duality.

\chap5{Summary and Conclusions}
Inclusive electron-nuclear scattering cross sections were measured in the nucleon resonance
region at $0.0 < W^{2} < 4.5$ GeV$^{2}$ for four-momentum transfer values between 
$0.5 < Q^{2} < 4.5$ GeV$^{2}$. 
The data were taken in May-July 2007 at Jefferson Laboratory (Newport News, Va, USA) 
at scattering angles between 12$^{\circ}$ and 76$^{\circ}$ degrees on 
$^{12}$C, $^{27}$Al, $^{56}$Fe, $^{64}$Cu targets with electron beam energy 2.097, 3.116, 3.269, 4.074, 4.134 and 5.150 GeV. 
The cross sections are estimated to have about 2\% systematic uncertainties.

An empirical fit to inelastic electron-nuclei scattering has been performed which describes 
available data reasonably well, within 3\% to 5\%, for 0.0 $<W^2<$ 4.5 GeV$^2$ and 0.5 $<Q^{2}<$ 5.0 GeV$^2$. 
The fit is useful in the evaluation of radiative corrections to experimental data, 
for extraction of spin structure functions from asymmetry measurements,
and for the evaluation of structure function moments.

The cross sections extracted from a deuteron target are used to study the nuclear dependence of the structure
function $R$. 
The cross section ratios of nuclear and deuteron targets showed no nuclear dependence of $R$.
This was further confirmed after extracting $R_{A}-R_{D}$ by performing Rosenbluth separations.
No significant nuclear dependence was found for $W^{2}>$ 2 GeV$^{2}$ for all targets.
The observed enhancement of $R_{A}-R_{D}$ near $W^{2} = $ 1.5 GeV$^2$ indicates some nuclear 
dependence which vanishes with increase of $Q^{2}$ and with nuclear number A. 
A final conclusion about the origins of the enhancement of $R_{A}-R_{D}$ can only be made after final analysis of experiment 
E06-009~\cite{E06009} which will provide an iterated model and precise cross sections for deuteron
in the same kinematic region as the current experiment. 
The results of $R_{A}-R_{C}$ indicate that there is no nuclear dependence of $R$ for heavy nuclei.

The $F_{2}^{A}$ structure function is extracted for all measured cross sections in a model 
dependent way by interpolating $R_{A}-R_{D}$ with a function and using a model for deuteron $R_{D}$.
The extracted $F_{2}^{A}$ shows that resonance structure is washed out by the Fermi motion of nucleons in nuclei
for $Q^{2} > $ 1 GeV$^2$ for nearly all targets and the scaling regime of $F_{2}^{A}$ versus Nachtmann $\xi$ is reached.
Since the Fermi momentum of copper is larger than that of other nuclei and copper data are taken at $Q^{2} > $ 1.5 GeV$^2$
scaling is observed even versus Bjorken $x$.
These results suggest that Fermi motion of nucleons inside nuclei are doing the local averaging of the structure 
function $F_{2}$ over $\xi$ for the nuclear resonances and produces a structure function consistent with the 
high $Q^{2}$ scaling limit of the $F_{2}^{A}$ structure function.

The structure functions $F_2$, $F_1$, $F_L$, and $R$ obtained by Rosenbluth separation 
have allowed us to study quark-hadron duality on nuclear targets in both the transverse 
and longitudinal channels. 
Comparison made to target mass corrected proton MRST (NNLO) parametrization~\cite{Martin:2004ir} with the 
EMC correction and isospin weighting applied, showed a similarity between the structure functions. 
Qualitatively, the nuclear effects in the resonance region appear to be similar to those in
the deep inelastic region. 
This may be viewed as another consequence of quark-hadron duality.

%

\appendix
\section{APPENDIX}
\subsection{Tables of Rosenbluth Separated \texorpdfstring{$R$}{R}, \texorpdfstring{$F_{1}$}{F1}, \texorpdfstring{$F_{2}$}{F2} and \texorpdfstring{$F_{L}$}{FL}}
\renewcommand{\arraystretch}{1.2}
\begin{table}[h]
\begin{center}
\begin{tabular}{|c|c|c|c|c|c|c|c|c|c|c|}
\hline
 $W^{2}$ & $Q^{2}$ & $x$ & $R$ & $\Delta R$ & $F_{1}$ & $\Delta F_{1}$ & $F_{2}$ & $\Delta F_{2}$ & $F_{L}$ & $\Delta F_{L}$ \\
\hline
 1.110 &  2.096 &  0.901 &  0.353 &  0.081 &  0.032 &  0.001 &  0.033 &  0.001 &  0.020 &  0.004\\
 1.110 &  3.773 &  0.943 &  0.155 &  0.097 &  0.009 &  0.000 &  0.010 &  0.001 &  0.003 &  0.002\\
 1.150 &  2.079 &  0.885 &  0.271 &  0.071 &  0.035 &  0.001 &  0.034 &  0.002 &  0.017 &  0.004\\
 1.150 &  3.753 &  0.933 &  0.170 &  0.097 &  0.009 &  0.001 &  0.011 &  0.001 &  0.003 &  0.002\\
 1.190 &  2.062 &  0.869 &  0.175 &  0.065 &  0.040 &  0.002 &  0.036 &  0.002 &  0.012 &  0.005\\
 1.190 &  3.734 &  0.923 &  0.212 &  0.100 &  0.009 &  0.001 &  0.012 &  0.001 &  0.004 &  0.002\\
 1.230 &  2.045 &  0.854 &  0.329 &  0.076 &  0.038 &  0.002 &  0.039 &  0.002 &  0.022 &  0.005\\
\hline
\end{tabular}
\end{center}
\caption{Structure functions of carbon.}
\end{table}

\begin{table}[p]
\begin{center}
\begin{tabular}{|c|c|c|c|c|c|c|c|c|c|c|}
\hline
 $W^{2}$ & $Q^{2}$ & $x$ & $R$ & $\Delta R$ & $F_{1}$ & $\Delta F_{1}$ & $F_{2}$ & $\Delta F_{2}$ & $F_{L}$ & $\Delta F_{L}$ \\
\hline
 1.710 &  1.905 &  0.697 &  0.325 &  0.050 &  0.070 &  0.002 &  0.068 &  0.002 &  0.032 &  0.004\\
 1.710 &  3.466 &  0.807 &  0.341 &  0.074 &  0.020 &  0.001 &  0.026 &  0.001 &  0.011 &  0.002\\
 1.750 &  1.889 &  0.685 &  0.367 &  0.050 &  0.072 &  0.002 &  0.072 &  0.002 &  0.036 &  0.004\\
 1.750 &  3.445 &  0.799 &  0.278 &  0.068 &  0.022 &  0.001 &  0.027 &  0.001 &  0.010 &  0.002\\
 1.790 &  1.804 &  0.665 &  0.318 &  0.046 &  0.084 &  0.002 &  0.079 &  0.002 &  0.036 &  0.004\\
 1.790 &  3.542 &  0.796 &  0.224 &  0.077 &  0.022 &  0.001 &  0.026 &  0.001 &  0.008 &  0.003\\
 1.830 &  1.786 &  0.653 &  0.299 &  0.043 &  0.090 &  0.002 &  0.083 &  0.002 &  0.035 &  0.004\\
\hline
\end{tabular}
\end{center}
\caption{Structure functions of carbon.}
\end{table}

\begin{table}[p]
\begin{center}
\begin{tabular}{|c|c|c|c|c|c|c|c|c|c|c|}
\hline
 $W^{2}$ & $Q^{2}$ & $x$ & $R$ & $\Delta R$ & $F_{1}$ & $\Delta F_{1}$ & $F_{2}$ & $\Delta F_{2}$ & $F_{L}$ & $\Delta F_{L}$ \\
\hline
 2.190 &  3.036 &  0.699 &  0.277 &  0.082 &  0.045 &  0.002 &  0.051 &  0.002 &  0.017 &  0.004\\
 2.190 &  3.683 &  0.738 &  0.228 &  0.053 &  0.031 &  0.001 &  0.036 &  0.001 &  0.010 &  0.002\\
 2.230 &  3.347 &  0.713 &  0.248 &  0.049 &  0.039 &  0.001 &  0.046 &  0.001 &  0.014 &  0.002\\
 2.270 &  1.597 &  0.535 &  0.374 &  0.040 &  0.149 &  0.003 &  0.134 &  0.002 &  0.060 &  0.005\\
 2.270 &  2.989 &  0.683 &  0.250 &  0.065 &  0.051 &  0.001 &  0.056 &  0.002 &  0.017 &  0.003\\
 2.270 &  3.648 &  0.724 &  0.180 &  0.059 &  0.035 &  0.001 &  0.039 &  0.001 &  0.009 &  0.003\\
 2.310 &  1.576 &  0.524 &  0.333 &  0.033 &  0.159 &  0.002 &  0.138 &  0.002 &  0.056 &  0.004\\
\hline
\end{tabular}
\end{center}
\caption{Structure functions of carbon.}
\end{table}

\begin{table}[p]
\begin{center}
\begin{tabular}{|c|c|c|c|c|c|c|c|c|c|c|}
\hline
 $W^{2}$ & $Q^{2}$ & $x$ & $R$ & $\Delta R$ & $F_{1}$ & $\Delta F_{1}$ & $F_{2}$ & $\Delta F_{2}$ & $F_{L}$ & $\Delta F_{L}$ \\
\hline
 2.630 &  1.538 &  0.468 &  0.395 &  0.037 &  0.198 &  0.003 &  0.172 &  0.003 &  0.073 &  0.005\\
 2.630 &  3.098 &  0.639 &  0.286 &  0.055 &  0.062 &  0.001 &  0.070 &  0.002 &  0.023 &  0.003\\
 2.670 &  1.000 &  0.359 &  0.430 &  0.038 &  0.332 &  0.006 &  0.234 &  0.004 &  0.102 &  0.008\\
 2.670 &  3.076 &  0.632 &  0.183 &  0.048 &  0.069 &  0.001 &  0.070 &  0.002 &  0.016 &  0.003\\
 2.710 &  0.921 &  0.335 &  0.494 &  0.038 &  0.356 &  0.006 &  0.249 &  0.004 &  0.118 &  0.007\\
 2.710 &  2.337 &  0.561 &  0.234 &  0.048 &  0.116 &  0.003 &  0.109 &  0.003 &  0.030 &  0.006\\
 2.710 &  3.054 &  0.625 &  0.271 &  0.051 &  0.069 &  0.001 &  0.076 &  0.002 &  0.023 &  0.003\\
\hline
\end{tabular}
\end{center}
\caption{Structure functions of carbon.}
\end{table}

\begin{table}[p]
\begin{center}
\begin{tabular}{|c|c|c|c|c|c|c|c|c|c|c|}
\hline
 $W^{2}$ & $Q^{2}$ & $x$ & $R$ & $\Delta R$ & $F_{1}$ & $\Delta F_{1}$ & $F_{2}$ & $\Delta F_{2}$ & $F_{L}$ & $\Delta F_{L}$ \\
\hline
 3.070 &  2.378 &  0.521 &  0.315 &  0.046 &  0.133 &  0.002 &  0.130 &  0.002 &  0.044 &  0.004\\
 3.070 &  2.937 &  0.573 &  0.235 &  0.057 &  0.096 &  0.003 &  0.097 &  0.003 &  0.026 &  0.006\\
 3.110 &  0.813 &  0.267 &  0.461 &  0.025 &  0.466 &  0.005 &  0.278 &  0.003 &  0.115 &  0.005\\
 3.110 &  1.749 &  0.440 &  0.220 &  0.061 &  0.229 &  0.007 &  0.177 &  0.006 &  0.044 &  0.010\\
 3.110 &  2.355 &  0.514 &  0.332 &  0.046 &  0.138 &  0.002 &  0.135 &  0.002 &  0.047 &  0.004\\
 3.110 &  3.642 &  0.620 &  0.300 &  0.088 &  0.062 &  0.002 &  0.073 &  0.003 &  0.023 &  0.004\\
 3.150 &  0.786 &  0.257 &  0.607 &  0.038 &  0.453 &  0.008 &  0.289 &  0.004 &  0.141 &  0.007\\
\hline
\end{tabular}
\end{center}
\caption{Structure functions of carbon.}
\end{table}

\begin{table}[p]
\begin{center}
\begin{tabular}{|c|c|c|c|c|c|c|c|c|c|c|}
\hline
 $W^{2}$ & $Q^{2}$ & $x$ & $R$ & $\Delta R$ & $F_{1}$ & $\Delta F_{1}$ & $F_{2}$ & $\Delta F_{2}$ & $F_{L}$ & $\Delta F_{L}$ \\
\hline
 3.350 &  3.589 &  0.592 &  0.249 &  0.080 &  0.075 &  0.002 &  0.083 &  0.003 &  0.022 &  0.005\\
 3.390 &  2.220 &  0.469 &  0.254 &  0.039 &  0.178 &  0.003 &  0.155 &  0.003 &  0.042 &  0.004\\
 3.390 &  2.792 &  0.527 &  0.139 &  0.048 &  0.131 &  0.003 &  0.117 &  0.004 &  0.019 &  0.006\\
 3.390 &  3.330 &  0.570 &  0.248 &  0.079 &  0.088 &  0.003 &  0.094 &  0.003 &  0.025 &  0.005\\
 3.430 &  2.198 &  0.463 &  0.251 &  0.039 &  0.186 &  0.003 &  0.160 &  0.003 &  0.043 &  0.005\\
 3.430 &  2.712 &  0.515 &  0.156 &  0.053 &  0.139 &  0.004 &  0.123 &  0.004 &  0.022 &  0.007\\
 3.470 &  2.175 &  0.456 &  0.274 &  0.038 &  0.189 &  0.003 &  0.164 &  0.003 &  0.047 &  0.004\\
\hline
\end{tabular}
\end{center}
\caption{Structure functions of carbon.}
\end{table}

\begin{table}[p]
\begin{center}
\begin{tabular}{|c|c|c|c|c|c|c|c|c|c|c|}
\hline
 $W^{2}$ & $Q^{2}$ & $x$ & $R$ & $\Delta R$ & $F_{1}$ & $\Delta F_{1}$ & $F_{2}$ & $\Delta F_{2}$ & $F_{L}$ & $\Delta F_{L}$ \\
\hline
 3.870 &  2.094 &  0.412 &  0.360 &  0.070 &  0.227 &  0.005 &  0.198 &  0.005 &  0.067 &  0.008\\
 3.870 &  2.666 &  0.471 &  0.189 &  0.047 &  0.171 &  0.004 &  0.148 &  0.004 &  0.030 &  0.006\\
 3.910 &  2.073 &  0.406 &  0.289 &  0.066 &  0.239 &  0.006 &  0.196 &  0.005 &  0.056 &  0.008\\
 3.910 &  2.646 &  0.466 &  0.285 &  0.051 &  0.166 &  0.004 &  0.155 &  0.004 &  0.044 &  0.006\\
 3.950 &  2.284 &  0.427 &  0.260 &  0.071 &  0.215 &  0.008 &  0.180 &  0.007 &  0.048 &  0.011\\
 3.950 &  2.988 &  0.493 &  0.310 &  0.052 &  0.137 &  0.003 &  0.138 &  0.003 &  0.042 &  0.004\\
 3.990 &  2.298 &  0.425 &  0.265 &  0.078 &  0.218 &  0.008 &  0.183 &  0.008 &  0.049 &  0.012\\
 3.990 &  2.935 &  0.486 &  0.269 &  0.055 &  0.146 &  0.003 &  0.140 &  0.003 &  0.038 &  0.005\\
 4.030 &  2.617 &  0.454 &  0.179 &  0.051 &  0.187 &  0.004 &  0.156 &  0.004 &  0.030 &  0.007\\
 4.030 &  3.322 &  0.513 &  0.257 &  0.063 &  0.121 &  0.003 &  0.122 &  0.003 &  0.032 &  0.005\\
 4.070 &  2.597 &  0.449 &  0.218 &  0.053 &  0.187 &  0.004 &  0.161 &  0.004 &  0.037 &  0.007\\
 4.070 &  3.295 &  0.508 &  0.242 &  0.055 &  0.127 &  0.002 &  0.126 &  0.003 &  0.031 &  0.004\\
 4.110 &  2.491 &  0.435 &  0.210 &  0.065 &  0.206 &  0.006 &  0.171 &  0.006 &  0.038 &  0.009\\
 4.110 &  2.903 &  0.473 &  0.190 &  0.051 &  0.163 &  0.003 &  0.144 &  0.003 &  0.029 &  0.005\\
 4.150 &  2.934 &  0.473 &  0.267 &  0.054 &  0.157 &  0.003 &  0.148 &  0.004 &  0.040 &  0.006\\
 4.190 &  2.911 &  0.468 &  0.309 &  0.056 &  0.158 &  0.003 &  0.153 &  0.004 &  0.046 &  0.006\\
 4.230 &  2.889 &  0.463 &  0.264 &  0.053 &  0.167 &  0.003 &  0.155 &  0.004 &  0.041 &  0.006\\
\hline
\end{tabular}
\end{center}
\caption{Structure functions of carbon.}
\end{table}
\cleardoublepage
\begin{table}[p]
\begin{center}
\begin{tabular}{|c|c|c|c|c|c|c|c|c|c|c|}
\hline
 $W^{2}$ & $Q^{2}$ & $x$ & $R$ & $\Delta R$ & $F_{1}$ & $\Delta F_{1}$ & $F_{2}$ & $\Delta F_{2}$ & $F_{L}$ & $\Delta F_{L}$ \\
\hline
 1.110 &  2.096 &  0.901 &  0.395 &  0.101 &  0.030 &  0.002 &  0.031 &  0.002 &  0.021 &  0.005\\
 1.110 &  3.733 &  0.942 &  0.274 &  0.137 &  0.008 &  0.001 &  0.011 &  0.001 &  0.004 &  0.002\\
 1.150 &  2.079 &  0.885 &  0.354 &  0.091 &  0.032 &  0.002 &  0.033 &  0.002 &  0.020 &  0.005\\
 1.150 &  3.714 &  0.932 &  0.272 &  0.133 &  0.009 &  0.001 &  0.012 &  0.001 &  0.005 &  0.002\\
 1.190 &  2.062 &  0.869 &  0.313 &  0.087 &  0.035 &  0.002 &  0.035 &  0.002 &  0.019 &  0.005\\
 1.190 &  3.695 &  0.923 &  0.290 &  0.131 &  0.009 &  0.001 &  0.012 &  0.001 &  0.005 &  0.002\\
 1.230 &  2.045 &  0.854 &  0.308 &  0.083 &  0.038 &  0.002 &  0.038 &  0.002 &  0.020 &  0.005\\
\hline
\end{tabular}
\end{center}
\caption{Structure functions of aluminum.}
\end{table}

\begin{table}[p]
\begin{center}
\begin{tabular}{|c|c|c|c|c|c|c|c|c|c|c|}
\hline
 $W^{2}$ & $Q^{2}$ & $x$ & $R$ & $\Delta R$ & $F_{1}$ & $\Delta F_{1}$ & $F_{2}$ & $\Delta F_{2}$ & $F_{L}$ & $\Delta F_{L}$ \\
\hline
 1.750 &  3.424 &  0.798 &  0.305 &  0.076 &  0.022 &  0.001 &  0.027 &  0.001 &  0.011 &  0.002\\
 1.790 &  1.804 &  0.665 &  0.232 &  0.046 &  0.087 &  0.002 &  0.077 &  0.002 &  0.027 &  0.005\\
 1.790 &  3.509 &  0.794 &  0.355 &  0.092 &  0.021 &  0.001 &  0.028 &  0.001 &  0.012 &  0.003\\
 1.830 &  1.786 &  0.653 &  0.287 &  0.046 &  0.090 &  0.002 &  0.082 &  0.002 &  0.034 &  0.004\\
 1.830 &  3.488 &  0.786 &  0.345 &  0.090 &  0.022 &  0.001 &  0.029 &  0.002 &  0.012 &  0.003\\
 1.870 &  1.769 &  0.641 &  0.223 &  0.043 &  0.098 &  0.002 &  0.084 &  0.002 &  0.028 &  0.005\\
 1.870 &  3.522 &  0.781 &  0.355 &  0.076 &  0.023 &  0.001 &  0.030 &  0.001 &  0.013 &  0.002\\
\hline
\end{tabular}
\end{center}
\caption{Structure functions of aluminum.}
\end{table}

\begin{table}[p]
\begin{center}
\begin{tabular}{|c|c|c|c|c|c|c|c|c|c|c|}
\hline
 $W^{2}$ & $Q^{2}$ & $x$ & $R$ & $\Delta R$ & $F_{1}$ & $\Delta F_{1}$ & $F_{2}$ & $\Delta F_{2}$ & $F_{L}$ & $\Delta F_{L}$ \\
\hline
 2.270 &  3.684 &  0.726 &  0.157 &  0.065 &  0.034 &  0.001 &  0.038 &  0.002 &  0.008 &  0.003\\
 2.310 &  1.576 &  0.524 &  0.334 &  0.035 &  0.159 &  0.002 &  0.138 &  0.002 &  0.056 &  0.004\\
 2.310 &  2.976 &  0.675 &  0.253 &  0.076 &  0.052 &  0.002 &  0.057 &  0.002 &  0.018 &  0.004\\
 2.310 &  3.663 &  0.719 &  0.222 &  0.068 &  0.034 &  0.001 &  0.040 &  0.002 &  0.011 &  0.003\\
 2.350 &  1.571 &  0.517 &  0.324 &  0.032 &  0.165 &  0.002 &  0.141 &  0.002 &  0.055 &  0.004\\
 2.350 &  2.953 &  0.668 &  0.235 &  0.073 &  0.056 &  0.002 &  0.060 &  0.002 &  0.017 &  0.004\\
 2.350 &  3.670 &  0.714 &  0.254 &  0.078 &  0.035 &  0.001 &  0.042 &  0.002 &  0.013 &  0.003\\
\hline
\end{tabular}
\end{center}
\caption{Structure functions of aluminum.}
\end{table}

\begin{table}[p]
\begin{center}
\begin{tabular}{|c|c|c|c|c|c|c|c|c|c|c|}
\hline
 $W^{2}$ & $Q^{2}$ & $x$ & $R$ & $\Delta R$ & $F_{1}$ & $\Delta F_{1}$ & $F_{2}$ & $\Delta F_{2}$ & $F_{L}$ & $\Delta F_{L}$ \\
\hline
 2.630 &  3.098 &  0.639 &  0.235 &  0.060 &  0.064 &  0.002 &  0.069 &  0.002 &  0.019 &  0.004\\
 2.670 &  0.977 &  0.353 &  0.418 &  0.039 &  0.343 &  0.007 &  0.237 &  0.005 &  0.101 &  0.009\\
 2.670 &  3.076 &  0.632 &  0.238 &  0.057 &  0.065 &  0.002 &  0.070 &  0.002 &  0.020 &  0.003\\
 2.710 &  0.907 &  0.331 &  0.491 &  0.041 &  0.362 &  0.007 &  0.251 &  0.005 &  0.118 &  0.008\\
 2.710 &  2.337 &  0.561 &  0.319 &  0.056 &  0.110 &  0.003 &  0.110 &  0.003 &  0.039 &  0.006\\
 2.710 &  3.054 &  0.625 &  0.301 &  0.058 &  0.066 &  0.002 &  0.074 &  0.002 &  0.025 &  0.003\\
 2.750 &  0.897 &  0.324 &  0.513 &  0.041 &  0.368 &  0.007 &  0.256 &  0.005 &  0.122 &  0.008\\
\hline
\end{tabular}
\end{center}
\caption{Structure functions of aluminum.}
\end{table}

\begin{table}[p]
\begin{center}
\begin{tabular}{|c|c|c|c|c|c|c|c|c|c|c|}
\hline
 $W^{2}$ & $Q^{2}$ & $x$ & $R$ & $\Delta R$ & $F_{1}$ & $\Delta F_{1}$ & $F_{2}$ & $\Delta F_{2}$ & $F_{L}$ & $\Delta F_{L}$ \\
\hline
 3.070 &  1.765 &  0.446 &  0.284 &  0.064 &  0.216 &  0.007 &  0.177 &  0.006 &  0.055 &  0.011\\
 3.070 &  2.378 &  0.521 &  0.327 &  0.049 &  0.133 &  0.003 &  0.131 &  0.003 &  0.045 &  0.005\\
 3.070 &  3.559 &  0.619 &  0.144 &  0.083 &  0.066 &  0.002 &  0.068 &  0.003 &  0.012 &  0.004\\
 3.110 &  0.810 &  0.266 &  0.520 &  0.027 &  0.458 &  0.006 &  0.284 &  0.003 &  0.127 &  0.005\\
 3.110 &  1.749 &  0.440 &  0.264 &  0.062 &  0.224 &  0.007 &  0.179 &  0.006 &  0.052 &  0.010\\
 3.110 &  2.355 &  0.514 &  0.319 &  0.051 &  0.137 &  0.003 &  0.133 &  0.003 &  0.045 &  0.005\\
 3.110 &  3.642 &  0.620 &  0.154 &  0.089 &  0.064 &  0.002 &  0.066 &  0.003 &  0.012 &  0.005\\
\hline
\end{tabular}
\end{center}
\caption{Structure functions of aluminum.}
\end{table}

\begin{table}[p]
\begin{center}
\begin{tabular}{|c|c|c|c|c|c|c|c|c|c|c|}
\hline
 $W^{2}$ & $Q^{2}$ & $x$ & $R$ & $\Delta R$ & $F_{1}$ & $\Delta F_{1}$ & $F_{2}$ & $\Delta F_{2}$ & $F_{L}$ & $\Delta F_{L}$ \\
\hline
 3.350 &  3.015 &  0.550 &  0.353 &  0.078 &  0.099 &  0.004 &  0.109 &  0.004 &  0.038 &  0.007\\
 3.350 &  3.585 &  0.592 &  0.333 &  0.093 &  0.072 &  0.002 &  0.084 &  0.003 &  0.028 &  0.005\\
 3.390 &  2.220 &  0.469 &  0.372 &  0.046 &  0.172 &  0.003 &  0.164 &  0.003 &  0.060 &  0.005\\
 3.390 &  2.792 &  0.527 &  0.218 &  0.062 &  0.122 &  0.004 &  0.116 &  0.004 &  0.028 &  0.007\\
 3.390 &  3.365 &  0.573 &  0.254 &  0.087 &  0.085 &  0.003 &  0.091 &  0.003 &  0.025 &  0.005\\
 3.430 &  2.198 &  0.463 &  0.322 &  0.045 &  0.180 &  0.003 &  0.164 &  0.003 &  0.053 &  0.005\\
 3.430 &  2.712 &  0.515 &  0.305 &  0.070 &  0.127 &  0.004 &  0.127 &  0.005 &  0.040 &  0.008\\
\hline
\end{tabular}
\end{center}
\caption{Structure functions of aluminum.}
\end{table}

\begin{table}[p]
\begin{center}
\begin{tabular}{|c|c|c|c|c|c|c|c|c|c|c|}
\hline
 $W^{2}$ & $Q^{2}$ & $x$ & $R$ & $\Delta R$ & $F_{1}$ & $\Delta F_{1}$ & $F_{2}$ & $\Delta F_{2}$ & $F_{L}$ & $\Delta F_{L}$ \\
\hline
 3.790 &  2.739 &  0.485 &  0.205 &  0.064 &  0.156 &  0.005 &  0.140 &  0.005 &  0.031 &  0.008\\
 3.790 &  3.337 &  0.534 &  0.288 &  0.063 &  0.104 &  0.003 &  0.110 &  0.003 &  0.032 &  0.005\\
 3.830 &  2.037 &  0.408 &  0.339 &  0.038 &  0.234 &  0.004 &  0.199 &  0.003 &  0.065 &  0.005\\
 3.830 &  2.663 &  0.475 &  0.382 &  0.074 &  0.152 &  0.005 &  0.154 &  0.005 &  0.055 &  0.008\\
 3.830 &  3.297 &  0.528 &  0.190 &  0.066 &  0.113 &  0.003 &  0.110 &  0.003 &  0.023 &  0.005\\
 3.870 &  2.112 &  0.414 &  0.386 &  0.047 &  0.219 &  0.005 &  0.195 &  0.004 &  0.070 &  0.007\\
 3.870 &  2.666 &  0.471 &  0.323 &  0.063 &  0.157 &  0.005 &  0.151 &  0.005 &  0.048 &  0.007\\
\hline
\end{tabular}
\end{center}
\caption{Structure functions of aluminum.}
\end{table}
\cleardoublepage
\begin{table}[p]
\begin{center}
\begin{tabular}{|c|c|c|c|c|c|c|c|c|c|c|}
\hline
 $W^{2}$ & $Q^{2}$ & $x$ & $R$ & $\Delta R$ & $F_{1}$ & $\Delta F_{1}$ & $F_{2}$ & $\Delta F_{2}$ & $F_{L}$ & $\Delta F_{L}$ \\
\hline
 1.310 &  2.064 &  0.828 &  0.167 &  0.069 &  0.041 &  0.002 &  0.037 &  0.002 &  0.011 &  0.005\\
 1.350 &  2.048 &  0.813 &  0.403 &  0.085 &  0.040 &  0.002 &  0.042 &  0.002 &  0.026 &  0.005\\
 1.390 &  2.032 &  0.799 &  0.280 &  0.075 &  0.044 &  0.002 &  0.043 &  0.002 &  0.020 &  0.005\\
 1.430 &  2.016 &  0.786 &  0.254 &  0.071 &  0.047 &  0.002 &  0.045 &  0.002 &  0.019 &  0.005\\
 1.470 &  2.000 &  0.772 &  0.277 &  0.071 &  0.050 &  0.002 &  0.048 &  0.002 &  0.021 &  0.005\\
 1.510 &  2.003 &  0.761 &  0.275 &  0.059 &  0.052 &  0.002 &  0.050 &  0.002 &  0.022 &  0.004\\
 1.550 &  1.984 &  0.748 &  0.327 &  0.059 &  0.054 &  0.002 &  0.054 &  0.002 &  0.026 &  0.004\\
\hline
\end{tabular}
\end{center}
\caption{Structure functions of iron.}
\end{table}

\begin{table}[p]
\begin{center}
\begin{tabular}{|c|c|c|c|c|c|c|c|c|c|c|}
\hline
 $W^{2}$ & $Q^{2}$ & $x$ & $R$ & $\Delta R$ & $F_{1}$ & $\Delta F_{1}$ & $F_{2}$ & $\Delta F_{2}$ & $F_{L}$ & $\Delta F_{L}$ \\
\hline
 2.630 &  1.538 &  0.468 &  0.296 &  0.043 &  0.200 &  0.004 &  0.161 &  0.004 &  0.055 &  0.007\\
 2.670 &  1.000 &  0.359 &  0.407 &  0.062 &  0.331 &  0.012 &  0.230 &  0.008 &  0.097 &  0.014\\
 2.710 &  0.921 &  0.335 &  0.539 &  0.069 &  0.345 &  0.012 &  0.249 &  0.008 &  0.124 &  0.014\\
 2.710 &  2.337 &  0.561 &  0.200 &  0.063 &  0.113 &  0.004 &  0.103 &  0.004 &  0.025 &  0.007\\
 2.750 &  0.909 &  0.327 &  0.500 &  0.065 &  0.360 &  0.012 &  0.250 &  0.008 &  0.118 &  0.014\\
 2.790 &  0.881 &  0.316 &  0.514 &  0.050 &  0.375 &  0.009 &  0.256 &  0.006 &  0.122 &  0.010\\
 2.830 &  0.925 &  0.322 &  0.557 &  0.067 &  0.358 &  0.012 &  0.257 &  0.008 &  0.128 &  0.013\\
\hline
\end{tabular}
\end{center}
\caption{Structure functions of iron.}
\end{table}
\cleardoublepage
\begin{table}[p]
\begin{center}
\begin{tabular}{|c|c|c|c|c|c|c|c|c|c|c|}
\hline
 $W^{2}$ & $Q^{2}$ & $x$ & $R$ & $\Delta R$ & $F_{1}$ & $\Delta F_{1}$ & $F_{2}$ & $\Delta F_{2}$ & $F_{L}$ & $\Delta F_{L}$ \\
\hline
 1.110 &  3.717 &  0.942 &  0.517 &  0.196 &  0.007 &  0.001 &  0.011 &  0.001 &  0.007 &  0.002\\
 1.150 &  3.696 &  0.932 &  0.455 &  0.185 &  0.007 &  0.001 &  0.011 &  0.001 &  0.006 &  0.002\\
 1.190 &  3.676 &  0.922 &  0.492 &  0.178 &  0.008 &  0.001 &  0.013 &  0.001 &  0.008 &  0.002\\
 1.230 &  3.655 &  0.913 &  0.244 &  0.141 &  0.010 &  0.001 &  0.012 &  0.001 &  0.004 &  0.002\\
 1.310 &  3.662 &  0.895 &  0.399 &  0.119 &  0.010 &  0.001 &  0.014 &  0.001 &  0.007 &  0.002\\
 1.350 &  3.640 &  0.886 &  0.226 &  0.100 &  0.012 &  0.001 &  0.014 &  0.001 &  0.005 &  0.002\\
 1.390 &  3.618 &  0.877 &  0.272 &  0.100 &  0.012 &  0.001 &  0.015 &  0.001 &  0.006 &  0.002\\
\hline
\end{tabular}
\end{center}
\caption{Structure functions of copper.}
\end{table}

\begin{table}[p]
\begin{center}
\begin{tabular}{|c|c|c|c|c|c|c|c|c|c|c|}
\hline
 $W^{2}$ & $Q^{2}$ & $x$ & $R$ & $\Delta R$ & $F_{1}$ & $\Delta F_{1}$ & $F_{2}$ & $\Delta F_{2}$ & $F_{L}$ & $\Delta F_{L}$ \\
\hline
 2.350 &  3.111 &  0.679 &  0.286 &  0.080 &  0.046 &  0.002 &  0.053 &  0.002 &  0.018 &  0.004\\
 2.390 &  3.086 &  0.672 &  0.221 &  0.074 &  0.051 &  0.002 &  0.056 &  0.002 &  0.015 &  0.004\\
 2.430 &  3.062 &  0.664 &  0.275 &  0.076 &  0.052 &  0.002 &  0.059 &  0.002 &  0.019 &  0.004\\
 2.470 &  3.038 &  0.656 &  0.159 &  0.070 &  0.056 &  0.002 &  0.056 &  0.002 &  0.012 &  0.004\\
 2.510 &  3.013 &  0.649 &  0.187 &  0.069 &  0.059 &  0.002 &  0.061 &  0.002 &  0.014 &  0.004\\
 2.550 &  2.989 &  0.642 &  0.321 &  0.076 &  0.058 &  0.002 &  0.066 &  0.002 &  0.024 &  0.004\\
 2.590 &  2.965 &  0.634 &  0.294 &  0.073 &  0.063 &  0.002 &  0.070 &  0.002 &  0.023 &  0.004\\
\hline
\end{tabular}
\end{center}
\caption{Structure functions of copper.}
\end{table}

\begin{table}[p]
\begin{center}
\begin{tabular}{|c|c|c|c|c|c|c|c|c|c|c|}
\hline
 $W^{2}$ & $Q^{2}$ & $x$ & $R$ & $\Delta R$ & $F_{1}$ & $\Delta F_{1}$ & $F_{2}$ & $\Delta F_{2}$ & $F_{L}$ & $\Delta F_{L}$ \\
\hline
 3.590 &  3.165 &  0.539 &  0.170 &  0.072 &  0.108 &  0.003 &  0.103 &  0.004 &  0.020 &  0.006\\
 3.630 &  2.470 &  0.473 &  0.187 &  0.067 &  0.170 &  0.005 &  0.145 &  0.006 &  0.030 &  0.009\\
 3.630 &  3.143 &  0.533 &  0.265 &  0.078 &  0.105 &  0.003 &  0.108 &  0.004 &  0.030 &  0.006\\
 3.670 &  2.402 &  0.463 &  0.211 &  0.048 &  0.179 &  0.004 &  0.153 &  0.004 &  0.035 &  0.006\\
 3.670 &  3.119 &  0.528 &  0.338 &  0.082 &  0.108 &  0.003 &  0.116 &  0.004 &  0.038 &  0.006\\
 3.710 &  2.804 &  0.498 &  0.191 &  0.061 &  0.141 &  0.004 &  0.127 &  0.004 &  0.027 &  0.007\\
 3.710 &  3.399 &  0.546 &  0.173 &  0.078 &  0.098 &  0.003 &  0.096 &  0.004 &  0.019 &  0.006\\
\hline
\end{tabular}
\end{center}
\caption{Structure functions of copper.}
\end{table}
\cleardoublepage

\subsection{HMS Kinematic Settings}

\begin{table}[ht]
\begin{center}
\begin{tabular}{|c|c|}
\hline
HMS Central Momentum P (GeV) & HMS central angle $\theta$\\
\hline
0.465 &  45.00 \\
0.535 &  45.00 \\
0.615 &  45.00 \\
0.707 &  45.00 \\
0.465 &  60.00 \\
0.535 &  60.00 \\
0.615 &  60.00 \\
0.707 &  60.00 \\
0.935 &  60.00 \\
0.465 &  75.00 \\
0.535 &  75.00 \\
0.615 &  75.00 \\
0.707 &  75.00 \\
0.813 &  75.00 \\
\hline
\end{tabular}
\end{center}
\caption{Beam energy is 2.097 GeV. }
\end{table}

\begin{table}[ht]
\begin{center}
\begin{tabular}{|c|c|c|}
\hline
HMS Central Momentum P (GeV) & HMS central angle $\theta$\\
\hline
1.184 & 15.00 \\
1.361 & 15.00 \\
1.029 & 25.00 \\
1.184 & 25.00 \\
1.361 & 25.00 \\
1.565 & 25.00 \\
0.778 & 35.00 \\
0.895 & 35.00 \\
1.029 & 35.00 \\
1.184 & 35.00 \\
1.361 & 35.00 \\
0.678 & 45.00 \\
0.778 & 45.00 \\
0.895 & 45.00 \\
1.029 & 45.00 \\
1.184 & 45.00 \\
1.361 & 45.00 \\
1.565 & 45.00 \\
0.505 & 60.00 \\
0.581 & 60.00 \\
0.678 & 60.00 \\
0.778 & 59.99 \\
0.895 & 60.00 \\
0.439 & 74.99 \\
0.439 & 74.99 \\
0.505 & 74.99 \\
0.581 & 75.00 \\
0.678 & 75.00 \\
0.778 & 75.00 \\
\hline
\end{tabular}
\end{center}
\caption{Beam energy is 3.116 GeV.}
\end{table}

\begin{table}[ht]
\begin{center}
\begin{tabular}{|c|c|}
\hline
HMS Central Momentum P (GeV) & HMS central angle $\theta$\\
\hline
0.535 & 76.00 \\
0.615 & 76.00 \\
0.749 & 60.00 \\
0.862 & 60.00 \\
0.991 & 60.00 \\
0.991 & 50.00 \\
1.140 & 50.00 \\
1.310 & 50.00 \\
1.140 & 43.00 \\
0.569 & 75.00 \\
0.569 & 75.00 \\
0.659 & 75.00 \\
0.659 & 75.00 \\
0.759 & 58.00 \\
0.879 & 58.00 \\
0.999 & 58.00 \\
0.879 & 48.00 \\
0.999 & 48.00 \\
1.148 & 48.00 \\
1.318 & 48.00 \\
1.318 & 48.00 \\
1.318 & 39.00 \\
\hline
\end{tabular}
\end{center}
\caption{Beam energy is 4.134 GeV.}
\end{table}

\begin{table}[ht]
\begin{center}
\begin{tabular}{|c|c|}
\hline
HMS Central Momentum P (GeV) & HMS central angle $\theta$\\
\hline
3.146 & 12.00 \\
3.615 & 12.00 \\
4.160 & 18.00 \\
3.146 & 18.00 \\
3.615 & 18.00 \\
2.380 & 23.00 \\
2.736 & 23.00 \\
3.146 & 23.00 \\
3.615 & 23.00 \\
3.615 & 23.00 \\
2.067 & 27.00 \\
2.377 & 27.00 \\
2.736 & 27.00 \\
3.150 & 27.00 \\
1.798 & 30.00 \\
2.067 & 30.00 \\
2.377 & 30.00 \\
2.740 & 30.00 \\
2.740 & 30.00 \\
1.798 & 33.00 \\
2.067 & 33.00 \\
2.377 & 33.00 \\
1.568 & 38.00 \\
1.798 & 38.00 \\
\hline
\end{tabular}
\end{center}
\caption{ Beam energy is 5.151 GeV.}
\end{table}

\cleardoublepage

\subsection{Cross Sections} \label{sec:CrossSections}
\begin{figure}[htp]
\begin{center}
\epsfig{file=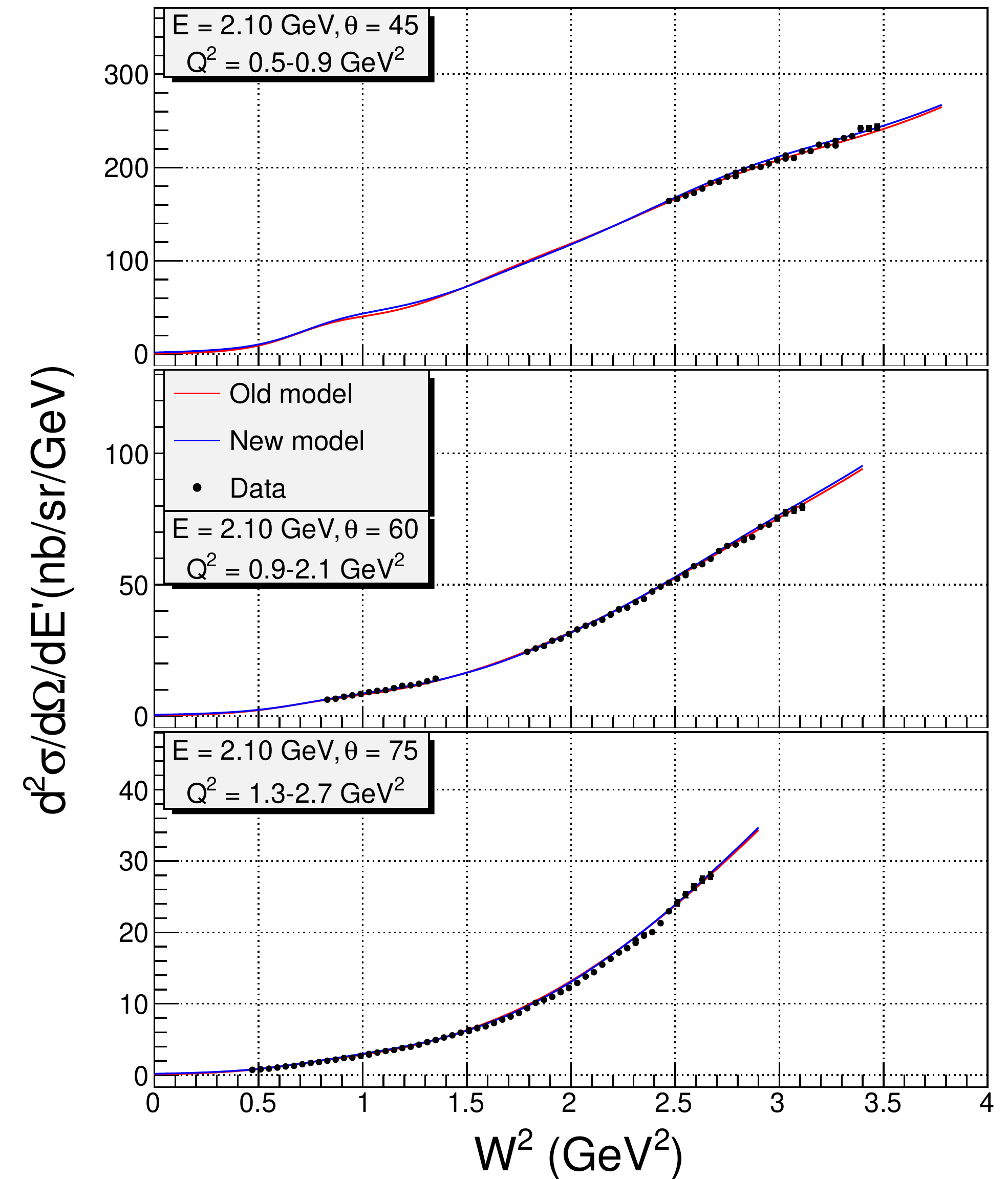,width=5.0in}
\end{center}
\caption { Extracted differential cross section for carbon compared to the model cross section. }
\label{fig:csC2097}
\end{figure}

\begin{figure}[p]
\begin{center}
\epsfig{file=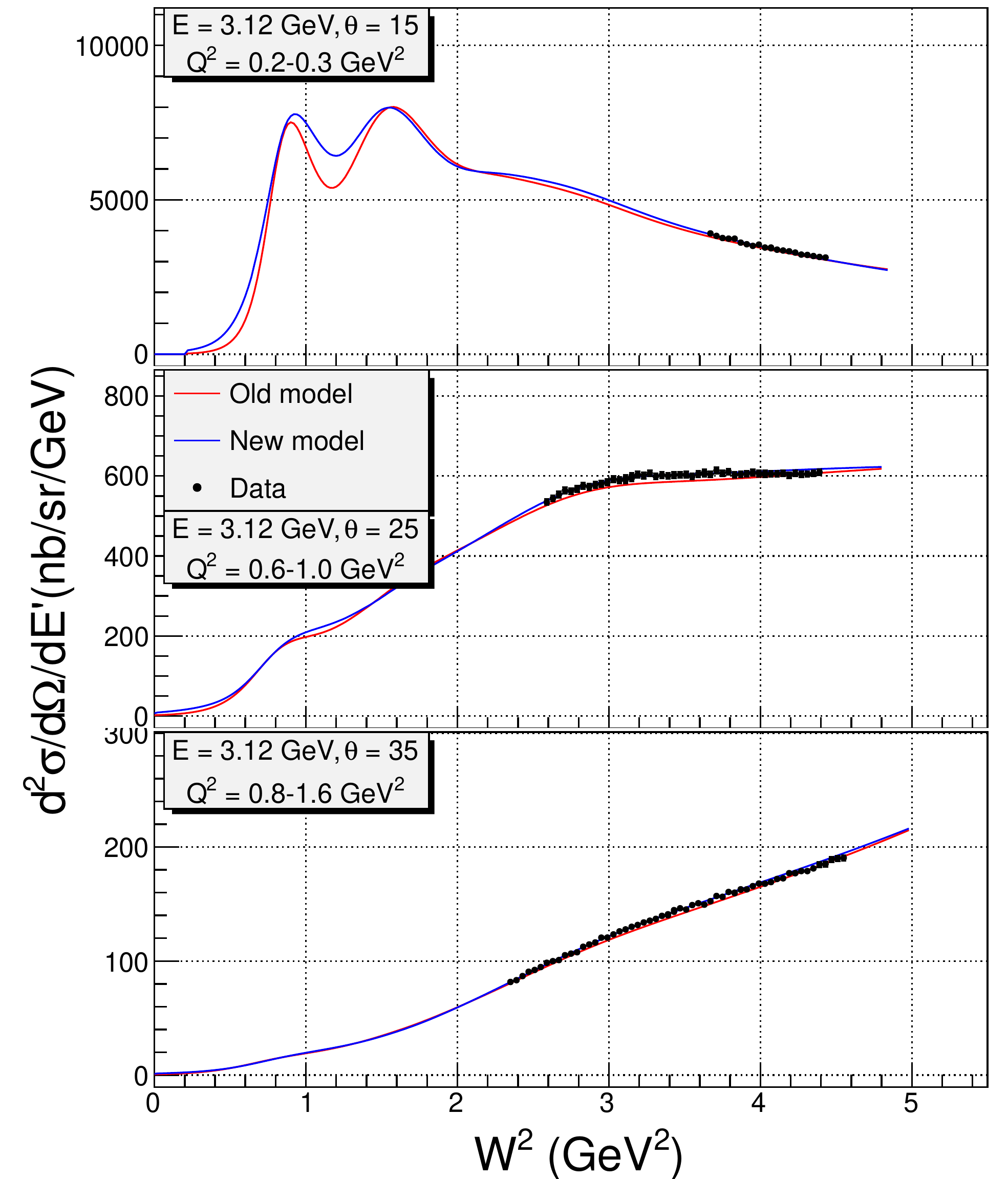,width=5.5in}
\end{center}
\caption { Extracted differential cross section for carbon compared to the model cross section.}
\label{fig:csC3116_0}
\end{figure}

\begin{figure}[p]
\begin{center}
\epsfig{file=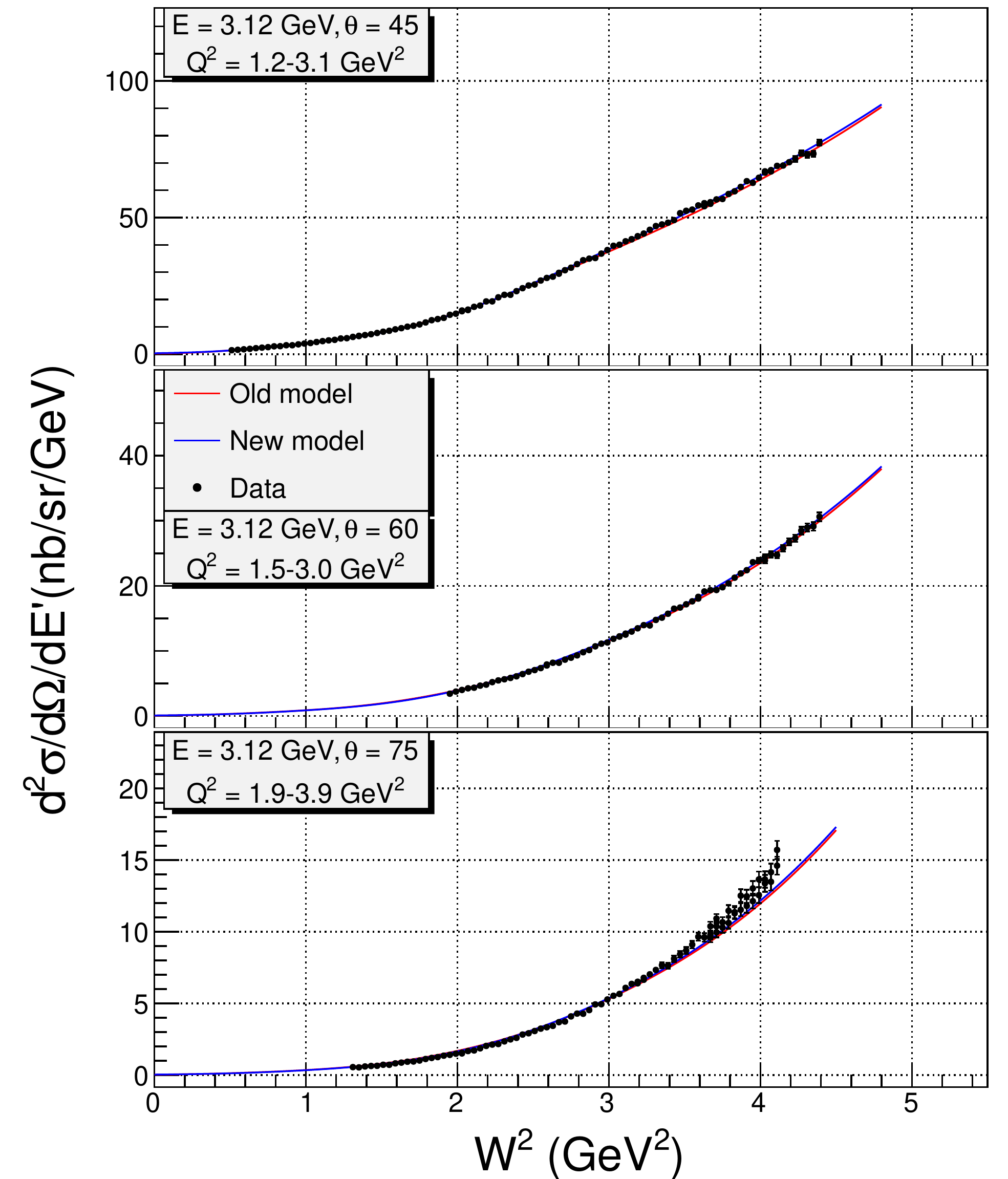,width=5.5in}
\end{center}
\caption { Extracted differential cross section for carbon compared to the model cross section. }
\label{fig:csC3116_1}
\end{figure}

\begin{figure}[p]
\begin{center}
\epsfig{file=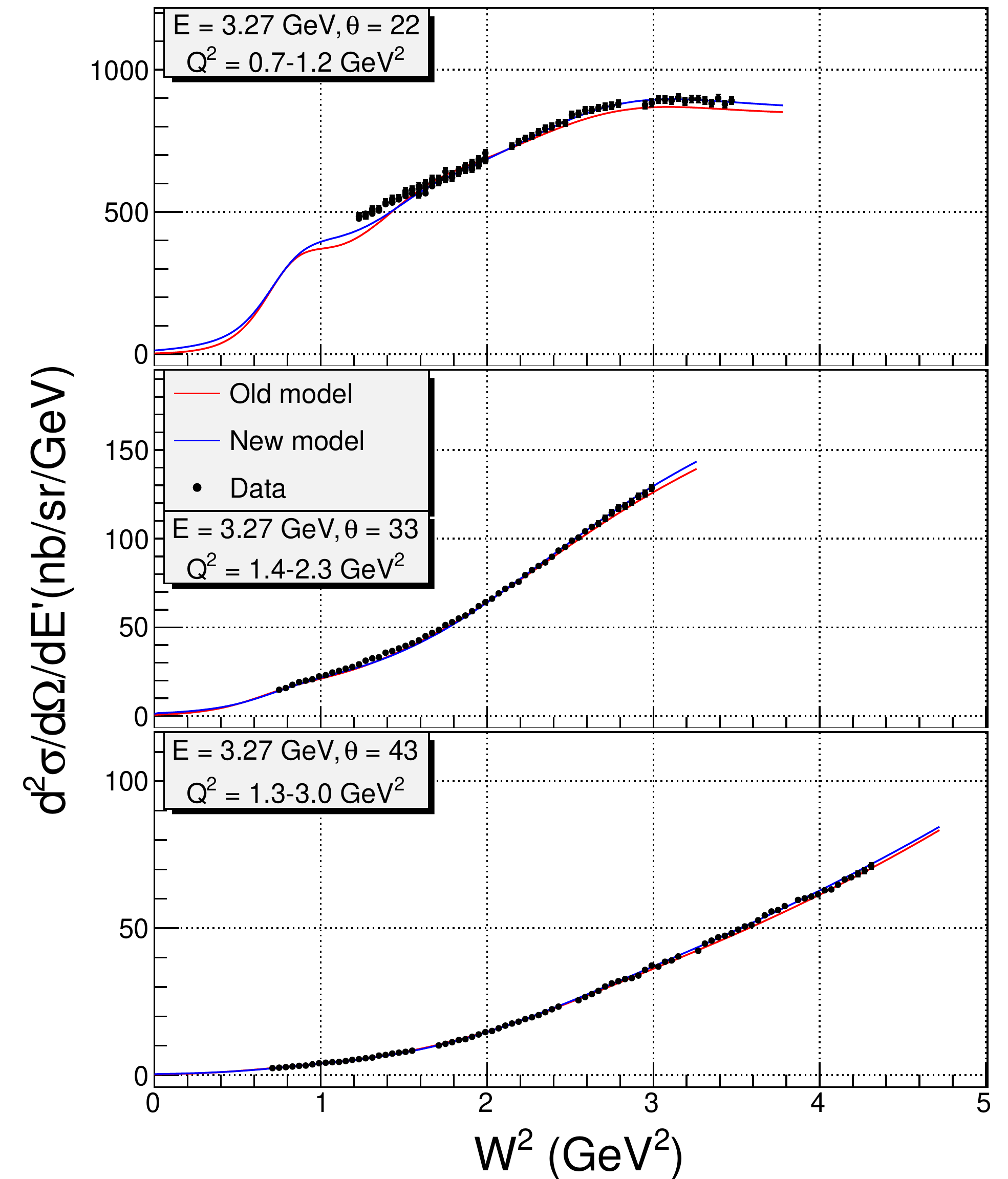,width=5.5in}
\end{center}
\caption { Extracted differential cross section for carbon compared to the model cross section. }
\label{fig:csC3270_0}
\end{figure}

\begin{figure}[p]
\begin{center}
\epsfig{file=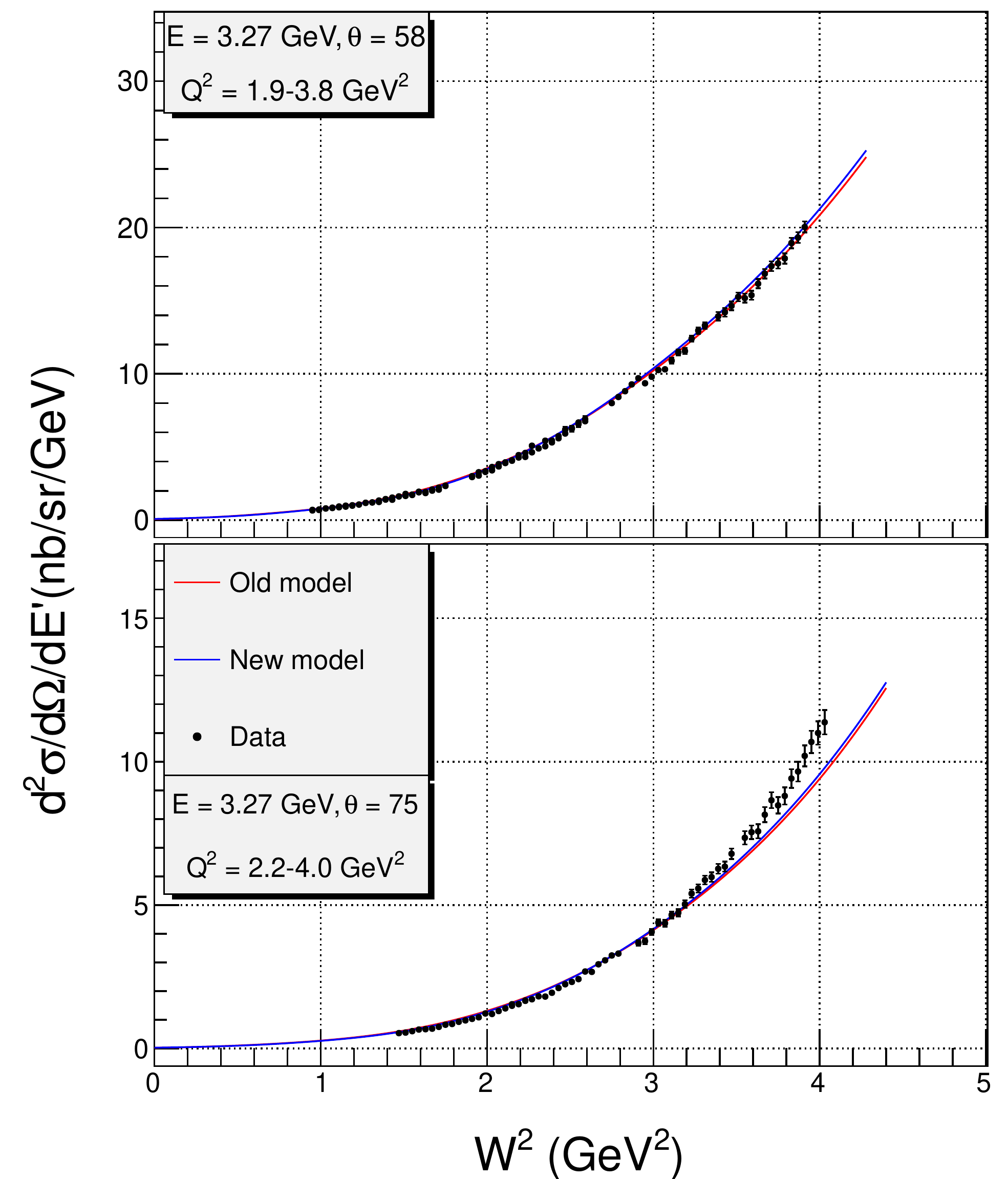,width=5.5in}
\end{center}
\caption { Extracted differential cross section for carbon compared to the model cross section. }
\label{fig:csC3270_1}
\end{figure}

\begin{figure}[p]
\begin{center}
\epsfig{file=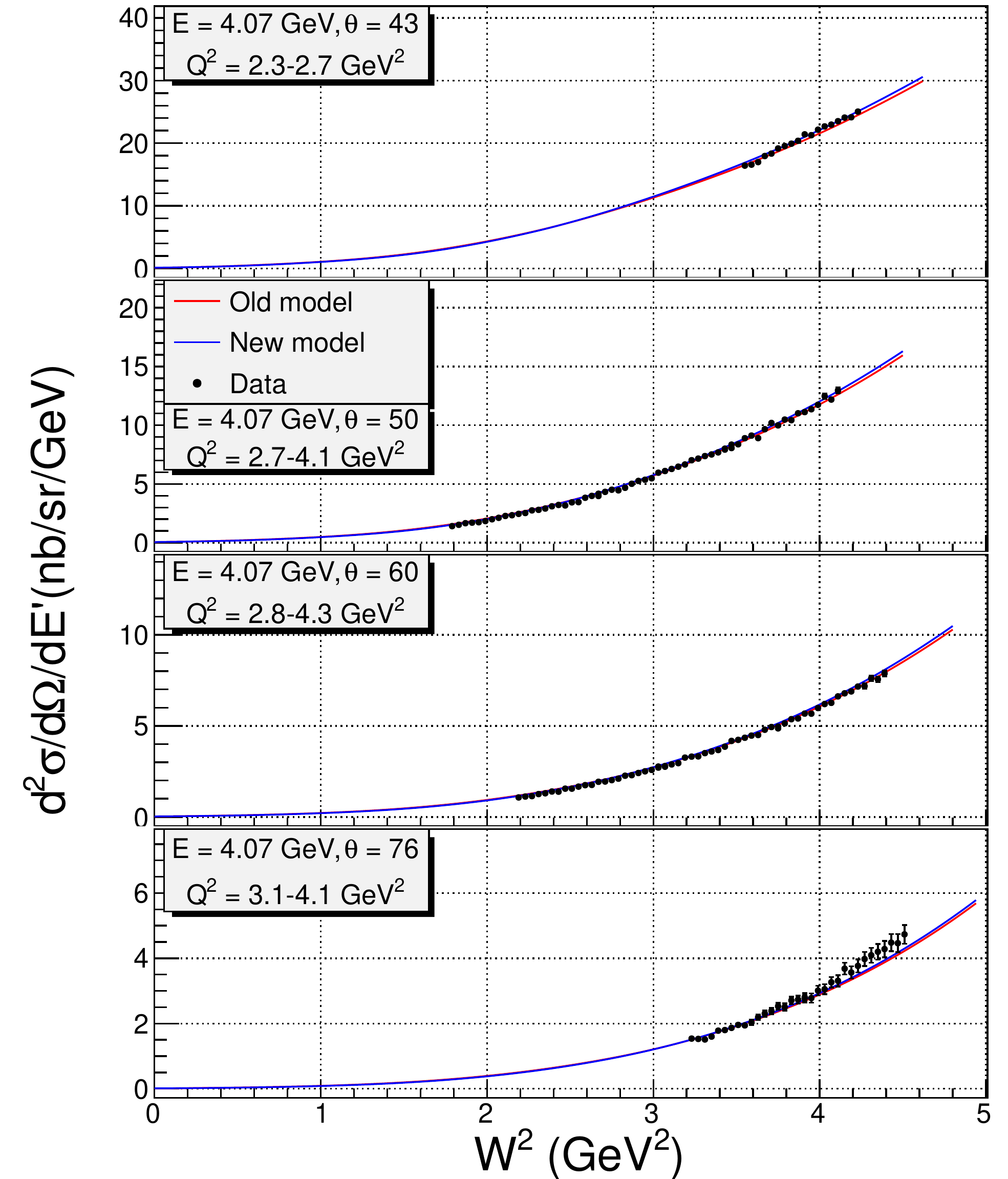,width=5.5in}
\end{center}
\caption {  Extracted differential cross section for carbon compared to the model cross section.}
\label{fig:csC4074_0}
\end{figure}

\begin{figure}[p]
\begin{center}
\epsfig{file=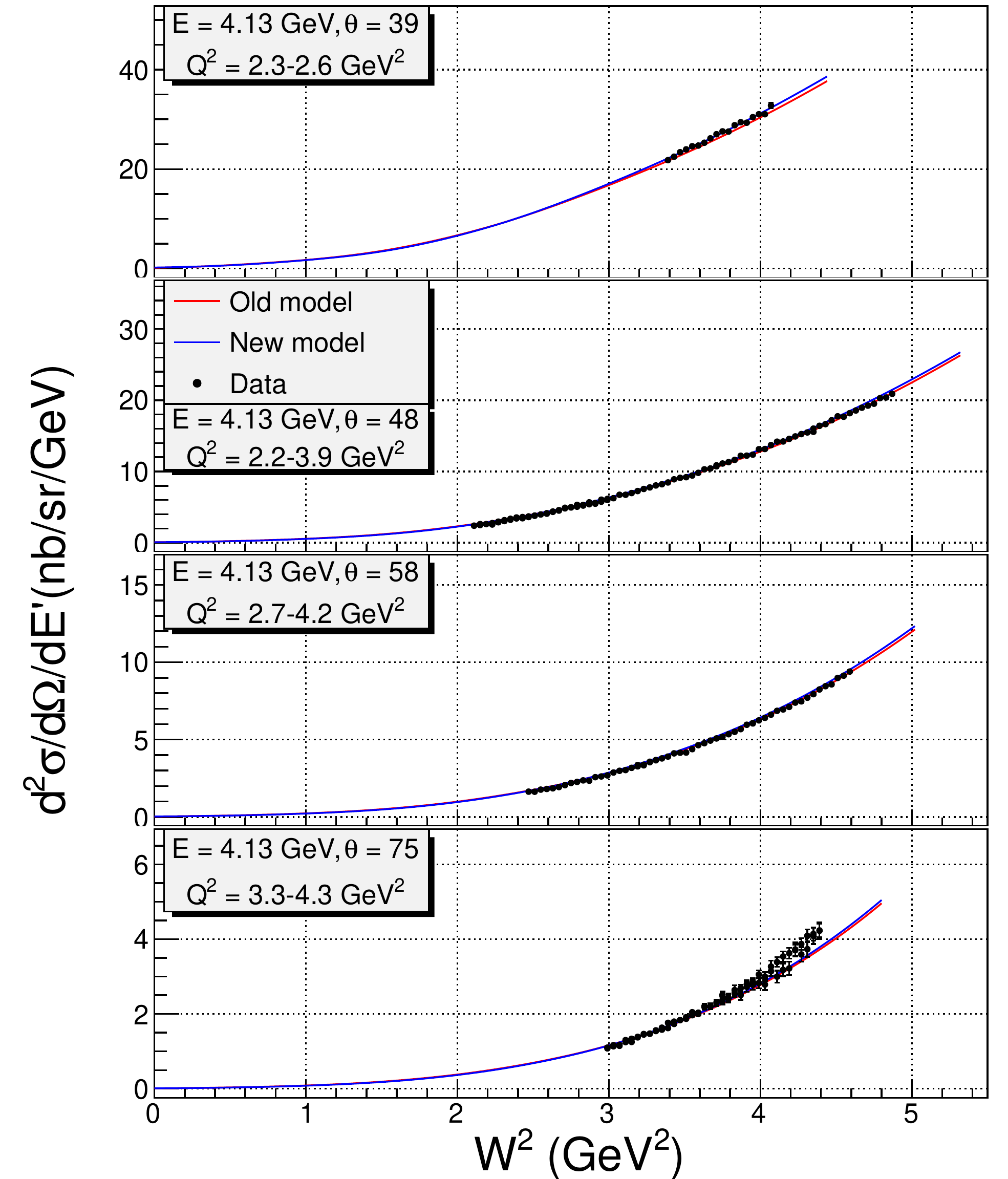,width=5.5in}
\end{center}
\caption {  Extracted differential cross section for carbon compared to the model cross section.}
\label{fig:csC4134_0}
\end{figure}

\begin{figure}[p]
\begin{center}
\epsfig{file=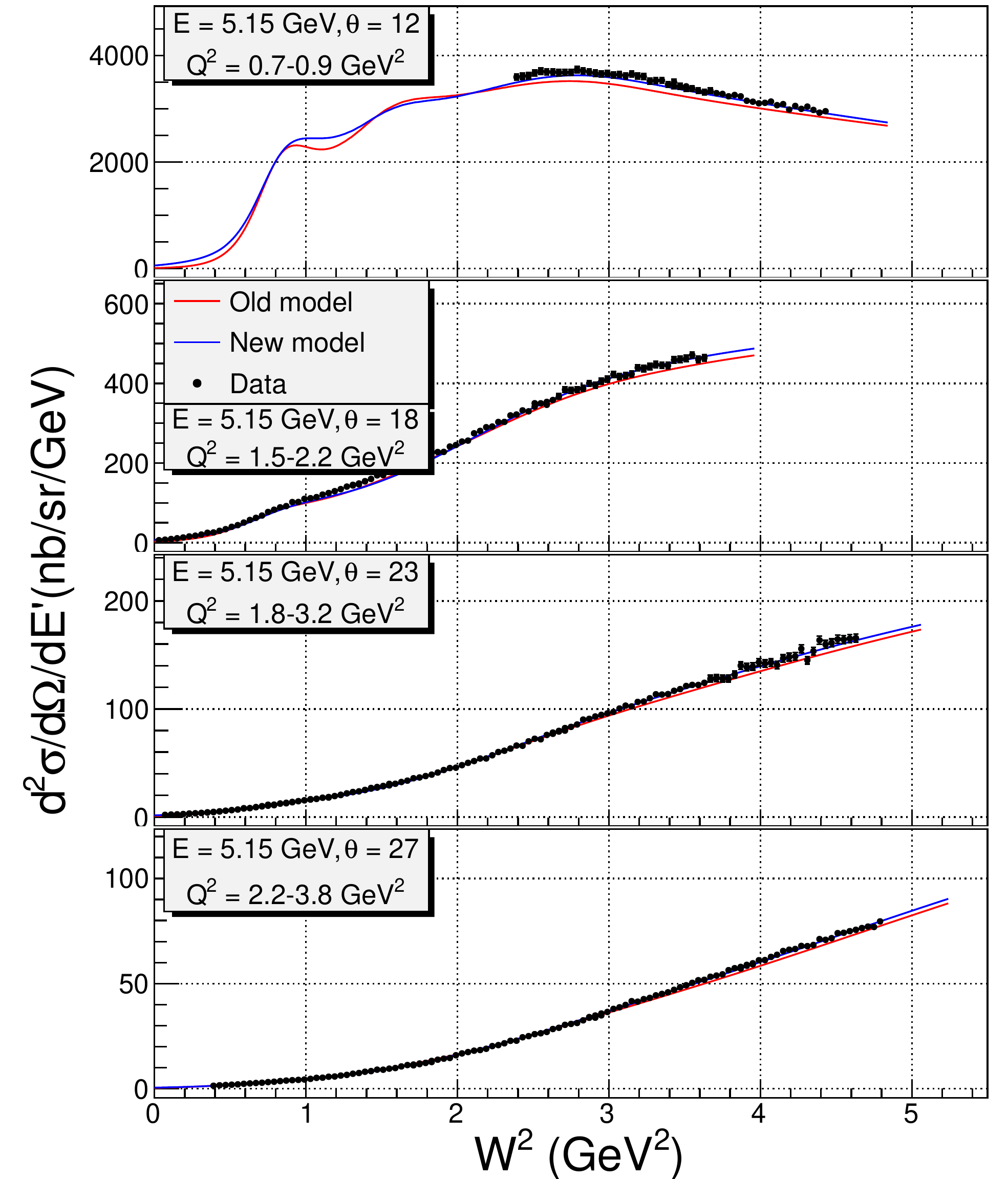,width=5.5in}
\end{center}
\caption { Extracted differential cross section for carbon compared to the model cross section. }
\label{fig:csC5151_0}
\end{figure}

\begin{figure}[p]
\begin{center}
\epsfig{file=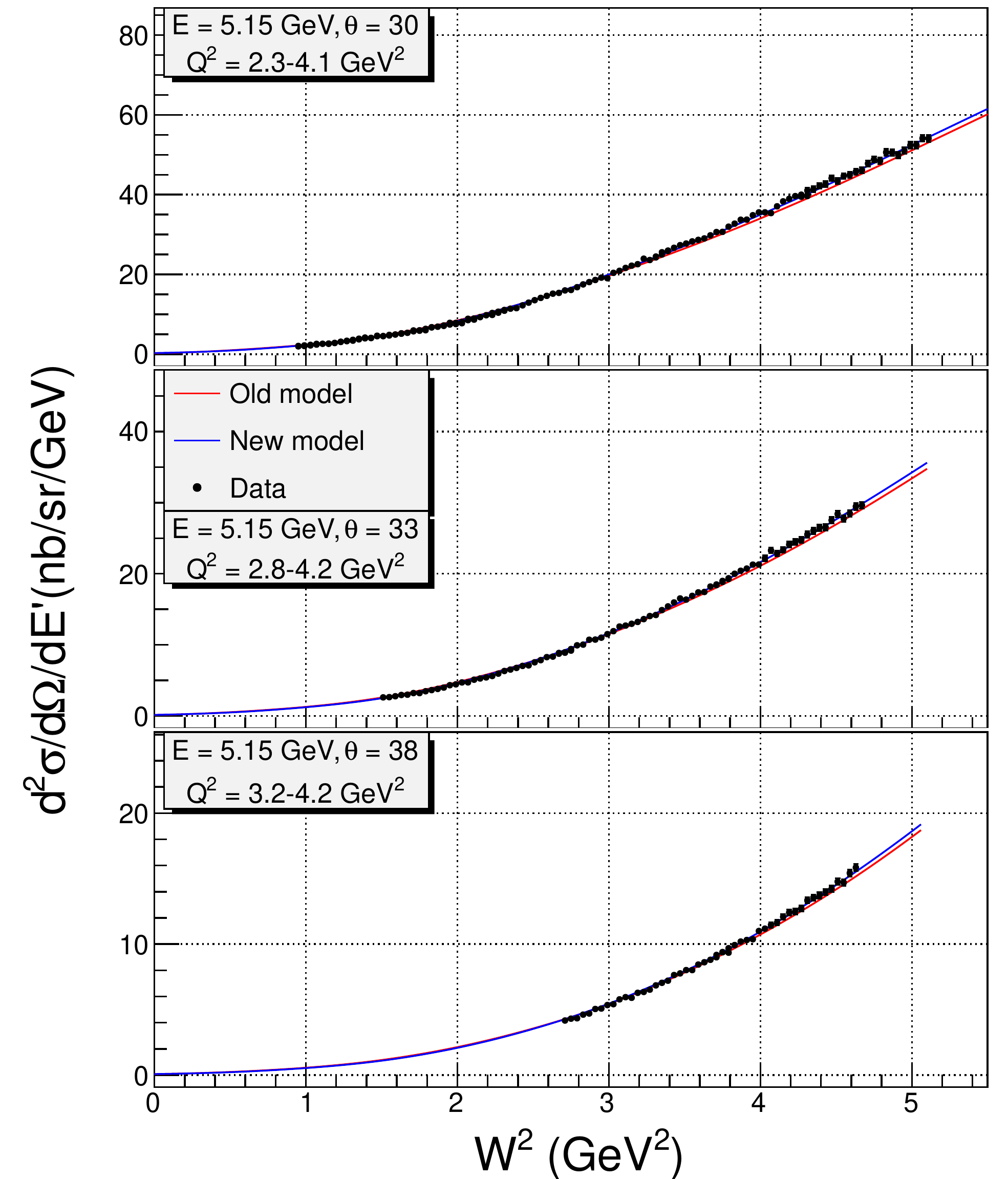,width=5.5in}
\end{center}
\caption { Extracted differential cross section for carbon compared to the model cross section. }
\label{fig:csC5151_1}
\end{figure}


\begin{figure}[htp]
\begin{center}
\epsfig{file=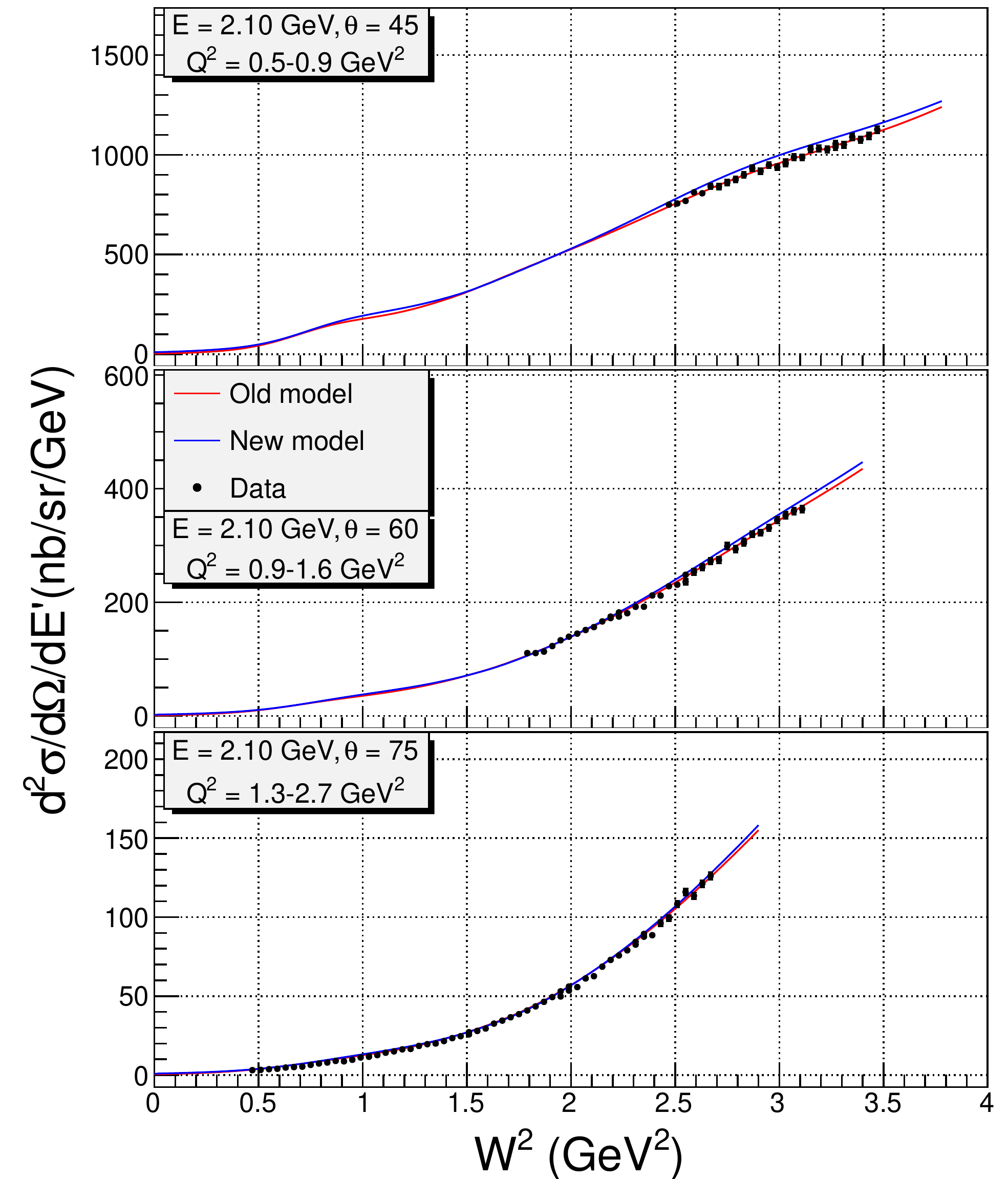,width=5.0in}
\end{center}
\caption { Extracted differential cross section for iron compared to the model cross section. }
\label{fig:csFe2097}
\end{figure}

\begin{figure}[p]
\begin{center}
\epsfig{file=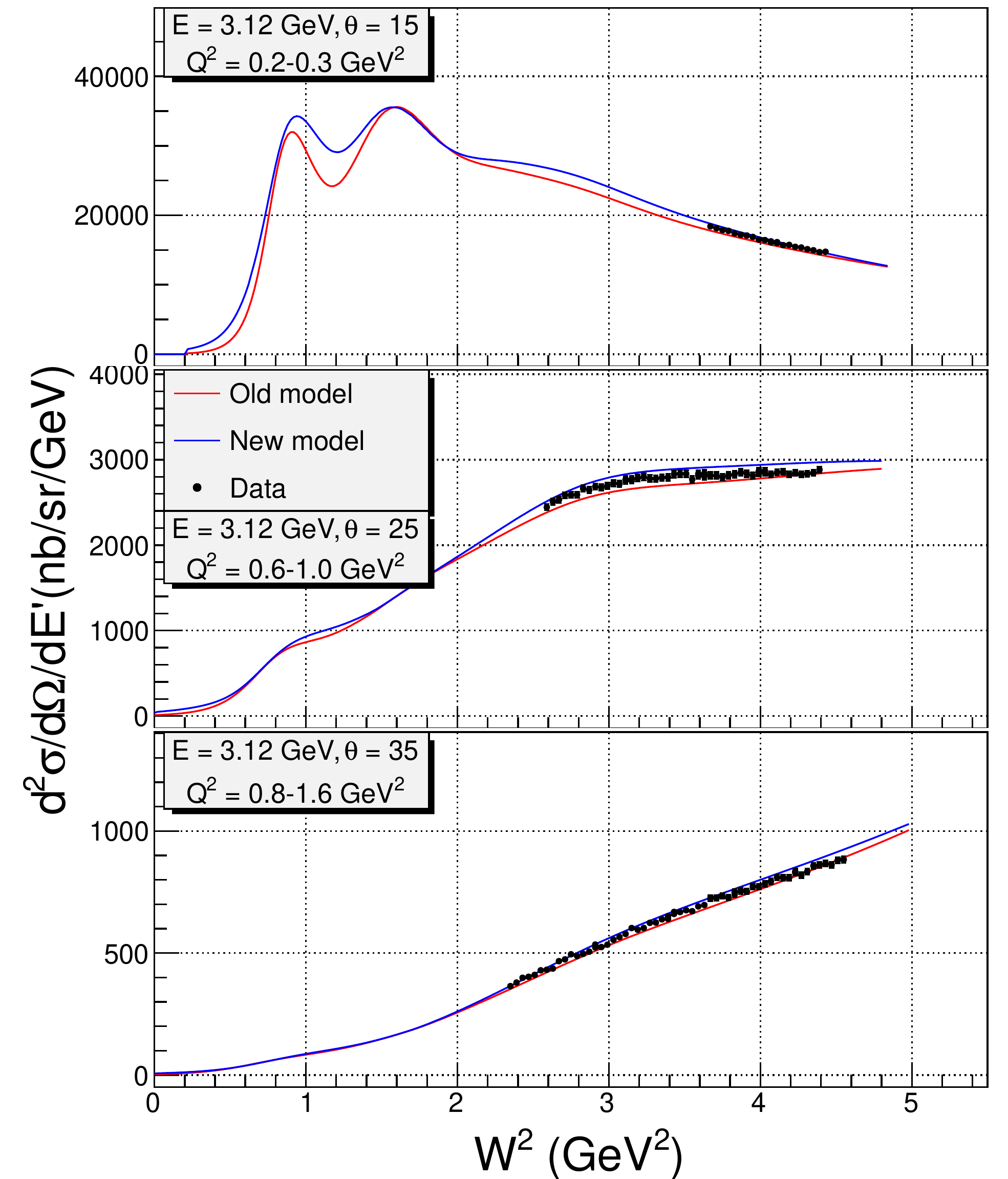,width=5.5in}
\end{center}
\caption { Extracted differential cross section for iron compared to the model cross section.}
\label{fig:csFe3116_0}
\end{figure}

\begin{figure}[p]
\begin{center}
\epsfig{file=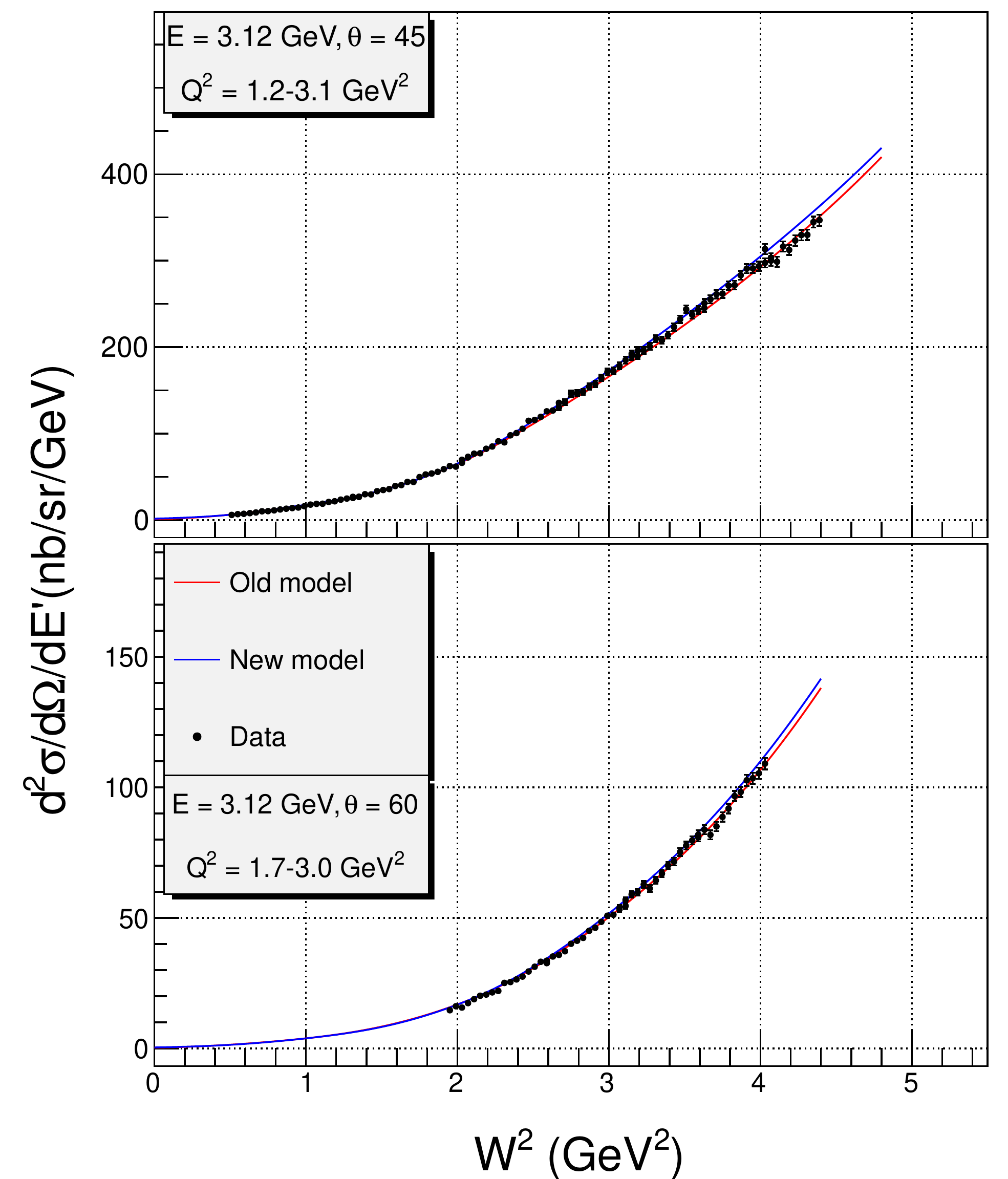,width=5.5in}
\end{center}
\caption { Extracted differential cross section for iron compared to the model cross section. }
\label{fig:csFe3116_1}
\end{figure}

\begin{figure}[p]
\begin{center}
\epsfig{file=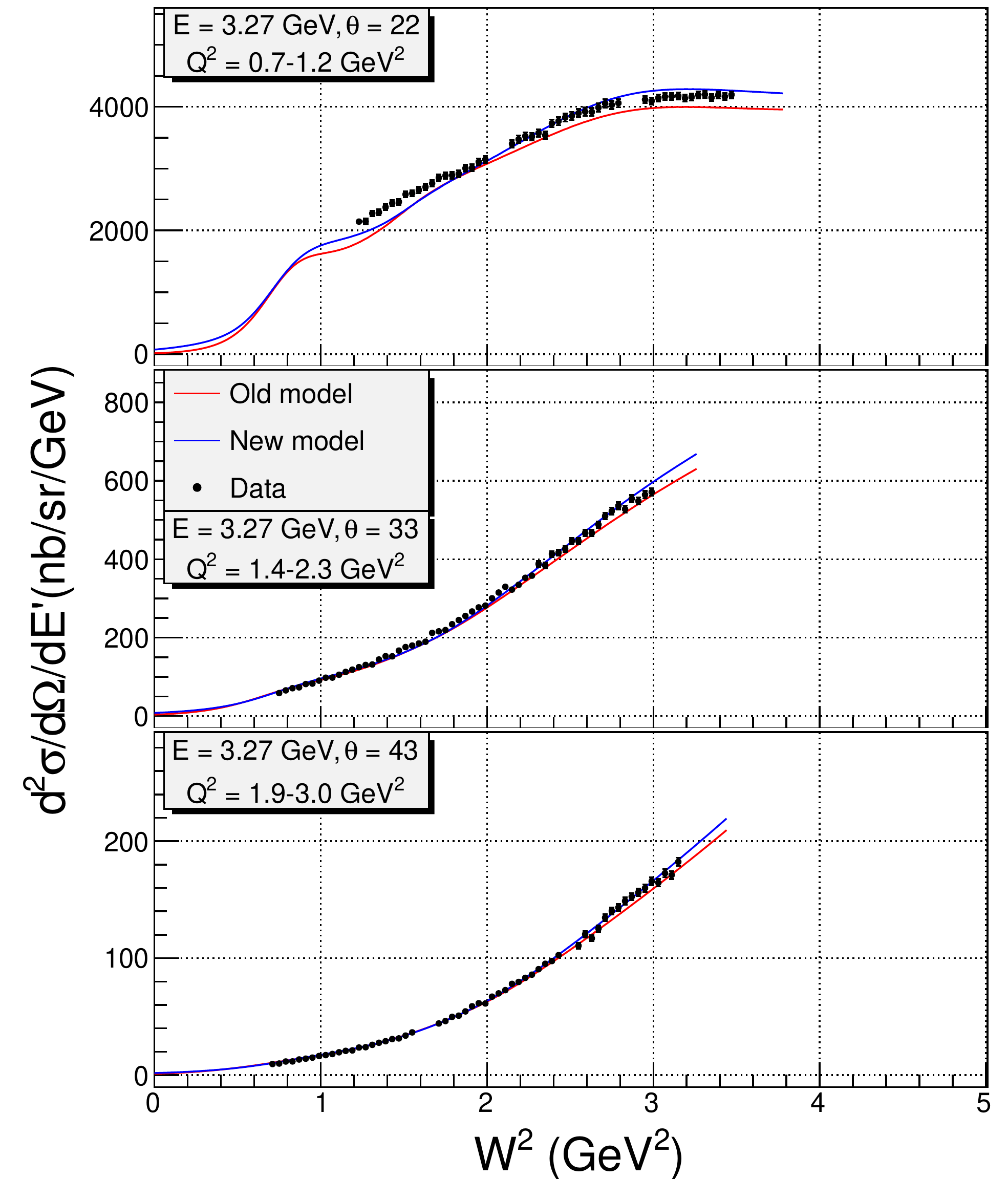,width=5.5in}
\end{center}
\caption { Extracted differential cross section for iron compared to the model cross section. }
\label{fig:csFe3270_0}
\end{figure}

\begin{figure}[p]
\begin{center}
\epsfig{file=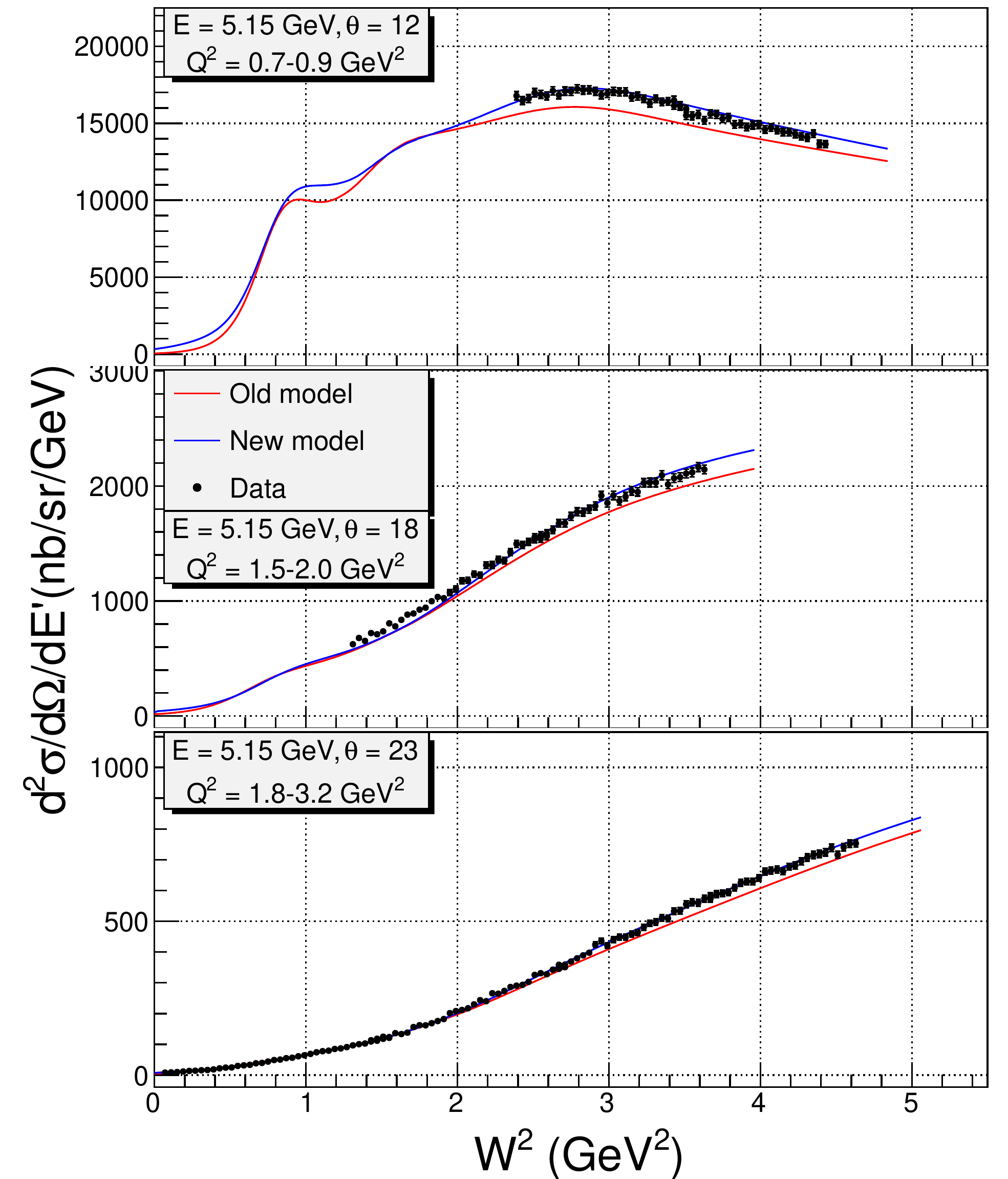,width=5.5in}
\end{center}
\caption { Extracted differential cross section for iron compared to the model cross section. }
\label{fig:csFe5151_0}
\end{figure}

\newpage

\begin{figure}[p]
\begin{center}
\epsfig{file=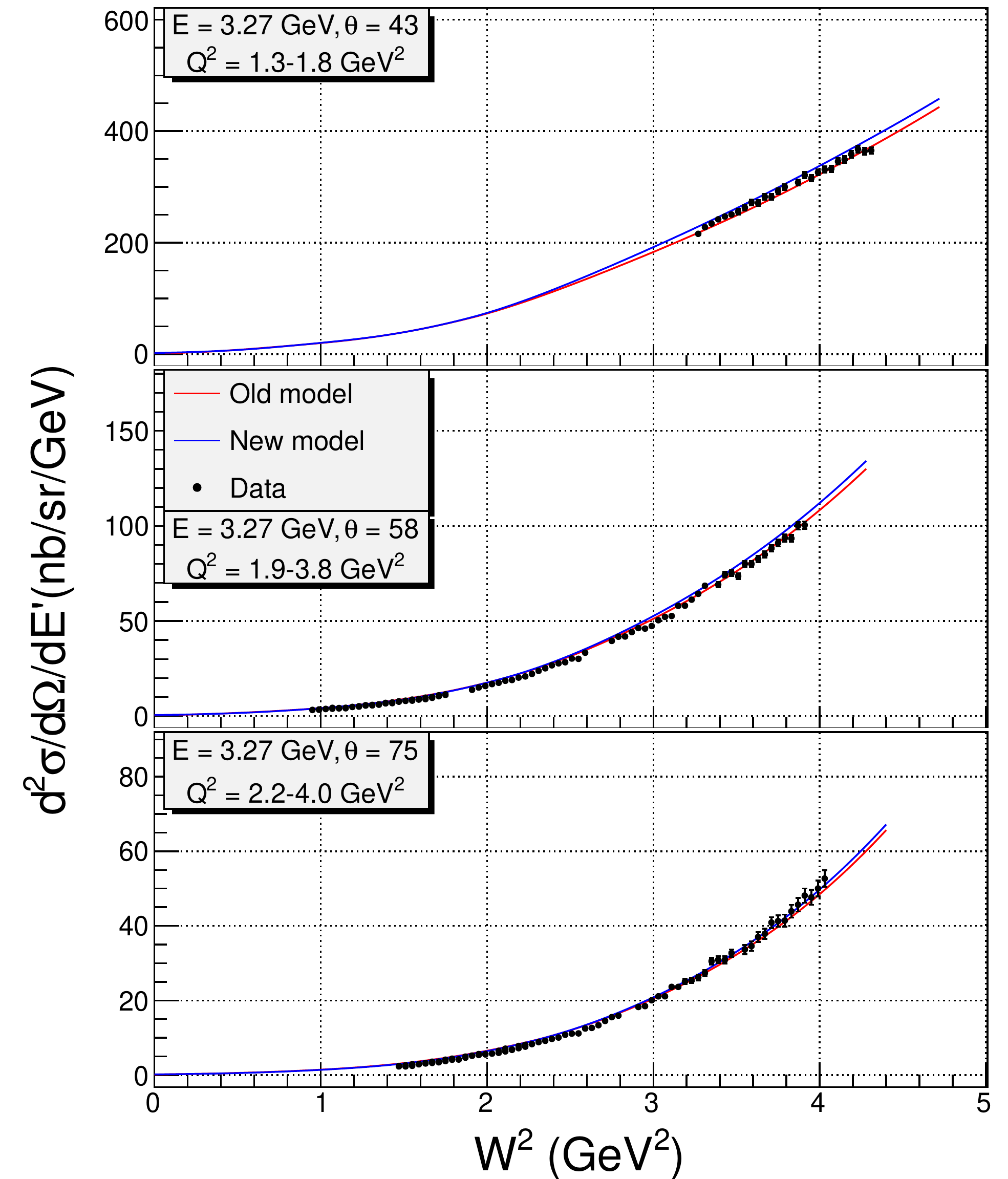,width=5.5in}
\end{center}
\caption { Extracted differential cross section for copper compared to the model cross section. }
\label{fig:csCu3270_0}
\end{figure}

\begin{figure}[p]
\begin{center}
\epsfig{file=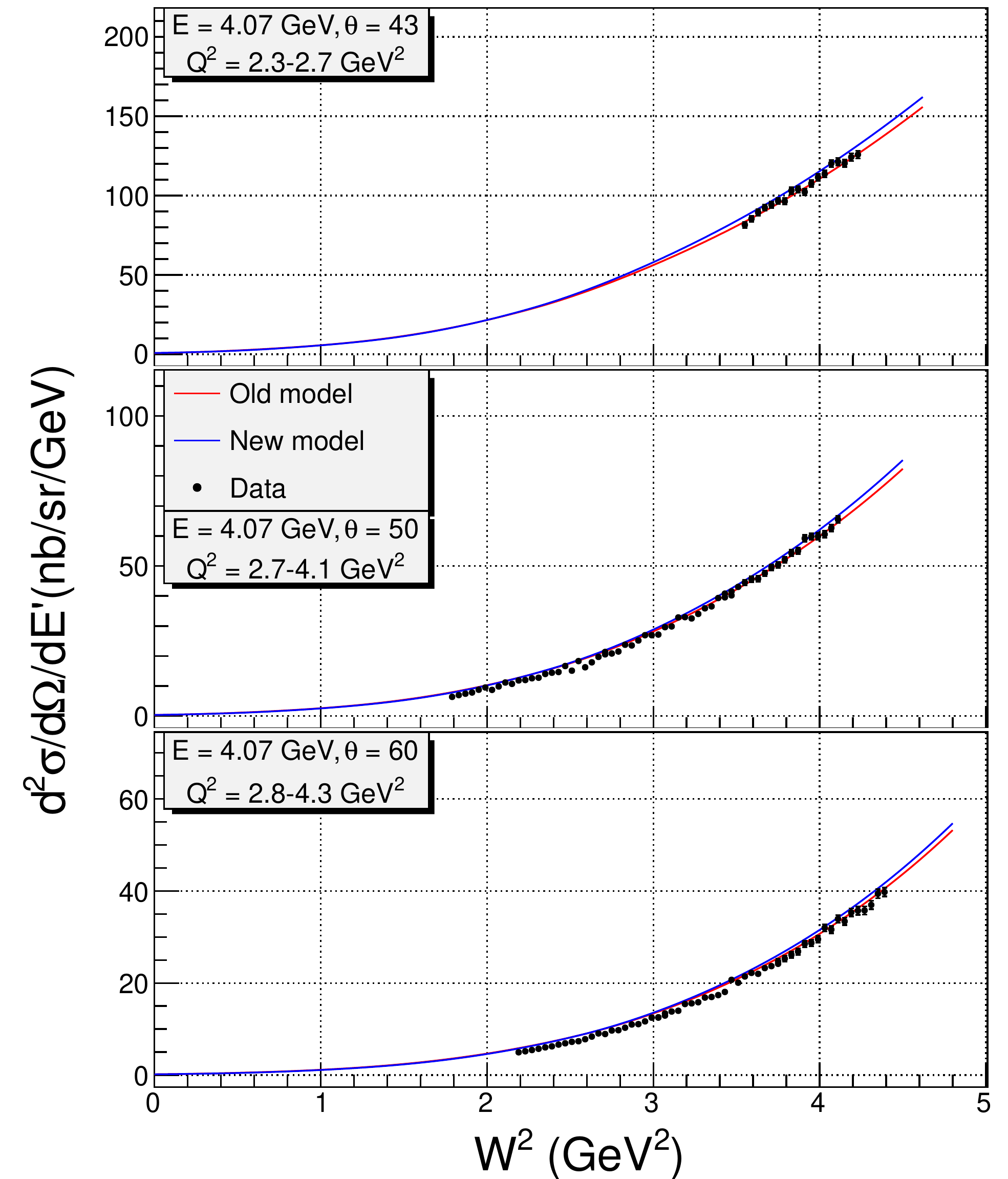,width=5.5in}
\end{center}
\caption {  Extracted differential cross section for copper compared to the model cross section.}
\label{fig:csCu4074_0}
\end{figure}

\begin{figure}[p]
\begin{center}
\epsfig{file=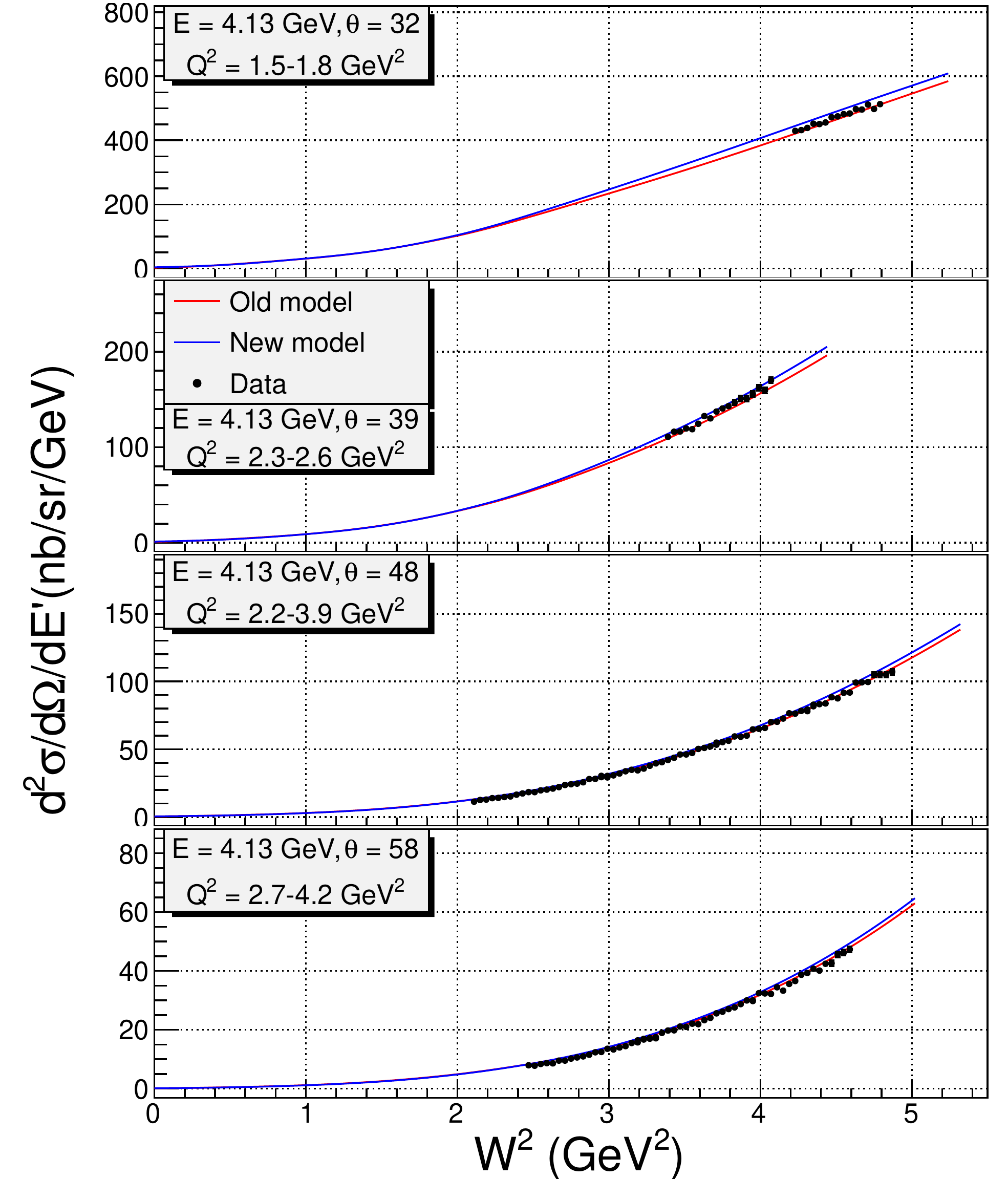,width=5.5in}
\end{center}
\caption {  Extracted differential cross section for copper compared to the model cross section.}
\label{fig:csCu4134_0}
\end{figure}

\begin{figure}[p]
\begin{center}
\epsfig{file=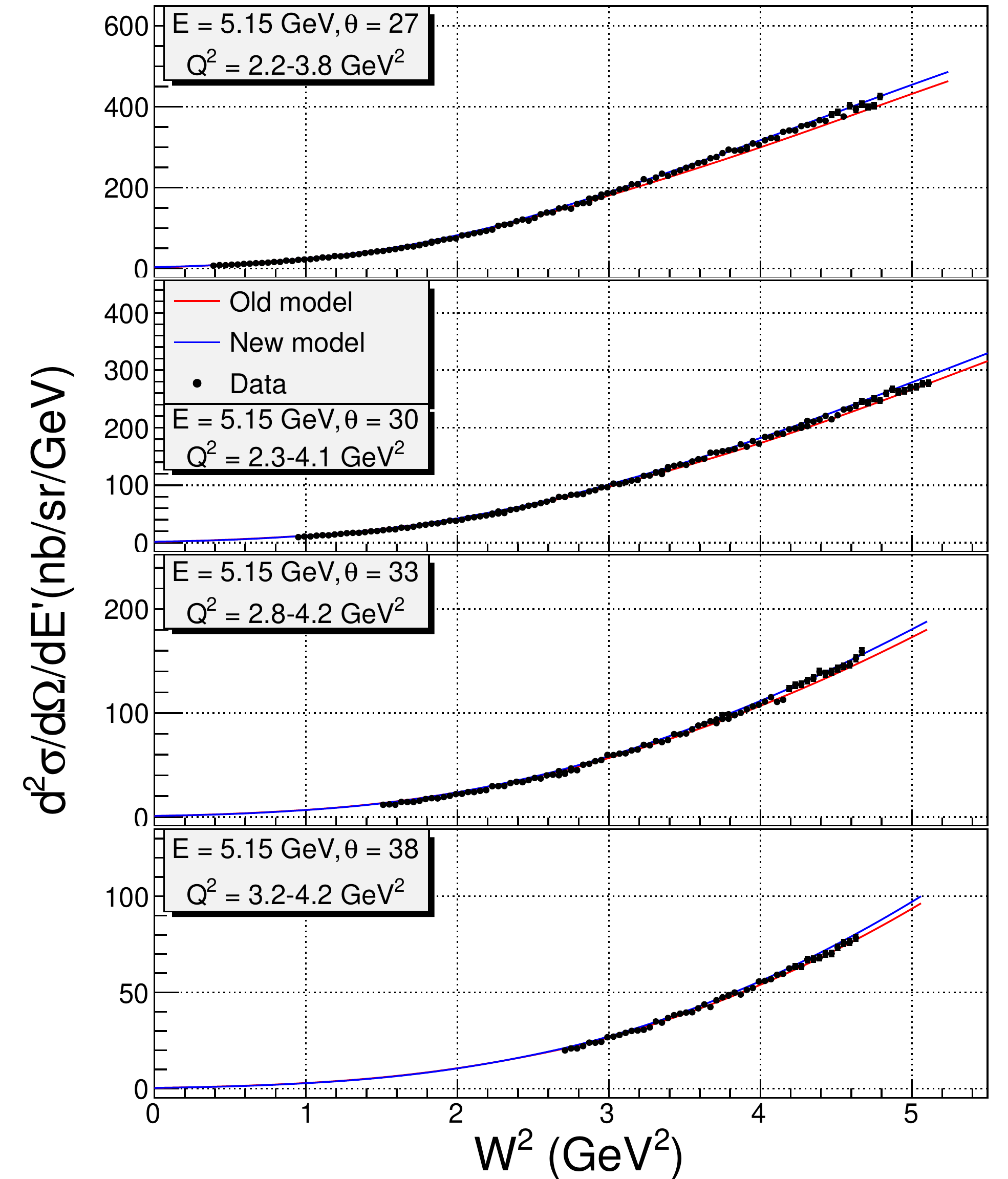,width=5.5in}
\end{center}
\caption { Extracted differential cross section for copper compared to the model cross section. }
\label{fig:csCu5151_0}
\end{figure}



\addcontentsline{toc}{section}{References} 
\begin{thebibliography} {99}



\bibitem{Ioffe:1983ju}
  B.~L.~Ioffe and A.~V.~Smilga,
  Nucl.\ Phys.\  B {\bf 232}, 109 (1984).

\bibitem{Balitsky:1983xk}
  I.~I.~Balitsky and A.~V.~Yung,
  Phys.\ Lett.\  B {\bf 129} (1983) 328.

\bibitem{Nesterenko:1984tk}
  V.~A.~Nesterenko and A.~V.~Radyushkin,
  JETP Lett.\  {\bf 39}, 707 (1984)
  [Pisma Zh.\ Eksp.\ Teor.\ Fiz.\  {\bf 39}, 576 (1984)].

\bibitem{Belyaev:1992xfa}
  V.~M.~Belyaev and I.~I.~Kogan,
  Int.\ J.\ Mod.\ Phys.\  A {\bf 8}, 153 (1993).

\bibitem{Radyushkin:2004mt}
  A.~Radyushkin,
  Annalen Phys.\  {\bf 13}, 718 (2004)
  [arXiv:hep-ph/0410153].

\bibitem{Coward:1967au}
  D.~H.~Coward {\it et al.},
  Phys.\ Rev.\ Lett.\  {\bf 20}, 292 (1968).

\bibitem{Bloom:1969kc}
  E.~D.~Bloom {\it et al.},
  Phys.\ Rev.\ Lett.\  {\bf 23}, 930 (1969).

\bibitem{Breidenbach:1969kd}
  M.~Breidenbach {\it et al.},
  Phys.\ Rev.\ Lett.\  {\bf 23}, 935 (1969).

\bibitem{Bloom:1970xb}
  E.~D.~Bloom and F.~J.~Gilman,
  Phys.\ Rev.\ Lett.\  {\bf 25}, 1140 (1970).

\bibitem{Collins:1977jy}
  P.~D.~B.~Collins,
  ``An Introduction To Regge Theory And High-Energy Physics'', Cambridge University Press, {\it  Cambridge 1977, 445p.}

\bibitem{Donnachie:2002en}
  S.~Donnachie, H.~G.~Dosch, O.~Nachtmann and P.~Landshoff,
  Camb.\ Monogr.\ Part.\ Phys.\ Nucl.\ Phys.\ Cosmol.\  {\bf 19} (2002) 1.

\bibitem{De Rujula:1976tz}
  A.~De Rujula, H.~Georgi and H.~D.~Politzer,
  Annals Phys.\  {\bf 103}, 315 (1977).

\bibitem{NeutrinoTargets} Fermilab MINERvA NUMI Neutrino proposal \url{http://nuint.ps.uci.edu/files/} 
The EOIs from the groups that combined to form the MINERvA collaboration, the NuMI On-axis and NuMI Off-axis
groups, can be found on the MINERvA home page at: \url{http://www.pas.rochester.edu/minerva/}

\bibitem{Filippone:1992iz}
  B.~W.~Filippone {\it et al.},
  Phys.\ Rev.\  C {\bf 45}, 1582 (1992).


\bibitem{Arrington:1998ps}
  J.~Arrington {\it et al.},
  Phys.\ Rev.\ Lett.\  {\bf 82}, 2056 (1999)
  [arXiv:nucl-ex/9811008].

\bibitem{Arrington:2001ni}
  J.~Arrington {\it et al.},
  Phys.\ Rev.\  C {\bf 64}, 014602 (2001)
  [arXiv:nucl-ex/0102004].

\bibitem{Miller:2000ta}
  G.~A.~Miller, S.~J.~Brodsky and M.~Karliner,
  Phys.\ Lett.\  B {\bf 481}, 245 (2000)
  [arXiv:hep-ph/0002156].

\bibitem{E02-109} { M. Christy, C. Keppel, Measurement of $R=\sigma_{L}/\sigma_{T}$ on Deuterium in the Nucleon Resonance Region,  
                  \url{http://www.jlab.org/exp_prog/proposals/02/PR02-109.ps}. }


\bibitem{Rosenbluth:1950yq}
  M.~N.~Rosenbluth,
  Phys.\ Rev.\  {\bf 79}, 615 (1950).



\bibitem{Thomas:2001kw}
  A.~W.~Thomas and W.~Weise,
  ``The Structure of the Nucleon,''
{\it  Berlin, Germany: Wiley-VCH (2001) 389 p}

\bibitem{Close:1979bt}
  F.~E.~Close,
  ``An Introduction To Quarks And Partons,''
{\it  Academic Press/london 1979, 481p}

\bibitem{Bjorken:1968dy}
  J.~D.~Bjorken,
  Phys.\ Rev.\  {\bf 179}, 1547 (1969).

\bibitem{Hagiwara:2002fs}
  K.~Hagiwara {\it et al.}  [Particle Data Group],
  Phys.\ Rev.\  D {\bf 66}, 010001 (2002).

\bibitem{Roberts:1990ww}
  R.~G.~Roberts,
  ``The Structure Of The Proton: Deep Inelastic Scattering,''
{\it  Cambridge, UK: Univ. Pr. (1990) 182 p. (Cambridge monographs on mathematical physics)}



\bibitem{Whitlow:1990gk}
  L.~W.~Whitlow, S.~Rock, A.~Bodek, E.~M.~Riordan and S.~Dasu,
  Phys.\ Lett.\  B {\bf 250}, 193 (1990).


\bibitem{Dasu:1988ru}
  S.~Dasu {\it et al.},
  Phys.\ Rev.\ Lett.\  {\bf 60}, 2591 (1988).

\bibitem{Aubert:1985fx}
  J.~J.~Aubert {\it et al.}  [European Muon Collaboration],
  Nucl.\ Phys.\  B {\bf 259}, 189 (1985).

\bibitem{Benvenuti:1989rh}
  A.~C.~Benvenuti {\it et al.}  [BCDMS Collaboration],
  Phys.\ Lett.\  B {\bf 223}, 485 (1989).

\bibitem{Berge:1989hr}
  J.~P.~Berge {\it et al.},
  Z.\ Phys.\  C {\bf 49}, 187 (1991).

\bibitem{Close:1978fk}
  F.~E.~Close,
{\it In the Proceedings of 19th International Conference on High-Energy Physics, Tokyo, Japan, 23-30 Aug 1978, pp 209}.
 The quark parton model, F. Close 1979 Rep. Prog. Phys. 42 1285.

\bibitem{Ashman:1989ig}
  J.~Ashman {\it et al.}  [European Muon Collaboration],
  Nucl.\ Phys.\  B {\bf 328}, 1 (1989).

\bibitem{COMPASS} { Celso Franco on behalf of the COMPASS collaboration, ``New COMPASS results on the gluon polarisation using D0 production asymmetries'', 
                    XVIII International Workshop on Deep-Inelastic Scattering and Related Subjects, DIS 2010 April 19-23, 2010 Firenze, Italy.}

\bibitem{Gribov:1972ri}
  V.~N.~Gribov and L.~N.~Lipatov,
  Sov.\ J.\ Nucl.\ Phys.\  {\bf 15}, 438 (1972)
  [Yad.\ Fiz.\  {\bf 15}, 781 (1972)].

\bibitem{Dokshitzer:1977sg}
  Y.~L.~Dokshitzer,
  Sov.\ Phys.\ JETP {\bf 46}, 641 (1977)
  [Zh.\ Eksp.\ Teor.\ Fiz.\  {\bf 73}, 1216 (1977)].

\bibitem{Altarelli:1977zs}
  G.~Altarelli and G.~Parisi,
  Nucl.\ Phys.\  B {\bf 126}, 298 (1977).

\bibitem{Nachtmann:1973mr}
  O.~Nachtmann,
  Nucl.\ Phys.\  B {\bf 63}, 237 (1973).

\bibitem{Gluck:1995yr}
  M.~Gluck, E.~Reya, M.~Stratmann and W.~Vogelsang,
  Phys.\ Rev.\  D {\bf 53}, 4775 (1996)
  [arXiv:hep-ph/9508347].




\bibitem{Itow:2001ee}
  Y.~Itow {\it et al.}  [The T2K Collaboration],
  arXiv:hep-ex/0106019.


\bibitem{NUMI} {The NUMI Low Energy Neutrino Program at Fermilab.}\url{http://www-nova.fnal.gov/nova_beam_anu.html}

\bibitem{Yang:2001xc}
  U.~K.~Yang {\it et al.}  [CCFR/NuTeV Collaboration],
  Phys.\ Rev.\ Lett.\  {\bf 87}, 251802 (2001)
  [arXiv:hep-ex/0104040].



\bibitem{Bloom:1971ye}
  E.~D.~Bloom and F.~J.~Gilman,
  Phys.\ Rev.\  D {\bf 4}, 2901 (1971).

\bibitem{Miller:1971qb}
  G.~Miller {\it et al.},
  Phys.\ Rev.\  D {\bf 5}, 528 (1972).

\bibitem{Benhar:1995rh}
  O.~Benhar and S.~Liuti,
  Phys.\ Lett.\  B {\bf 358}, 173 (1995)
  [arXiv:hep-ph/9505213].


\bibitem{E89008} D. Day, B. Filippone, ``Inclusive Scattering for Nuclei at $x>1$ and High $Q^2$'', \url{http://www.jlab.org/exp\_prog/proposals/89/PR89-008.pdf}

\bibitem{Cornwall:1968cx}
  J.~M.~Cornwall and R.~E.~Norton,
  Phys.\ Rev.\  {\bf 177}, 2584 (1969).

\bibitem{Wilson:1969zs}
  K.~G.~Wilson,
  Phys.\ Rev.\  {\bf 179}, 1499 (1969).


\bibitem{Feynman} { R.P. Feynman, ``Partons'', Prepared for Symposium on the Past Decade in Particle Theory, Austin, Tex., 14-17 Apr 1970.
 Published in Austin 1970, ``The past decade in particle theory'', p.773-813.}        


\bibitem{Georgi:1976vf}
  H.~Georgi and H.~D.~Politzer,
  Phys.\ Rev.\ Lett.\  {\bf 36}, 1281 (1976)
  [Erratum-ibid.\  {\bf 37}, 68 (1976)].





\bibitem{Arneodo:1992wf}
  M.~Arneodo,
  Phys.\ Rept.\  {\bf 240}, 301 (1994).

\bibitem{Bertsch:1993vx}
  G.~F.~Bertsch, L.~Frankfurt and M.~Strikman,
  Science {\bf 259} (1993) 773.

\bibitem{Miller}
  G.~A.~Miller,
  Phys.\ Rev.\  C {\bf 64}, 022201 (2001)
  [arXiv:nucl-th/0104025].


\bibitem{Piller:1999wx}
  G.~Piller and W.~Weise,
  Phys.\ Rept.\  {\bf 330}, 1 (2000)
  [arXiv:hep-ph/9908230].

\bibitem{Frankfurt:1988nt}
  L.~L.~Frankfurt and M.~I.~Strikman,
  Phys.\ Rept.\  {\bf 160}, 235 (1988).


\bibitem{Thomas:1983fh}
  A.~W.~Thomas,
  Phys.\ Lett.\  B {\bf 126}, 97 (1983).

\bibitem{Frankfurt:1989wq}
  L.~L.~Frankfurt, M.~I.~Strikman, L.~Mankiewicz, A.~Schafer, E.~Rondio, A.~Sandacz and V.~Papavassiliou,
  Phys.\ Lett.\  B {\bf 230}, 141 (1989).



\bibitem{Carlson:1993wy}
  C.~E.~Carlson and N.~C.~Mukhopadhyay,
  Phys.\ Rev.\  D {\bf 47}, 1737 (1993).

\bibitem{Gran:2008zz}
  R.~Gran  [MINERvA Collaboration],
  J.\ Phys.\ Conf.\ Ser.\  {\bf 136} (2008) 042040.

\bibitem{Cabibbo:1963yz}
  N.~Cabibbo,
  Phys.\ Rev.\ Lett.\  {\bf 10}, 531 (1963).


\bibitem{Burkert:2002zz}
  V.~D.~Burkert, R.~De Vita, M.~Battaglieri, M.~Ripani and V.~Mokeev,
  Phys.\ Rev.\  C {\bf 67}, 035204 (2003)
  [arXiv:hep-ph/0212108].

\bibitem{Aznauryan:2004jd}
  I.~G.~Aznauryan, V.~D.~Burkert, H.~Egiyan, K.~Joo, R.~Minehart and L.~C.~Smith,
  Phys.\ Rev.\  C {\bf 71}, 015201 (2005)
  [arXiv:nucl-th/0407021].

\bibitem{Lalakulich:2006sw}
  O.~Lalakulich, E.~A.~Paschos and G.~Piranishvili,
  Phys.\ Rev.\  D {\bf 74}, 014009 (2006)
  [arXiv:hep-ph/0602210].




\bibitem{e04001}{ A. Bodek, C. Keppel, ``Measurements of F$_{2}$ and R on Nuclear Targets in Resonance Region'', 
        Jefferson Lab Experiment E04-001, \url{http://www.jlab.org/exp_prog/proposals/04/PR04-001.pdf}.}

\bibitem{Leemann:2001dg}
  C.~W.~Leemann, D.~R.~Douglas and G.~A.~Krafft,
  Ann.\ Rev.\ Nucl.\ Part.\ Sci.\  {\bf 51}, 413 (2001).


\bibitem{Kazimi:2000gg}
  R.~Kazimi, C.~K.~Sinclair and G.~A.~Krafft,
in {\it Proc. of the 20th Intl. Linac Conference LINAC 2000 } ed. Alexander W. Chao,
{\it In the Proceedings of 20th International Linac Conference (Linac 2000), Monterey, California, 21-25 Aug 2000, pp MOB14}.

\bibitem{Yan:1995uu}
  C.~Yan, P.~Adderley, R.~Carlini, C.~Cuevas, W.~Vulcan and R.~Wines,
  Nucl.\ Instrum.\ Meth.\  A {\bf 365}, 46 (1995).

\bibitem{gue2} P. Gueye, Status of the Actual Beam Position Monitors in the Hall C Beamline, CEBAF Internal Report.

\bibitem{uncer1} { K. B. Unser, The Parametric Current Transformer, a Beam Current Monitor Developed for LEP, CERN SL/91-42 (unpublished).)}

\bibitem{Unser:1981fh}
  K.~Unser,
  IEEE Trans.\ Nucl.\ Sci.\  {\bf 28}, 2344 (1981).

\bibitem{E06009} { M. Christy, C. Keppel, ``Measurement of R=$\sigma_{L}/\sigma_{T}$ on Deuterium in the Nucleon Resonance Region and Beyond'', 
\url{http://www.jlab.org/exp\_prog/proposals/06/PR06-009.pdf}.}

\bibitem{Baker:1995ky}
  O.~K.~Baker {\it et al.},
  Nucl.\ Instrum.\ Meth.\  A {\bf 367}, 92 (1995).


\bibitem{Arring} J. Arrington, Inclusive Electron Scattering From Nuclei at x $>$1 and High Q$^2$, Ph.D. Thesis, California Institute of Technology (1998). \url{http://www1.jlab.org/Ul/Publications/documents/ACF1155.pdf}

\bibitem{coda} { G. Heyes et al. 
``The CEBAF online data acquisition system'',
Given at Computing in High-energy Physics (CHEP 94), San Francisco, CA, 21-27 Apr 1994.
In San Francisco 1994, Computing in High Energy Physics 94, p.122-126.}                      

\bibitem{Dalesio:1994qp}
  L.~R.~Dalesio {\it et al.},
  Nucl.\ Instrum.\ Meth.\  A {\bf 352}, 179 (1994).

\bibitem{tsup} {JLab CODA group. Trigger Supervisor Manual. Unpublished internal document, 1996.}




\bibitem{liang} Yongguang Liang, Ph.D. Thesis, Measurement of $R=\sigma_{L}/\sigma_{T}$ in the Nucleon Resonance Region, 
The American University (2003). \url{http://www1.jlab.org/Ul/Publications/documents/liang_Rproton_thesis.pdf}.

\bibitem{vlad} V. Tvaskis, Longitudinal-Transverse Separation of Deep-Inelastic Scattering at low $Q^{2}$ on Nucleons and Nuclei, Ph.D. Thesis, Vrije Universiteit (2004).

\bibitem{THorn} T. Horn, E. Christy, E. Segbefia, Rate Dependence of the HMS Tracking Efficiency, (2007).

\bibitem{cosy} M. Berz, COSY Infinity Version and Reference Manual, NSCL Technical Report MSUCL-977, Michigan State University, (1995). 

\bibitem{Bardin} A. A. Akhundov, D. Yu. Bardin and N. M. Shumeiko, Sov. J. Nucl. Phys., 26, (1977); 
D. Yu. Bardin and N. M. Shumeiko, Sov. J. Nucl. Phys., 29, (1979); and
A. A. Akhundov et al., Sov. J. Nucl. Phys., 44, (1986).

\bibitem{Dasu} Sridhara Rao Dasu, Precision Measurement of $x$, $Q^{2}$ and A-dependence of \mbox{$R=\sigma_{L}/\sigma_{T}$} and $F_{2}$
in Deep Inelastic Scattering, Ph.D. Thesis, The University of Rochester (1988).


\bibitem{Mo:1968cg}
  L.~W.~Mo and Y.~S.~Tsai,
  Rev.\ Mod.\ Phys.\  {\bf 41}, 205 (1969).


\bibitem{BostedPCom} Peter Bosted, private communication.


\bibitem{Jones:1999rz}
  M.~K.~Jones {\it et al.}  [Jefferson Lab Hall A Collaboration],
  Phys.\ Rev.\ Lett.\  {\bf 84}, 1398 (2000)
  [arXiv:nucl-ex/9910005].

\bibitem{Gayou:2001qd}
  O.~Gayou {\it et al.}  [Jefferson Lab Hall A Collaboration],
  Phys.\ Rev.\ Lett.\  {\bf 88}, 092301 (2002)
  [arXiv:nucl-ex/0111010].


\bibitem{Bosted:1994tm}
  P.~E.~Bosted,
  Phys.\ Rev.\  C {\bf 51}, 409 (1995).


\bibitem{Arrington:2003df}
 J.~Arrington,
 Phys.\ Rev.\  C {\bf 68}, 034325 (2003)
 [arXiv:nucl-ex/0305009].

\bibitem{Blunden:2003sp}
 P.~G.~Blunden, W.~Melnitchouk and J.~A.~Tjon,
 Phys.\ Rev.\ Lett.\  {\bf 91}, 142304 (2003)
 [arXiv:nucl-th/0306076].


\bibitem{Rekalo:2003xa}
  M.~P.~Rekalo and E.~Tomasi-Gustafsson,
  Eur.\ Phys.\ J.\  A {\bf 22}, 331 (2004)
  [arXiv:nucl-th/0307066].

\bibitem{Chen:2004tw}
  Y.~C.~Chen, A.~Afanasev, S.~J.~Brodsky, C.~E.~Carlson and M.~Vanderhaeghen,
  Phys.\ Rev.\ Lett.\  {\bf 93}, 122301 (2004)
  [arXiv:hep-ph/0403058].





\bibitem{Fancher:1976ea}
  D.~L.~Fancher {\it et al.},
  Phys.\ Rev.\ Lett.\  {\bf 37}, 1323 (1976).


\bibitem{Aste:2005wc}
  A.~Aste, C.~von Arx and D.~Trautmann,
  Eur.\ Phys.\ J.\  A {\bf 26}, 167 (2005)
  [arXiv:nucl-th/0502074].

\bibitem{Gueye:1999mm}
  P.~Gueye {\it et al.},
  Phys.\ Rev.\  C {\bf 60}, 044308 (1999).

\bibitem{Day:1990mf}
  D.~B.~Day, J.~S.~McCarthy, T.~W.~Donnelly and I.~Sick,
  Ann.\ Rev.\ Nucl.\ Part.\ Sci.\  {\bf 40}, 357 (1990).
\bibitem{Donnelly:1998xg}
  T.~W.~Donnelly and I.~Sick,
  Phys.\ Rev.\ Lett.\  {\bf 82}, 3212 (1999)
  [arXiv:nucl-th/9809063].
\bibitem{Donnelly:1999sw}
  T.~W.~Donnelly and I.~Sick,
  Phys.\ Rev.\  C {\bf 60}, 065502 (1999)
  [arXiv:nucl-th/9905060].
\bibitem{Maieron:2001it}
  C.~Maieron, T.~W.~Donnelly and I.~Sick,
  Phys.\ Rev.\  C {\bf 65}, 025502 (2002)
  [arXiv:nucl-th/0109032].
%


\bibitem{Cenni:1996zh}
  R.~Cenni, T.~W.~Donnelly and A.~Molinari,
  Phys.\ Rev.\  C {\bf 56}, 276 (1977)
  [arXiv:nucl-th/9606058].

\bibitem{Amaro:2004bs}  J.~E.~Amaro, M.~B.~Barbaro, J.~A.~Caballero, T.~W.~Donnelly, A.~Molinari and I.~Sick,
  Phys.\ Rev.\  C {\bf 71}, 015501 (2005)
  [arXiv:nucl-th/0409078].



\bibitem{Tsai:1973py}
  Y.~S.~Tsai,
  Rev.\ Mod.\ Phys.\  {\bf 46}, 815 (1974).

\bibitem{Aubert:1983xm}
  J.~J.~Aubert {\it et al.}  [European Muon Collaboration],
  Phys.\ Lett.\  B {\bf 123}, 275 (1983).

\bibitem{Eric:proton} M. E. Christy and P. E. Bosted, arXiv:0712.3731 [hep-ph]

\bibitem{Bosted:2007xd}
  P.~E.~Bosted and M.~E.~Christy,
  Phys.\ Rev.\  C {\bf 77}, 065206 (2008)
  [arXiv:0711.0159 [hep-ph]].

\bibitem{Lacombe:1980dr}
  M.~Lacombe, B.~Loiseau, J.~M.~Richard, R.~Vinh Mau, J.~Cote, P.~Pires and R.~De Tourreil,
  Phys.\ Rev.\  C {\bf 21}, 861 (1980).

\bibitem{Amaudruz:1991cca}
  P.~Amaudruz {\it et al.}  [New Muon Collaboration],
  Z.\ Phys.\  C {\bf 51}, 387 (1991).

\bibitem{Gomez:1993ri}
  J.~Gomez {\it et al.},
  Phys.\ Rev.\  D {\bf 49}, 4348 (1994).

\bibitem{NadiaThesis} N. Fomin, Inclusive Electron Scattering from Nuclei at $x > $ 1 and High $Q^2$ with a 5.75 GeV Beam, Ph.D. Thesis, University of Virginia (2008).
\url{http://www1.jlab.org/Ul/Publications/documents/JLAB-PHY-08-888.pdf}

\bibitem{Benhar:2006er}
  O.~Benhar, D.~Day and I.~Sick,
  arXiv:nucl-ex/0603032.

\bibitem{BostedVaheFit} P.Bosted, V. Mamyan, Empirical Fit to Inclusive Electron-Nuclear Cross Section in Resonance Region, to be published.


%

\bibitem{Christy:2004rc}
  M.~E.~Christy {\it et al.}  [E94110 Collaboration],
  Phys.\ Rev.\  C {\bf 70}, 015206 (2004)
  [arXiv:nucl-ex/0401030].

\bibitem{Abramowicz:1997ms}
  H.~Abramowicz and A.~Levy,
  arXiv:hep-ph/9712415.


\bibitem{Donnachie:1993it}
  A.~Donnachie and P.~V.~Landshoff,
  Z.\ Phys.\  C {\bf 61}, 139 (1994)
  [arXiv:hep-ph/9305319].


\bibitem{Gluck:2007ck}
  M.~Gluck, P.~Jimenez-Delgado and E.~Reya,
  Eur.\ Phys.\ J.\  C {\bf 53}, 355 (2008)
  [arXiv:0709.0614 [hep-ph]].


\bibitem{Martin:2004ir}
  A.~D.~Martin, R.~G.~Roberts, W.~J.~Stirling and R.~S.~Thorne,
  Phys.\ Lett.\  B {\bf 604}, 61 (2004)
  [arXiv:hep-ph/0410230].

\bibitem{EricChr} {M.E. Christy, (private communication).}

\bibitem{Arneodo:1995cq}
  M.~Arneodo {\it et al.}  [New Muon Collaboration.],
  Phys.\ Lett.\  B {\bf 364}, 107 (1995)
  [arXiv:hep-ph/9509406].
















































%
%




%



%
%

%
%






\end{thebibliography}
\end{document}